%% file: my_thesis.tex
\documentclass[10pt,b5paper]{phdthesis}
\pdfoutput=1

\usepackage{fancyhdr} 
\usepackage{xspace}
\usepackage{caption}
\def\red#1 {\textcolor{red}{#1}\ }   
\usepackage[referable]{threeparttablex}
\usepackage{setspace}
\usepackage{enumitem}
\usepackage{chapterbib}
\usepackage[toc,page]{appendix}
\usepackage[usenames, dvipsnames]{color}
\usepackage[dvipsnames]{xcolor}
\usepackage{palatino}
\usepackage{quoting}

\usepackage[toc,page]{appendix}

\usepackage{makeidx}     
\makeindex

\input{epsf}
\usepackage{multirow}
\usepackage{deluxetable}
\usepackage{amsmath}
\usepackage{amssymb}
\usepackage{amsthm}
\usepackage{epsfig}
\usepackage{graphics}
\usepackage{float}
\usepackage{here}
\usepackage{sectionapp}
 \usepackage{aas_macros} 
\usepackage{comment}
\usepackage{epigraph}
\usepackage{natbib}
\usepackage[english]{babel}
\usepackage[T1]{fontenc}
\usepackage[applemac]{inputenc}
\usepackage{longtable}
\usepackage{wallpaper}
\usepackage{subfigure}
\usepackage{wrapfig}
\usepackage{url}
\usepackage{tabularx,booktabs}
\usepackage[outer=10mm,inner=20mm,top=15mm,bottom=20mm]{geometry}

\def\Mstar{\mbox{M$_*$}\xspace}
\def\Mgas{\mbox{M$_{\rm gas}$}\xspace}

\def\aCO{\mbox{$\alpha_{\rm CO}$}\xspace}

\def\pct{\mbox{\,$\%$}\xspace}

\def\Xc{\mbox{$X_{\rm{C}}$}\xspace}

\def\Tk{\mbox{$T_{\rm{k}}$}\xspace}
\def\Te{\mbox{$T^{\rm{e}}$}\xspace}
\def\Td{\mbox{$T_{\rm{dust}}$}\xspace}
\def\Pe{\mbox{$P_{\rm{ext}}$}\xspace}

\def\Hpe{\mbox{$\Gamma_{\rm{PE}}$}\xspace}
\def\Hpi{\mbox{$\Gamma_{\rm{PI}}$}\xspace}
\def\Hcrhi{\mbox{$\Gamma_{\rm{CR,H}\textsc{i}}$}\xspace}
\def\Hcrh2{\mbox{$\Gamma_{\rm{CR,H}_2}$}\xspace}

\def\Cions{\mbox{$\Lambda_{\rm{ions+atoms}}$}\xspace}
\def\Crec{\mbox{$\Lambda_{\rm{rec}}$}\xspace}
\def\Cff{\mbox{$\Lambda_{\rm{f-f}}$}\xspace}
\def\Ch2{\mbox{$\Lambda_{\rm{H}_2}$}\xspace}
\def\Cco{\mbox{$\Lambda_{\rm{CO}}$}\xspace}
\def\Cgd{\mbox{$\Lambda_{\rm{gas-dust}}$}\xspace}
\def\Ccii{\mbox{$\Lambda_{\rm{C}\rm{\textsc{ii}}}$}\xspace}

\def\Coi{\mbox{$\Lambda_{\rm{O}\rm{\textsc{i}}}$}\xspace}

\def\nH{\mbox{$n_{\rm{H}}$}\xspace}
\def\nHplus{\mbox{$n_{\rm{H}^+}$}}
\def\nhi{\mbox{$n_{\rm{H}\textsc{i}}$}\xspace}
\def\ncii{\mbox{$n_{\rm{C}\textsc{ii}}$}\xspace}
\def\ne{\mbox{$n_{\rm{e}}$}\xspace}
\def\nav{\mbox{$\langle n \rangle$}\xspace}
\def\xe{\mbox{$x_{\rm{e}}$}\xspace}
\def\next{\mbox{$n_{\rm{H,ext}}$}\xspace}
\def\nh2{\mbox{$n_{\rm{H}_2}$}\xspace}
\def\colH{\mbox{$N_{\rm{H}}$}\xspace}
\def\NH2{\mbox{$N_{\rm{H}_2}$}\xspace}
\def\nhi{\mbox{$n_{\rm{H}\textsc{i}}$}\xspace}
\def\xco{\mbox{$X_{\rm{CO}}$}\xspace}
\def\aco{\mbox{$\alpha_{\rm{CO}}$}\xspace}
\def\fh2{\mbox{$f'_{\rm mol}$}\xspace}
\def\fmol{\mbox{$f'_{\rm mol}$}\xspace}

\def\Mh2{\mbox{$M_{\rm{H}_2}$}\xspace}
\def\mH{\mbox{$m_{\rm{H}}$}\xspace}
\def\Mgmc{\mbox{$m_{\rm{GMC}}$}\xspace}
\def\mgmc{\mbox{$m_{\rm{GMC}}$}\xspace}

\def\mgas{\mbox{$m_{\rm{gas}}$}\xspace}
\def\msph{\mbox{$m_{\rm{SPH}}$}\xspace}
\def\Mneu{\mbox{$M_{\rm{neutral}}$}\xspace}
\def\mneu{\mbox{$m_{\rm{neutral}}$}\xspace}
\def\mmol{\mbox{$m_{\rm{mol}}$}\xspace}

\def\g0{\mbox{$G_0$}\xspace}
\def\ga{\mbox{$G_{0{\rm ,att}}$}\xspace}
\def\cri{\mbox{$\zeta_{\rm{CR}}$}\xspace}
\def\crimw{\mbox{$\zeta_{\rm{CR,MW}}$}\xspace}
\def\crihi{\mbox{$\zeta_{\rm{CR,H}\textsc{i}}$}\xspace}
\def\crimol{\mbox{$\zeta_{\rm{CR,H}_2}$}\xspace}
\def\Av{\mbox{$A_{\rm{v}}$}\xspace}
\def\Z{\mbox{$Z'$}\xspace}
\def\Zn{\mbox{$Z'$}}

\def\Avtr{\mbox{$A_{\rm{v}}^{\rm{(tr)}}$}\xspace}
\def\sv{\mbox{$\sigma_{\rm{v}}$}\xspace}
\def\Re{\mbox{$R_{\rm{eff}}$}\xspace}

\def\SFRsd{\mbox{$\Sigma_{\rm{SFR}}$}\xspace}
\def\gassd{\mbox{$\Sigma_{\rm{gas}}$}\xspace}
\def\starssd{\mbox{$\Sigma_{\rm{*}}$}\xspace}
\def\gd{\mbox{$\Sigma_{\rm{gas}}$}\xspace}
\def\Sh2{\mbox{$\Sigma_{\rm{mol}}$}}
\def\Shihii{\mbox{$\Sigma_{\rm H{\sc I}+H{\sc II}}$}}
\def\rh2{\mbox{$R_{\rm{H}_2}$}\xspace}
\def\rh2{\mbox{$R_{\rm{H}_2}$}\xspace}
\def\rrh2{\mbox{$r_{\rm{H}_2}$}\xspace}

\def\rci{\mbox{$R_{\rm{C}\textsc{i}}$}\xspace}
\def\rrci{\mbox{$r_{\rm{{\sc C\textsc{i}}}}$}\xspace}
\def\rcl{\mbox{$R_{\rm{cloud}}$}\xspace}
\def\rl10{\mbox{$R_{\rm{L10}}$}\xspace}
\def\rp06{\mbox{$R_{\rm{P06}}$}\xspace}
\def\fcii{\mbox{$f_{\rm{\rm{C}\textsc{ii}}}$}\xspace}
\def\oi{\mbox{\rm{O}{\sc i}}\xspace}
\def\oil{\mbox{[\rm{O}{\sc i}]}\xspace}
\def\CIIsd{\mbox{$\Sigma_{\rm{[C\textsc{ii}]}}$}\xspace}
\def\Lcii{\mbox{$L_{\rm{[C}\rm{\textsc{ii}]}}$}\xspace}

\def\Lir{\mbox{$L_{\rm{IR}}$}\xspace}
\def\Lfir{\mbox{$L_{\rm{FIR}}$}\xspace}
\def\Ltir{\mbox{$L_{\rm{TIR}}$}\xspace}

\def\hu{\mbox{ergs\,cm$^{-3}$\,s$^{-1}$}\xspace}
\def\Zsun{\mbox{$Z_\odot$}\xspace}
\def\lsun{\mbox{$L_\odot$}\xspace}
\def\cms{\mbox{cm$^{2}$}\xspace} 
\def\cmps{\mbox{cm$^{-2}$}\xspace} 
\def\cmpc{\mbox{cm$^{-3}$}\xspace} 
\def\kms{\mbox{km~s$^{-1}$}\xspace}
\def\ps{\mbox{s$^{-1}$}\xspace}
\def\pc{\mbox{\rm pc}\xspace}

\def\msun{\mbox{$\rm{M}_\odot$}\xspace}
\def\sfru{\mbox{$\rm{M}_\odot$~yr$^{-1}$}\xspace}
\def\arcsec{\mbox{''}\xspace}
\def\arcmin{\mbox{'}\xspace}

\def\h2{\mbox{{\sc H}$_2$}\xspace}
\def\hi{\mbox{\rm{H}\textsc{i}}\xspace}
\def\hii{\mbox{\rm{H}\textsc{ii}}\xspace}
\def\cii{\mbox{[\rm{C}\textsc{ii}]}\xspace}
\def\nii{\mbox{[\rm{N}\textsc{ii}]}\xspace}
\def\cplus{\mbox{\rm{C}$^{\rm{+}}$}\xspace}
\def\ci{\mbox{\rm{C}\textsc{i}}\xspace}
\def\cil{\mbox{[\rm{C}\textsc{i}]}\xspace}

\def\sigame{\texttt{S\'IGAME}\xspace}
\def\gadget{{\small GADGET-3}\xspace}
\def\gadgettwo{{\small GADGET-2}\xspace}
\def\cloudy{\texttt{CLOUDY}\xspace}
\def\lime{\texttt{LIME}\xspace}

\def\ls{\mbox{$\lesssim$}}
\def\gs{\mbox{$\gtrsim$}}

\newcommand{\e}  [1]{\ensuremath{\times10^{#1}}}

\newcommand{\ave}[1]{\ensuremath{\langle #1 \rangle}}

\DeclareRobustCommand{\ion}[2]{%
\relax\ifmmode
\ifx\testbx\f@series
{\mathbf{#1\,\mathsc{#2}}}\else
{\mathrm{#1\,\mathsc{#2}}}\fi
\else\textup{#1\,{\mdseries\textsc{#2}}}%
\fi}

\begin{document}

\pagenumbering{arabic}

\include{thesisfront}

\frontmatter

\leavevmode
\vspace{2cm}

\begin{center}
{\Huge \scshape Observing }\\[2mm]
{\Huge \scshape and simulating }\\[2mm]
{\Huge \scshape galaxy evolution }\\[3mm]
{{ \LARGE  					
- from X-ray to millimeter wavelengths
}}
\end{center}
\newpage
\leavevmode
\vspace{2cm}
\newpage

\include{abstr}
\newpage

\thispagestyle{empty}
\addtocontents{toc}{\protect\thispagestyle{empty}}
\tableofcontents

\pagestyle{fancy}
\rfoot{\thepage}
\lhead{}
\rfoot{}
\rhead{}
\renewcommand{\headrulewidth}{0pt}


\pagenumbering{arabic}

\mainmatter	

\include{bkg}

\include{intro}

\include{CO}

\include{CII}

\include{Part_II}

\include{summary}

\include{akn}

\include{appendix}


\end{document}

%% file: thesisfront.tex
\pagestyle{empty}
\ThisCenterWallPaper{1.05}{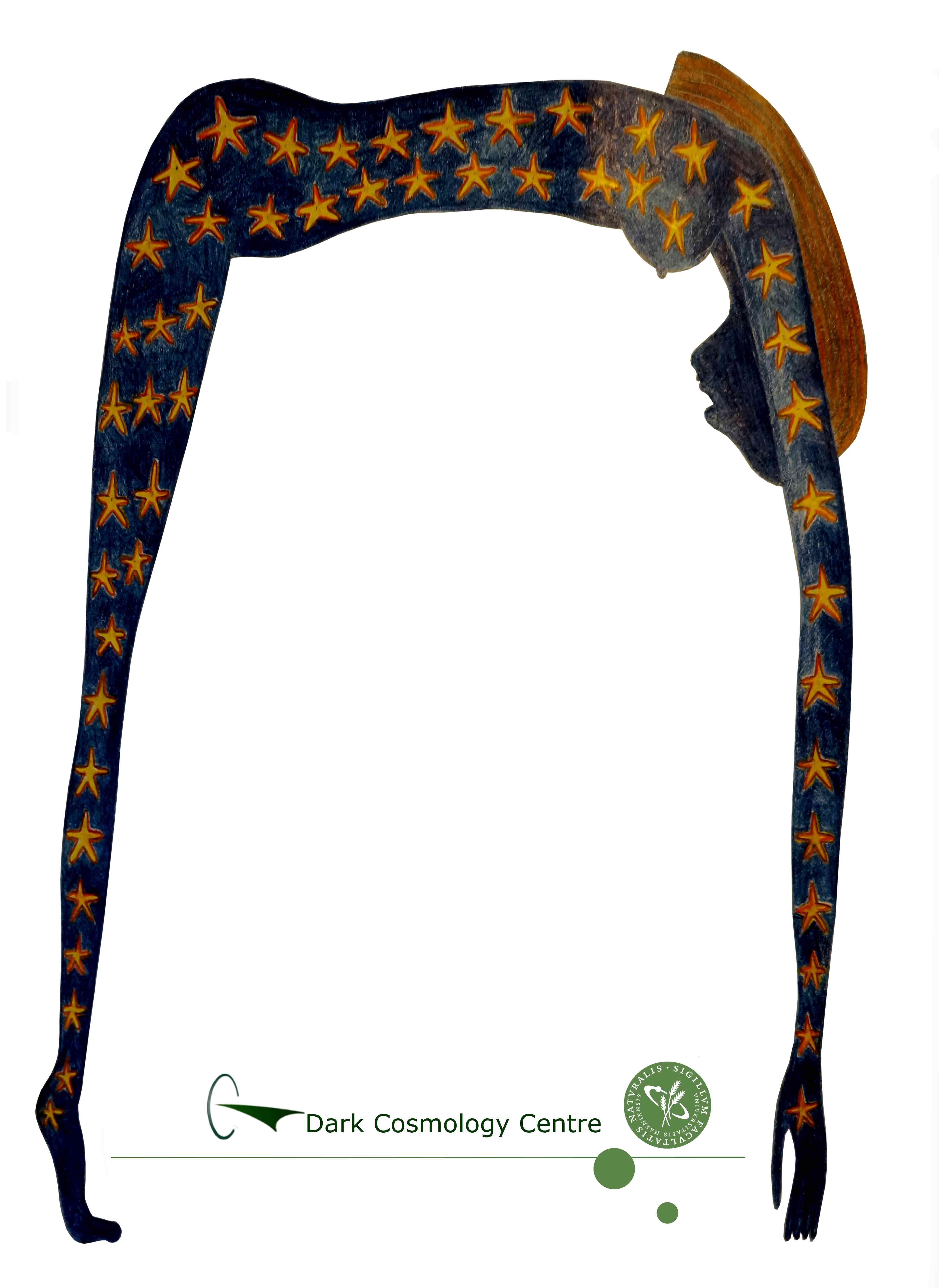}

\vspace*{4cm}
\begin{center}
\colorlet{myblue}{RoyalBlue!90!Black!200!}
{\color{myblue}\Huge \scshape Observing }\\[2mm]
{\color{myblue}\Huge \scshape and simulating }\\[2mm]
{\color{myblue}\Huge \scshape galaxy evolution }\\[3mm]
{{\color{myblue} \LARGE  					
- from X-ray to millimeter wavelengths
}}\\[3mm]

\vspace{4cm}

\color{myblue}{Dissertation submitted for the degree of\\[1mm]
{\bf\huge\scshape Philosophi\ae{} Doctor}}\\[2mm]
\color{myblue}{to the PhD School of The Faculty of Science, University of Copenhagen}\\[5mm]
\color{myblue}{on April 10 2015, by}\\[5mm]
\color{Bittersweet}{\large{\bf Karen Pardos Olsen}}\\[1mm]
\color{myblue}{\large Supervisors:} {\large{\it Sune Toft} and {\it Thomas Greve}} \\[5mm]

\end{center}

\newpage 
\leavevmode
\vspace{18cm}

{\it Cover art:}\\
Nut, goddess of the sky in ancient Egyptian religion. 
Contrary to most other religions, the sky was feminine and her brother, Geb, 
personified the Earth. 
At dusk, Nut would swallow the sun god, Ra, who 
would pass through her belly during the night and be 
reborn at dawn. 
Sometimes she was depicted as a cow, but most of the time as a star-covered nude woman arching over, 
and protecting, the Earth.









%% file: abstr.tex
\chapter*{Abstract}
It remains a quest for modern astronomy to answer what main mechanisms set the star formation rate (SFR) of galaxies. 
Massive galaxies present a good starting point for such a quest due to their relatively easy detection at every redshift. 
Since stars form out of cold and dense gas, a comprehensive model for galaxy evolution should explain any observed connection 
between SFR and the amount and properties of the molecular gas of the interstellar medium (ISM). 
In proposed models of that kind, an active galactic nucleus (AGN) phase is often invoked as the cause for the decrease or 
cease of star formation. 
This thesis consists of models and observations of gas and AGNs in massive galaxies at $z\sim2$, 
and how they may affect the overall SFR and the subsequent evolutionary trajectory of massive galaxies to $z=0$.

In this work, a new method and code is presented; SImulator of GAlaxy Millimeter/submillimeter Emission 
(\sigame), the aim of which is to improve our understanding of 
the connection between emission lines from the gas and the underlying ISM physics. 
By post-processing the outputs of cosmological simulations of 
galaxy formation with sub-grid physics recipes, \sigame divides the ISM into different gas phases and derives density and 
temperature structure, with locally resolved radiation fields. 
In the {\bf first study}, \sigame is combined with the radiative 
transfer code \lime to model the spectral line energy distribution (SLED) of the CO molecule. 
A CO SLED close to that of the Milky Way is found for normal star-forming massive galaxies at $z\sim2$, but $50\,\%$ 
smaller CO-\h2 conversion factors, with the latter decreasing towards the center of each model galaxy. 
In a {\bf second study}, \sigame is adapted to model the fine-structure line of singly ionized carbon, 
\cii at $158\,\mu$m, 
the most powerful emission line of neutral ISM. 
Most \cii emission in the same type of $z\sim2$ galaxies is revealed to trace mainly the molecular part of their ISM. 
The observed relation between \cii luminosity and SFR at $z>0.5$ is reproduced 
and a similar relation is established on kpc scales for the first time theoretically.

A {\bf third study} uncovers the presence of AGNs among massive galaxies at $z\sim2$, and sheds light on the AGN-host 
co-evolution by connecting the fraction and luminosity of AGNs with galaxy properties. 
By analyzing a large survey in X-ray, AGNs of high and low X-ray luminosity are extracted among massive galaxies at $z\sim2$ 
via AGN classification methods and stacking techniques in X-ray. It is found that about every fifth massive galaxy, 
quenched or not, contain an X-ray luminous AGN. Interestingly, an even higher fraction of low-luminosity AGNs reside 
in the X-ray undetected galaxies, and preferentially in the quenched ones, lending support to the importance of AGNs 
in impeding star formation during galaxy evolution.

%% file: bkg.tex
\chapter{Background}\label{intro}

\begin{quoting}[font=itshape,leftmargin=5cm,rightmargin=1cm]
Do not feel lonely, the entire Universe is inside you. - Rumi\footnote{Persian poet and Sufi mystic of the 13th century}
\end{quoting}


While Europeans were taking their medieval nap, Arabs carried astronomy forward. 
Not only did they collect and combine the astronomy from e.g. Greece, Egypt and India 
into one mathematical language, they also made contributions to the Ptolemaic system 
and created instruments such as celestial globes, astrolabs and large observatories. 
Their motivation was mostly a practical one of wanting to determine 
prayer times and the direction to Mecca to high accuracy. 
But what might the Arabs have thought for themselves while looking up at the stars? 
While this will remain unknown, we now know that most of that twinkling 
in the sky comes from our home, the Milky Way. 
Stars in the Milky Way are organized in a slowly rotating disk of spiral arms, 
embraced by an additional spheroidal component, the bulge. 
If you are lucky, you can even see a diffuse and `milky' band going across the night sky, 
and that is the combined light 
from millions of unresolved stars in the Milky Way disk.
What you cannot see with your naked eye, is all the gas and dust 
out of which new stars are born. 
Nor can you see the dark matter, 
emitting no light at all and believed to inhabit a large sphere embracing the Milky Way entirely.
Such assemblies of baryonic and dark matter we call galaxies, and the Universe is full of them. 
Understanding how these galaxies form and evolve while 
making room for life\footnote{In at least one case}, is one of the fundamental aims of modern astronomy.
This thesis is about the gas and the black hole component in galaxies. 
A theoretical model is developed to simulate the amount and state of gas in galaxies as well 
as its light emission in the infrared. 
A separate study uncovers the fraction of massive galaxies dominated by the powerful energetics 
associated with their central black hole, during an important cosmic epoch. 
Together, these projects offer a better view of galaxy evolution and will be useful for predicting and interpreting 
future observations.

\subsection{Setting the scale}
\begin{figure}[!htbp] 
\centering
\hspace{.0em}\raisebox{0.cm}{\includegraphics[width=0.8\columnwidth]{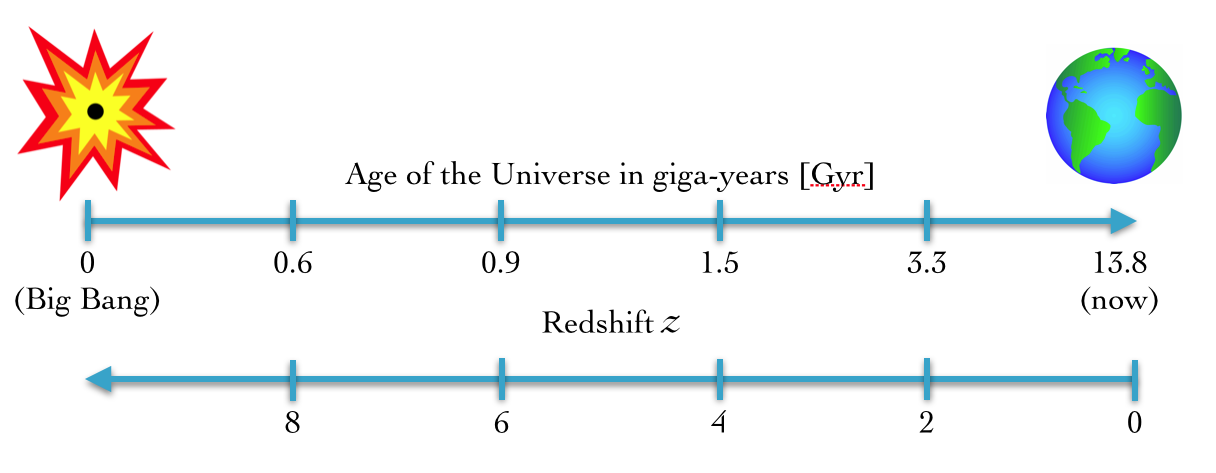}}
\caption{\footnotesize{A scale comparing age of the Universe with redshift, across its total 
lifespan of $\sim13.8$\,Gyr. Note that the redshift drops with age in a exponential-like behaviour, 
meaning that the Universe reached half its age at quite low redshift, $z\approx0.7$. Photons decoupled 
from matter at $z=1{,}100$, only $\sim378{,}000$\,yr after the Big Bang, and that is as far back as we can observe.
The ages have been calculated according to the now generally accepted spatially flat $\Lambda$CDM cosmology model, 
that describes the composition of the Universe with just six parameters 
\citep[fixed here to the most recent results by the][]{planck15}.
}}
\label{f:scales}
\end{figure}

The finite speed of light is of great value to astronomers, because it allows us to look back in time. 
In fact, astronomical distances, both within and between galaxies, are measured in light years (ly; 
the distance light can travel in 1 year) or parsecs ($1\,\pc=3.26\,$ly). 
On top of that, the expansion of the Universe causes light to `stretch' as 
it traverses the great expanses between galaxies. The amount of 
`stretching' serves as a label on every photon, saying how far it has been traveling. 
We call this `the redshift of light', $z$, 
and often use it as a measure of distance from us or age of the Universe. 
Fig.\,\ref{f:scales} provides a quick callibration between age of the Universe and redshift. 

\subsection{The cosmic web of galaxies}
\begin{figure}[!htbp] 
\centering
\hspace{.0em}\raisebox{0.cm}{\includegraphics[width=0.8\columnwidth]{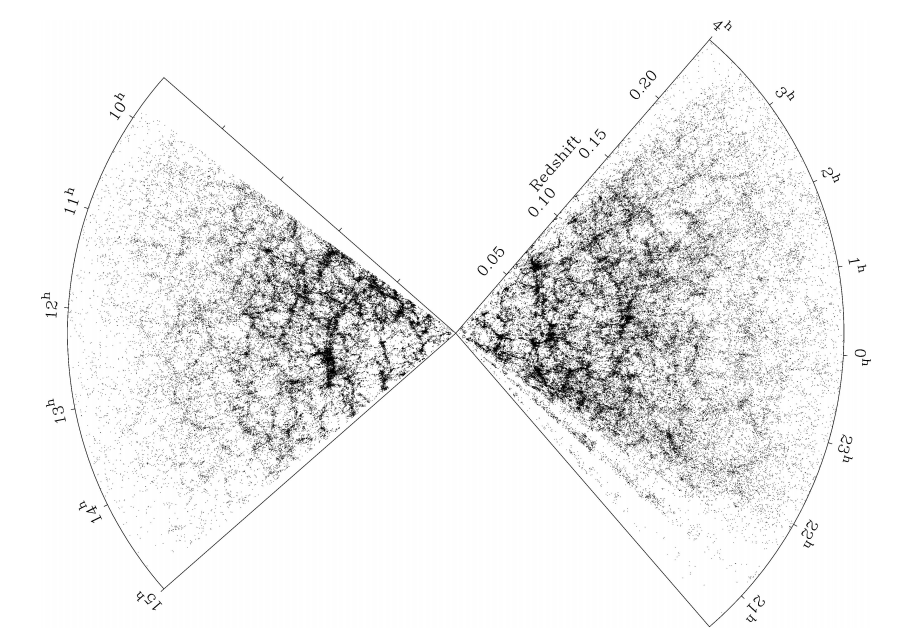}}
\caption{\footnotesize{Map of galaxies around us and out to $z\approx 0.3$ from the 
2dF Galaxy Redshift Survey by \citep{colless01} who studied two strips on the sky, 
one in the northern (left) and one in the southern (right) hemisphere.}}
\label{f:web}
\end{figure}

Galaxies are not evenly distributed in space, but rather follow the underlying cosmic web of 
dark matter that attracts the gas via gravitation. 
Fig.\,\ref{f:web} is a map carried out by \cite{colless01} of galaxies close to the MW and out to  
$z\approx0.3$ over a total sky area of 2000\,deg$^2$ or almost 5\pct of the total sky.
Like a Swiss cheese, there are regions with very low density of galaxies (so-called voids) 
and other regions with many galaxies living close together (clusters) connected by 
filaments and sheets. 
It has been confirmed by observations, that in the latter, highly concentrated regions, 
galaxies are prone to interact and possibly merge into even bigger ones.
Galaxies are therefore not isolated systems that go about their own business, 
but are constantly influenced by their surroundings via e.g. mergers, tidal strippings 
and gas inflow from the Intra Cluster Medium (ICM). 
In some of the following sections we will describe their internal components and inner workings, 
treating them as isolated systems, but one should never forget that galaxies in general 
interact with their surroundings.

\section{`Normal star-forming galaxies'}\label{intro:MS}

\begin{figure}[!htbp] 
\centering
\hspace{.0em}\raisebox{0.cm}{\includegraphics[width=0.7\columnwidth]{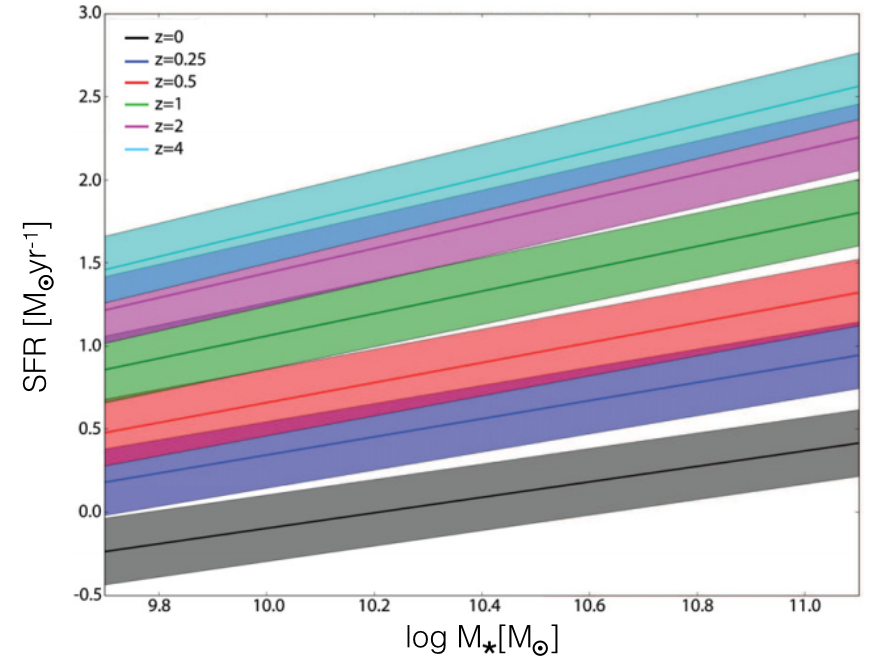}}
\caption{\footnotesize{Main sequence evolution. Power law fits to star-forming galaxies in different redshift bins 
from the compilation study of \cite{speagle14}. Shaded areas show an estimated `true' scatter of $0.2\,$dex.}}
\label{f:MS_z}
\end{figure}

Galaxies come in a great variety from containing few stars (dwarf galaxies) 
to containing many stars (massive galaxies), from compact to extended, 
from star-forming to inactive (quiescent in the following) and with different compositions of stars, gas, dust and dark matter. 
That said, most of them follow a basic rule: 
The more stars they have, the more stars they form. 
In other words, plotted in a star formation rate (SFR) vs. stellar mass (\Mstar) diagram, most 
galaxies fall on a power law relation called the `main sequence' (MS). 
The evolution of this sequence with cosmic time has been investigated recently by \cite{speagle14} 
who compiled 25 studies from the existing literature to find that the MS evolves towards higher SFR 
as we look back in time. This is shown in Fig.\,\ref{f:MS_z} with best fits to the observations 
at different redshifts, without plotting individual galaxies for the sake of simplicity.

Hence, like in crazy adolescent years, back in the day galaxies used to produce stars at much higher rates. 
The slope of the SFR-\Mstar relations shown in Fig.\,\ref{f:MS_z} is also called the specific star formation rate, 
SSFR~$={\rm SFR}/\Mstar$, and with very few galaxies at high-redshift, it is still hard to say whether the SSFR continues to 
rise with redshift or reaches a `plateau' \citep{behroozi13,speagle14}.

\subsection{The turning point at $z\sim2$}
\begin{figure}[!htbp] 
\centering
\hspace{.0em}\raisebox{0.cm}{\includegraphics[width=0.7\columnwidth]{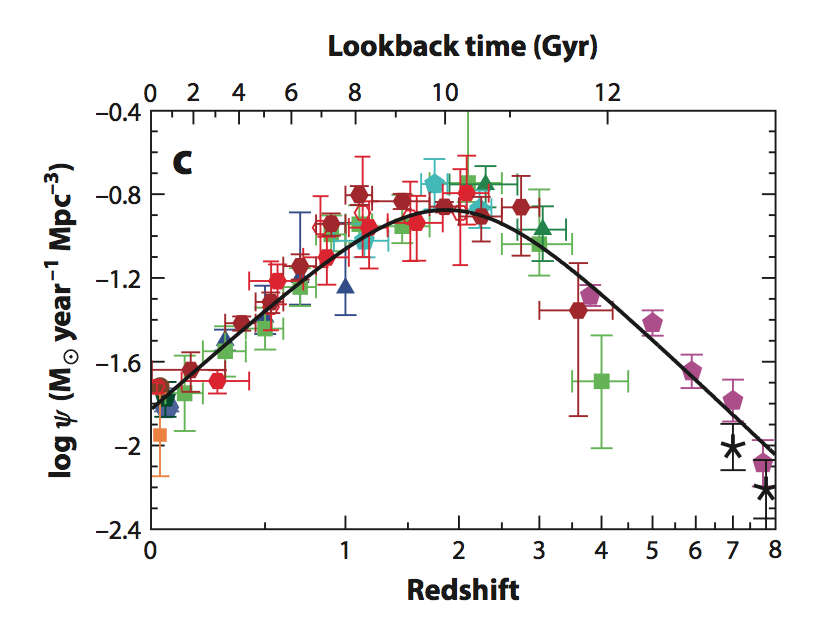}}
\caption{\footnotesize{The evolution of cosmic SFR density (SFRD) with redshift from the review of \cite{madau14}.}}
\label{f:SFRD_history}
\end{figure}

If one measures SFR per volume, there is a maximum in cosmic SFR density, SFRD, at 
$z\sim2$ after which it starts to decline again as shown in Fig.\,\ref{f:SFRD_history}.
The epoch around $z\sim2$, or about 10 billion years ago, is therefore an interesting one of 
phase change and one that we are just now beginning to uncover with modern telescopes (see Section \ref{intro:tele}).

\subsection{The case of `massive galaxies'} \label{intro:massive}
To understand how galaxies evolved since their formation some million years after the Big Bang, 
we need observations of all types of galaxies from high redshift until now. 
This task is impeded by the difficulty of resolving a galaxy very far away, because 
galaxies start to appear very dim and small.
The degree at which a large survey (such as the 2dF Survey in Fig.\,\ref{f:web}) 
is capturing all galaxies of a certain mass, is referred to as `completeness'. 
The state-of-the-art Cosmic Assembly Near-IR Deep Extragalactic Legacy 
Survey (CANDELS), 
carried out with the Hubble Space Telescope (HST),
looks all the way back to the high-redshift universe ($z\sim8$) 
with observations of different exposure time (depth). 
But even in the deep fields of GOODS-South/North, a typical lower mass limit of $10^{10}\,$\msun 
must be adopted to ensure $100\,\%$ completeness for studies going out to $z\sim2.5$ \citep[see e.g.][]{wuyts12}. 
In the smaller Ultra Deep Fields (UDFs), CANDELS can reach the same mass out to $z\sim3$ 
due to longer exposure times\footnote{see \url{http://candels.ucolick.org/survey/files/ferguson_STUC1102_v2.pdf}} 
\citep{hartley13}. 
But galaxies with stellar masses below these limits may 
exist that these surveys are incapable of detecting with available telescopes and integration time.
For this reason, massive (i.e. with high stellar mass) galaxies represent a convenient sample, 
because they are relatively easy to detect at all redshifts. 
My work concerns mainly massive galaxies, so this denomination will occur frequently in the present dissertation 
and the exact mass range will always be defined for each particular study.

\section{The Interstellar Medium (ISM)} \label{intro:ism}
\begin{wrapfigure}[19]{r}{7cm}
\centering
\includegraphics[width=6cm]{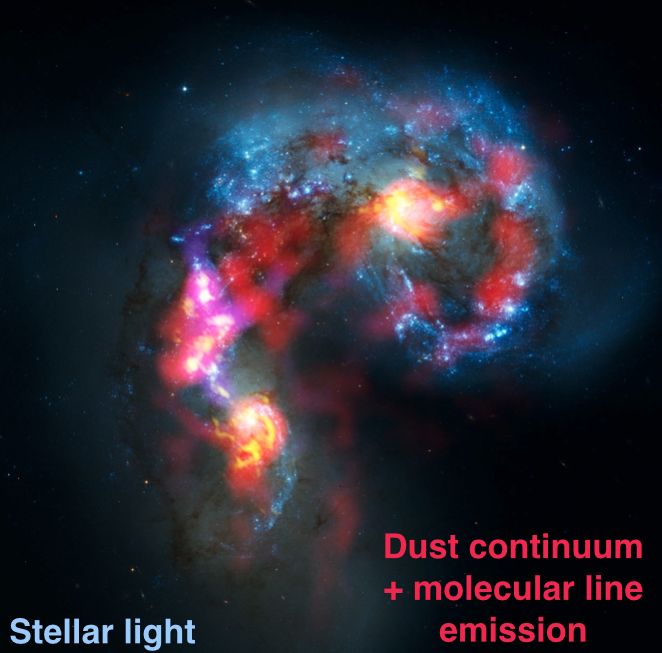}
\caption{\footnotesize{The Antennae Galaxies, credit: NASA/ESA HST for the optical image, 
and ALMA (ESO/NAOJ/NRAO) first science results for the observations of gas and dust at sub-mm wavelengths.}}
\label{f:antennae}
\end{wrapfigure}
Let's return for a moment to the sparkling stars that you may observe on a clear night. 
What you can't see with your naked eye is all the material filling out the vast spaces in between those stars, 
the Interstellar Medium or ISM. 
But it is out of this mixture of gas and dust that stars are born, 
the stars that expel heavier elements during their explosive deaths, thereby providing the universe with the 
atoms that we are ultimately made of. 
Carl Sagan once said that {\it `we are made of starstuff'}, 
yet stars would never have been formed had it not been for the gas that was already there. 
Fig.\,\ref{f:antennae} shows an example of what surprises lie in wait once telescopes are tuned to the 
frequencies at which the gas lights up in the ISM. The blue image is made of stellar light emission from the 
Antennae Galaxies -- two colliding spiral galaxies about 70 million ly away from us -- and the red image 
is a composite of observations with the interferometric array ALMA (to be described in section \ref{intro:tele}). 

Whether a cloud of gas in a galaxy can fragment and collapse to a density that allows fusion, i.e. the ignition of a new star, 
depends on the properties of that gas. The efficiency by which the gas forms stars is characterised by the gas depletion time, $t_{\rm dep}$:
\begin{align}
	t_{\rm dep}	=	\mgas/{\rm SFR}
\end{align}
that is, the time it would take to use up a cloud of mass \mgas\ if the current SFR stayed constant. 
Naively, one might expect that $t_{\rm dep}$ would be close to the time it takes a cloud to collapse onto itself under the influence of 
gravity only, also called the free-fall time, $t_{\rm ff}$. 
But observations show that $t_{\rm dep}$ is typically 1-3 orders of magnitude longer than $t_{\rm ff}$, 
requiring the existence of processes that slow down the star formation rate resulting from pure gravitation. 
These processes are broadly gathered under the term `feedback' from either nearby star formation \citep[see review by][]{krumholz14} 
or the powerful radiation from an Active Galactic Nucleus (see Section\,\ref{intro:agn}).

\begin{figure}[!htbp] 
\centering
\hspace{.0em}\raisebox{0.cm}{\includegraphics[width=0.9\columnwidth]{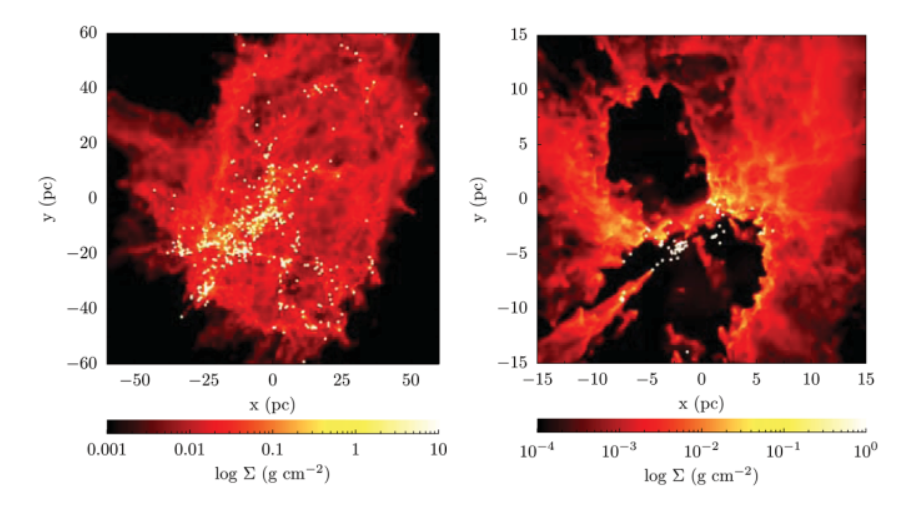}}
\caption{\footnotesize{{\it Left: }Column density map of a simulation of star cluster formation in a gas cloud with escape velocity 
$>10$\,\kms. White dots represent individual stars. 
{\it Right: }The same but for an escape velocity of $<10$\,\kms, in which case star formation is supressed. Adapted from \cite{krumholz14}.}}
\label{f:dale}
\end{figure}

Feedback on star formation is a whole topic in itself under fierce exploration with observations and theoretical models. 
An example of one strong type of feedback, is the ionizing field from young stars. 
Stars of masses $>10\,$\Mstar\ emit large quantities of ionizing photons creating large 
bubbles of expanding hot, ionized gas, that can hit nearby cold clouds and prevent them from 
collapsing in a type of explosive feedback \citep{krumholz14,dale12}. 
However, recent simulations by \cite{dale12} show that the effect of expanding regions of ionized gas in 
turbulent clouds depends strongly on the escape velocities in the gas. 
Fig.\,\ref{f:dale} illustrates, with snapshots from their simulations, how photoionization only has a significant impact 
in clouds of escape velocities, as set by the cloud mass and size, below about $10$\,\kms. 
This example shows that feedback from star formation involves many intercoupled processes.
Other types of stellar feedback include those of protostellar outflows, radiation pressure 
from massive stars, winds from hot stars, supernovae explosions and the mere thermal energy injected into the 
gas as gravitational potential energy is released during the collapse. In addition, nature most likely combines these 
mechanisms with the influence of magnetic fields \cite[e.g.][]{price09}. 

In principle, being able to observe the composition, turbulence, temperature, density and magnetic fields of the same gas cloud 
to high angular precision, would provide the building blocks for uniquely defining star formation on small ($<50$\,pc) scales. 
Therefore, {\it being able to characterise the gas in terms of these properties will help explain why some massive galaxies at 
$z\sim2$ are quiescent while others are even more star-forming than their local counterparts.}

\subsection{Chemical composition}
Before galaxies formed, the universe was pervaded by a primordial gas of only 4 atomic species; 
hydrogen (H), helium (He) and a trace of lithium (Li) and beryllium (Be), as well as a few of their isotopes. 
These were created from about 10 seconds to 20 minutes after the Big Bang\footnote{see \url{http://www.astro.ucla.edu/~wright/BBNS.html}}, 
but any elements heavier than 
Lithium and Beryllium had to wait for the formation of the first stars a few hundreds of million years later, in order to be produced.
Today, hydrogen and helium continue to be the dominant species found in the local ISM of the MW, 
with gas mass fractions of $\sim71.5$\,\% and $\sim27.1$\,\% respectively, compared to only $1.4$\,\% of heavier elements 
denoted `metals' \citep[e.g.][]{przybilla08}.

All atoms in the ISM can exist in atomic form or be ionized to several degrees or be combined in molecules of 
increasing complexity (see \citealt{tielens13} for an overview of just how complicated these molecules can get).
For the general life of a galaxy however, only a subset of these species needs to be considered for the 
structure and thermal state of the ISM relevant for star formation, as we shall see in the next subsections. 


\subsection{Distribution and density} \label{intro:dis}

\begin{figure}[!htbp] 
\centering
\includegraphics[width=0.4\columnwidth]{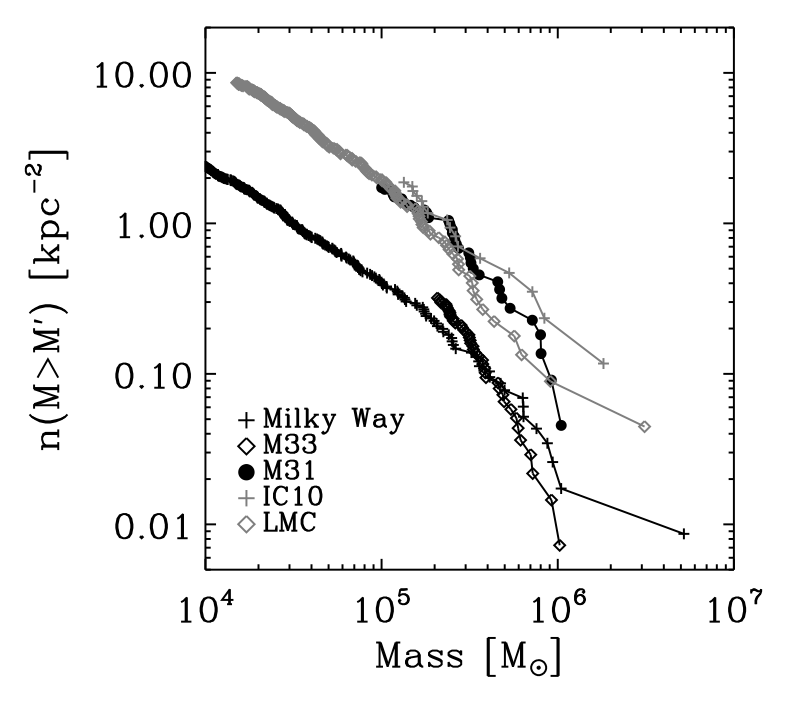}
\includegraphics[width=0.5\columnwidth]{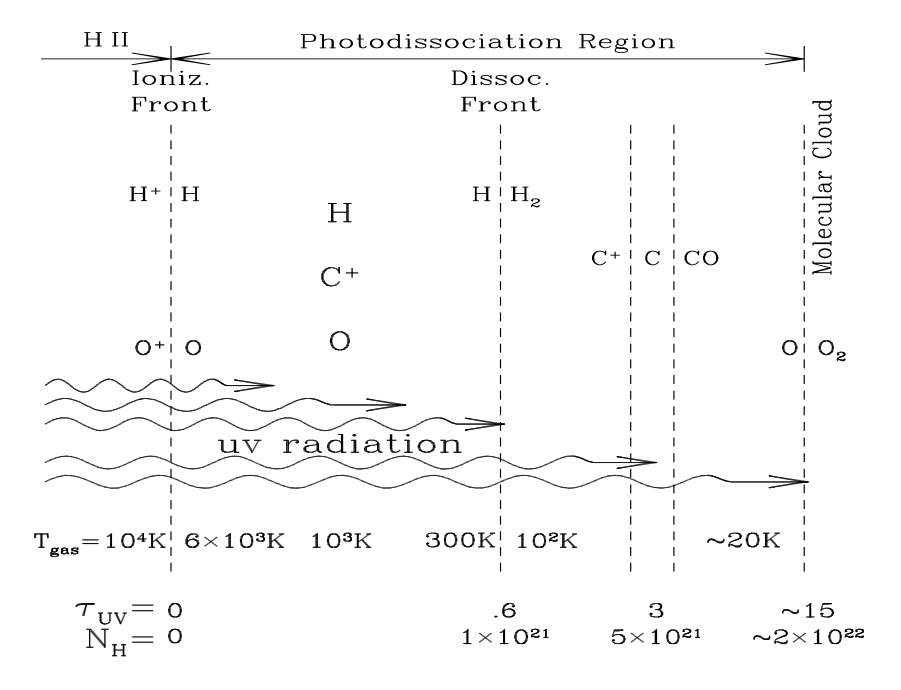}
\caption{\footnotesize{{\it Left:} Mass spectrum of GMCs in the MW and local galaxies \citep{blitz07}. 
{\it Right:} Schematic overview of the chemical stratification taking place in the PDR of a GMC, from \cite{papa13}.}}
\label{f:blitz07}
\end{figure}

Stars form generally in clusters as the simulated one depicted in Fig. \ref{f:dale} \citep{krumholz14}. 
Such constructions of dense knots and long filaments are collectively called Giant Molecular Clouds (GMCs) 
and are observed to have radii from about 5 to $\sim100$\,pc in the MW and local galaxies \citep{blitz07}. 
From measuring the broadening of emission lines from the gas and converting those into velocity dispersions, 
one can get an estimate of the mass, by assuming the cloud to be in virial equilibirum \citep{roman-duval10}:
\begin{align}
	M_{\rm vir} = \frac{1.3\sv^2 R}{G} ~ \Rightarrow ~  M_{\rm vir}[\msun] = 905 \cdot (\sigma_{v,1D}[\kms])^2 \cdot R[{\rm pc}]
	\label{eq:vir}
\end{align}
This was done by for example \cite{blitz07}, who found the mass spectrum shown to the left in Fig.\,\ref{f:blitz07}, 
the data of which can be fitted with a power law of slope $-1.71$ in the outer MW.

The reason why all gas in the galaxy is not neutral and forming molecules, is mainly the 
strong far-ultraviolet (FUV) radiation field from young stars, which ionizes the gas and 
creates a stratification of the ISM as illustrated on the right-hand side of Fig.\,\ref{f:blitz07}.
The region of stratified layers is also referred to as the Photodissociation Region (PDR). 
In addition to ionizing, the FUV radiation penetrates into the gas and determines its 
thermal and chemical state. 
This results in a set of distinct phases of the ISM, 
that have been observationally identified and are listed in Table \ref{t:phases}.
After the dense, molecular gas, comes the warm and cold neutral medium, often collectively named `WCNM', 
and finally, the least dense, ionized gas (\hii).

\begin{table}[htbp]
\centering
\begin{tabular*}{1\columnwidth}{p{4cm} | p{4 cm} p{2 cm} p{2 cm}} \toprule
Chemical state			&	Gas phase 			&	$n$ [\cmpc]	&	\Tk\ [K] 	\\ 	\hline	
Molecular gas			&	Diffuse \h2			&	$\sim100$	&	$\sim50$	\\
(17\% of total mass)	&	Dense \h2			&	$10^3-10^6$	&	$10-50$		\\ 	\hline
Atomic gas				&	Warm \hi\ (WNM)		&	$\sim0.6$	&	$\sim5000$	\\
(60\% of total mass)	&	Cold \hi\ (CNM)		&	$30$		&	$\sim100$	\\ 	\hline
Ionized gas				&	Coronal gas (HIM) 	&	$0.004$		&	$10^{5.5}$	\\
(23\% of total mass)	&	\hii\ gas			&	$0.3-10^4$	&	$10^4$		\\	\hline
\end{tabular*}
\caption{\footnotesize{Gas mass fractions, densities and temperatures of the different 
phases of interstellar gas, adapted from \cite{draine11}.}}
\label{t:phases}
\end{table}

If we take a look at the densities listed in Table \ref{t:phases}, 
it is clear that the ISM is in general {\it very} diffuse. 
You would need a cube of $700\times700\times700$\, meters of the densest \h2\ gas phase 
(with density $\nH\sim10^6$\,\cmpc) in order to have the same 
gas mass as present in 1\,cm$^3$ of air at sea level on Earth (with density $\rho\approx1.2$\,kg\,m$^{-3}$). 

Gas mass and SFR are related via the Kennicutt-Schmidt (KS) law, 
which is an observed tight relation between surface density of SFR, \SFRsd, and gas, \gd. 
As will be explained in further detail in Part \ref{part1}, the line emission from the 
CO molecule is particularly well suited for estimating the total gas mass. 
The KS-law for low and high-$z$ star-forming galaxies was estimated by \cite{genzel10} using a large 
data base of CO line observations and SFRs based on several SFR indicators:
\begin{align}
			\log \SFRsd\,[\sfru\,{\rm kpc}^{-2}] = 1.17\times \log\gd\,[\msun\,{\rm pc}^{-2}] - 3.48	
\end{align}
However, recent observations show that this correlation might break down already on scales of $\sim100\,$pc \citep{xu15}. 

\subsection{Thermal state}
The thermal state of the ISM is complicated by the fact that the ISM is never in thermodynamic equilibrium, 
but rather always subject to a flow of energy. It is constantly being heated, primarily by FUV radiation from 
young stars, while radiating away the heat via dust emission in the infrared and (mostly FIR) gas emission lines 
(see Section \ref{intro:obs} for more on the spectrum of a galaxy). 
To derive the gas kinetic temperature, \Tk, one must therefore consider all relevant heating and cooling 
mechanisms, calculate their energy rates and search for an equilibrium temperature. 

Out in the hot, ionized \hii\ regions (see Table\,\ref{t:phases}), gas is heated by photo-ionization 
of \hi\ gas. FUV photons ionize the hydrogen atoms and the free electrons convert their kinetic energy 
to heat via collisions with other gas particles. 
Cosmic rays are very energetic protons, most likely produced in supernovae \citep{ackermann13},  
and they represent a second radiation field that can heat the gas and penetrate even further into it than FUV photons 
due to their high energies ($\sim$\,GeV). 
Like FUV photons, cosmic rays can ionize the atomic gas and release electrons with kinetic energy, 
but cosmic rays can also interact with the free electrons, transferring kinetic energy directly via 
Coulomb interactions. 

\begin{figure}[!htbp] 
\centering
\hspace{.0em}\raisebox{0.cm}{\includegraphics[width=0.85\columnwidth]{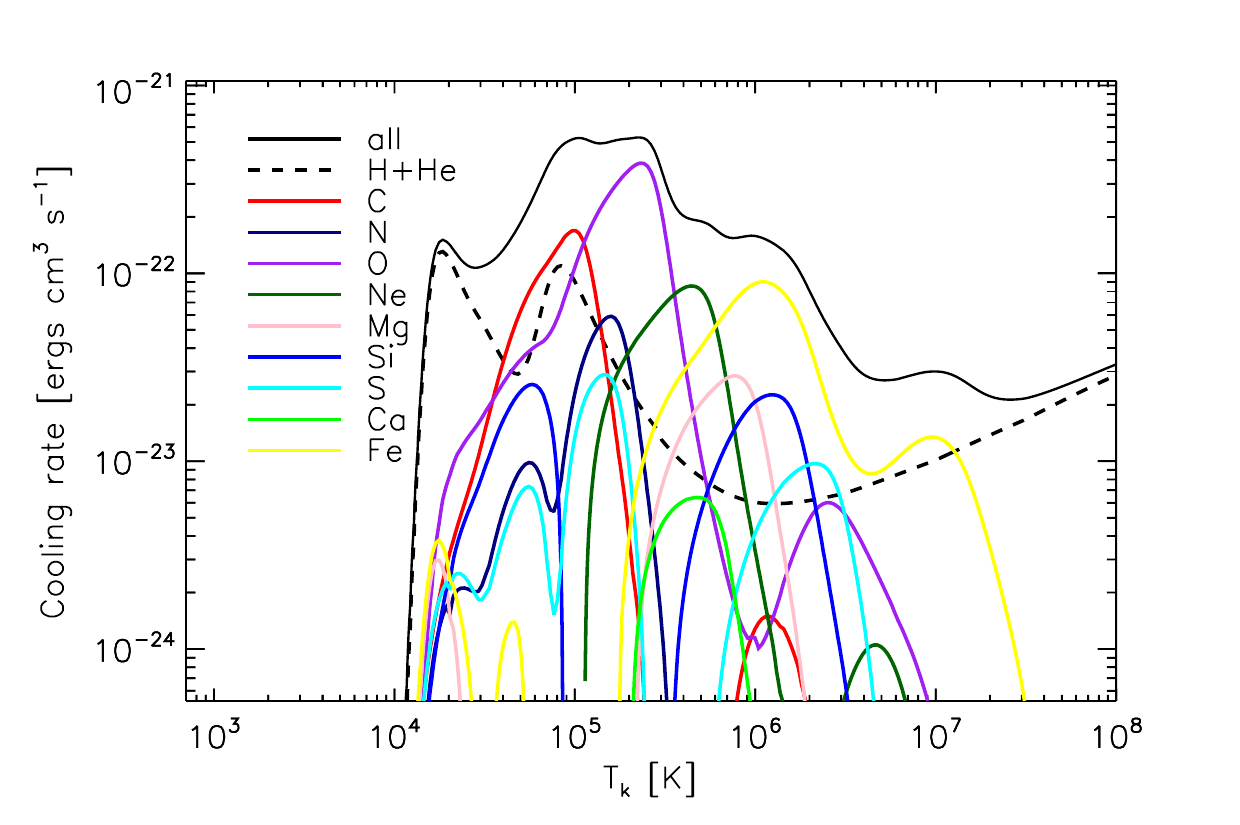}}
\caption{\footnotesize{Cooling rates of all (black) and individual atoms and ions 
in the hot ISM as function of temperature. Made with the publicly available code of 
\cite{wiersma09} which includes both collisional- and photo-ionization, though collisional ionization equilibrium (CIE) 
has been used for this plot. 
The code takes \Tk\ and \nH\ as input of which \nH is set to $0.01$\,\cmpc here, and 
solar abundances are adopted.}}
\label{f:metals}
\end{figure}

As for cooling, the gas has several options. First of all, emission lines from hydrogen and 
heavier elements, in their atomic and ionized states, can remove a great deal of energy, especially at 
$\Tk>10^4$\,K. The cooling rates as function of \Tk\ for the most important elements are shown 
in Fig.\,\ref{f:metals}. The sudden rise in cooling rate at $\Tk\sim10^4$\,K makes sure that the 
temperature of the \hii\ gas hardly ever exceeds $10^4$\,K.

In addition to these emission lines, electron
recombination with ions can cool the gas, as recombining electrons take away
kinetic energy from the plasma, a process which is important at temperatures $ >
10^3$\,K \citep{wolfire03}. At similar high temperatures another important
cooling mechanism is the scattering of free electrons off of other free ions,
whereby free-free emission removes energy from the gas \citep{draine11}.

Combining all of the above heating and cooling mechanisms, we arrive at one equation 
that describes the balance between the energy rates of heating and cooling:
\begin{align}
	\Hpi + \Hcrhi = \Cions + \Crec + \Cff
\end{align}
where \Hpi\ and \Hcrhi\ are the heating rates for photo-ionization and cosmic ray ionization, 
respectively, whereas \Cions\ is the combined cooling rate from the elements shown in 
Fig.\,\ref{f:metals}. Finally, \Crec\ and \Cff\ are the cooling rates due to recombination and free-free 
interactions as described above.

In the denser and colder ($\Tk<10^4$\,K) WCNM, other heating and cooling mechanisms take over. 
Rather than photo-ionizing the gas, FUV photons now heat the gas more efficiently via the 
photo-electric effect consisting of a FUV photon knocking loose an electron from the surface of 
a dust grain, with the same net result: The escaping electron can deposit its kinetic energy as heat in the 
surrounding gas. 
The dominating cooling mechanisms in neutral gas are fine-structure line emission from ioinised carbon, 
\cii, to be discussed in more detail in Section \ref{1:CII} in Part \ref{part1}, and from neutral 
atomic oxygen, \oil, leading to the following energy rate equation:
\begin{align}
	\Hpe+\Hcrhi = \Ccii+\Coi   
\end{align}
where \Hpe\ is the heating rate due to the photo-electric effect on dust grains, \Ccii\ is the cooling rate 
due to \cii\ line emission and \Coi\ is the cooling rate due to \oil line emission. 
In addition, the Lyman alpha (Ly-$\alpha$) line of hydrogen at $1216$\,\AA ~is important in the warm part of the WCNM.

In molecular gas, most of the FUV radiation will be attenuated by dust and self-shielding molecules, leaving 
in some cases cosmic rays as the dominant source of heating. 
Indeed, it has been suggested that cosmic rays control the initial conditions for star formation in regions of 
high SFR density where the sites for star formation might be completely UV-shielded \citep{papa11,papa13}. 
The prescription for cosmic ray heating per hydrogen atom in molecular gas is in principle different from its 
form in neutral gas, but the two turn out to be very similar \citep{stahler05} \citep[though see][who find 
higher heating rates in molecular gas by a factor 2-3]{glassgold12}. 
At increasingly higher temperature, \cii\ line cooling gives way for 
CO line cooling, and at very high densities ($>10^4$\,\cmpc), 
cooling by interactions between gas and dust particles can become dominating, as the gas approaches the dust temperature.
In molecular gas, line cooling from molecular hydrogen and atomic oxygen also contributes 
to the cooling rate with typically smaller amounts.
Summarized in one equation, the thermal equilibrium in molecular gas can hence be approximated by:
\begin{align}
	\Hpe+\Hcrh2 = \Ch2+\Cco+\Coi+\Ccii+\Cgd
\end{align}

\begin{figure}[!htbp] 
\centering
\hspace{.0em}\raisebox{0.cm}{\includegraphics[width=0.6\columnwidth]{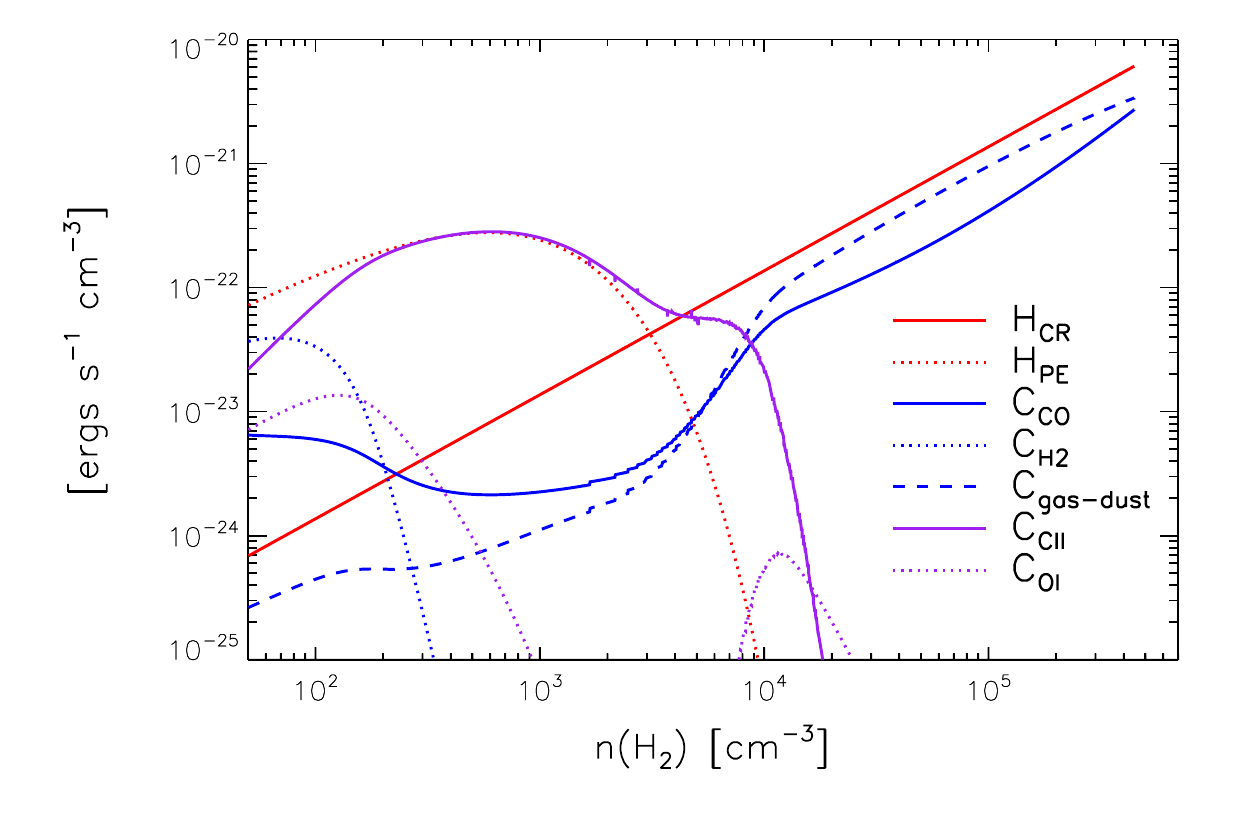}}
\caption{\footnotesize{Heating and cooling rates in molecular gas. 
The energy rates shown are calculated at the equilibrium temperature for that density. 
Calculated for a GMC of mass $\Mgmc=10^4$\,\msun, solar metallicity, 
an interstellar FUV field of $16$ Habing units and a cosmic ray ionization rate of $8\e{-16}$\,\ps. 
In the outskirts (low density) of the cloud, \cii line cooling and photo-electric heating dominate, 
while in the inner regions (high density), gas-dust cooling and CR heating take over, of which the latter 
is proportional to the density and hence a straight line of slope 1 \citep{stahler05}. 
Made with the \sigame code (to be described in Part \ref{part1}).}}
\label{f:hc_gmc}
\end{figure}

Fig.\,\ref{f:hc_gmc} gives an example of the dominant heating and cooling mechanisms mentioned above 
as functions of density for the interior of a model GMC of mass $\Mgmc=10^4$\,\msun and solar metallicity, 
immersed in an interstellar FUV field of $16$ Habing units and 
a cosmic ray ionization rate of $8\e{-16}$\,\ps.

\section{The Active Galactic Nucleus (AGN)} \label{intro:agn}

Apart from the gas, dust, stars and dark matter that I mentioned in the beginning of this introduction, 
there is a fifth component in the MW: black holes. 
A particularly big one or `supermassive black hole' (SMBH) is hiding at the very center of the MW, 
and most likely every galaxy, as it became clear in the 90's \citep{magorrian98}. 

The strong gravitational pull from a SMBH creates a rotating accretion disk of matter (yellow part in Fig.\,\ref{f:torus}) 
from its host galaxy. 
As gravitational energy is converted into mechanical and electromagnetic energy, gas in the accretion disk 
is heated to very high temperatures and starts to radiate fiercefully over a broad range of energies, 
from radio to X-ray \citep[see][for a review of the inner workings of these central engines]{krawczynski13}. 

Galaxies in which the spectra are dominated by this type of central energy source, are called Active Galactic Nuclei (AGNs). 
Observationally, AGNs have been classified into two main groups; (Seyfert) type I and type II based 
on the width of their optical emission lines. 
Type I AGNs show strong broad emission lines, while type II AGNs have relatively narrow emission lines and softer X-ray emission. 
In some cases, relativistic jets are produced perpendicularly to the accretion disk and their strong synchroton radiation in radio 
has led to the classification `radio-loud' AGNs. 
AGNs for which the optical light from the host galaxy is outshined by the central disk itself 
are dubbed quasars (QSOs). 

\subsection{Unification models} \label{intro:agn}

\begin{wrapfigure}[22]{L}{7.5cm}
\centering
\includegraphics[width=7.5cm]{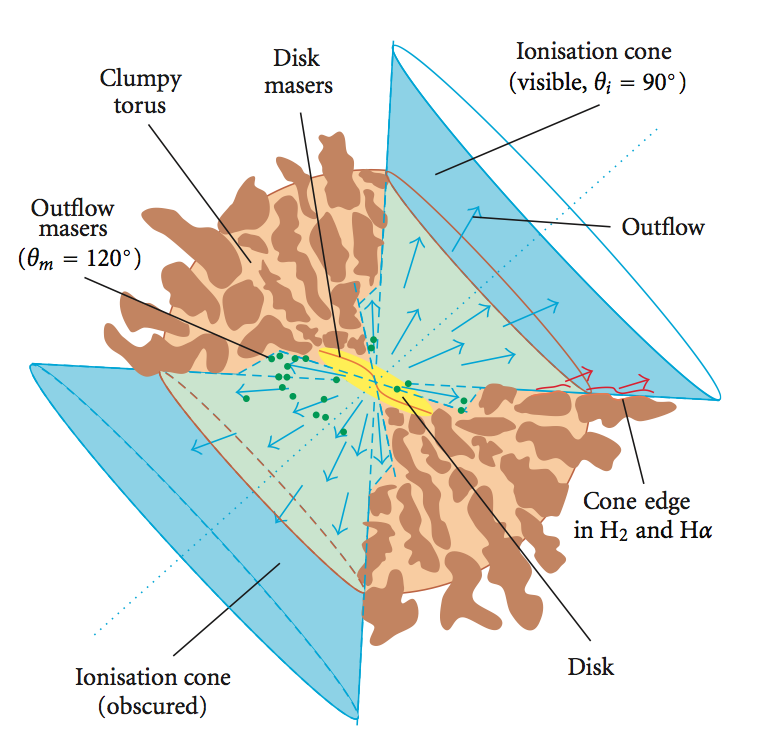}
\caption{\footnotesize{Drawing, from the review of \citep{bianchi12}, of the inner engine emitting the strong radiation of an AGN, 
including the disk (yellow) of matter accreting onto the SMBH, the surrounding clumpy torus (brown) 
and ionization cones (blue), also containing the relativistic jets.
}}
\label{f:torus}
\end{wrapfigure}

Over the past two decades, observations have led to the 
Standard Unified Model suggesting that type I and type II AGNs are intrinsically the same class of objects, 
but viewed from different angles, as pictured in Fig.\,\ref{f:torus} \citep{bianchi12}.
In this picture, the accretion disk is surrounded by a clumpy torus of gas and dust, 
and outflow is mainly allowed within the ionization cone 
encircling the relativistic jets.
A type I AGN is a galaxy inclined so that we see the inner parts of the accretion disk 
where high orbital speeds result in relatively broad lines. 
A type II AGN is orientated so that the inner accretion disk is obscured by the clumpy torus and only narrow lines 
from more distant cold material can be observed.

A tight correlation between optical and X-ray emission of unobscured quasars suggest that 
all luminous AGNs are intrinsically also X-ray bright \citep[e.g.][]{steffen06,gibson08}. 
But high amounts of obscuring gas and dust in the torus and elsewhere in the host galaxy 
can decrease the X-ray luminosity or change the spectral slope of the X-ray emission 
that actually escapes the galaxy, to be discussed further in Section \ref{intro:agn}. 
Alternative sources for strong X-ray emission in a galaxy are 
High-Mass X-ray Binary systems (HMXBs), which may be significant in case of high SFR \citep{ranalli03}, 
and halos of hot gas \citep{mulchaey10}. 

\section{The `spectral fingerprint' of a galaxy}\label{intro:obs}
\begin{figure}[!htbp] 
\centering
\hspace{.0em}\raisebox{0.cm}{\includegraphics[width=1.08\columnwidth]{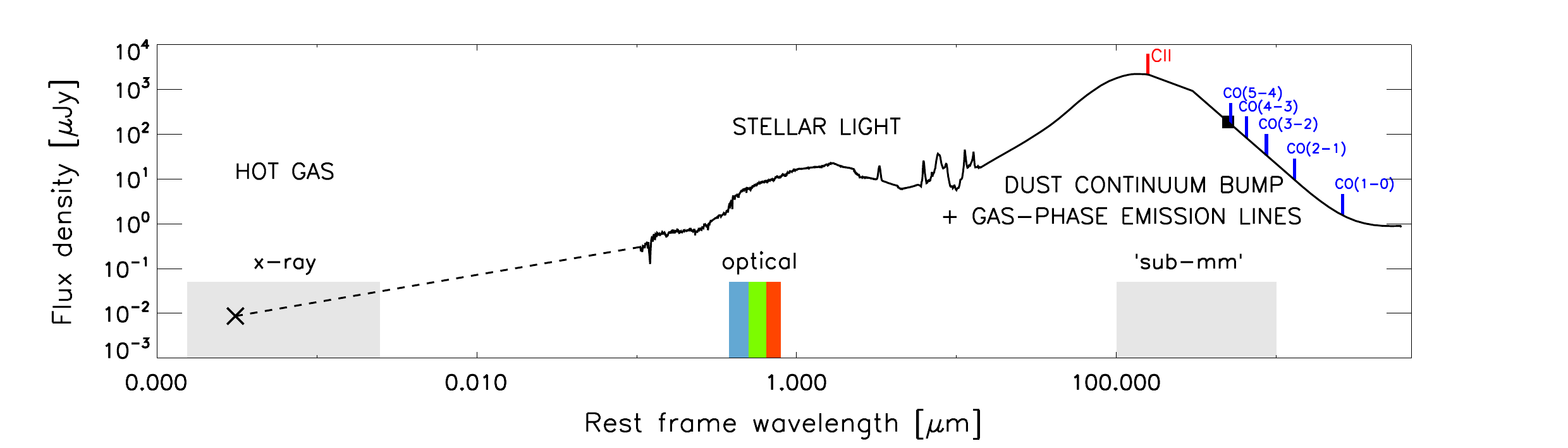}}
\caption{\footnotesize{The SED of a typical star-forming galaxy at $z=0$ with indications 
of the different components contributing to its shape. 
Also shown are the positions of some important emission lines from the gas, to be described in 
Part \ref{part1}. 
The flux has been scaled to the observed $500\,\mu$m luminosity of local galaxies for a mass of $10^{10}\,\msun$ 
as marked with a black square \citep{groves15}.
SED template of \cite{kirkpatrick12}: \protect\url{http://www.astro.umass.edu/~pope/Kirkpatrick2012/}.}}
\label{f:SED_z0}
\end{figure}

The best way to measure the conditions of gas in the ISM is by taking a `spectral fingerprint'. 
Fig.\,\ref{f:SED_z0} is a cartoon version of the spectral energy distribution (SED) of 
a galaxy if we were able to measure its emission from the smallest wavelengths (x-ray) 
to the longest ($\sim1$\,cm) of interest for the ISM. 
Furthermore, the SED is shown in the `rest frame' of the galaxy rather than in the observers frame, 
in which the SED would be red-shifted to longer wavelengths if the galaxy happened to be at high redshift. 
The unit is Jansky ($={\rm Jy}=10^{-26}\frac{{\rm W}}{{\rm m}^2\cdot {\rm Hz}}$) 
which is a typical flux unit in radio astronomy.
The SED shown in Fig.\,\ref{f:SED_z0} is that of a typical star-forming spiral galaxy, taken from the publicly available SED templates 
of \cite{kirkpatrick12}, with an additional imaginative X-ray flux measurement corresponding to 
an X-ray luminosity of $L_{\rm 0.5-8{\,\rm keV}}=10^{44}$\,ergs/s.

First noticeable are the two `bumps' at wavelengths around 1\,$\mu$m and $100$\,$\mu$m. 
The first one is dominated by direct stellar light at ultraviolet (UV) to near-infrared (NIR) wavelengths. 
The UV light is mainly produced by young massive stars recently formed 
and therefore traces the current SFR well, since massive stars have relatively short lifetimes.
On the other hand, the NIR flux is primarily consisting of light from old, less massive stars, making 
NIR luminosity a good tracer of total stellar mass. 
Good estimates of SFR and \Mstar can be made simultaneously by fitting the entire SED with 
stellar population synthesis models, typically only requiring assumptions about the initial mass function (IMF) 
of stars as they are born, the SFR history and the metallicity. 
For example, the SFRs in Fig.\,\ref{f:SFRD_history} were created assuming a Salpeter IMF \citep{salpeter55}.

\begin{figure}[!htbp] 
\centering
\hspace{.0em}\raisebox{0.cm}{\includegraphics[width=1\columnwidth]{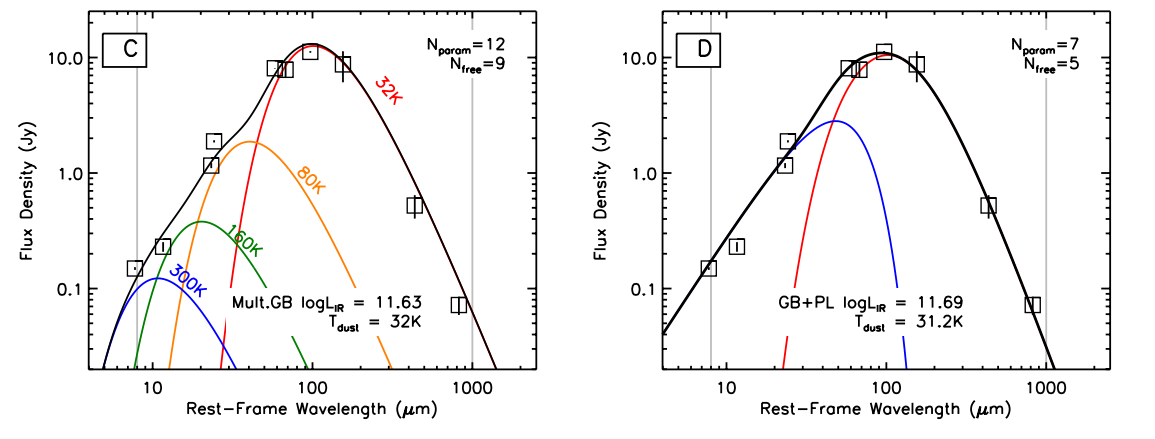}}
\caption{\footnotesize{Examples of how the spectra in MIR to FIR, caused by dust in a galaxy, can be 
approximated by either several `greybodies' of different dust temperature (left) or one greybody combined with 
a power law at high energies (right). Adapted from \cite{casey12} who presented the latter, simplified FIR SED fitting 
technique.}}
\label{f:dust_em}
\end{figure}

The second bump is caused by dust that absorbs light at many wavelengths, but it is a particularly powerful absorber of 
far-ultraviolet (FUV; $\sim100$-$200\,$nm) radiation, which is re-emitted again in the infrared (IR; $\sim1$-$1000\,\mu$m), 
thus `reprocessing' the emitted starlight. 
The shape of the dust `bump' can most of the time be approximated by a combination of a modified blackbody (a `greybody'), 
fixed at the temperature of the dust, 
and a power law in the mid-infrared (MIR) as described in detail by \cite{casey12} and shown with an example in Fig.\,\ref{f:dust_em}. 
On top of the dust peak a few important gas emission lines are shown, namely the CO (blue) and \cii (red) lines, that will 
be described in detail in Part \ref{part1}.

In order to analyze the gas component of a galaxy, there are basically two relevant regimes in wavelength, 
both of which are indicated with grey areas in the figure:\\
{\bf X-ray regime:} Thermal emission from galactic and extra-galactic hot, ionized gas. \\ 
{\bf Sub-mm regime:} Gas emission lines from colder, more neutral gas. \\
This thesis implements observational and theoretical techniques in both of these regimes, as will be summarized in 
Section \ref{intro:th}.

\clearpage
\section{An exciting time for observations}\label{intro:tele}

Facilities for observing gas and dust in galaxies accross cosmic time are 
experiencing a true revolution, that is gradually opening our eyes towards normal galaxies at high redshifts. 
Here, I will briefly go through the most important radio and X-ray telescopes in use or preparation.

\subsection{Radio Telescopes}

\begin{wrapfigure}[17]{r}{7cm}
\centering
\includegraphics[width=7cm]{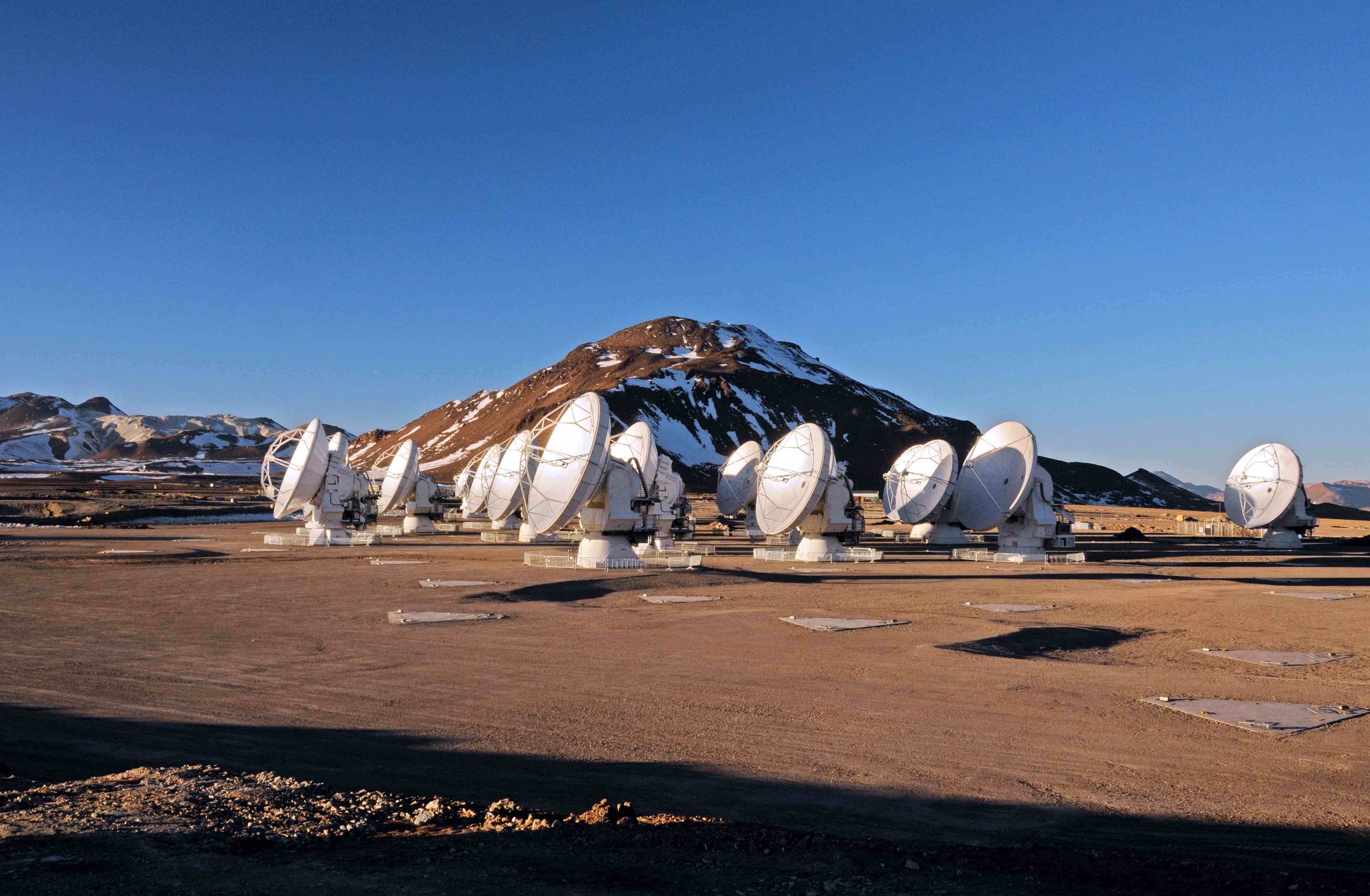}
\caption{\footnotesize{A photo of 19 ALMA antennas as they stood ready for the ALMA Early Science cycle. 
The array is located on the Chajnantor Plateu at 5000\,m above sea level in Atacama desert 
of northern Chile.
Credit: ALMA (ESO/NAOJ/NRAO)/W. Garnier (ALMA).}}
\label{f:alma}
\end{wrapfigure}

Probably the most impressive effort in this regard has been the 
Atacama Large Millimeter Array (ALMA), even if still under contruction. 
For an idea of its capabilities already in 2011 when the first scientific observations with ALMA began, 
see the image in Fig.\,\ref{f:antennae} made with only 12 antennas. 
When completed, ALMA will comprise 50 antennas of 12 metres in diameter, acting as a single 
telescope and with a very high spatial resolution thanks to the art of interferometry. 
An additional 4 12-m and 12 7-m antennas will create a compact array in the centre. 
Together, the 66 antennas (each weighing $>100$\,tons!) can be moved with special designed vehicles into different configurations 
with maximum antenna distances from 150\,m to an astonishing 16\,km. 
A more extended configuration corresponds to a higher spatial resolution, but also a lower sensitivity towards 
larger (more extended) sources on the sky, a trade-off that must be considered when planning observations. 
With bands of high spectral resolution from $312\,\mu$m to $3.56\,$mm (and 2 extra bands going down to 
about $7.4\,$mm possibly added in the future), ALMA will be able to detect \cii\ and CO lines in a normal galaxy at $z=3$ 
in a matter of hours.
For the next cycle of observations (Cycle 3, starting in October 2015), 
ALMA is anticipated to encompass a total of 48 antennas 
with most receiver bands available.

In the Northern Hemisphere, 
The NOrthern Extended Millimeter Array 
(NOEMA\footnote{\url{http://www.iram.fr/IRAMFR/GILDAS/doc/html/noema-intro-html/noema-intro.html}}, former PdBI) 
will be, upon completion in 2019, 
the facility in the Northen Hemisphere that comes closest to the capabilities of ALMA, 
with bands for observing in the $808\,\mu$m-$3.7\,$mm range. 
At longer wavelengths, the Jansky Very Large Array (JVLA) is currently the superior 
radio interferometer, covering from $0.6$ to $30\,$cm with 27 antennas of diameter 25\,m \citep{napier06}.  
A number of single-dish radio telescopes can also target gas and dust emission at high 
redshifts; the LMT, IRAM-30m, GBT, Onsala 20+25m, ARO, NMA and APEX to name a few. 

\begin{figure}[!htbp] 
\centering
\hspace{.0em}\raisebox{0.cm}{\includegraphics[width=1\columnwidth]{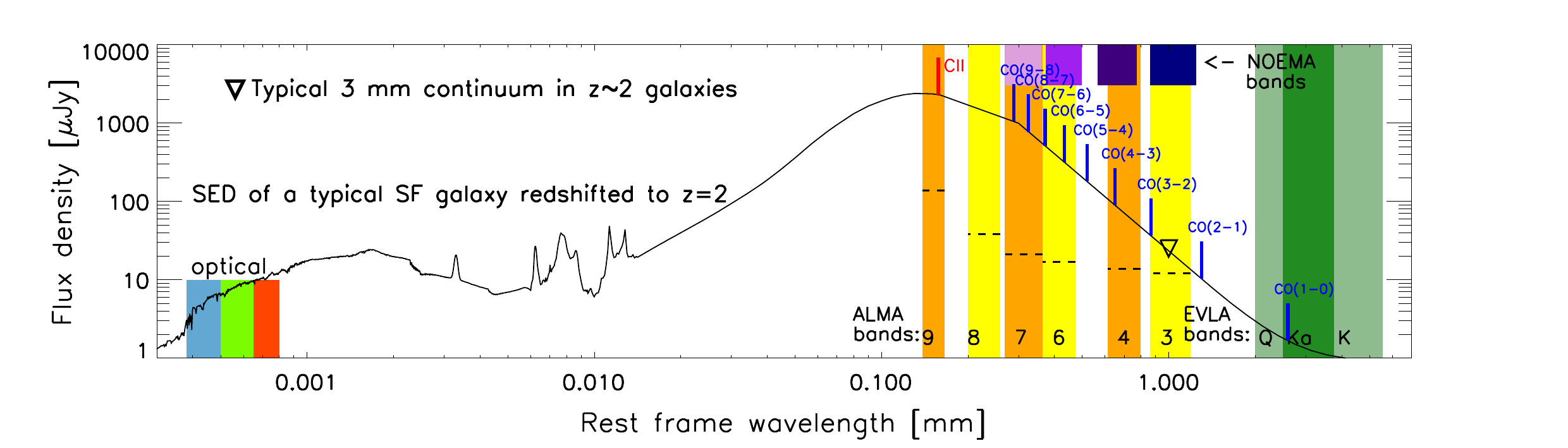}}
\caption{\footnotesize{The star-forming galaxy SED also used in Fig.\,\ref{f:SED_z0}, with 
the frequency bands of some of the worlds best radio interferometers shown in color. 
The SED has been scaled to $30\,\mu$Jy at 3\,mm, corresponding to typical massive disk galaxies at 
$z\sim2$ \citep{dannerbauer09}.
The dashed lines in the ALMA bands correspond to 3$\sigma$ after 5\,hrs integration time with the full ALMA. }}
\label{f:SED_z2}
\end{figure}

An overview of ALMA, EVLA and NOEMA is intended with 
Fig.\,\ref{f:SED_z2}, comparing the position of their frequency bands  
with the SED (including \cii\ and CO lines) of a typical star-forming galaxy at $z=2$.

The James Webb Space Telescope (JWST), scheduled to launch in 2018, will be an improved version of the 
famous Hubble Space Telescope (HST, expected to stay in operation until 2020), 
covering $0.6$ to $28\,\mu$m with the same resolution in NIR as HST has in the optical. 

Between the long wavelength range of EVLA/ALMA and the shorter of JWST, 
ground-based observations are hampered by atmospheric absorption, but 
new space missions are awaiting approval for adventuring into the MIR $28$-$312\,\mu$m regime. 
FIRSPEX is a satellite mission to continue the success of the Herschel Space Observatory 
({\it Herschel}), that ran out of fuel in 2013. The primary goal of FIRSPEX 
is to carry out a wide survey of the MW in \cii, \nii, \cil and 
CO$(6-5)$\footnote{FIRSPEX webpage: \url{http://astroweb1.physics.ox.ac.uk/~dar/FIR/firspex4.html}}, 
possibly detecting nearby galaxies as well that are redshifted into these bands. 
SPICA is a similar mission, of $\sim2$ orders of magnitude higher sensitivity in the FIR than {\it Herschel} 
and wavelength coverage from $5$ to $\sim210\,\mu$m \citep{goicoechea11}.

\subsection{X-ray Telescopes}

\begin{wrapfigure}[19]{l}{5.5cm}
\centering
\includegraphics[width=5cm]{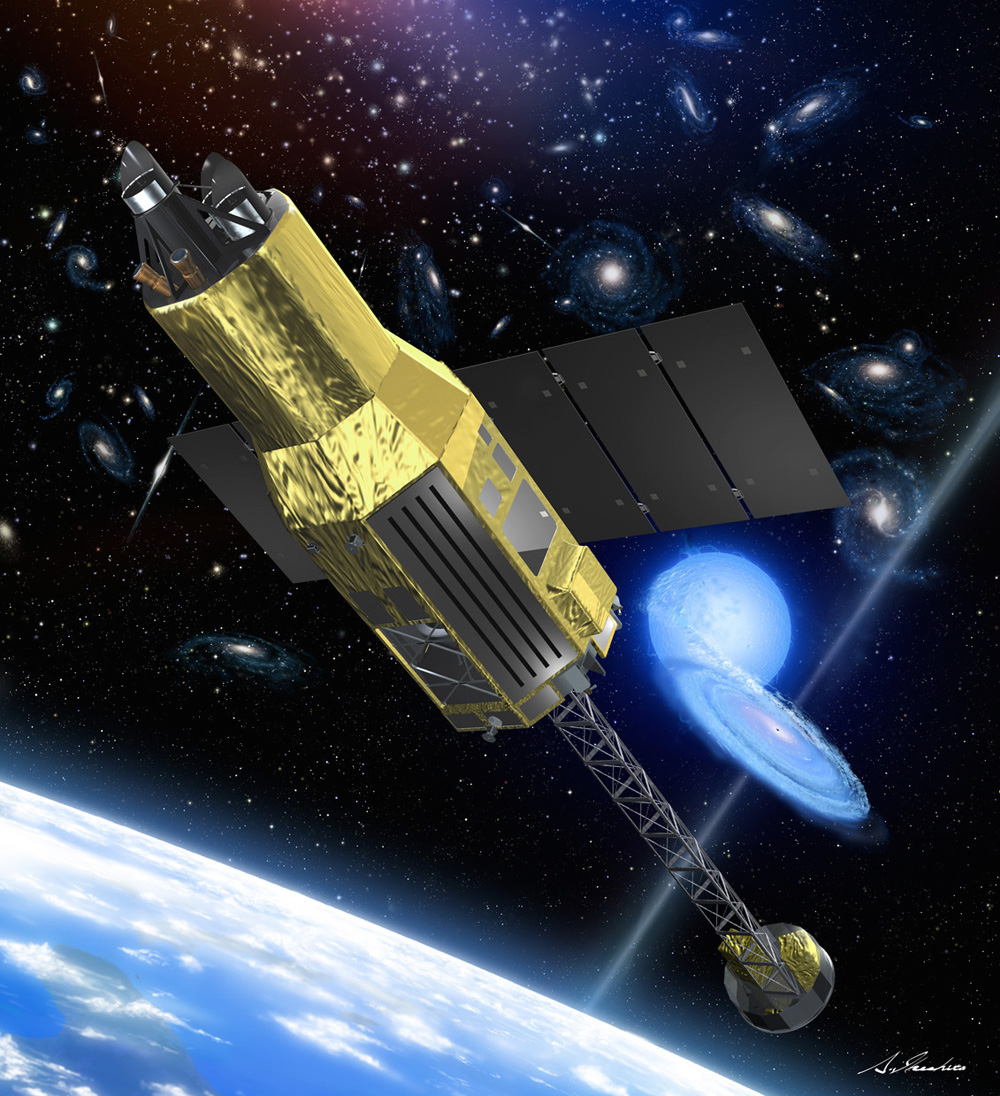}
\caption{\footnotesize{An illustration of the upcoming ASTRO-H when in orbit at an altitude of about 5-600\,km. 
From the mission home page: \protect\url{http://astro-h.isas.jaxa.jp/en/}.}}
\label{f:astro-h}
\end{wrapfigure}
To observe an AGN in X-ray, one has to escape the obscuring atmosphere of the Earth, 
which has led to many space missions since the beginnings of the 1980's.

The {\it Chandra} X-ray ($0.1-10$\,keV) observatory has been in orbit around the Earth since 1999, and has dedicated 
2 and 4\,Ms to observing the {\it Chandra} Deep Field North and South respectively (CDF-N and CDF-S). 
The large covering areas of roughly $450\,$arcmin$^2$ for each have proven ideal for detecting 
AGNs at redshifts $z\approx 0.1-5.2$ \citep{bauer04,xue11}. 
Similar missions, though of lower spatial resolution, include the XMM-Newton, ROSAT and Suzaku. 
At higher energies, the nuclear spectroscopic telescope array Nustar has been observing from space at $3-79$\,keV since 2012.  

The ASTRO-H mission (see Fig.\,\ref{f:astro-h}) is the latest in a long row of Japanese X-ray satellites, though ASTRO-H is being 
developed in a collaboration between Japanese (ISAS/JAXA) and US (NASA/GSFC) institutions.
It will among other things investigate the X-ray reflection signatures of AGN, with increased 
sensitivity for the soft X-rays as compared to {\it Chandra} \citep{reynolds14}.
Astrosat is India's first dedicated astronomy satellite and has the advantage of allowing 
simultaneous multi-wavelength observations at UV and X-ray wavelengths. 
It will have two bands for detecting X-rays (3-80\,keV and 10-150\,keV), 
and a large effective area in the hard X-ray band to facilitate studies of highly variable 
sources such as AGN \citep{biswajit13}. 

\clearpage

\section{Galaxy simulations}\label{intro:sim}
\begin{wrapfigure}[36]{r}{7cm}
\centering
\hspace{.0em}\raisebox{0.cm}{\includegraphics[width=6cm]{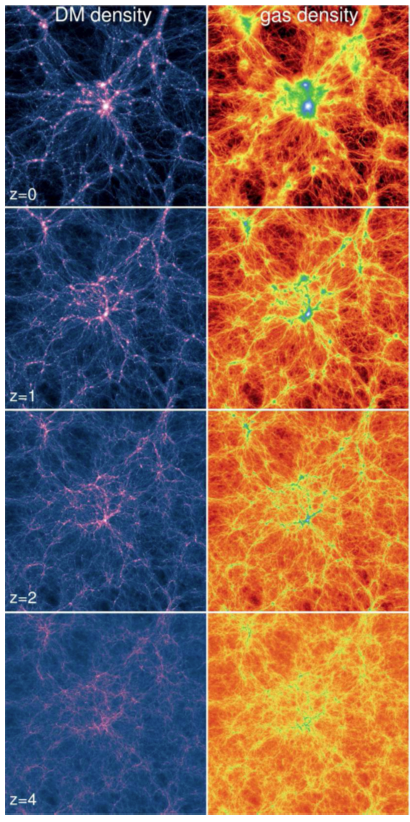}}
\caption{\footnotesize{$106.5\,$Mpc wide on each side, column density maps are shown 
of simulated dark matter (left) and gas column densities (right) from $z=4$ until now (bottom to top). 
Brightest colors in the column density plot 
correspond to a density of $\sim10^4\,\msun/$kpc$^3$. 
The images are from the Illustris Project -- one 
of the most comprehensive cosmological hydrodynamical simulations of 
galaxy formation to date \citep{vogelsberger14}. 
The recently developed moving-mesh code \texttt{AREPO} \citep{springel10} was 
used to follow DM and gas dynamics. 
}}
\label{f:illustris}
\end{wrapfigure}
With so many coupled mechanisms in a galaxy on scales from sub-parsecs to kilo-parsecs, 
the advantage of running numerical simulations of galaxy evolution has proven tremendous. 
Galaxy simulations can be roughly divided into two kinds: 
Cosmological simulations that start out with 
the small density pertubations of dark matter shortly after the Big Bang, 
and more idealized disk simulations that start out with an analytical density profile for the 
galaxy and allow gravitational and electromagnetic forces to perturb the galaxy in time. 
I will focus on the former kind as that is what my theoretical 
projects build on (see Parts \ref{part1} and \ref{part2}).

The series of images in Fig.\,\ref{f:illustris} illustrate how a cosmological simulation works, 
with snapshots, from high redshift ($z=4$) and until now ($z=0$), of the dark matter structures 
and the resulting gas mass distribution. 
In the present $\Lambda$CDM cosmological model, dark matter formed structures (or `minihalos') 
before baryonic gas could cool and collapse. 
Cosmological simulations therefore focus on modeling an underlying dark matter structure 
(left-hand side of Fig.\,\ref{f:illustris}) at the same time as 
the baryonic component of the universe is allowed to condense in these structures via 
gravity, dissipation and radiation (right-hand side of Fig.\,\ref{f:illustris}). 
As the universe ages, filaments and knots in the dark matter density distribution become more pronounced 
and infalling gas cools enough to form stars and galaxies in regions of high concentration.

The universe is then believed to have proceeded via hierarchical structure formation, in which small gravitationally 
bound structures, such as stars and galaxies, form first, followed by groups, clusters and superclusters of galaxies, 
not to mention the cosmic web in between them (c.f. Fig\,\ref{f:web}).

\begin{figure}[!htbp] 
\centering
\hspace{.0em}\raisebox{0.cm}{\includegraphics[width=0.8\columnwidth]{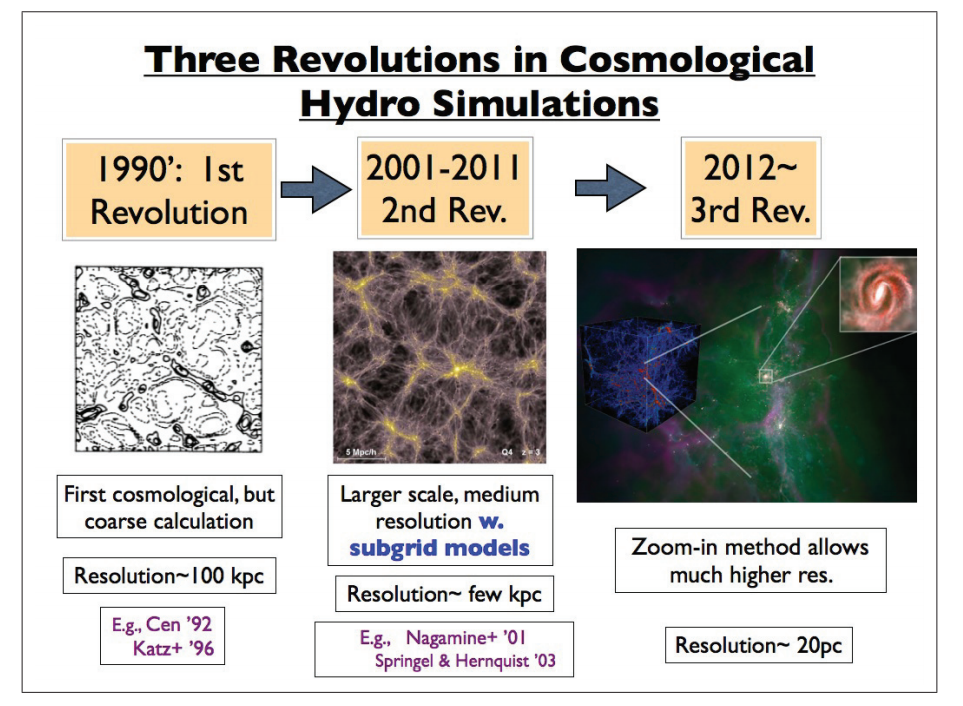}}
\caption{\footnotesize{Schematic overview of the history of cosmological galaxy simulations and their 
improvements over the past two decades, with references, as presented in the review by \cite{nagamine14}.}}
\label{f:sims}
\end{figure}

It was soon realized though, that these large-scale simulations with resolutions going down to kpc-scales, 
could not capture the actual physics inside galaxies, where the most important processes take place on $<1\,$pc scales.
In particular, simulations on smaller scales proved that localized feedback is needed in order 
to avoid overpredicting the star formation rate and to agree with the observed IMF of star clusters  
\citep{tasker11,krumholz11,hopkins11}.
Made possible by an increasing computing power and technology, cosmological galaxy simulations 
have seen two big `revolutions' since their beginning in the early 90's as illustrated in Fig.\,\ref{f:sims}, 
of which we are in the middle of the last one, the `zoom-in' simulations.

\subsection{`Zoom-in' simulations}
The concept behind `zoom-in' simulations is to select a galaxy at low redshift in a cosmological simulation of 
low resolution ($\gtrsim1\,$kpc), and trace the particles (or cells) comprising it back in time to the beginning of 
the simulation. A re-simulation is then initiated with much higher resolution, but only following the particles/cells 
that end up in the galaxy of interest. This re-simulation may take days or months, but provides the opportunity to 
follow star formation on realistic scales, while staying in agreement with the overall large scale structure.
Two computational techniques exist for treating the gas hydrodynamically at the necessary resolution:\\
{\bf Smoothed Particle Hydrodynamic (SPH) simulations:} The gas is represented by particles with properties such as mass, 
density and temperature in addition to a smoothing kernel that roughly speaking smears the particles out 
across a volume in space. Resolution is increased simply by increasing the number 
of particles while reducing their mass in order to keep mass conservation. \\
{\bf Adaptive Mesh Refinement (AMR) simulations:} The gas is divided into cells that can be subdivided in areas where higher 
resolution is needed. All interactions between gas cells take place as mass/temperature/pressure exchange through 
the cell partition walls.\\ 
Examples of SPH codes include \texttt{\textsc{Gadget}} and \texttt{\textsc{Gasoline}}, whereas 
\texttt{\textsc{Art}}, \texttt{\textsc{Enzo}} and \texttt{\textsc{Ramses}} are AMR codes. 
\cite{kim14} presented, with the AGORA project, a comparison of these in terms of resolution, physics of the ISM, 
feedback and galactic outflows. But see also \cite{springel10} for a novel combination of the two methods.

\subsection{The calibration to $z=0$}
While the initial conditions for galaxy formation and evolution are relatively strict, 
by e.g. measurements of the Cosmic Microwave Background (CMB), 
and the final endpoint is anchored with $z=0$ observations, what goes on in between is less well-defined 
by observations. 
Common tests for the evaluation of a cosmological simulation focus on how well it can reproduce the 
observed main sequence and luminosity functions at low- to high-$z$. 
Although simulations on the whole agree with observations, a discrepancy still persists at the time of 
maximum cosmic star formation ($1\lesssim z\lesssim3$), where the simulated main sequence is a factor 
$\sim2-3$ below the observed one \citep[e.g.][]{furlong14}. 
Possible explanations include an IMF that changes with redshift and galaxy properties 
(rather than staying fixed to the locally observed one as most simulations assume), or feedback from 
stellar winds, supernovae and AGN that is not properly accounted for in simulations 
\citep[e.g.][]{dave08,narayanan12b,sparre15}.
Until more resolved observations of ISM at $z\gtrsim2$ are obtained, 
{simulations of galaxy evolution are forced to make, and be inhibitted by, 
extrapolations to high-$z$ of local ISM physics and star-formation laws}. 
Some of these restrictions will be discussed in Part \ref{part1}.

\begingroup
\section{References}
\def\chapter*#1{}
\bibliographystyle{apj} 
	\setlength{\bibsep}{1pt}
	\setstretch{1}
\bibliography{bibs0}
\endgroup

%% file: intro.tex

\chapter{Introduction to this thesis}

\section{Evolution of massive galaxies across cosmic time}

With the previous brief description of the internal composition of galaxies and 
the observational and theoretical tools used for characterising them, 
we are now ready to take a look at the populations of massive galaxies so far uncovered and 
the current status on creating a consistent picture for their evolution with cosmic time. 

\subsection{High-redshift massive galaxy populations}\label{intro:sel}
An important fact to bear in mind when studying galaxy evolution, 
is that we are never observing {\it the same} galaxy at various epochs in the life of the universe. 
At each redshift, the most we can do is observe galaxies in a snapshot of their lifetime, 
and often any evolutionary connection between galaxies observed at different redshifts is hampered by 
different selection techniques and incompleteness (we cannot detect all galaxies at every redshift, cf. Section \ref{intro:massive}). 

The ideal way of classifying galaxies at all redshifts would be to measure their entire SED (like the one shown in Fig.\,\ref{f:SED_z0}) 
via spectroscopy. But since this would be very time-consuming, clever methods have been developed to 
first select a candidate group of the galaxies that might be of interest, and then perform high-resolution spectroscopy on these. 
Such methods typically build on photometry -- measuring the total flux through a limited set of photometric bands by applying filters 
that only allow certain ranges in wavelength.
For selecting galaxies of a certain mass, the $K$-band (centered on $2.2\,\mu$m in the NIR) is particularly useful since it captures 
light from primarily old stellar populations that dominate the mass budget of most galaxies. 
Therefore, a lower cut in $K$-band luminosity directly translates into a minimum stellar mass 
\cite{cimatti03}.
Follow-up spectroscopy can then determine the redshift, stellar mass and SFR to higher precision with SED fitting methods 
\citep[see][for applications of this method]{man12,cimatti04}.
This method works out to $z\sim4$ where the $K$-band starts to trace rest-frame UV light, which unlike NIR or optical, 
is dominated by the strong radiation from young stars, and is therefore a tracer of the amount of current star formation 
rather than stellar mass.

At $z\gtrsim3$, another way to classify galaxies is typically via their current amount of star formation. 
This star formation can be either un-obscured, as revealed in rest-frame UV, or -- in galaxies of high dust amount -- obscured 
resulting in high FIR luminosities (dust continuum `bump' in Fig.\,\ref{f:SED_z0}), or a combination of the two. 
Two examples of the first approach are the Lyman Break Galaxies (LBGs) and BzK-selected galaxies, 
whereas the second approach has revealed a group of very dusty galaxies, the Sub-mm Galaxies (SMGs) or nowadays, 
Dusty Star-forming Galaxies \citep[DSFGs;][]{casey14}. 

LBGs are selected to have a strong break and reduced flux at the wavelength $<\lambda=912$\,\AA. 
This break is caused by neutral hydrogen gas residing inside the galaxy and in the Inter Galactic Medium (IGM), 
absorbing light at wavelengths below $912\,$\AA, the threshold for ionizing hydrogen. 
The LBG technique requires that the galaxy be UV bright (relatively star-forming) and not dust-obscured \citep{steidel03}. 
The advantage of the LBG technique is that it makes a preselection of not only SFR but also redshift, by restricting the 
location of the break. 
\cite{carilli08} used a method called `stacking' (see application of this in X-ray in Part \ref{part2}), 
to find that LBGs at $z\sim3$ have SFRs of $31\pm7$\,\sfru, that is, not very high on the MS at that redshift (c.f. Fig.\,\ref{f:MS_z}). 

The BzK-selection takes advantage of absorption in stellar atmospheres creating a strong break at $\sim4000$\,\AA\, 
so that three photometric bands ($B$, $z$ and $K$) can effectively be used to select galaxies at 
$1.4\lesssim z\lesssim 2.5$ and roughly estimate 
their SFR \citep{daddi04}. 
Similar groups of color-selected galaxies are those of `BX' and `BM' galaxies, selected via their color in 
the three UV and optical filters $U_{n}$, $G$ and $\mathcal{R}$ \citep{adelberger04,steidel04}.
The color criteria for `BX' galaxies restricts them to lie at $2.0\leq z\leq 2.5$, while the 
`BM' galaxies lie at $1.5\leq z\leq 2.0$. 
Both techniques give rise to slightly higher SFRs than the BzK-selection would, as shown by \cite{reddy05} 
using X-ray inferred bolometric SFRs.

\begin{wrapfigure}[18]{r}{7cm}
\centering
\includegraphics[width=7cm]{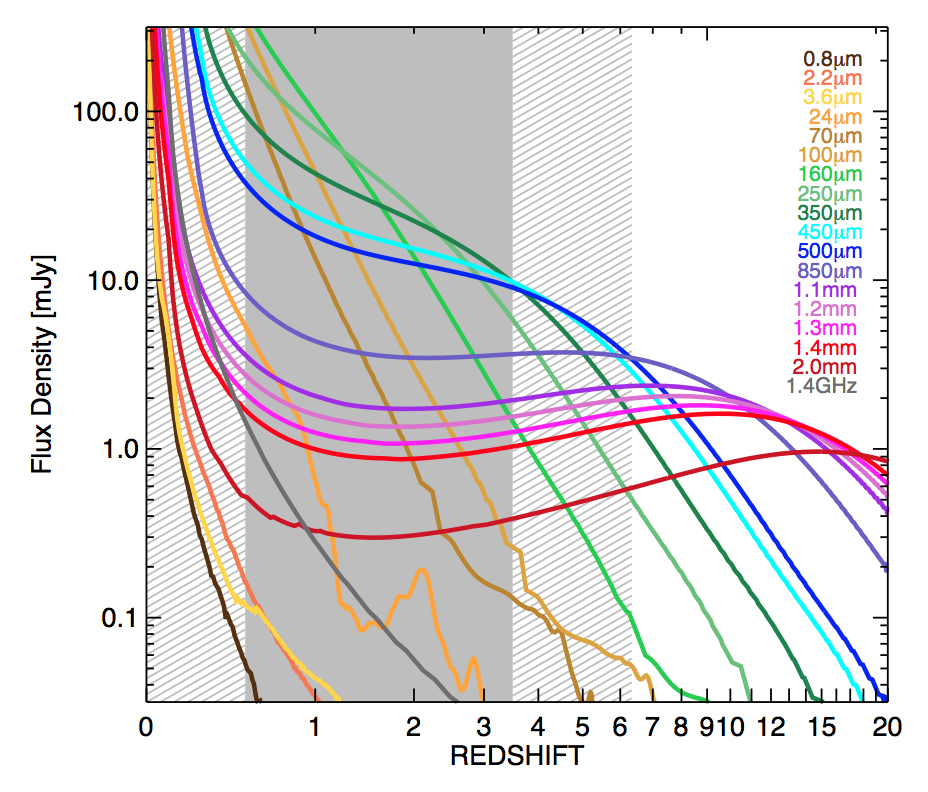}
\caption{\footnotesize{Observed flux densities for a typical dusty, star-forming galaxy as presented 
in the review by \cite{casey14}. Note the nearly-constant flux density in the $\sim1$\,mm bands accross 
$1\lesssim z\lesssim 10$ redshift range.}}
\label{f:negk}
\end{wrapfigure}

SMGs on the other hand, are selected for having a strong rest-frame NIR emission, and were first 
detected with the ground-based instrument SCUBA at 850\,$\mu$m 
\citep[e.g.][]{smail97,hughes98,barger99}, soon followed by surveys including the use of instruments such as MAMBO at 
1.2\,mm and AzTEC at 1.1\,mm \citep{coppin05,chapman05,greve08,perera08}. 
The advantage of looking in the rest-frame $1-3$\,mm wavelength range at high redshift, is that 
the flux stays roughly the same owing to the effect of `negative $K$-correction' as illustrated in Fig.\,\ref{f:negk}.
Due to this effect, SMGs have been identified mainly at  $z\sim2$ but with a tail extending up to $z\sim5$ 
\citep{chapman05,wardlow11,simpson14}. 
Converting the high IR luminosities to dust-obscured SFRs, implies SFRs of $\sim100-1000$\,\sfru\ \citep{karim13}. 
For the most part, SMGs display disturbed stellar morpohologies and/or kinematics as compared to 
normal non-interacting rotating disks, 
suggesting recent or ongoing merger events \citep[e.g.][]{conselice03,chen15}.
The disturbed signatures are backed up by molecular emission line observations of clumps of 
dense and highly star-forming gas in SMGs \citep[e.g.][]{tacconi08,riechers11,hodge12}.


\subsection{Observing galaxies at $z>4$ via their gas emission lines}

In the early ($z>4$) universe, very few normal star-forming galaxies have so far been detected, 
typically via the LBG technique \citep[see][for a compilation]{behroozi13}, 
and even fewer quiescent galaxies have been discovered at $z\gtrsim2.5$, with the most distant ones at $z\sim3$ 
\citep{gobat12,fan13}. 
Only one galaxy at $z>7$ has so far had its redshift determined from spectroscopy of its 
stellar continuum. 
But the galaxy A1689-zD1 at $z=7.5\pm0.2$, has now been detected in dust emission 
with ALMA resulting in a combined (UV+IR) SFR of $\sim12\,\sfru$ \citep{watson15}, i.e. a normal star-forming galaxy. 
In addition to dust continuum, ALMA can also detect line emission from the gas, thereby 
determining redshifts and characterizing the ISM at the same time for normal star-forming galaxies 
(see more on this in the Outlook section of Part \ref{part1}).
An example of such an emission line, is \cii from singly ionized carbon, which is expected to be the most 
luminous cooling line in neutral gas. 
We return with a more detailed description of \cii in Chapter \ref{paper2}, 
but mention here that at $z\sim5$-$7$, the \cii line falls in bands 6-7 of ALMA, 
currently the most sensitive millimeter interferometer by an order of magnitude,  
and therefore a promising tool for detecting \cii in normal star-forming galaxies at $z\sim6$.
Indeed, it has been detected in and around normal star-forming galaxies at $z\sim5-7$ \citep{maiolino15,capak15}.
{\it Our only way of characterizing the ISM of high-$z$ galaxies might very well be 
to observe emission lines from the gas.}

\subsection{The red and blue galaxies} \label{redblue}

\begin{figure}[!htbp] 
\centering
\hspace{.0em}\raisebox{0.cm}{\includegraphics[width=0.7\columnwidth]{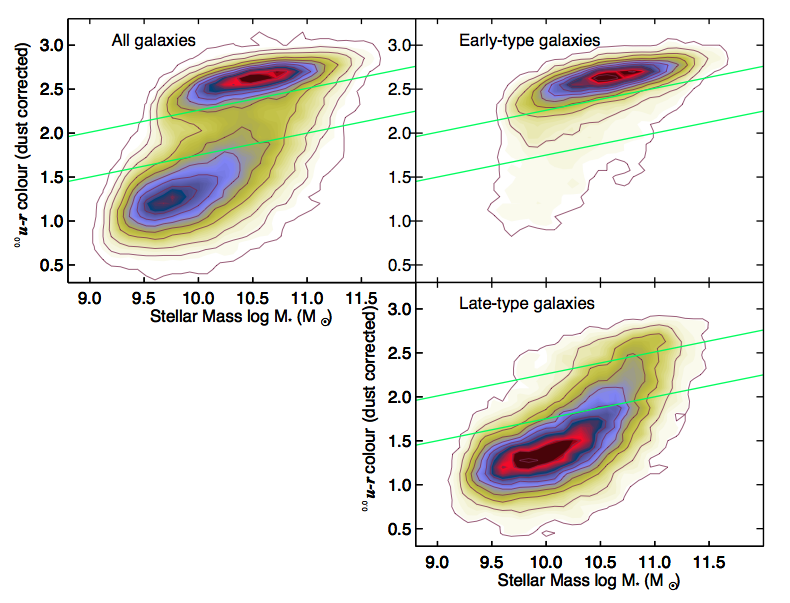}}
\caption{\footnotesize{The UV-optical color bimodality of 
$0.02<z<0.05$ galaxies, as measured by \cite{schawinski14} 
using the SDSS+{\it GALEX}+Galaxy Zoo data. The `$0.0$' in $^{0.0}u$ refers 
to the $K$-correction of the observed color to $z=0$ (see Section \ref{intro:sel}). 
Green lines indicate the `green valley' in between the two galaxy populations.}}
\label{f:bimo}
\end{figure}

The local galaxy population is bimodal in terms of UV-to-optical color, with a distinct group of 
blue, star-forming galaxies and another of red, quiescent galaxies 
\citep{strateva01,schawinski14}. 
This is shown with color-mass diagrams in Fig.\,\ref{f:bimo} for low redshift galaxies. 
The connection between color and star formation activity 
is a result of younger stars having relatively bluer colors than old stars, which emit their 
light predominantly at longer (optical and NIR) wavelengths. 
Historically, the red and blue galaxies are also called `early' and `late'-type galaxies 
from the morphological classification of \cite{hubble26}. 
In general, late type galaxies display stellar disks with spiral arms and exponential light 
profiles whereas the stellar light from early-type galaxies (or sometimes named ellipticals) is more concentrated 
with even structures and no clear sign of spiral arms. 
But detailed observations or their stellar populations have later shown that this naming makes sense, 
since the nearby red galaxies most likely formed much earlier than the blue ones \citep[e.g.][]{kartaltepe14}. 
As always, dust plays an important role and can make star-forming galaxies appear red by absorbing 
the blue light and re-emitting it in the FIR. 
Indeed \cite{sodre13} showed that some local galaxies with extremely red colors, are reddened by 
dust more than by older (redder) stellar populations.
But the diagrams in Fig.\,\ref{f:bimo} have been corrected for reddening by dust, by measuring the dust 
extinction in optical light. 

In between the red and the blue group, lies the `green valley' with very few galaxies. 
There is a discussion about how many of these green galaxies are actually late-type galaxies 
being quenched \citep{pan13,schawinski07} and how many are 
early-type galaxies with star formation initiated recently in a sort of `rejuvenation' \citep{fang12}.
In Section \ref{intro:MS} the main sequence (MS) of star formation was introduced, and Fig.\,\ref{f:MS_z} showed 
how the MS grows towards higher SFR with redshift. 
Blue galaxies lie on the MS relation, whereas red galaxies lie below and typically  at high stellar mass. 
One of the biggest quest in astronomy for understanding galaxy evolution is; 
{\it Why did blue galaxies form more stars at higher redshift?}

By now, the division into quiescent or star-forming has been observationally confirmed out to $z\sim3$, 
\citep{kriek08,williams09,viero13,man14}, leading to a second, equally important, question: 
{\it When did the onset of the red sequence take place and what caused it?}

\subsection{Quenching star formation} \label{intro:quench}
AGNs are one of the preferred hypothesized mechanisms for quenching star formation, 
as radiation, winds and jets from the central accreting black hole can remove the gas or heat it so that 
contraction of molecular clouds cannot take place (see \citealt{fabian12} and \citealt{mcnamara07}
for a reviews on models and observational evidence). 
A distinction is made between `quasar mode', by which feedback from an AGN ejects 
all gas from the galaxies \citep{granato04}, and `radio mode', during which powerful jets heat 
the circumgalactic medium to such high temperatures that further accretion of cold gas 
onto the galaxy is prevented \citep{croton06}. 
Semi-analytical simulations, by e.g. \cite{dimatteo05}, show how reasonable descriptions 
of the AGN feedback on star formation in elliptical galaxies can reproduce the observed local mass function. 


In addition to quenching by AGNs (possibly ignited by a major merger), the environment in which the galaxy 
lives is believed to raise further quenching mechanisms. 
This suspicion came from observations showing 
that galaxies in clusters are in general more quenched and passive than field galaxies \citep{verdugo08}. 
Similar to AGNs, most of these proposed internal/external mechanisms have to do 
with the removal, distribution or heating of gas, i.e. 
the fuel for further star formation. 
This can happen via galaxy-galaxy interactions or influence by the Intra Cluster Medium (ICM) itself. 
The following are some of the more popular environmental quenching mechanisms, but see 
reviews by e.g. \cite{boselli06} and \cite{blanton09} for a much more complete overview: 
i) Harassment, when during close encounters or direct mergers, 
galaxies alter each others' morphology \citep{moore98}; 
ii) Ram pressure stripping, where 
galaxies traveling through the ICM at high speed have their ISM shock-heated and 
eventually stripped off, ending up in the cluster as seen in observations of individual cases 
and proposed by several simulations \citep[e.g.][]{steinhauser12,bosch13,ebeling14}; 
iii) Tidal stripping, by which galaxies passing through the ICM feel the global tidal field of their host, significantly altering their 
morphologies \citep{villalobos14};
iv) Cosmological `starvation' or `strangulation', when the hot gas halo around a galaxy is removed 
due to hydrodynamical interactions with the ICM, and further infall of gas is prevented causing the 
galaxy to slowly run out of fuel on time scales of a few Gyr \citep{balogh00,bekki02,feldmann15}; 
v) Compactness quenching, as suggested by observations of satellites in the outskirts of haloes, 
that show a strong correlation between quenching and compactness of the galaxy \citep{woo15}.

These last types of quenching mechanisms are also referred to collectively as `satellite quenching', 
because they primarily happen to satellite galaxies in a cluster. 
However, it has been hypothesized that satellite quenching and that of AGN (also called `mass-quenching'), 
are actually manifestations of the same underlying process \citep{carollo14,knobel15}.

\subsection{Connecting the dots} \label{intro:dots}

The fact that the red sequence is already in place at $z\sim2$, means that we have 
to push even further out in redshift in order to find the onset of this 
`dead' population, and with that, hopefully also the `murder weapon'.
In addition, half of the most massive galaxies ($\Mstar>10^{11}$\,\msun) at $z\approx2$ are already 
compact and quiescent galaxies (CQGs) with old stellar populations \citep[e.g.][]{toft07,vandokkum08,szomoru12}. 
The stellar spectra suggest that major starburst events took place in CQGs some 1-2 Gyr 
prior to the time of observation, but were quenched by mechanisms either internal or external (or both) to the galaxy 
\citep[e.g.][]{gobat12,whitaker13,vandesande13}.
So far, very few have CQGs have been found in the present-day universe \citep[e.g.][]{trujillo09,taylor10}, though see 
also \cite{graham15} who suggest that CQGs could be hiding in the local universe as the bulges of spheroidal 
galaxies having grown an additional (2D) disk over the past 9-11\,Gyr.

\begin{figure}[!htbp] 
\centering
\hspace{.0em}\raisebox{0.cm}{\includegraphics[width=0.8\columnwidth]{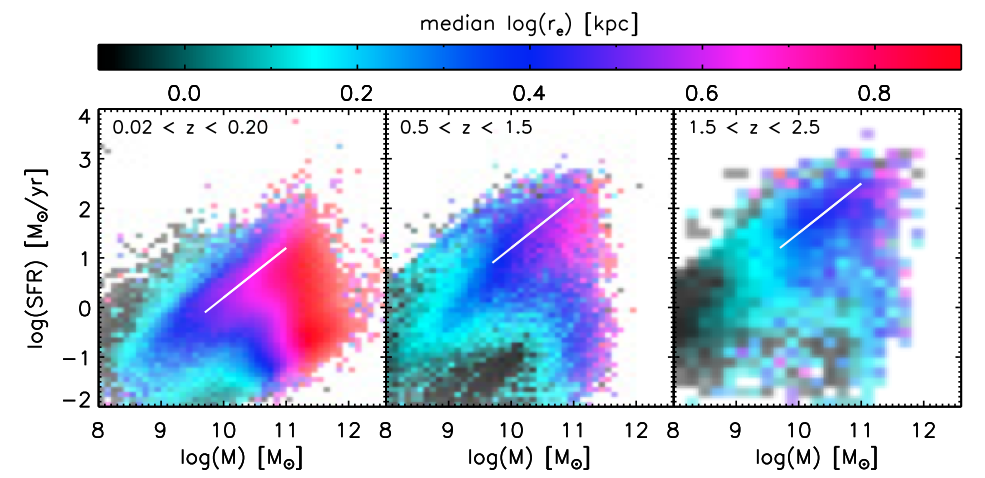}}
\caption{\footnotesize{Galaxy SFR-mass diagrams in three redshift bins, colorcoded by 
size which is expressed as a circularized effective radius from fitting a S\'ersic 
profile to the rest frame visible or NIR depending on redshift. From \cite{wuyts11}, 
who study galaxies from various large samples, including CANDELS (see Section \ref{intro:massive}).}}
\label{f:wuyts}
\end{figure}

It is still an open question what kind of galaxy evolution leads to the CQGs observed at $z\sim2$, 
and how they increased in size, without increasing their SFR significantly, down to $z\sim0$, for 
which a size increase by a factor of $\sim3-6$ is required \citep[e.g.][]{trujillo06,toft07,szomoru12}. 
Basically, three formation scenarios have been put forth: 
i) Massive CQGs formed in a monolithic way, assembling most of their mass 
at $z>2-3$ and then `puffed up' (by adiabatic expansion) into 
the big, red ellipticals we see today \citep[e.g.][]{bezanson09,damjanov09}. \\
ii) A similar scenario to i), but CQGs enhanced their size from $z\sim2$ to $0$ by 
inside-out growth, most likely due to merging with other small galaxies 
(minor merging), following a type of hierarchical structure formation like the dark matter \citep{naab09,oser12}. 
The minor merging scenario is in good agreement with the observed 
metallicity gradients \citep{kim13} and stellar kinematics \citep[e.g.][]{vandesande13,arnold14} 
in $z=0$ quiescent galaxies. \\
iii) The third possibility, requires a bit more thought. 
Compared to star-forming galaxies, local quiescent galaxies are more massive at the same SFR but 
also slightly smaller at the same mass, for redshifts out to $z\sim2$. 
This can be seen in Fig.\,\ref{f:wuyts} which shows the main sequence of galaxies, 
including the quiescent ones below it, from $z\sim0$ to $z\sim2$. 
The average size of the quiescent population could grow with time simply if 
recently quenched (larger) star-forming galaxies are constantly being added to the sample as 
time moves forward \citep[e.g.]{trujillo12,poggianti13,krogager14}. 
This hypothesis also goes under the name `progenitor bias'. 
However, \cite{belli15} argued that, for at least half of all CQGs at $z=2$, the increase in size until 
$z=1.5$ must have taken place via growth rather than additions of increasingly larger galaxies to the sample. 


\begin{wrapfigure}[18]{r}{5.7cm}
\centering
\includegraphics[width=6cm]{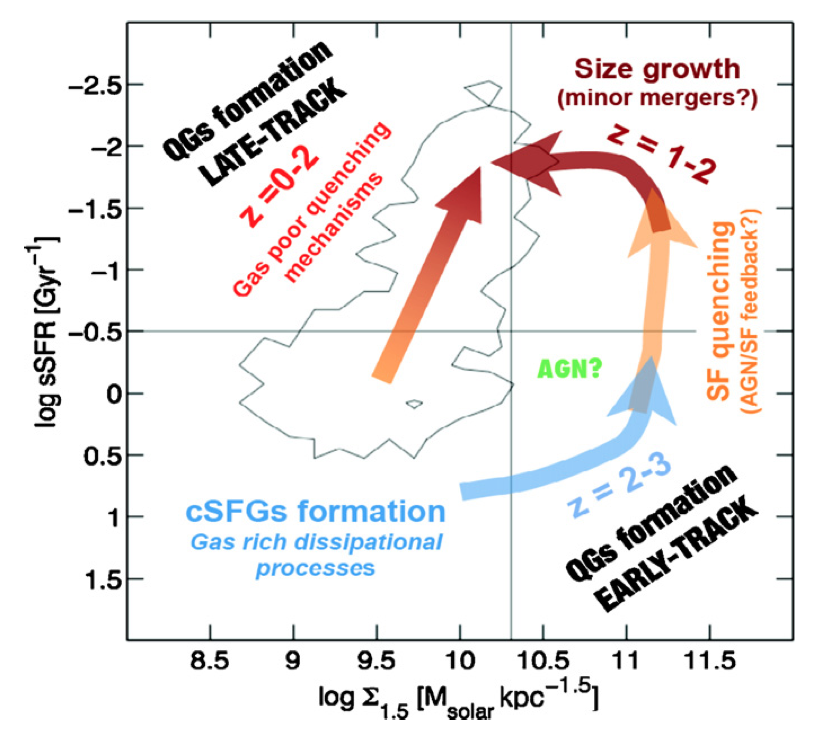}
\caption{\footnotesize{Two suggested evolutionary paths from star-forming galaxies at $z=2-3$ to 
quiescent galaxies at $z=0$ \citep{barro13}.}}
\label{f:barro13}
\end{wrapfigure}

The potential quenching by AGNs has led to the idea of a fast evolutionary path from 
compact star-bursting galaxy at $z\sim3$ to a galaxy quenched by AGN feedback some 
$1-2$\,Gyr later and, via size-growth, to a larger, however still quiescent galaxy at $z=0$. 
This is illustrated in Fig.\,\ref{f:barro13} as an `early-track' route \citep{barro13}. 
In addition, \cite{barro13} suggest a late-track evolutionary path, by which larger $z\sim2$ star-forming galaxies slowly 
turn off star formation between $z=2$ and $z=0$, without passing through a compact phase.  
This idea was expanded upon by \cite{williams14}, who suggest that massive, compact galaxies 
take the fast route and quench sooner than the normally-sized LBGs at $z\sim2-3$.

Galaxy mergers present the kind of violent events that are required for the `early-track' route. 
In particular, simulations have shown that gas-rich (wet) mergers can lead to bursts of star formation, 
but that gas is driven towards the center via dissipation and may ignite an AGN. 
Wet mergers can increase the stellar mass but leave a more compact remnant, 
whereas gas-poor (dry) mergers can build up a galaxy in mass and size without altering 
star formation too much \citep[see][and references therein]{lin08}. 
In addition, major mergers are those between galaxies of similar mass whereas minor mergers are those in which  
the stellar masses differ by a factor of at least $\sim4$ \citep{man14}.

\begin{figure}[!htbp] 
\centering
\hspace{.0em}\raisebox{0.cm}{\includegraphics[width=1\columnwidth]{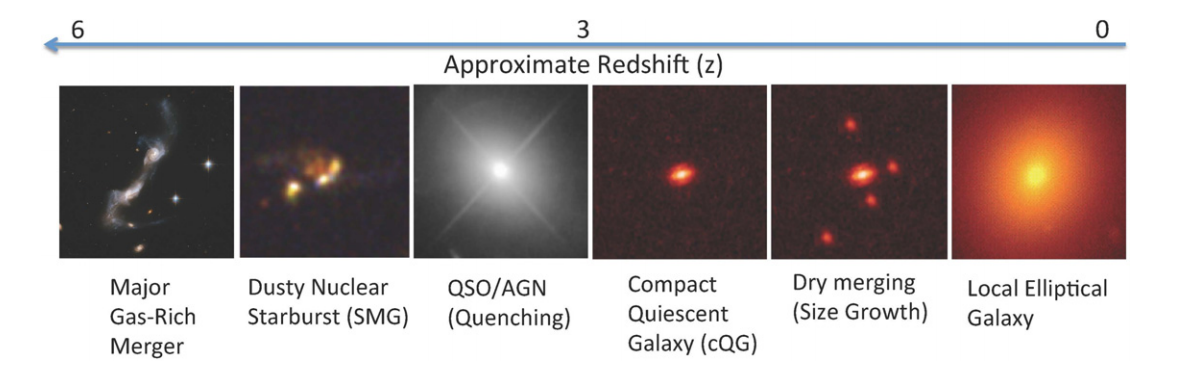}}
\caption{\footnotesize{The proposed evolutionary sequence by \cite{toft14} going from major gas-rich mergers at 
high redshift (left) to local elliptical quiescent galaxies (right).}}
\label{f:smg-elliptical}
\end{figure}

In order to find the actual progenitors (and descendants for that matter) of the CQGs, 
we have to look outside the MS, consisting of normal galaxies.
Fig.\,\ref{f:smg-elliptical} illustrates an evolutionary sequence connecting $z\gtrsim3$ SMGs 
through compact quiescent $z\sim2$ galaxies to local elliptical ones 
\citep{hopkins06,ricciardelli10,toft14,marchesini14}. 
Here, gas-rich major mergers in the early universe trigger a short time 
period ($42^{+40}_{29}$\,Myr) of high SFR in the form of a nuclear dust-enshrouded starbursts, that we observe as an SMG. 
That star formation is subsequently quenched either due to gas exhaustion, feedback from the 
starburst itself or the ignition of an AGN. What is left behind, is a compact stellar remnant 
that is observed as a CQG about a Gyr after, at $z\sim2$. 
Finally, the CQGs grow gradually, and primarily through minor merging, into the elliptical galaxies 
that we observe today. 

But observations of merger rates in massive galaxies  at $z\sim3-0$ show that major and minor merging 
is not enough to bring the CQGs to the sizes of locally observed early-type galaxies, 
suggesting that other mechanisms are needed for the size growth \citep{man14}.
And, as summarized in Section \ref{intro:quench}, there are many alternative ways to quench star formation in addition to the 
powerful feedback of an AGN. 
In order to understand evolution of massive galaxies at high redshifts, it is fundamental 
to find out which of these mechanisms dominate and when.

\section{This thesis}\label{intro:th}

The two fundamental questions about galaxy evolution, that were posed in the beginning of this chapter, 
in Section \ref{redblue}, could both be related to the observed main sequence  
of star formation (MS; see  Section\,\ref{intro:MS}): 

1) {\it Why does the MS evolve towards higher SFR with redshift?}

2) {\it What causes some galaxies to be outliers to the MS, or `quiescent' galaxies?}

\begin{figure}[!htbp] 
\centering
\includegraphics[width=\columnwidth]{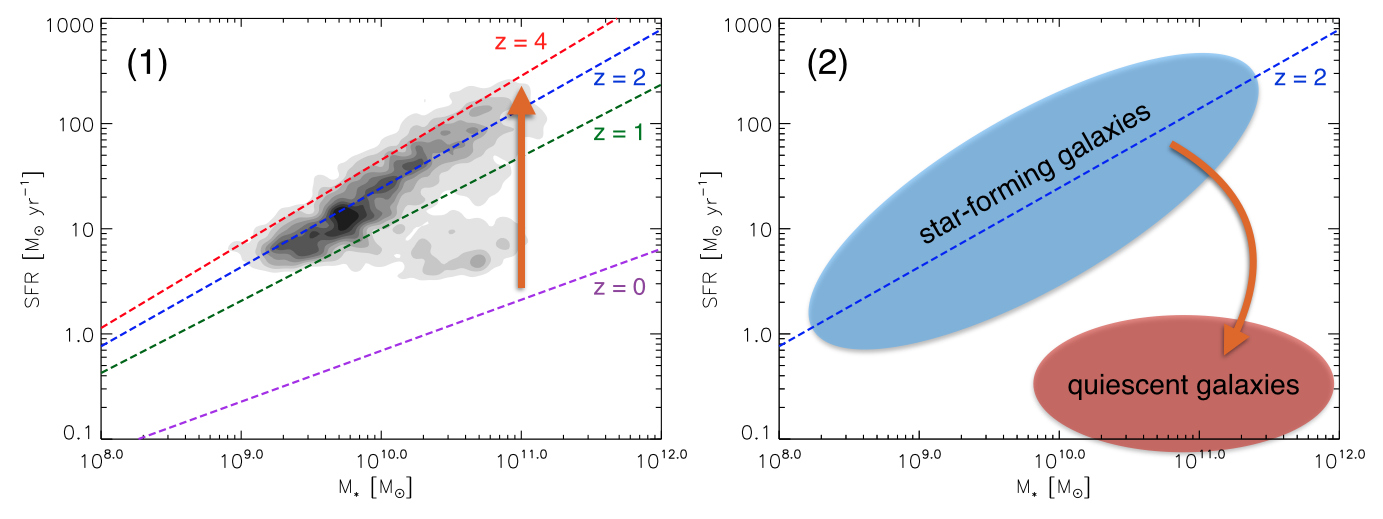}
\caption{\footnotesize{Plots showing the MS at various redshifts from \cite{speagle14}, 
illustrating the two open questions that need to be answered in order to understand galaxy 
evolution across cosmic time: 1) Why the MS evolves towards higher SFR with redshift and, 2) Why some galaxies 
are hardly forming stars, creating a group of red, quiescent galaxies observed all the way out to 
redshift 3. Contours in the left-hand plot show the position of star-forming galaxies 
at $1.4<z2.5$ from the NEWFIRM Medium-Band Survey \citep{whitaker11}.  
}}
\label{ms1}
\end{figure}

These issues are illustrated in Fig.\,\ref{ms1}, and compose the focus of this thesis.
Stars form out of gas, and it would therefore seem natural that these questions translate into:
{\it At what rate did galaxies acquire and use up their gas mass?}
The larger part of my thesis focuses on the gas content of galaxies, on which the following 
subsection expands upon.

\subsection{All eyes on the gas}

Most of the above-mentioned mechanisms depend in one way or another on gas amount, 
as summarized below:\\
$\bullet$ The gas content together with the amount of in/outflow of gas sets the fuel available 
for star formation and thereby a 1st order estimate of how long the galaxy can continue producing 
stars at the current SFR.\\
$\bullet$ Removal or heating of this gas is what quenches star formation, and is an important 
milestone in the life of massive galaxies.\\
$\bullet$ Mergers and the resulting triggering of an AGN, are often invoked as a quenching mechanism, 
but the outcome of a merger depends critically on how gas-rich the merging galaxies are.
 
\begin{figure}[!htbp] 
\centering
\includegraphics[width=7cm]{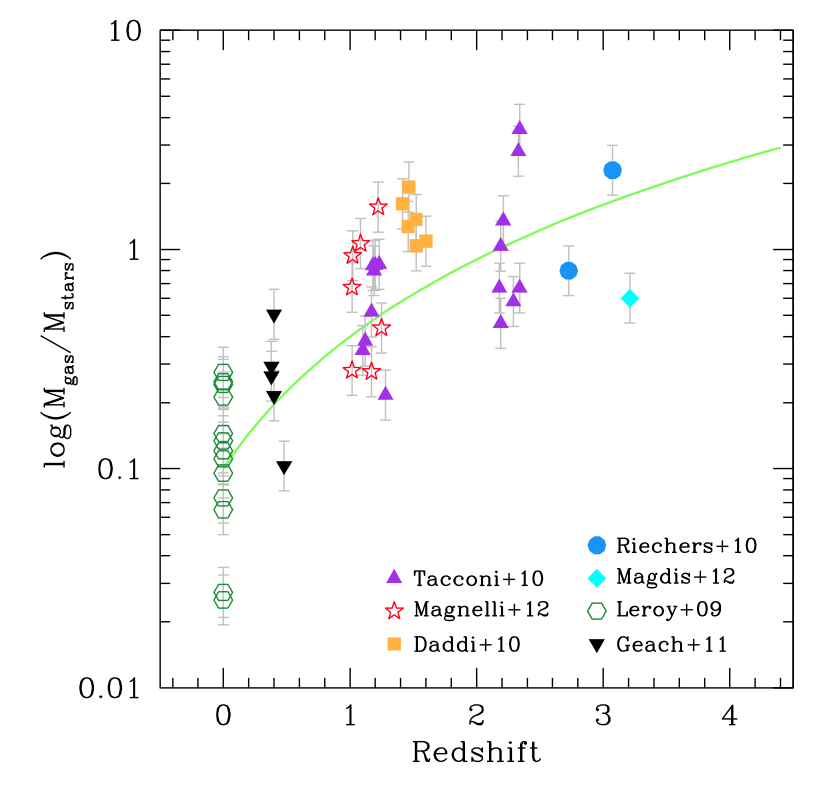}
\caption{\footnotesize{Evolution of the gas to stellar mass ratio, \Mgas/\Mstar, with redshift for 
star-forming disk galaxies with $\Mstar>10^{10}\,$\msun \citep{carilli13}.}}
\label{f:fgas}
\end{figure}

As mentioned in section \ref{intro:MS}, typical galaxies spend 95\pct of their lifetime since $z=1$ 
within a factor 4 of their average SFR at a given $\Mstar$ \citep{noeske07}. 
In light of this, most of the evolution in the MS, where galaxies spend most of their lives, 
might be due to a rising gas mass fraction with redshift. 
The general consensus is that galaxies at higher redshift were definitely more gas-rich, but 
the very few observations at $z>3$ make this statement hard to quantify, as Fig.\,\ref{f:fgas} illustrates.
This figure shows the gas mass fraction defined as $f_{\rm gas}=\Mgas/(\Mgas+\Mstar)$ 
as a function of redshift and reveals the uncertainty in its evolution at $z\gtrsim2$.

\begin{figure}[!htbp] 
\centering
\includegraphics[width=0.7\columnwidth]{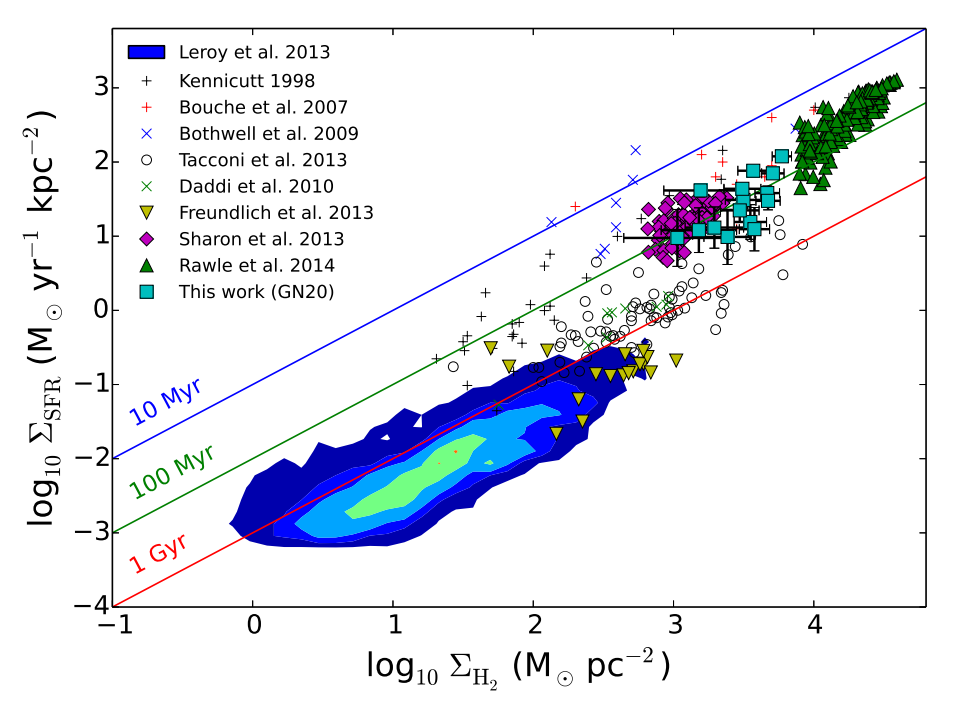}
\caption{\footnotesize{The KS law for different galaxy types at high and low redshift from the work of \cite{hodge15} 
who carried out resolved CO(2-1) observations of the $z=4.05$ SMG GN20, shown with cyan squares. 
GN20 lies significantly above the relation for local MS galaxies (contours), 
together with two other strongly lensed, resolved SMGs (magenta diamonds and green triangles), 
unresolved SMGs (blue crosses and red plus-signs) and local ultraluminous infrared galaxies 
(ULIRGs; black plus-signs).
Close to the local MS galaxies are $z\sim1.2$ massive star-forming galaxies 
(yellow upside-down triangles), $z\sim1-3$ color-selected galaxies (black open circles) 
and $z\sim1.5$ BzK galaxies (green crosses). See references for these observations in \cite{hodge15}.}}
\label{f:sfrsd}
\end{figure}

Another topic related to the gas and under current debate, is whether the KS law (introduced in Section \ref{intro:dis}) 
might be different in normal galaxies compared to starburst galaxies or mergers, as shown 
in Fig.\,\ref{f:sfrsd} (though such conclusions depend critically on how the gas mass is determined, a topic we return to 
in Part \ref{part1}). 

A huge missing piece to the puzzle is therefore knowledge of the gas (and dust) in massive galaxies, 
especially at $z\gtrsim2$. 
The premonition, shared by several research groups, is that 
future observations of the gas and dust content, morphology and dynamics in high-$z$ MS and quiescent 
galaxies, will solve the puzzle \citep{carilli13,blain15}. 
As put by \cite{glazebrook13}: {\it `...we need to consider the fuel as well as the fire'}.
As mentioned in \ref{intro:tele}, it is now possible with radio telescopes such as 
ALMA, JVLA and NOEMA to access the high-redshift domain where the red sequence must have formed. 

To answer the above questions, we not only need to observe the gas at high redshift in more detail, 
we also need to know how to interpret those observations better. 


\subsection{Quick summary of my projects}
I have worked on improving our knowledge of the evolutionary state of massive galaxies at $z=2$, 
with regard to 3 aspects:
\begin{itemize}
\item Amount and distribution of molecular gas in the ISM\\
A key species for studying the gas in the ISM was already introduced in Chapter \ref{intro} 
(Section \ref{intro:dis}) and will be described in detail in Part \ref{part1}: carbonmonoxide (CO). 
I used galaxy simulations and developed sub-grid procedures that could be combined with radiative transfer 
calculations in order to improve predictions for the use of CO lines to 
estimate the amount and distribution of molecular gas in the ISM of 
normal star-forming galaxies at $z\sim2$ (\sigame, see Part \ref{part1} Chapter \ref{gas1}).
\item Tracing SFR on local and global scales\\
The line emission of \cplus has been used to trace SFR and various phases of the ISM, including neutral and ionized 
gas, but it holds many mysteries with its exact origin still unknown.
Combining galaxy simulations with sub-grid procedures, I developed a method for modeling \cii 
line emission in order to study its origin in the ISM, and make 
predictions for its use as a star formation tracer at $z\sim2$ (\sigame, see Part \ref{part1} Chapter \ref{gas2}).
\item Importance of AGNs in massive galaxies at $z\sim2$\\
By analyzing {\it Chandra} X-ray data from a sample of galaxies at $1.5\lesssim z\lesssim2.5$, 
including the stacked signal from low-luminosity sources, I derive the fraction of luminous and 
low-luminosity AGNs in star-forming as well as in quiescent massive galaxies (see Part \ref{part2}).
\end{itemize}

At the beginning of the following chapters some additional background is provided on the specific techniques 
used in the above listed project, and each chapter will be followed by a list of references.

\begingroup
	\section{References}
	\def\chapter*#1{}
	\bibliographystyle{apj} 
	\setlength{\bibsep}{1pt}
	\setstretch{1}
	\bibliography{bibs0}
\endgroup

%% file: CO.tex
\part{Modeling the ISM of $z\sim2$ galaxies with \textnormal{\sigame}}
\label{part1}

\chapter{CO emission lines from galaxies} \label{gas1}

\section{Probing the molecular gas} \label{1:CO}
Due to the high abundance of hydrogen, the most common molecule to find in the ISM, is by far \h2. 
But in terms of observing cold molecular gas, \h2 is practically invisible, 
because the lack of an electric dipole moment means that the lowest possible rotational transitions 
of \h2 are electric quadrupole moments with very high energy. 
The first rotational transition, S(0), is at $510$\,K, meaning that even the lowest roational 
transitions can only be excited in shock waves, FUV-rich PDRs or regions affected by 
turbulent dissipation \citep{ingalls11}.

\begin{wrapfigure}[21]{r}{7cm}
\centering
\includegraphics[width=7cm]{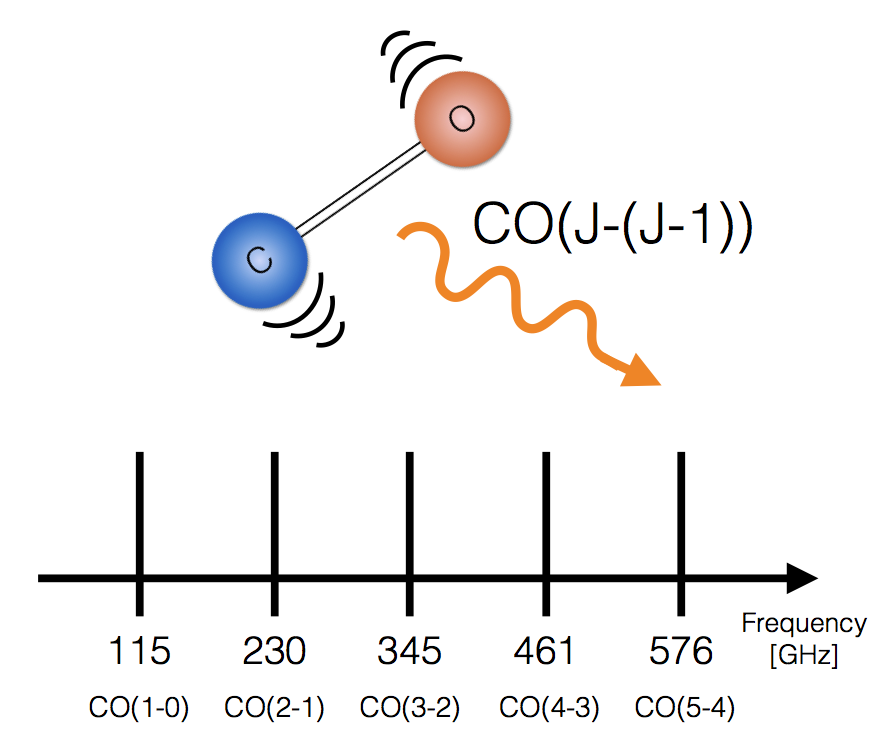}
\caption{\footnotesize{Illustration of the rotating CO molecules together with the 
frequencies of the first 5 rotational lines of the CO molecular 
are shown. Below are the corresponding critical densities 
($n_{\rm crit}=A_{ul}\Sigma_{i\neq u}C_{ui}$) as calculated by \cite{greve14} for a kinetic temperature 
of $\Tk=40$\,K and assuming \h2 to be the main collision partner.}}
\label{f:COillu}
\end{wrapfigure}

The second most abundant molecule is carbonmonoxide, CO, consisting of one carbon and one oxygen atom. 
The abundance relative to \h2, $[\rm{CO}/\h2]$, is roughly $10^{-4}$ in MW ISM as has been determined 
via observations of early cold cores in the MW \citep{glover11,liu13} 
and simulations of molecular clouds with $\nH\gtrsim1000\,\cmpc$ \citep{glover11}.

CO is the most used molecule for observing molecular gas, since its rotational transitions 
conveniently sample the typical densities and temperatures inside GMCs. 
To prope the very densest parts of GMCs, other molecular tracers exist with higher critical 
densities (again see \citealt{tielens13} for an overview of most detected molecules, and \citealt{aalto13} 
for a subset of the denser gas tracers), but the low CO lines have proven excellent for capturing 
the mayority of the molecular gas.

Fig.\,\ref{f:COillu} provides an illustration of the rotating CO molecule, together 
with the frequencies and critical densities of the first 5 transitions.
Also, the positions of the first 5 transitions on a typical SED can be seen in Fig.\,\ref{f:SED_z0}.
In GMCs, the excitation of CO molecules to higher rotational levels is believed to come 
mainly from collisions with \h2 molecules.
The population of the various rotional levels are therefore determined by density and temperature 
(how fast the molecules are moving) of the gas.
The amount of flux in each rotational transition creates a CO Spectral Line Energy Distribution (SLED) or 
`CO ladder' which can take on different 
shapes depending on the gas properties. 

\section{Observations of CO line emission from different types of galaxies at $z\gtrsim2$} \label{1:CO_obs}
At $z\sim2$, the vast majority of CO observations have been of extreme galaxies,
such as submillimeter galaxies (SMGs) with very high SFRs or quasi-stellar
objects (QSOs) containing powerful AGNs. Only a few galaxies on the ``main
sequence of star formation'' \citep[MS, e.g.][]{wuyts11}, have been observed in
CO.  These are typically massive ($10^{10}-10^{11}$\,\msun) galaxies selected by
the BzK technique to be moderately star-forming with SFR$\,\simeq10-500$\,\sfru,
i.e. specific SFR (SSFR) of around $0.5$\,Gyr$^{-1}$.  The molecular gas of SMGs
and QSOs is highly excited with peak fluxes at the CO(5-4) and CO(6-5)
transitions, which is evidence of dense and possibly warm gas \citep{papa12},
but little is known for the normal star-forming galaxies observed in CO to date.
The only such galaxies, with $4$ observed CO transitions, are the BzK-selected
galaxies BzK$-$4171, BzK$-$16000 and BzK$-$21000 \citep{daddi15}.  
These galaxies were previously believed to contain mainly gas reminiscent to that of 
the Milky Way (MW) \citep{dannerbauer09,aravena14}, 
but new CO$(5-4)$ observations by \cite{daddi15} require 
a second component of higher density and possibly higher temperature. 

With instruments of high spatial resolution, it has recently become possible to
resolve galaxies at $z\gtrsim2 $ on kiloparsec scales, as shown with the VLA by
e.g. \cite{walter07} when observing low-$J$ CO transitions.  Scales of $1$\,kpc
require a resolution of $\sim0.1$\arcsec at $z=2$, achievable with ALMA for
higher CO lines that are not redshifted below its bands.

\begin{figure}[!htbp] 
\centering
\includegraphics[width=0.7\columnwidth]{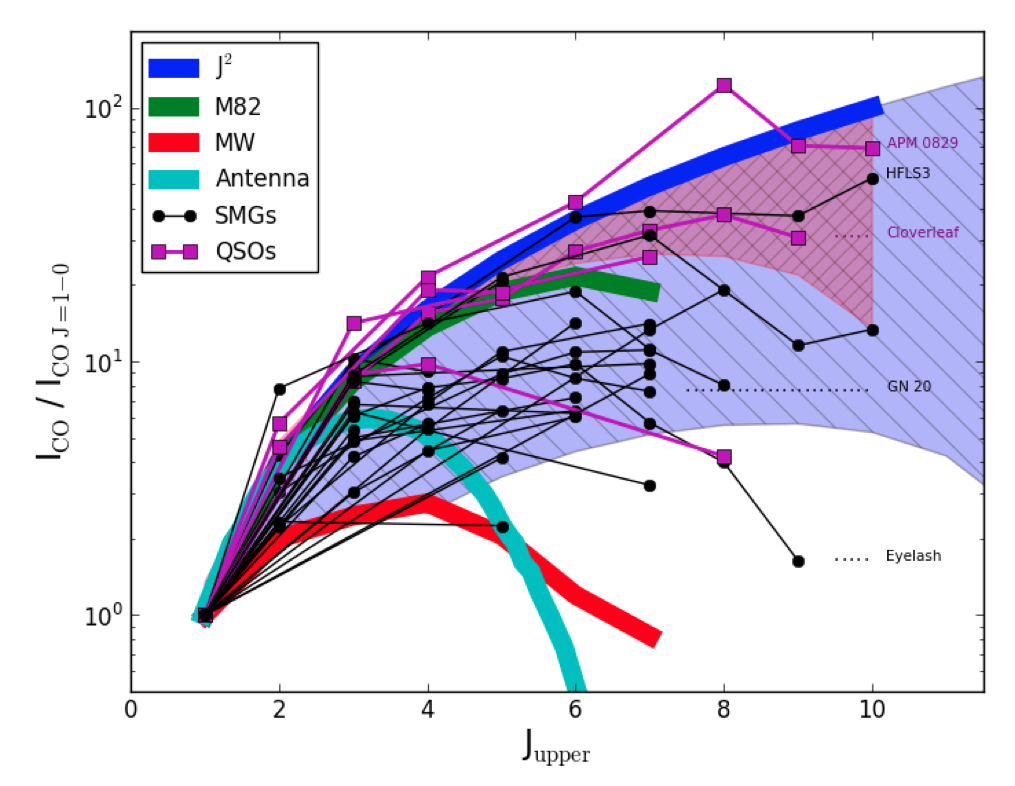}
\caption{\footnotesize{Observed CO ladders for different galaxies, normalized to the CO(1-0) transition. From the review of \cite{casey14}.}}
\label{f:sleds}
\end{figure}

When measuring the CO ladder of galaxies with different compositions, 
SFR and AGN activity, a huge variety emerges. Fig.\,\ref{f:sleds} compares the rather low-excitation ladder of 
the MW to the CO ladders of SMGs and QSOs which peak at increasingly higher $J$-values. 
An upper limit as to how steep the ladder can be, is reached if the gas is in Local Thermal Equilibrium (LTE), in 
which case the level populations just follow a Boltzmann distribution set by the gas kinetic temperature 
and no radiative transitions are considered. 
For gas in LTE and of kinetic temperature above that of the upper level transition level, 
the CO ladder would grow as $J^2$ as indicated in Fig.\,\ref{f:sleds} in blue.
But in general, this upper limit is not reached by the local kinetic temperature 
and the high $J$-levels are less populated.


\subsection{The \xco factor}
Probably the widest use of CO emission lines, is the estimation of total molecular gas mass. 
The low critical density ($\sim300\,$\cmpc) of the first CO(1-0) transition makes it a tracer of 
most gas in a GMC. In order to convert from CO(1-0) line intensity and column density of \h2, the CO-\h2\ 
conversion factor (\xco\ or `$X$-factor') gives the \h2 column density per CO(1-0) intensity unit: 
\begin{align}
	\xco \,[\rm{cm}^{-2}/(\rm{K\,km}\,\ps)^{-1}]  = 	\rm{N}(\h2)/W_{\rm{(CO(1-0))}}	
\end{align}
A similar conversion factor exists between total \h2 gas mass amount and CO line luminosity: 
\begin{align}
	\aco \,[\msun/(\rm{K\,km\,pc}^{-2})^{-1}]  = 	M(\h2)/L_{\rm{CO(1-0)}}	
\end{align}
The most common way of inferring their values, is to estimate the gas mass from either 
(i) the CO line width assuming the GMCs are in virial equilibrium (cf. eq.\,\ref{eq:vir}), 
(ii) the dust mass assuming a dust-to-gas mass ratio, 
or (iii) $\gamma$-ray emission from cosmic ray interactions with \h2 \citep[see references in][]{narayanan12}.
In the inner disk of the MW, they are relatively constant at 
$\xco\approx 2\e{20}\,$cm$^{-2}$/(K\,km\,\ps)$^{-1}$ 
and $\aco\approx4.3\,$\msun/(K\,km\,pc$^{-2}$)$^{-1}$ 
\citep[from the comprehensive review on the $X$-factor by][]{bolatto13}. 

\begin{figure}[!htbp] 
\centering
\includegraphics[width=0.7\columnwidth]{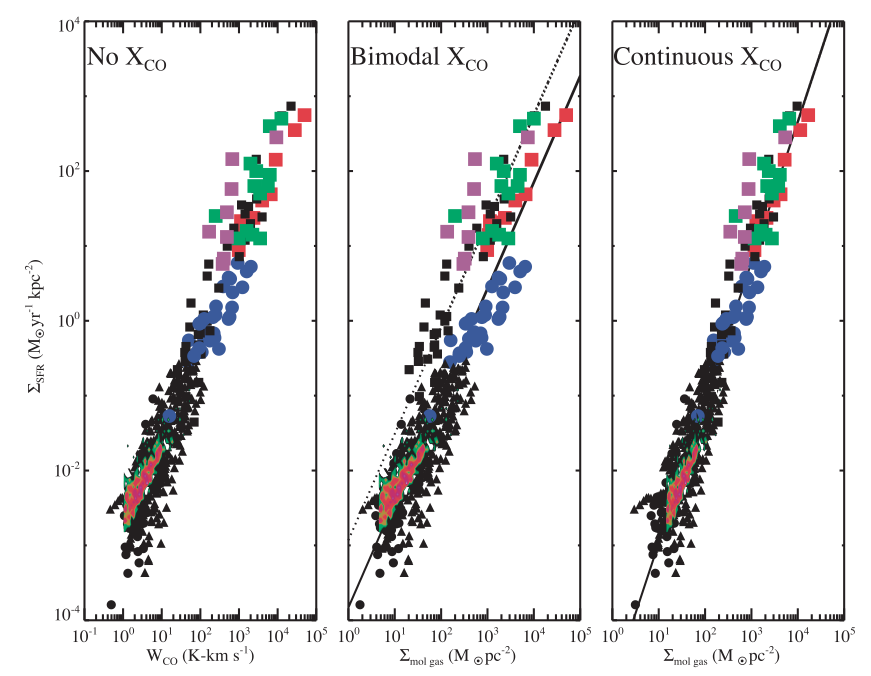}
\caption{\footnotesize{The effect of using an effectively bimodal \xco factor 
(mid panel; $\aco=4.5$ for local disks, 3.6 for high-$z$ disks and 
0.8\,\msun/(K\,km\,pc$^{-2}$)$^{-1}$ for mergers) 
or a continuous one (right panel) 
on the KS law \citep{narayanan12}.}}
\label{f:xco}
\end{figure}

But recent observations, at low and high redshift, show that \xco must be lower 
in galaxies of high-surface density environments, 
and larger in low-metallicity environments.
This has led to the belief that maybe there exists two versions of the KS star formation relation, one 
for gas-rich mergers and another for star-forming normal disk galaxies \citep{genzel10}.
Based on detailed modeling, \cite{narayanan12} came up with a function for \xco\ 
that depends on CO line luminosity and metallicity. With this prescription for \xco\ 
a single continous KS star formation 
relation can be drawn for starbursts and normal disk galaxies at low and high redshift as shown in Fig.\,\ref{f:xco}. 
However, see the work of \cite{hodge15}, implying that SMGs really can have more efficient star formation 
on small scales, regardless of the choice of conversion factor.

\section{Modeling of CO emission lines} 
The CO ladder of a galaxy can be modeled by combining galaxy simulations 
(either zoom-in versions of cosmological simulations or isolated galaxy models, cf. Section \ref{intro:sim}) 
with sub-grid prescriptions for the GMC structure and radiative transfer codes or LVG models for 
the transport of CO line emission through the ISM. 
\begin{figure}[!htbp] 
\centering
\includegraphics[width=1\columnwidth]{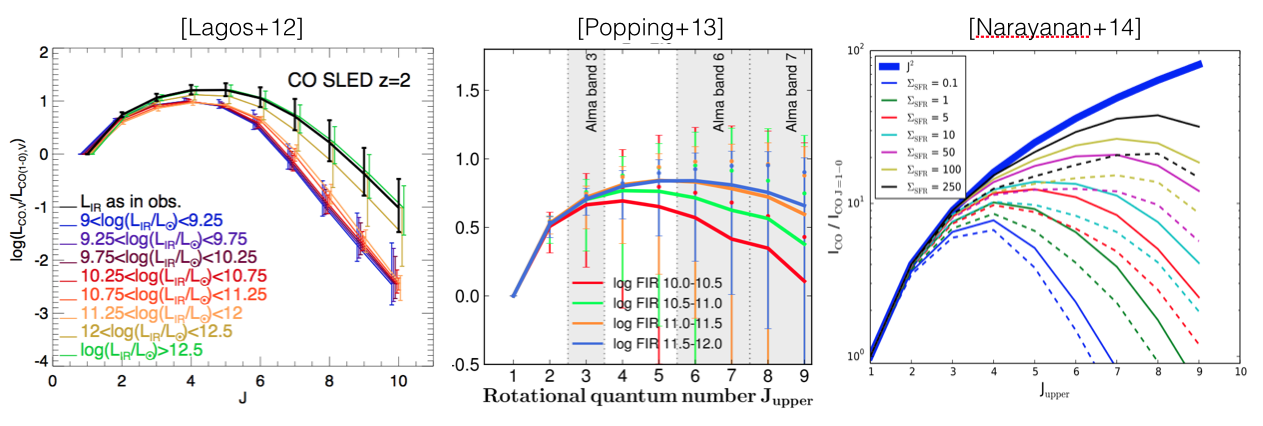}
\caption{\footnotesize{Examples of simulated CO ladders for different galaxy types. 
{\it Left:} The shape of CO ladders for $z=2$ galaxies, binned according to infrared ($8-1000\,\mu$m) luminosity, \Lir, 
as predicted by \cite{lagos12}, 
{\it Middle:} The same as in the left panel according to the method of \cite{popping14}, 
{\it Right:} CO ladders parameterized by \SFRsd for simulations of resolved (solid) and unresolved (dashed) cases, 
as derived by \cite{narayanan14}. }}
\label{f:COsims}
\end{figure}
Over the past decade in particular, detailed modeling of CO emission has been developed,
the basic steps of which can be summarised as follows: (1) Simulate a galaxy
with dark matter, stars and atomic gas (if possible, with a multiphase molecular
gas), (2) Apply semi-analytical physical recipes to estimate the amount and
state of dust and molecular gas, (3) Calculate the radiative transfer of
selected CO lines. A full description of such a process can be found in
\cite{narayanan08}, who investigated the signatures in CO morphology and line
profile of outflows created by active galactic nuclei (AGNs) and starbursts. The
same method has been used to study other aspects of the ISM, such as the
Kennicutt-Schmidt relation and the \xco factor which relates
velocity-integrated CO line intensity to \h2 column density, in low- and
high-redshift galaxies \citep[e.g.,][]{narayanan11,narayanan13}. By running
smoothed particle hydrodynamical (SPH) simulations with the GADGET-3 code,
\cite{narayanan11} were able to follow the evolution with time of a multiphase
ISM in isolated disk galaxies undergoing a merger, allowing the gas to cool down
to temperatures of $\sim10$\,K. \cite{lagos12} simulated the CO emission of
quiescently star-forming as well as starburst galaxies at redshifts ranging from
$z=0$ to $z=6$ by coupling cosmological galaxy formation simulations with the
tabulated output of a code describing photon dominated regions (PDRs). \cite{lagos12}
found that the global velocity-integrated CO line intensity peaks at higher
rotational transitions with increasing infrared (IR) luminosity ($J$-level $4$ to
$5$ when median $L_{\rm IR}$ of the sample goes from about $1.3\sim10^{9}$ to
$8\times10^{12}$\,\lsun). 
In a similar study, \cite{popping14} found that models of normal star-forming galaxies 
at high redshift have much higher rotational transitions than their local counterparts. 
Finally, the modeling of CO lines from a large set of simulated disk galaxies, 
recently carried out by \cite{narayanan14} showed that the CO line energy distribution 
can be parameterised better with respect to SFR surface density compared to total SFR. 
Simulated CO SLEDs in in Fig.\,\ref{f:COsims} provide examples of these previous works. 
However simulations of resolved CO line emission have been lacking.

\chapter{Simulating CO line emission from massive star-forming galaxies at $z=2$ (Paper I)} \label{paper1}

\section{Aim of this project}
We wanted to expand on previous modeling done by other groups of CO emission lines from galaxies at high redshift by 
developing a method that focuses on 
radial gradients of the \xco factor and CO line ratios. 
The aim of the project was therefore to serve as a preview of what astronomers might one day 
observe in `normal' star-forming galaxies at $z\sim2$ with new and upcoming instruments.

For that purpose, we created a new numerical framework for simulating the 
line emission of normal star-forming galaxies. The code -- 
SImulator of GAlaxy Millimeter/submillimeter Emission (\sigame\footnote{\sigame in 
Spanish translates as `follow me' - which in this context refers to the pursue of line emission 
through a galaxy.}) 
-- combines (non-)cosmological simulations of galaxy formation with subgrid prescriptions for the
\h2/\hi fraction and thermal balance throughout the ISM, down to parsec scales.
\sigame accounts for a FUV and cosmic ray intensity field that vary
with local SFR density within the galaxy. \sigame can be applied to
any galaxy simulated in the SPH formalism, 
though currently restricted to galaxies dominated by star formation processes 
rather than AGN and 
with mean metallicities above about $0.01\Zsun$. 
Here, we adapt the code to cosmological SPH simulations of three 
massive, normal star-forming galaxies at $z=2$ (i.e., so-called main-sequence 
galaxies), and model their CO rotational line spectrum using a publically 
available 3D radiative transfer code. 
We show that \sigame reproduces observed low-$J$ CO line luminosities and provides 
new estimates of the \xco factor for main-sequence galaxies at $z\sim2$, 
while at the same time predicting their CO line luminosities at high-$J$ 
($J_{\rm up}>6$) transitions where observations are yet to be made. 

The structure of this chapter is as follows. 
Section \ref{cosmological_simulations} describes the cosmological
SPH simulations used, along with the basic properties of the three star-forming
galaxies extracted from the simulations. 
A detailed description of
\sigame is presented in Section \ref{modeling}. 
The CO emission maps and spectra
obtained after applying \sigame to the three simulated galaxies are
presented in Section \ref{results}, where we also compare to actual CO observations of
similar galaxies at $z\sim 2$.  Section \ref{dis_models} discusses the strengths and
weaknesses of \sigame in the context of other molecular line (CO)
simulations and recent observations. Finally, in Section \ref{CO:con} we summarise the main steps of
\sigame, and list its main findings and predictions regarding the CO
line emission from massive, star-forming galaxies at $z\simeq 2$.  We adopt a
flat cold dark matter ($\Lambda$CDM) scenario with $\Omega_m=0.3$,
$\Omega_\Lambda=0.7$ and $h=0.65$. 

\section{Cosmological Simulations}\label{cosmological_simulations}
\subsection{SPH simulations}
We employ a cosmological TreeSPH code for simulating galaxy formation and
evolution, though in principle, grid-based hydrodynamic simulations could be
incorporated equally well.  The TreeSPH code used for the simulations is in most
respects similar to the one described in \cite{somm05} and \cite{romeo06}. A
cosmic baryon fraction of $f_{\rm{b}}=0.15$ is assumed, and simulations are
initiated at a redshift of $z=39$.  The implementation of star formation and
stellar feedback, however, has been manifestly changed. 

Star formation is assumed to take place in cold gas ($\Tk \lesssim 10^4$\,K) at
densities $\nH > 1$\,\cmpc. The star formation efficiency (or probability that a
gas particle will form stars) is formally set to 0.1, but is, due to effects of
self-regulation, considerably lower.
Star formation takes place in a stochastic way, and in a star formation event,
$1$ SPH gas particle is converted completely into $1$ stellar SPH particle,
representing the instantaneous birth of a population of stars according to a
Chabrier (2003) stellar initial mass function (IMF; \citealt{chabrier03}) -- see
further below.

The implementation of stellar feedback is based on a sub-grid super-wind model,
somewhat similar to the `high-feedback' models by \cite{stinson06}. These
models, though, build on a supernova blast-wave approach rather than super-wind
models. Both types of models invoke a Chabrier (2003) IMF, which is somewhat
more top-heavy in terms of energy and heavy-element feedback than, e.g., the
standard Salpeter IMF. The present models result in galaxies characterised by
reasonable $z=0$ cold gas fractions, abundances and circum-galactic medium
abundance properties. They also improve considerably on the ``angular momentum
problem'' relative to the models presented in, e.g., \cite{somm03}. The models
will be described in detail in a forthcoming paper. 

\subsection{The model galaxies}\label{model_galaxies}
Three model galaxies, hereafter referred to as G1, G2 and G3 in order of
increasing SFR, were extracted from the above SPH simulation and re-simulated
using the `zoom-in' technique described in \citep[e.g.,][]{somm03}. The emphasis
in this study is on massive ($M_* \gtrsim 5\times 10^{10}\,\msun$) galaxies, and the
three galaxies analysed are therefore larger, rescaled versions of galaxies
formed in the 10/$h$\,Mpc cosmological simulation described in \cite{somm03}.
The linear scale-factor is of the order 1.5, and since the CDM power spectrum is
fairly constant over this limited mass range the rescaling is a reasonable
approximation. 

Galaxy G1 was simulated at fairly high resolution, using a total of $1.2\times
10^6$ SPH and dark matter particles, while about $9\times 10^5$ and $1.1\times
10^6$ particles were used in the simulations of G2 and G3, respectively.  For
the G1 simulation, the masses of individual SPH gas, stellar and dark matter
particles are $m_{\rm{SPH}}=m_*\approx6.3\e{5}$ $h^{-1}$M$_{\odot}$ and
$m_{\rm{DM}}$=$3.5\e{6}$ $h^{-1}$M$_{\odot}$, respectively.  Gravitational
(cubic spline) softening lengths of 310, 310 and 560 $h^{-1}$pc, respectively,
were employed. Minimum gas smoothing lengths were about $50\,h^{-1}$pc.  For the
lower resolution simulations of galaxies G2 and G3, the corresponding particle
masses are $m_{\rm{SPH}}=m_*\approx4.7\e{6}$\,$h^{-1}$M$_{\odot}$ and
$m_{\rm{DM}}= 2.6\e{7}$ $h^{-1}$M$_{\odot}$, respectively, and the gravitational
softening lengths were 610, 610 and 1090 $h^{-1}$pc. Minimum gas smoothing
lengths were about 100 $h^{-1}$pc.  
\begin{figure}
\centering
\includegraphics[width=0.6\columnwidth]{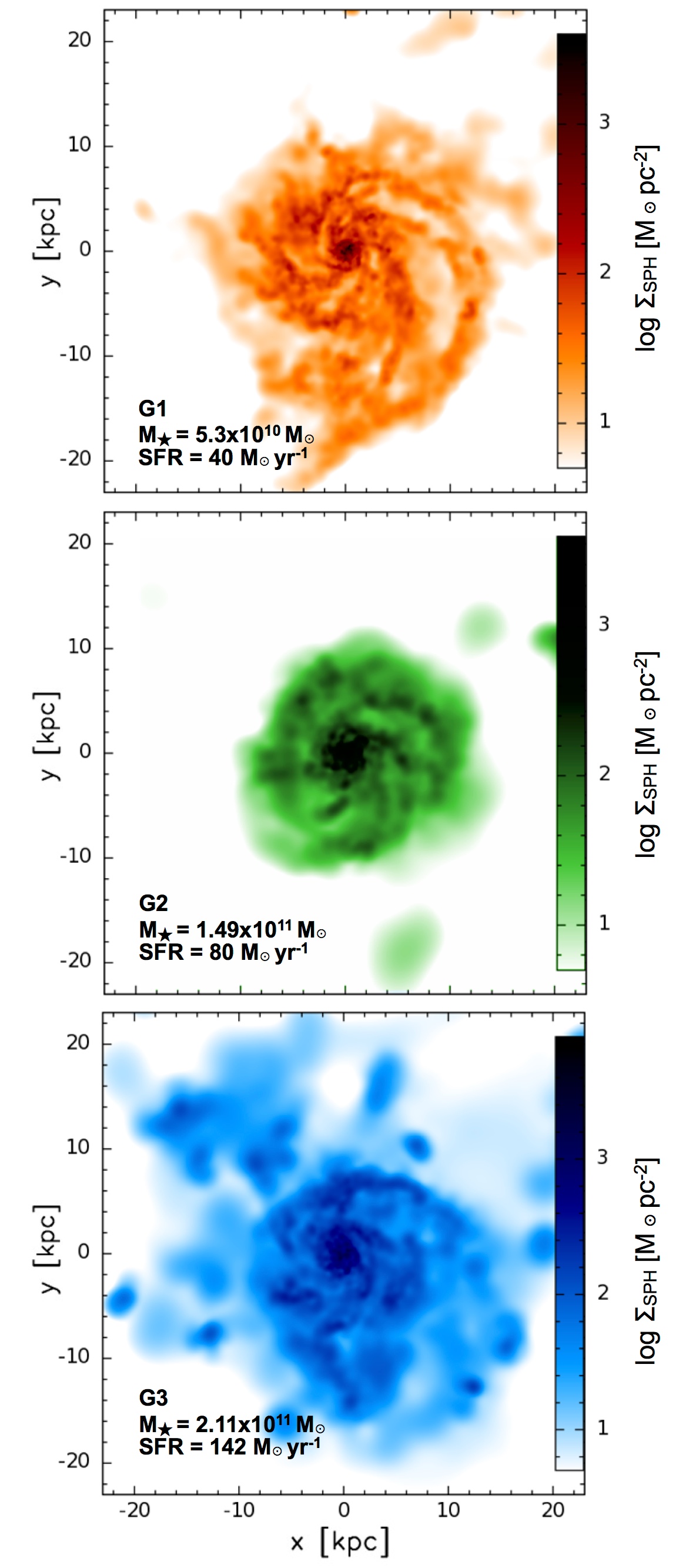}
\caption{\footnotesize{SPH gas surface density maps of the three model galaxies G1 (top), G2
(middle), and G3 (bottom) viewed face-on.  The stellar masses and SFRs of each
galaxy are indicated  (see also Table\,\ref{model_galaxies_t1}).  The maps have
been rendered with the visualization tool SPLASH version 2.4.0 \citep{price07}
using the gas smoothing lengths provided by the simulations. 
}}
\label{model_galaxies_f1}
\end{figure}

Due to effects of gravitational softening, typical velocities in the innermost
parts of the galaxies (typically at radii less than about
2$\epsilon_{\rm{SPH}}$, where $\epsilon_{\rm{SPH}}$ is the SPH and star particle
gravitational softening length) are somewhat below dynamical values \citep[see,
e.g.][]{somm98}. The dynamical velocities will be of the order
$v_{\rm{dyn}}=\sqrt{G M(R)/R}$, where $G$ is the gravitational constant, $R$ is
the radial distance from the centre of the galaxy and $M(R)$ is the total mass
located inside of $R$. Indeed, it turns out that for the simulated galaxies
considered in this project SPH particle velocities inside of
2$\epsilon_{\rm{SPH}}$ are only about 60-70\% of what should be expected from
dynamics. To coarsely correct for this adverse numerical effect, for SPH
particles inside of 2$\epsilon_{\rm{SPH}}$ the velocities are corrected as
follows: For SPH particles of total velocity less than $v_{\rm{dyn}}$, the
tangential component of the velocity is increased such that the total velocity
becomes equal to $v_{\rm{dyn}}$. Only the tangential component is increased in
order not to create spurious signatures of merging. With this correction
implemented, the average ratio of total space velocity to dynamical velocity of
all SPH particles inside of 2$\epsilon_{\rm{SPH}}$ equals unity.

\bigskip

Figure \ref{model_galaxies_f1} shows surface density maps of the SPH gas in G1,
G2, and G3, i.e., prior to any post-processing by \sigame.  The gas is seen to
be strongly concentrated towards the centre of each galaxy and structured in
spiral arms containing clumps of denser gas.  The spiral arms reach out to a
radius of about $20$\,kpc in G1 and G3, with G2 showing a more compact
structure that does not exceed $R\sim15$\,kpc.  Table \ref{model_galaxies_t1}
lists key properties of the simulated galaxies, namely their SFR, stellar mass
($M_*$), SPH gas mass ($M_{\rm SPH}$), SPH gas mass fraction ($f_{\rm
SPH}=M_{\rm SPH}/(M_*+M_{\rm SPH})$), and metallicity \Z. These quantities were
measured within a radius ($R_{\rm cut}$, also given in
Table\,\ref{model_galaxies_t1}) corresponding to where the radial cumulative
stellar mass function has flattened out. The metallicity is in units of solar
metallicity and is calculated from the abundances of C, N, O, Mg, Si, S, Ca and
Fe in the SPH simulations, and adjusted for the fact that not all heavy metals
have been included according to the solar element abundancies measured by
\cite{asplund09}.  
\begin{center}
\begin{table}
\centering
\caption{Physical properties of the three simulated galaxies G1, G2, and G3} 
\begin{tabular}{@{}lllllll@{}} 
\hline
\hline
 	 	&	SFR 		& 	$M_{\ast}$	 		 & $M_{\rm SPH}$ 		&	$f_{\rm SPH}$	&	$\Z$	& 	$R_{\rm cut}$  \\ 
		&[$\rm{M}_\odot$~yr$^{-1}$]& 	[$10^{11}\,\rm{M}_\odot$]	 & [$10^{10}\,\rm{M}_\odot$]	&					&		& 	[kpc]  \\ 
\hline
G1				&	40	& 	$0.53$		&	$2.07$		 &	28\%		&	1.16	&	$20$		\\ 
G2				& 	80	&	$1.49$		&	$2.63$		 &	15\%		&	1.97	&	$15$		\\ 
G3 				& 	142	&	$2.11$		&	$4.66$		 &	18\%		&	1.36	&	$20$		\\ 
\hline
\end{tabular}
\\
	{\bf Notes.} \footnotesize{All quantities are determined within a radius $R_{\rm cut}$,
	which is the radius where the cumulative radial stellar mass function of
	each galaxy becomes flat. The gas mass ($M_{\rm SPH}$) is the total SPH gas
	mass within $R_{\rm cut}$.  The metallicity ($\Z=Z/Z_{\odot}$) is the mean
	of all SPH gas particles within $R_{\rm cut}$.}
\label{model_galaxies_t1}
\end{table}
\end{center}

\begin{figure*}
\centering
\includegraphics[width=0.8\columnwidth]{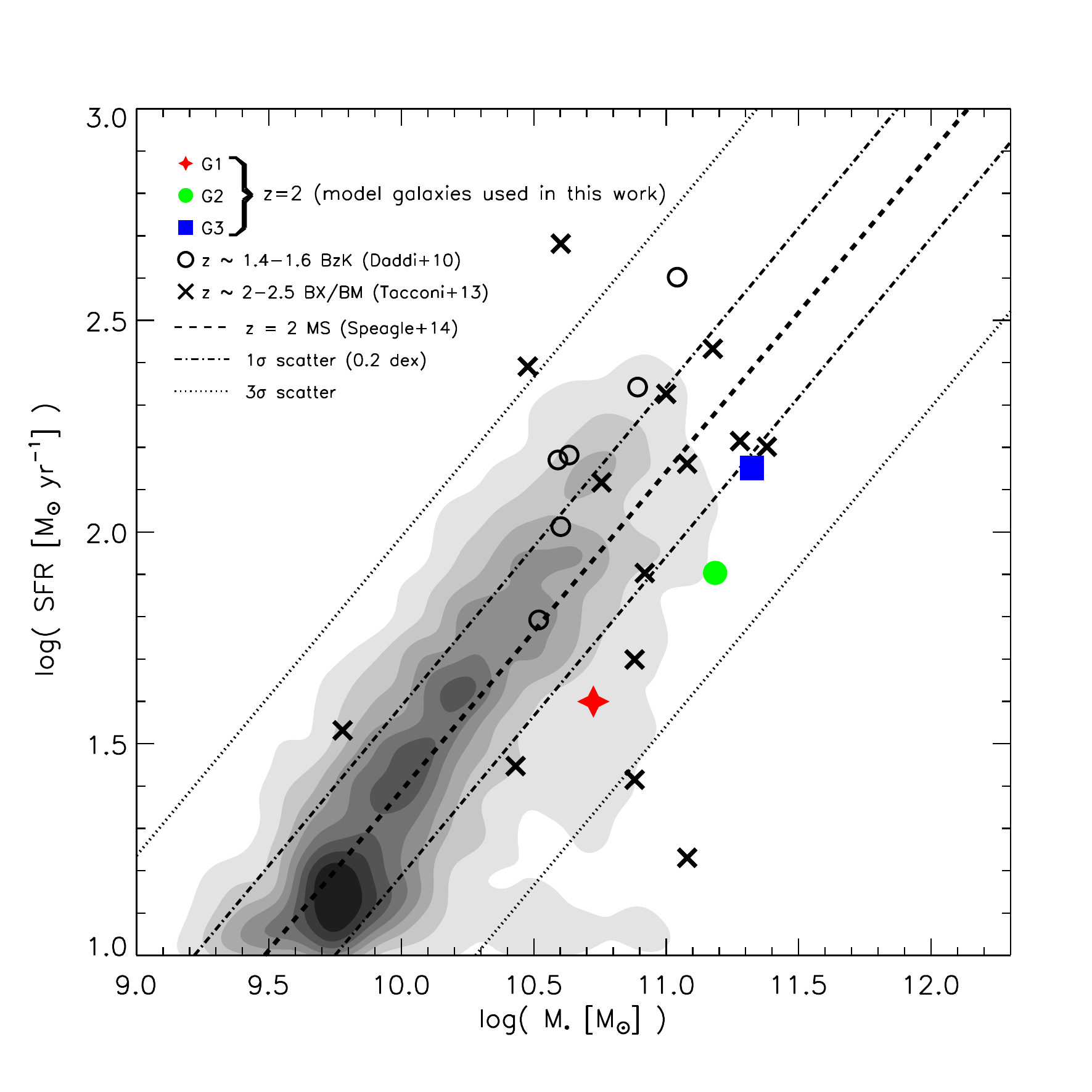}
\caption{\footnotesize{Position of the three model galaxies studied here (G1, G2 and G3 with
filled a red star, a green circle and a blue square respectively), on a SFR--M$_{\ast}$ diagram. The grey
filled contours show the $z\sim 2$ number density of $3754$
$1.4<z<2.5$ galaxies from the NEWFIRM Medium-Band Survey \citep{whitaker11}.
The ${\rm SFR-M_{\ast}}$ relation at $z\simeq 2$ as determined by \citet{speagle14} is
indicated by the dashed line, with the $1\sigma$ and $3\sigma$ scatter of the
relation shown by the dot-dashed and dotted lines, respectively.  
Also shown are six $z\sim1.4-1.6$ BzK galaxies (black circles; \citealt{daddi10}) and 14
$z\sim2-2.5$ Bx/BM galaxies (black crosses; \citealt{tacconi13}). The BzK
galaxies are from top to bottom: BzK$-$12591, BzK$-$21000, BzK$-$16000,
BzK$-$17999, BzK$-$4171 and BzK$-$2553 (following the naming convention of
\citealt{daddi10}).
}}
\label{model_galaxies_f_M_SFR}
\end{figure*}

The location of our three model galaxies in the SFR-$M_*$ diagram is shown in
Figure \ref{model_galaxies_f_M_SFR} along with a sample of 3754 $1.4 < z < 2.5$
main-sequence galaxies selected in near-IR from the NEWFIRM Medium-Band Survey
\citep{whitaker11}.  The latter used a Kroupa IMF but given its similarity with
a Chabrier IMF no conversion in the stellar mass and SFR was made (cf.,
\citet{papovich11} and \citet{zahid12} who use conversion factors of 1.06 and
1.13, respectively).  Also shown is the determination of the main-sequence
relation at $z\simeq2$ by \cite{speagle14}, and the 1 and $3\sigma$ scatter
around it.  G1, G2, and G3 are seen to lie within the $3\sigma$ scatter around the $z\simeq 2$ main
sequence, albeit offset somewhat towards lower SFRs. This latter tendency is
also found among a subset of CO-detected BX/BM galaxies at $z\sim2-2.5$
\citep{tacconi13}, highlighted in Figure \ref{model_galaxies_f_M_SFR} along
with a handful of $z\sim1.5$ BzK galaxies also detected in CO \citep{daddi10}.
The BX/BM galaxies are selected by a UGR colour criteria \citep{adelberger04},
while the BzK galaxies are selected by the BzK colour criteria \citep{daddi04}.
Based on the above we conclude that, in terms of stellar mass and SFR, our
three model galaxies are representative of the star-forming galaxy population
detected in CO at $z\sim2$.

\section{Modeling the ISM with \sigame}\label{modeling}
\subsection{Methodology overview}\label{subsection:methodology}
Here we give an overview of the major steps that go into \sigame,
along with a brief description of each. The full details of each step are given
in subsequent sections and in appendices \ref{apB} through \ref{apD}. We stress,
that \sigame operates entirely in the post-processing stage of an SPH
simulation, and can in principle easily be adapted to any given SPH galaxy
simulation as long as certain basic quantities are known for each SPH particle
in the simulation, namely: position ($\boldsymbol{r}$), velocity ($\vec{v}$), atomic
hydrogen density (\nH), metallicity (\Z), kinetic temperature (\Tk), smoothing
length ($h$), and star formation rate (SFR). The key steps involved in
\sigame are:

\begin{enumerate}
\item Cooling of the SPH gas. The initially hot ($T_{\rm k}\sim 10^{3-7}\,{\rm
K}$) SPH gas particles are cooled to temperatures typical of the warm neutral
medium ($\lesssim 10^4\,{\rm K}$) by atomic and ionic cooling lines primarily. 

\item Inference of the molecular gas mass fraction ($\fh2 = m_{\rm H_2}/m_{\rm
SPH}$) of each SPH particle after the initial cooling in step 1.  \fh2 
for a given SPH particle is calculated by taking into account its temperature,
metallicity, and the local CR and FUV radiation field impinging on it.

\item Distribution of the molecular gas into GMCs. Cloud masses and sizes are
obtained from random sampling of the observed GMC mass-spectrum in nearby
quiescent galaxies and applying the local GMC mass-size relation.

\item GMC thermal structure. A radial density profile is adopted for each GMC
and used to calculate the temperature structure throughout individual clouds,
taking into account heating and cooling mechanisms relevant for neutral and
molecular gas, when exposed to the local CR and attenuated FUV fields. 

\item Radiative transport of CO lines. Finally, the CO line spectra are calculated 
separately for each GMC with a radiative transfer code and accumulated on a common 
velocity axis for the entire galaxy.
\end{enumerate}

Determining the temperature (i) and the molecular gas mass fraction
(ii) of a warm neutral gas SPH particle cannot be done independently of
each other, but must be solved for simultaneously in an iterative fashion (see
Sections \ref{WCNM} and \ref{HI_to_H2}).  As already mentioned, we shall apply
\sigame to the SPH simulations of galaxies G1, G2, and G3 described in Section
\ref{cosmological_simulations}, and in doing so we will use them to illustrate
the workings of the code.

\subsection{The Warm and Cold Neutral Medium}\label{WCNM}
\begin{figure} 
\hspace{1.2cm}
\includegraphics[width=0.8\columnwidth]{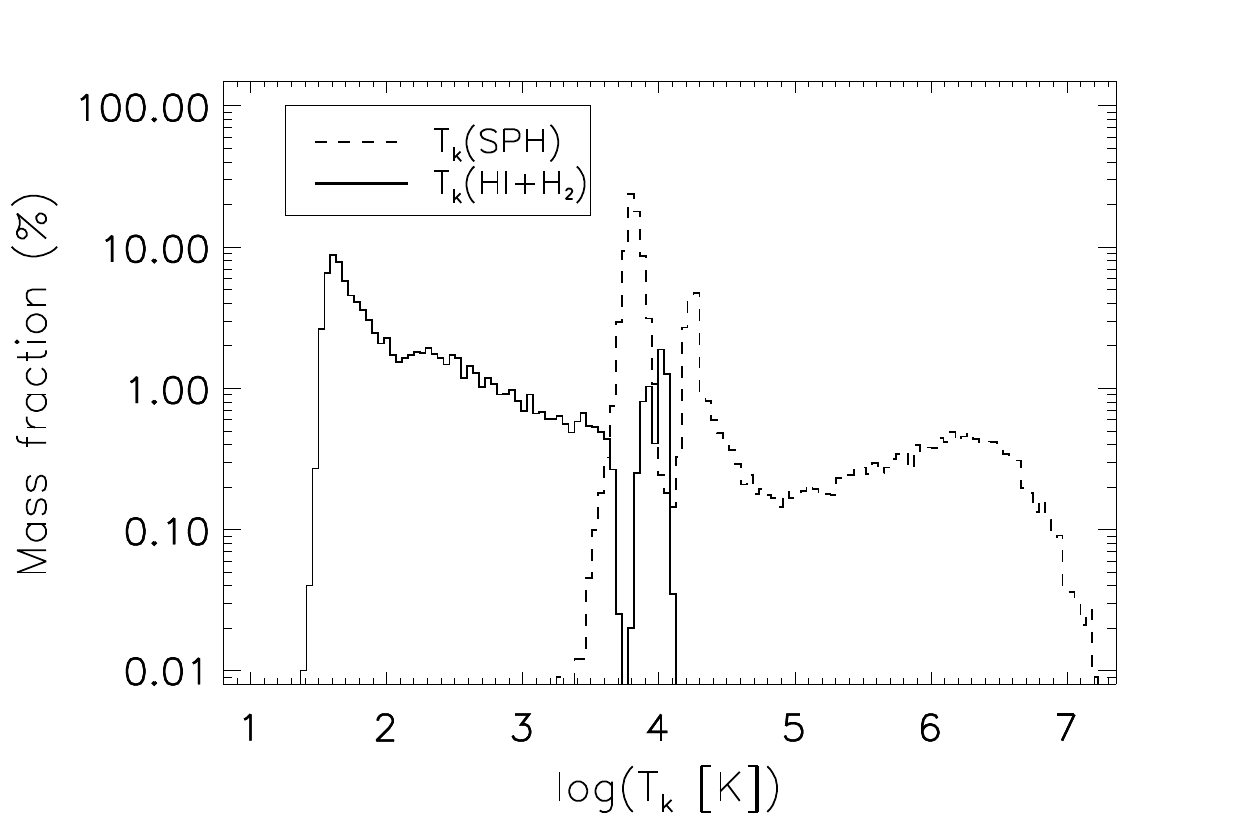}
\caption{\footnotesize{The distributions of gas kinetic temperature before (dashed histogram)
and after (solid histogram) applying the heating and cooling mechanisms of
eq.\,\ref{WCNM_e1} to galaxy G1. The original hot SPH gas is seen to span a
temperature range from about $10^3$\,K up to $\sim10^7$\,K, while once the gas
has been cooled the temperature distribution only barely exceeds $\sim10^4$\,K.
}}
\label{WCNM_f1}
\end{figure}

In SPH simulations of galaxies the gas is typically not cooled to temperatures
below a few thousand Kelvin \citep{springel03}. This is illustrated in Figure
\ref{WCNM_f1}, which shows the SPH gas temperature distribution (dashed
histogram) in G1 (for clarity we have not shown the corresponding temperature
distributions for G2 and G3, both of which are similar to that of G1). Minimum
SPH gas temperatures in G1, G2 and G3 are about 1200\,K, 3100\,K and 3200\,K,
respectively, and while temperatures span the range $\sim10^{3-7}\,{\rm K}$, the
bulk ($\sim 80-90\%$) of the gas mass in all three galaxies is at $\Tk \lesssim
10^5\,{\rm K}$.

At these temperatures the gas will be in atomic or ionised form, and H atoms that
attach to dust grain surfaces via chemical bonds (chemisorbed) will evaporate from
the grains before \h2 can be formed. \h2 can effectively only exist at
temperatures below $\sim 10^3$\,K, assuming a realistic desorption energy of
$3\times 10^4$\,K for chemisorbed H atoms \citep[see][]{cazaux04}.  The first
step of \sigame is therefore to cool some portion of the hot SPH gas
down to $\Tk \lesssim 10^3\,{\rm K}$, i.e., a temperature range characteristic of a warm
and cold neutral medium for which we can meaningfully employ a prescription for
the formation of H$_2$. 

\sigame employs the standard cooling and heating mechanisms pertaining to a
hot, partially ionised gas. Cooling occurs primarily via emission lines from H,
He, C, O, N, Ne, Mg, Si, S, Ca, and Fe in their atomic and ionised states, with
the relative importance of these radiative cooling lines depending on the
temperature \citep{wiersma09}. In addition to these emission lines, electron
recombination with ions can cool the gas, as recombining electrons take away
kinetic energy from the plasma, a process which is important at temperatures $ >
10^3$\,K \citep{wolfire03}. At similar high temperatures another important
cooling mechanism is the scattering of free electrons off other free ions,
whereby free-free emission removes energy from the gas \citep{draine11}.
Working against the cooling is heating caused by cosmic rays via the expulsion
of bound electrons from atoms or the direct kinetic transfer to free electrons
via Coulomb interactions \citep{Simnett69}.  
By ignoring heating of the gas by photo-ionization, we shall adopt an 
assumption typically used for galactic ISM \citep{gnat07,smith08}, 
though see \cite{wiersma09} who demonstrate how cooling rates are 
reduced when including photoionization in the intergalactic medium 
and proto-galaxies.

We arrive at a first estimate of the temperature of the neutral medium by
requiring energy rate equilibrium between the above mentioned heating and
cooling mechanisms:
\begin{equation}
\Hcrhi = \Cions + \Crec + \Cff,
\label{WCNM_e1} 
\end{equation}
where \Cions is the cooling rate due to atomic and ionic emission lines, \Crec
and \Cff are the cooling rates from recombination processes and free-free
emission as described above, and \Hcrhi is the CR heating rate in
atomic, partly ionised, gas.  The detailed analytical expressions employed by
\sigame for these heating and cooling rates are given in appendix \ref{apB}.

The abundances of the atoms and ions included in \Cions either have to be
calculated in a self-consistent manner as part of the SPH simulation or set by
hand. For our set of galaxies, the SPH simulations follow the abundances of H,
C, N, O, Mg, Si, S, Ca and Fe, while for the abundances of He and Ne, we adopt
solar mass fractions of $0.2806$ and $10^{-4}$, respectively, as used in
\cite{wiersma09}.

\Hcrhi depends on the {\it primary} CR ionization rate (\cri), a
quantity that is set by the number density of supernovae since they are thought
to be the main source of CRs \citep{ackermann13}. In \sigame, this is accounted
for by parameterizing \cri as a function of the {\it local} star formation rate
density (SFRD) as it varies across the simulated galaxy. The details of this
parameterization are deferred to Section \ref{Tk_GMC}.

In eq.\,\ref{WCNM_e1}, all terms but \Cions depend on the ionization fraction
($x_e = n_e/n_{\rm HI}$) of the gas, which in turn depends on the {\it
total} CR ionization rate (i.e., \cri corrected for secondary
ionizations of H and He), the gas temperature (\Tk), and the H\,{\sc i} density
(\nhi) (see appendix \ref{apB}).  The ionization fraction is calculated taking
into account the ionization of H and He (with a procedure kindly provided by
I.\ Pelupessy; see also \cite{pelupessy05}). Since, \nhi is set by the
molecular gas mass fraction (\fh2), which in turn also depends on \Tk (see
Section \ref{HI_to_H2} on how \fh2 is calculated), eq.\ \ref{WCNM_e1} has to be
solved in an iterative fashion until consistent values for \Tk, \ne, and \fh2
are reached. Example solutions are given in Figure \ref{apB1} in Appendix
\ref{apB}.  

\smallskip

The temperature distribution that results from solving eq.\ \ref{WCNM_e1} for
every SPH particle in the G1 simulation is shown in Figure \ref{WCNM_f1} (very
similar distributions are obtained for G2 and G3). The gas has been cooled to
$\Tk\lesssim 10^4$\, K, with temperatures extending down to $\sim 25\,{\rm K}$.
This new gas phase represents both the warm neutral medium (WNM) and the cold
neutral medium (CNM), and from it we derive the molecular gas phase.

\subsection{\hi to \h2 conversion}\label{HI_to_H2}
For the determination of the molecular gas mass fraction associated with each
SPH gas particle, a prescription of \cite{pelupessy06} is used, inferred by
equating the formation rate of \h2 on dust grains with the photodissociation
rate of \h2 by Lyman-Werner band photons, and taking into account the
self-shielding capacity of \h2 and dust extinction. We ignore \h2 production in
the gas phase \citep[cf.,][]{christensen12} since only in diffuse low
metallicity ($\lesssim0.1Z_{\odot}$) gas is this thought to be the dominant
formation route \citep{norman97}, and so should not be relevant in our model
galaxies that have mean metallicities $\Z > 1$ (see
Table\,\ref{model_galaxies_t1}) and very little gas with $\Z<0.1$ (see
Figure \ref{apD1} in Appendix \ref{apD}). We adopt a steady-state for the
H\,{\sc i}$\rightarrow$H$_2$ transition, meaning that we ignore any time
dependence owing to temporal changes in the UV field strength and/or disruptions
of GMCs, both of which can occur on similar time scales as the \h2 formation.
This has been shown to be a reasonable assumption for environments with
metallicities $\gtrsim 0.01\,Z_{\odot}$ \citep{narayanan11,krumholz11}.

\smallskip

The first step is to derive the FUV field strength, \g0, which sets the H\,{\sc
i}$\rightarrow$H$_2$ equilibrium. In \sigame, \g0 consists of a
spatially varying component that scales with the local SFRD (${\rm
SFRD_{local}}$) in different parts of the galaxy on top of a constant component
set by the total stellar mass of the galaxy.  This is motivated by
\citet{seon11}  who measured the average FUV field strength in the MW
($G'_{\rm 0,MW}$) and found that about half comes from star light directly with
the remainder coming from diffuse background light.  We shall assume that in the
MW the direct stellar contribution to $G'_{\rm 0,MW}$ is determined by the
average SFRD ($\rm SFRD_{MW}$), while the diffuse component is fixed by the
stellar mass (${\rm M}_{*,\rm{MW}}$). From this assumption, i.e., by calibrating
to MW values, we derive the desired scaling relation for \g0 in our simulations: 
\begin{equation}
\g0 = G'_{\rm 0,MW} \left( 0.5\frac{{\rm SFRD}_{\rm{local}}}{{\rm SFRD}_{\rm{MW}}} + 0.5\frac {{\rm M}_*}{{\rm M}_{\rm *,MW}}\right), 
\label{equation:g0} 
\end{equation}
where $G'_{\rm 0,MW} = 0.6$\,Habing \citep{seon11}, and ${\rm
M}_{*,\rm{MW}}=6\e{10}$ \,\msun \citep{mcmillan11}.  For ${\rm SFRD}_{\rm{MW}}$
we adopt $0.0024$\,\sfru\,kpc$^{-3}$, inferred from the average SFR within the
central 10\,kpc of the MW \citep[$0.3$\,\sfru; ][]{heiderman10} and within a
column of height equal to the scale height of the young stellar disk
\citep[$0.2\,{\rm kpc}$; ][]{bovy12} of the MW disk. ${\rm SFRD_{local}}$ is the
SFRD ascribed to a given SPH particle, and is calculated as the volume-averaged
SFR of all SPH particles within a $5\,{\rm kpc}$ radius. Note, that the stellar mass
sets a lower limit on \g0, which for G1, G2, and G3 are $0.22$, $0.62$, and
$0.88$\,Habing, respectively. 

\smallskip

Next, the gas upon which the FUV field impinges is assumed to reside in
logotropic clouds, i.e., clouds with radial density profiles given by
$n(r)=\next\left( r/R \right)^{-1}$, where \next is the density at the cloud
radius $R$. 
For a logotropic
density profile, the external density is given by $n_{\rm{H,ext}} = 2/3
\ave{\nH}$, where we approximate $\ave{\nH}$ with the original SPH gas density.
From \cite{pelupessy06} we then have that the molecular gas mass
fraction, including heavier elements than hydrogen, of each cloud is given
by\footnote{We will use lower case $m$ when dealing with individual SPH
particles. Furthermore, \fh2 is not to be confused with $f_{\rm mol}$
describing the total molecular gas mass fraction of a galaxy and to be
introduced in Section \ref{results}.}:
\begin{equation}
\fh2\equiv \frac{m_{\rm mol}}{m_{\rm SPH}}=\exp{\left[ -4 \frac{\Avtr}{\ave{\Av}} \right]}.
\label{HI_to_H2_e1}
\end{equation}
Here \ave{\Av} is the area-averaged visual extinction of the cloud and \Avtr is
the extinction through the outer layer of neutral hydrogen.  The area-averaged
extinction, \ave{\Av}, is calculated from the metallicity and average cloud
density, $\ave{\nH}$: 
\begin{equation} 
	\ave{\Av} = 7.21\e{-22}\Z \langle n_{\rm H}\rangle R,
\end{equation}
where $\ave{\nH} R$ is set by the well known density-size scaling
relation for virialised clouds, normalised by the external boundary pressure, \Pe, i.e.:
\begin{equation}
	\ave{\nH} = n_{\rm ds}\left( \frac{\Pe/k_{\rm B}}{10^4\,\cmpc\,{\rm K}}\right)^{1/2} \left ( \frac{R}{\rm pc}\right )^{-1}.
	\label{nR_Pe}
\end{equation}
For the normalization constant we adopt $n_{\rm ds}=10^3$\,\cmpc, as inferred
from studies of molecular clouds in the MW \citep{larson81,wolfire03,heyer04},
although we note that \cite{pelupessy06} uses 1520\,\cmps.  The external
hydrostatic pressure for a rotating disk of gas and stars is calculated at
mid-plane following \cite{swinbank11}:
\begin{equation}
	P_{\rm tot}\approx\frac{\pi}{2}\rm{G}\gassd \left[ \gassd+\left( \frac{\sigma_{\rm gas\perp}}{\sigma_{\rm *\perp}} \right) \starssd\right], 
	\label{eq:Pe}
\end{equation}
where $\sigma_{\rm gas\perp}$ and $\sigma_{\rm *\perp}$ are the local vertical
velocity dispersions of gas and stars respectively, and $\Sigma$ denotes surface
densities of the same.  These quantities are all calculated directly from the
simulation output, using the neighbouring SPH particles within 5\,kpc, 
weighted by mass, density
and the cubic spline kernel \citep[see also][]{monaghan05}. 
The external cloud pressure that enters in eq.\,\ref{nR_Pe}, is assumed to be
equal to $P_{\rm tot}/(1+\alpha_0+\beta_0)$ for relative cosmic and magnetic
pressure contributions of $\alpha_0=0.4$ and $\beta_0=0.25$
\citep{elmegreen89,swinbank11}.  For the MW, $\Pe/k_{\rm B}\sim10^4\,{\rm \cmpc\,K}$
\citep{elmegreen89}, but as shown in Figure\,\ref{apD1} in Appendix \ref{apD}, our
model galaxies span a wide range in $\Pe/k_{\rm B}$ of
$\sim10^2-10^7\,{\rm \cmpc\,K}$.  

For \Avtr, the following expression is provided by \cite{pelupessy06}:
\begin{equation} 
\Avtr = 1.086 \nu \xi_{\rm{FUV}}^{-1} \times \ln \left[ 1+\frac{\frac{G_0'}{\nu\mu S_\text{H} (T_{\rm k})} \sqrt{ \frac{\xi_{\text{FUV}}}{Z' T_{\rm k}} }}{\next/135~\cmpc } \right], 
\label{HI_H2_e2}
\end{equation}
where $\xi_{\text{FUV}}$ is the ratio between dust FUV absorption cross section
($\sigma$) and the effective grain surface area ($\sigma_d$), and is set to
$\xi_{\text{FUV}} = \sigma/\sigma_{\rm d} = 3$. $S_\text{H} = (1+0.01 T_{\rm
k})^{-2}$ is the probability that \hi atoms stick to grain surfaces, where
$T_k$ is the kinetic gas temperature (determined in an iterative way as
explained in Section \ref{WCNM}). Furthermore, $\nu=\next R \sigma (1+\next R
\sigma)^{-1}$, and $\mu$ \citep[set to $3.5$ as suggested by][]{pelupessy06} is a
parameter which incorporates the uncertainties associated with primarily
$S_\text{H}$ and $\sigma_{\rm d}$.

\smallskip

\begin{figure}
\hspace{1.2cm}
\includegraphics[width=0.8\columnwidth]{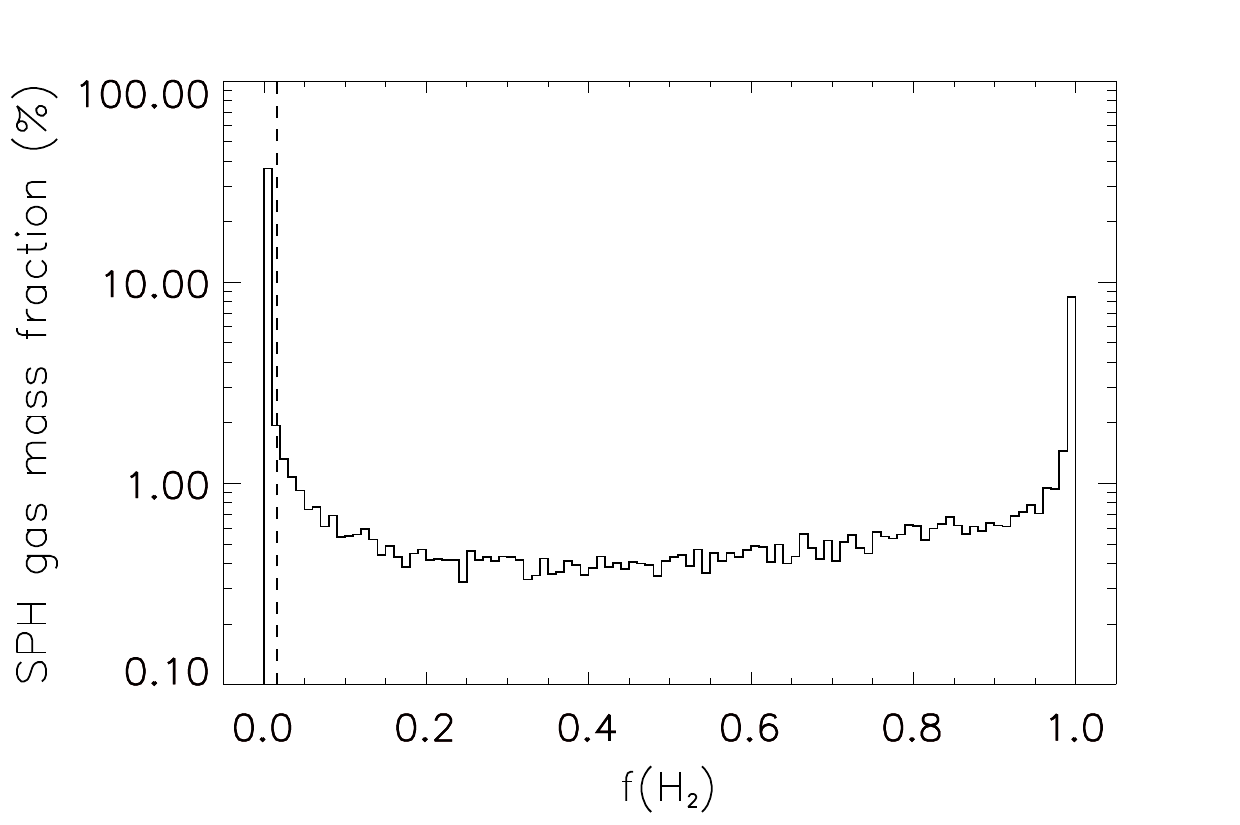}
\caption{\footnotesize{The distribution of the \h2 gas mass fraction of the SPH particles in
G1 calculated according to eq.\ \ref{HI_to_H2_e1} (solid line histogram).
Similar distributions are found for G2 and G3. The lower limit on \fh2, defined
as described in Section \ref{split}, is indicated by the dashed vertical line.}}
\label{figure:f_H2}
\end{figure}

Following the above prescription \sigame determines the molecular gas mass
fraction and thus the molecular gas mass ($m_{\rm mol}=\fh2 m_{\rm SPH}$)
associated with each SPH particle. Depending on the environment (e.g., \g0,
\Z, \Tk, \Pe,...), the fraction of the SPH gas converted into \h2 can take
on any value between 0 and 1, as seen from the \fh2 distribution of G1 in Figure
\ref{figure:f_H2}.  Overall, the total mass fraction of the SPH gas in G1, G2,
and G3 (i.e., within $R_{\rm cut}$) that is converted to molecular gas is
$29$\,\%, $52$\,\% and $34$\,\%, respectively. 

\subsection{Structure of the molecular gas}\label{structure}
Having determined the molecular gas mass fractions, \sigame proceeds by
distributing the molecular gas into GMCs, and calculates their masses and sizes,
along with internal density and temperature structures, as
described in the following. 

\subsubsection{GMC masses and sizes}\label{split}
\begin{figure}
\hspace{1.2cm}
\includegraphics[width=0.8\columnwidth]{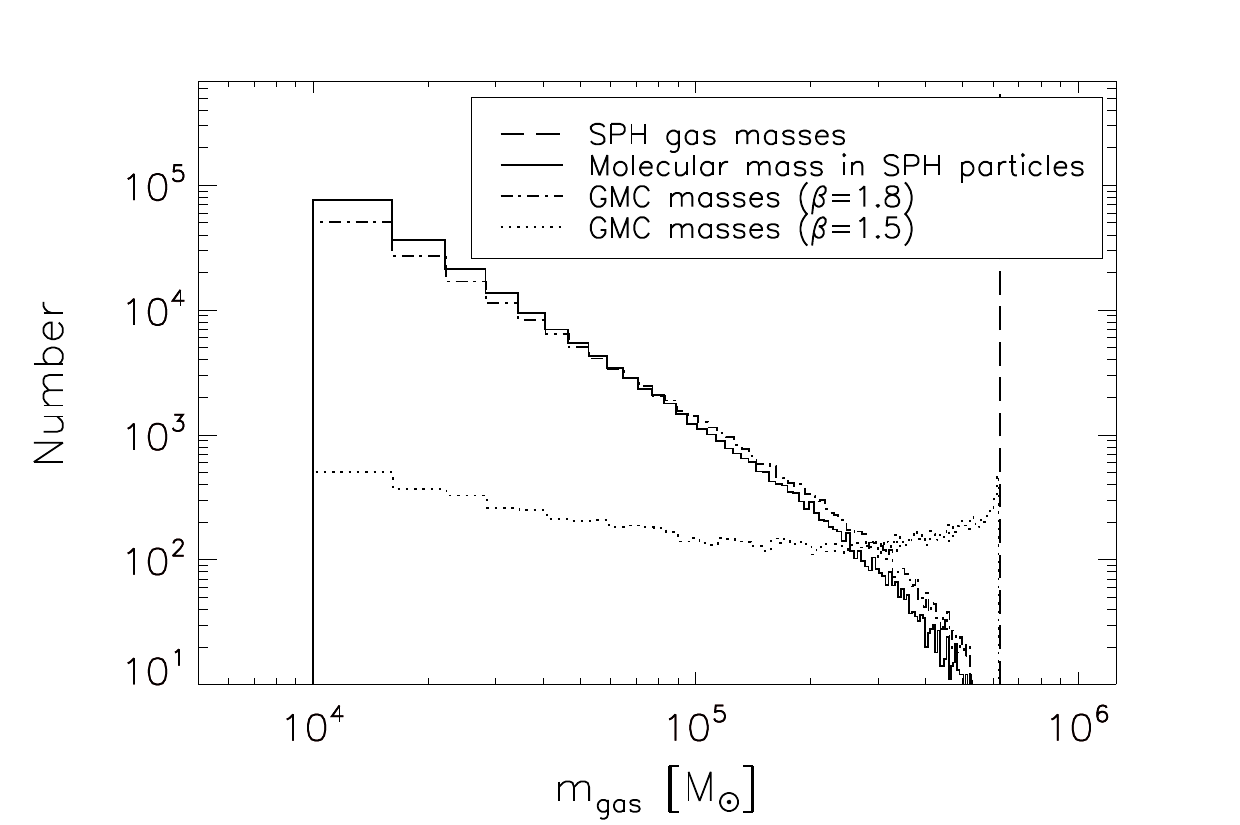}
\caption{\footnotesize{The distribution of GMC masses in G1 obtained by applying eq.\
\ref{equation:mass-spectrum}, with $\beta = 1.8$ (solid histogram) and $\beta =
1.5$ (dash-dotted histogram), compared to the total molecular gas masses associated with
SPH particles (distribution shown as dotted histogram). The dashed vertical line
indicates the SPH gas mass resolution of the simulation ($=6.3\times
10^5\,h^{-1}\,{\rm \msun}$ in the case of G1).}}
\label{structure_f1}
\end{figure}

The molecular gas associated with a given SPH particle is divided into 
GMCs by randomly sampling a power-law mass spectrum of the form:
\begin{equation}
	\frac{dN}{dm_{\rm GMC}}\propto \Mgmc^{-\beta}.
	\label{equation:mass-spectrum}
\end{equation}
For GMCs in the MW disk and Local Group galaxies $\beta\simeq1.8$ \citep{blitz07}, and
is the value adopted by \sigame in this work unless otherwise stated. Lower and upper mass
cut-offs at $10^4\,\msun$ and $10^6\,\msun$, respectively, are enforced in order
to span the mass range observed by \cite{blitz07}. A similar approach was
adopted by \citet{narayanan08,narayanan08a}. Note that, in the case of G1 the upper
cut-off on the GMC masses is in fact set by the mass resolution of the SPH
simulation ($6.3\times 10^5\,h^{-1}\,{\rm \msun}$). For G1, typically $\lesssim
30$ GMCs are created in this way per SPH particle, while for G2 and G3, which
were run with SPH gas particles masses almost an order of magnitude higher, as
much as $\sim 100$ GMCs can be extracted from a given SPH particle for
$\beta=1.8$. Figure \ref{structure_f1} shows the resulting mass distribution of
all the GMCs in G1, along with the distribution of molecular mass associated
with the SPH gas particles prior to it being divided into GMCs. The net effect
of re-distributing the \h2 mass into GMCs is a mass distribution dominated by
relatively low cloud masses, which is in contrast to the relatively flat SPH \h2
mass distribution. Note, the lower cut-off at $\Mgmc=10^4\,\msun$ implies that
if the molecular gas mass associated with an SPH particle (i.e., $m_{\rm
mol}=\fh2 m_{\rm SPH}$) is less than this lower limit it will not be
re-distributed into GMCs. Since $m_{\rm SPH}$ is constant in our simulations
($6.3\times 10^5\,h^{-1}\,\msun$ for G1 and $4.7\times 10^6\,h^{-1}\,\msun$ for
G2 and G3) the lower limit imposed on $m_{\rm GMC}$ translates directly into a
lower limit on \fh2 ($0.016$ for G1 and 0.002 for G2 and G3, shown as a dashed
vertical line for G1 in Figure \ref{figure:f_H2}).  As a consequence, 0.2, 0.005
and 0.01\,\% of the molecular gas in G1, G2, and G3, respectively, does not end
up in GMCs. These are negligible fractions and the molecular gas they represent
can therefore be safely ignored.

\begin{figure}
\hspace{1.2cm}
\includegraphics[width=0.8\columnwidth]{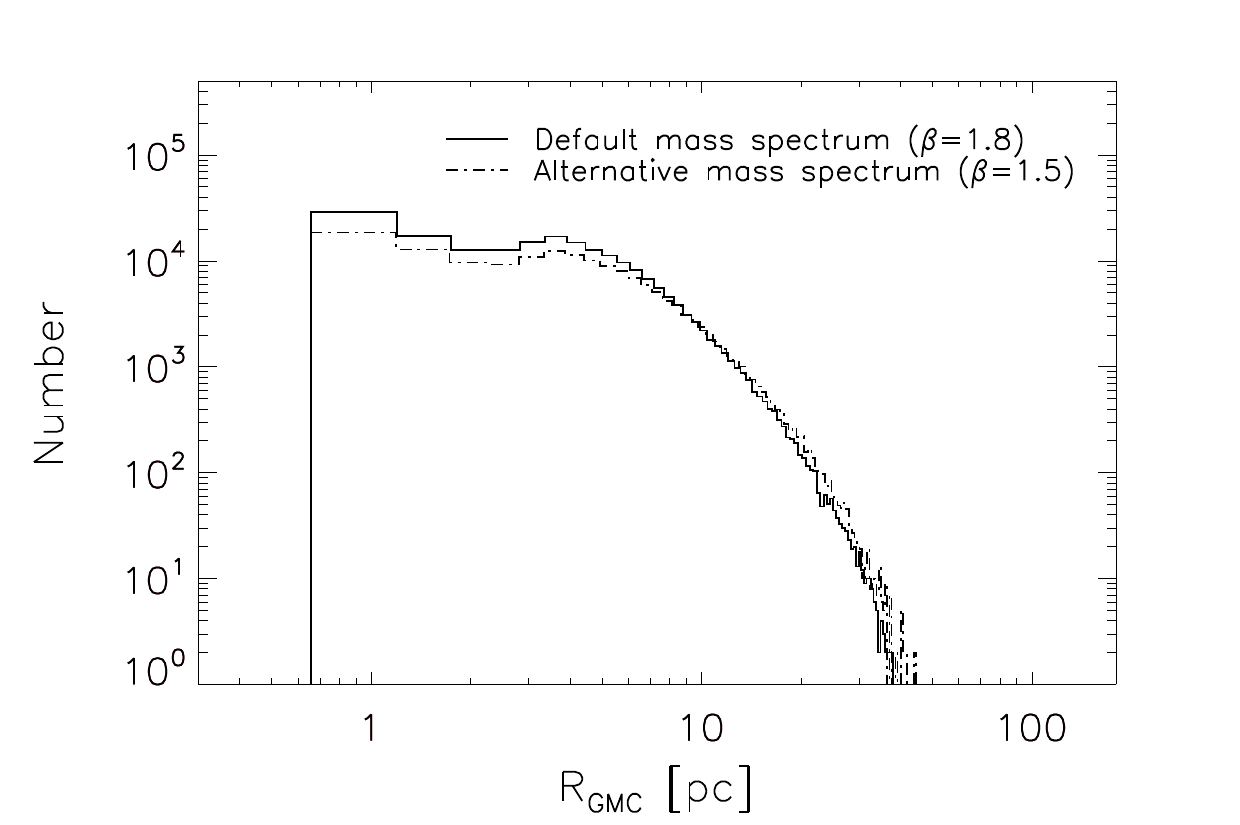}
\caption{\footnotesize{The distribution of GMC radii in G1 for the adopted cloud mass
spectrum with $\beta = 1.8$ (solid histogram) and for a slightly modified
spectrum with $\beta = 1.5$ (dash-dotted histogram).  Similar distributions are
found for G2 and G3.}}
\label{structure_f2}
\end{figure}

The GMC sizes are derived from the virial theorem, which relates the radius of a GMC
(\Re) to its mass and external pressure (\Pe) according to:
\begin{equation} 
	\frac{\Re}{\text{pc}} = \left( \frac{\Pe/k_{\rm B}}{10^4\,\cmpc\,{\rm K}}\right)^{-1/4} \left (\frac{\Mgmc}{290\,\msun}\right )^{1/2}.
	\label{structure_e1}
\end{equation}
Similarly, the internal velocity dispersion ($\sigma_v$) of the clouds is given by:  
\begin{equation} 
	\frac{\sigma_v}{{\rm km\,s^{-1}}} = 1.2 \left( \frac{\Pe/k_{\rm B}}{10^4\,\cmpc\,{\rm K}}\right)^{1/4}\left ( \frac{\Re}{\rm{pc}} \right )^{1/2}
	\label{structure_e2}
\end{equation}
\citep[e.g.,][]{elmegreen89,swinbank11}.
Figure \ref{structure_f2} shows the resulting distribution of GMC radii in G1
(solid histogram). The minimum and maximal cloud radii found in G1 are $\sim 0.07\,{\rm
pc}$ and $\sim 100\,{\rm pc}$, respectively, and are set by the pressure and the 
imposed limits on $m_{\rm GMC}$.

Observations have indicated that the shape of the GMC mass spectrum might be
different in gas-rich systems where a high-pressure ISM leads to the
characteristic mass of star-forming clumps of molecular gas being much higher
than what is observed in normal spirals \citep[e.g.][]{swinbank11,leroy15}.  In
Section \ref{dif_ISM} we therefore examine the effects of adopting a more
top-heavy GMC mass distribution, corresponding to $\beta=1.5$ (shown as
dot-dashed histogram in Figure \ref{structure_f1}). 
The total amount of molecular gas in our galaxies does not change significantly
between $\beta = 1.8$ and $\beta = 1.5$, and the CO simulation results are
robust against (reasonable) changes in the mass-spectrum.

The GMCs are placed randomly around the position of their `parent' SPH
particle, albeit with an inverse proportionality between radial displacement
and mass of the GMC. The latter is done in order to retain the mass
distribution of the original galaxy simulation as best as possible. The GMCs
are assigned the same bulk velocity, $\bar{v}$, \Z, \g0 and \cri as their
`parent' SPH particle.

\subsubsection{GMC density structure}\label{plummer}
In order to ensure a finite central density in our GMCs, \sigame adopts a
Plummer radial density profile \citep[e.g.,][]{plummer11,whitworth01}: 
\begin{equation} 
	\frac{n_{\text{H}_2} (R)}{\cmpc} = 3.55 \left (\frac{m_{\rm GMC}}{\msun}\right ) \left (\frac{R_p}{{\rm pc}}\right )^{-3}
        \left( 1+\frac{R^2}{R_p^2} \right)^{-5/2}, 
	\label{equation:plummer-profile}
\end{equation}
where $R_p$ is the so-called Plummer radius, which we set to $R_p=0.1 \Re$.
The latter allows for a broad range in gas densities throughout the clouds,
from $\sim 10^{5}$\,\cmpc in the central parts to a few $10\,\cmpc$ further
out. Note, eq.\ \ref{equation:plummer-profile} accounts for the presence of
helium in the GMC.  

\begin{figure}
\hspace{1.2cm}
\includegraphics[width=0.8\columnwidth]{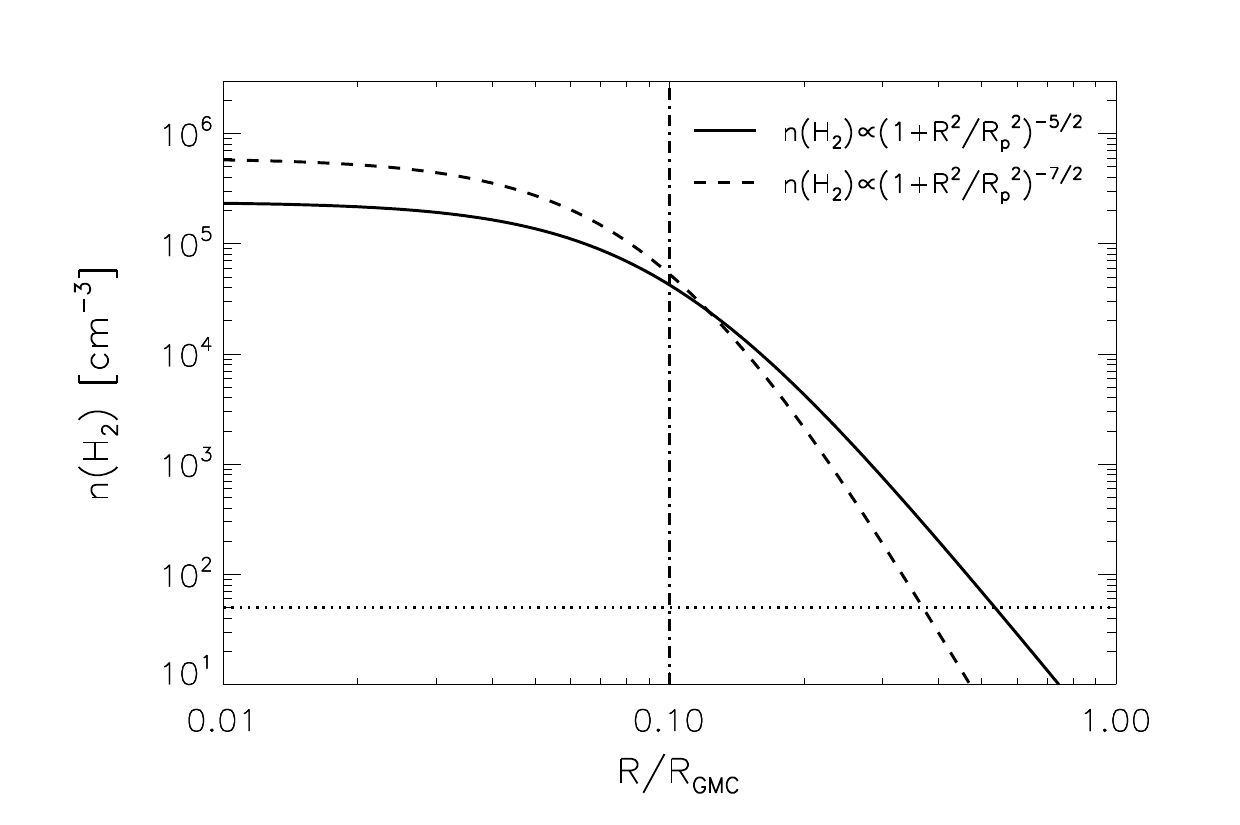}
\caption{\footnotesize{Plummer density profiles (eq.\ \ref{equation:plummer-profile}) with
exponents $-5/2$ (solid line) and $-7/2$ (dashed line) for a GMC mass of
$10^4$\,\msun and an external pressure of $\Pe/k_{\rm B}=10^4\,$K\,cm$^{-3}$.
The corresponding cloud radius is $\rcl=5.9$\,pc (eq.\ \ref{structure_e1}).
The vertical dash-dotted line marks the Plummer radius ($=0.1R_{\rm GMC}$), and
the horizontal dotted line the \nh2 value ($50\,{\rm cm^{-3}}$) below
which CO is assumed to photo-dissociate in our simulations.}}
\label{fig:plummer}
\end{figure}

In Section \ref{dif_ISM} we examine the impact on our simulation results
if a steeper GMC density profile with an exponent of $-7/2$ is adopted instead
of the default $-5/2$ in eq.\ \ref{equation:plummer-profile}.
This modified Plummer profile, as well as the default profile, are shown in
Figure \ref{fig:plummer} for a GMC with mass of $10^4$\,\msun and external
pressure $10^4\,{\rm K\,cm^{-3}}$.

\subsubsection{GMC thermal structure}\label{Tk_GMC}
Having established the masses, sizes, and density structure of the GMCs,
\sigame solves for the kinetic temperature throughout the clouds by balancing
the relevant heating and cooling mechanisms as a function of cloud radius. 

GMCs are predominantly heated by FUV photons (via the photo-electric effect) 
and cosmic rays. For the strength of the FUV field impinging on the GMCs, we
take the previously calculated values at the SPH gas particle position (eq.\
\ref{equation:g0}).  We shall assume that the CR ionization rate scales in a
similar manner, since CRs have their origin in supernovae and are therefore
related to the star formation and the total stellar mass of the galaxy: 
\begin{align}
\cri &= \zeta_{\rm CR,MW} \left( 0.5\frac{{\rm SFRD}_{\rm{local}}}{{\rm SFRD}_{\rm{MW}}} + 0.5\frac {{\rm M}_*}{{\rm M}_{\rm *,MW}}\right) 
\label{equation:CR}
\end{align}
where ${\rm SFRD}_{\rm{local}}$, ${\rm SFRD}_{\rm{MW}}$ and ${\rm M}_{\rm
*,MW}$ are as described in Section \ref{HI_to_H2}, and \cri is scaled to the
`canonical' MW value of \cri$_{\rm ,MW}=3\e{-17}$\,\ps
\citep[e.g.][]{webber98}.  While the FUV radiation is attenuated by dust and
therefore does not heat the GMC centres significantly, cosmic rays can
penetrate these dense regions and heat the gas there \citep{papa11}.  For this
reason \sigame attenuates the FUV field throughout the GMCs, while the CR
ionization rate remains constant for a given cloud. The extinction of the FUV
field at a certain point within a GMC is derived by integrating the Plummer
density profile from that point and out to \Re. This H column density is
converted into a visual extinction ($A_{\rm V} = \colH/2.2\e{21}\,\cmps$) so
that the attenuated FUV field becomes:
\begin{align}
\ga=\g0 e^{-1.8A_{\rm V}},
\label{ga}
\end{align}
where the factor of $1.8$ is the adopted conversion from visual to FUV
extinction \citep{black77}. 

\sigame calculates the gas kinetic temperature throughout each GMC via the
following heating and cooling rate balance:
\begin{equation}
\Hpe+\Hcrh2 = \Ch2+\Cco+\Coi+\Ccii+\Cgd.
\label{Tk_GMC_e1}
\end{equation}
\Hpe is the photo-electric heating by FUV photons, and \Hcrh2 is the cosmic ray
heating in molecular gas.  \Ch2 is the cooling rate of the two lowest \h2
rotational lines (S(0) and S(1)), and \Cco is the cooling rate of the combined
CO rotational ladder. \Cgd is the cooling rate due to interactions between gas
molecules and dust particles, which only becomes important at densities above
$10^4$\,\cmpc \citep[e.g.,][]{goldsmith01,glover12}.  \Ccii and \Coi are the
cooling rates due to \cii and \oi line emission, respectively.  The abundances
of carbon and oxygen used in the prescriptions for their cooling rates, scale
with the cloud metallicity, while the CO cooling scales with the relative CO to
neutral carbon abundance ratio, set by the molecular density (see Appendix
\ref{apC}).

\Cgd depends on the temperature difference between the gas and the dust (see
eq.\ \ref{equation:dust-gas-cooling}). The dust temperature, \Td, is set by the
equilibrium between the absorption of FUV and the emission of IR radiation by
the dust grains. We adopt the approximation given by \citet{tielens05}:
\begin{equation}
\frac{\Td}{{\rm K}} \simeq 33.5 \left( \frac{a}{1\mu {\rm m}}\right)^{-0.2} \left( \frac{\ga}{10^4\,{\rm Habing}} \right)^{0.2},
\label{Tk_GMC_e2}
\end{equation}
where \ga is the dust-attenuated FUV field (eq.\ \ref{ga}) and $a$ is the grain
size, which we set to 1\,$\mu$m for simplicity.  Values for \Td using eq.\
\ref{Tk_GMC_e2} range from 0 to 8.9\,K, but we enforce a lower limit on \Td
equal to the $z=2$ CMB temperature of $8.175$\,K. \Td is therefore essentially
constant ($\sim8-9$\,K) throughout the inner region of the GMC models, similar
to the value of $\Td=8$\,K adopted by \cite{papa13} for CR-dominated cores.
Analytical expressions for all of the above heating and cooling rates are given
in Appendix \ref{apC}, which also shows their relative strengths as a function
of density for two example GMCs (Figure \ref{apC1}).

\bigskip

Figure \ref{apD3} in Appendix \ref{apD} shows the resulting \Tk versus \nh2
behaviour for 80 GMCs spanning a broad range of GMC masses, metallicities, and
star formation rate densities.  As seen in the displayed GMC models, some
general trends can be inferred from the $\Tk-\nh2$ diagrams. At fixed
metallicity and mass, an increase in \g0 (and therefore also in \cri), leads to
higher temperatures throughout the models.  In the outer regions this is due
primarily to the increased photoelectric heating, while in the inner regions,
heating by the unattenuated cosmic rays takes over as the dominating heating
mechanism (Figure\,\ref{apC1}). Keeping \g0 (and \cri) fixed, lower \Tk-levels
and shallower $\Tk-\nh2$ gradients are found in GMCs with higher metallicities.
Both these trends are explained by the fact that the \cii and \oi cooling rates
scale linearly with \Z (see Appendix\,\ref{apC}). 

Moving from the outskirts and inward towards the GMC centres, \Tk drops as the
attenuation of \g0 reduces the photoelectric heating.  However, the transition
from cooling via \cii to the less efficient cooling mechanism by CO lines,
causes a local increase in \Tk at $\nh2\sim10^3-10^{4.5}$\,\cmpc, as also seen
in the detailed GMC simulations by \cite{glover12}. The exact density at which
this `bump' in \Tk occurs depends strongly on the mass of the GMC.  Our choice
of the Plummer model for the radial density profile means that the extinction,
$A_{\rm V}$, at a certain density increases with GMC mass. This in turn
decreases the FUV heating, and as a result the $\Tk-\nh2$ curve moves to lower
densities with increasing GMC mass.

As the density increases towards the cloud centres (i.e., $n_{\rm H_2} \gtrsim
10^{4-5}$\,\cmpc) molecular line cooling and also gas-dust interactions become
increasingly efficient and start to dominate the cooling budget.  The $\Tk-\nh2$
curves are seen to be insensitive to changes in \Z, which is expected since the
dominant heating and cooling mechanisms in these regions do not depend on \Z.
Eventually, in the very central regions of the clouds, the gas reaches
temperatures close to that of the ambient CMB radiation field, irrespective of
the overall GMC properties and the conditions at the surface (Figure
\ref{apC1}). 

\subsubsection{GMC grid models} \label{gmcgrid}
The $\Tk-\nh2$ curve for a given GMC is determined by the following quantities: 
\begin{itemize}
\item \g0 and \cri, which govern the gas heating and are set by the local star
formation rate density (SFRD$_{\rm local}$) and the total stellar mass ($M_*$)
according to eqs.  \ref{equation:g0} and \ref{equation:CR}.
\item \Mgmc and \Pe which determine the effective radius of a cloud (eq.\
\ref{structure_e1}) and thus its density profile (eq.\
\ref{equation:plummer-profile}). 
\item The local metallicity (\Z), which influences the fraction of \h2 gas and
plays an important role in cooling the gas. 
\end{itemize}

These local parameters together with the most important global parameters used by \sigame 
are listed in Table \ref{sigame_par}. 
The GMC ensemble distributions of \g0, \cri, \Mgmc, $\Pe/k_{\rm B}$, and \Z for
each of the galaxies G1, G2 and G3 are shown in Figure \ref{apD1}. 

There are more than $100,000$ GMCs in a single model galaxy and, as Figure
\ref{apD1} shows, they span a wide range in \g0, \cri, \Mgmc, $\Pe/k_{\rm B}$,
and \Z.  Thus, in order to shorten the computing time, we calculated $\Tk-\nh2$
curves for a set of 630 GMCs, chosen to appropriately sample the distributions
at certain grid values (listed in Table \ref{grid_par}, and marked by vertical
black lines in Figure \ref{apD1}). 
Every GMC in our simulations was
subsequently assigned the $\Tk-\nh2$ curve of the GMC grid model closest to it
in the (\g0, $m_{\rm GMC}$, \Z) parameter space.
In the default GMC grid, we keep $\Pe/k_{\rm B}$ fixed 
to a MW-like value of $10^4\,\rm{K}\,\cmpc$, thereby also anchoring the scaling relations 
in eq. \ref{structure_e1} and \ref{structure_e2} for size and velocity dispersion. 
This is done to minimise the number of radiative transfer calculations, 
effectively keeping the amount of molecular gas mass dependent on local pressure, 
but removing local pressure as a free parameter in, and hence simplifying, 
the calculation of CO emission.

In addition to the default grid described above, 
we made two separate tests to explore the GMC 
parameter space more fully. 
These tests are described in the following and results of their 
application are presented in Section \ref{dif_ISM}. 
The external cloud pressure, $\Pe/k_{\rm B}$, spans a wide range from $\sim10^3$ to 
$10^9\,\rm{K}\,\cmpc$ as shown in Figure \ref{apD1}. 
In a separate test, we therefore constructed the same GMC grid, but for
fixed pressures of $\Pe/k_{\rm B}=10^{5.5}\,$ and $10^{6.5}\,$K\,\cmpc 
and interpolated among the resulting three values of $\Pe/k_{\rm B}$ 
($[10^{4},10^{5.5},10^{6.5}]\,$K\,\cmpc) as 
a demonstration of how to incorporate local pressure in \sigame. 
We also test the effect of adopting the alternative radial GMC 
density profile defined in Section \ref{plummer}.
$\Tk-\nh2$ curves were calculated for each possible combination of the
parameter grid values listed in Table\,\ref{grid_par} for both density profiles, 
giving us a total of
$3\times2\times630=3780$ GMC grid models.

\begin{center}
\begin{table*}
\centering
\caption{\footnotesize{Parameters, global and local, used by \sigame, together with relevant equations in this work.}}
\begin{tabular}{p{3cm}|   p{6cm} p{4cm}} 
\hline
\hline
Global parameters 				&	[CO/\h2], $\beta$ (GMC mass spectrum), GMC density profile & Eqs.\,\ref{equation:mass-spectrum} and \ref{equation:plummer-profile} \\
Local parameters 				&	\Mgmc, \g0, \cri, \Z, \Pe, SFRD	&	Eqs.\,\ref{equation:mass-spectrum}, \ref{equation:g0}, \ref{equation:g0} and \ref{eq:Pe} \\
Derived internal GMC parameters	&	\nH, \Tk, \xe, \sv				&	Eqs.\,\ref{equation:plummer-profile} and \ref{Tk_GMC_e1}\\
\hline
\end{tabular}
\label{sigame_par}
\end{table*}
\end{center}

\begin{center}
\begin{table}
\centering
\caption{\footnotesize{Grid parameter values}}
\begin{tabular}{l |   l} 
\hline
\hline
\g0	[Habing]  			&	0.5, 1, 4, 7, 10, 13, 16, 19, 23, 27	 \\ 
$\log(\Mgmc\,[\msun])$		&	4.0, 4.25, 4.5, 4.75, 5.0, 5.25, 5.5, 5.75, 6.0	\\ 
$\log(Z/Z_{\odot})$		& 	-1, -0.5, 0, 0.5, 1, 1.4, 1.8	\\ 
$\log(\Pe/k_{\rm B}\,{\rm [K\,\cmpc]})$ 	&	4.0, 5.5, 6.5	 \\ 
\hline
\end{tabular}
\label{grid_par}
\end{table}
\end{center}

\subsection{Radiative transfer of CO lines}\label{radiative_transfer}

\sigame assumes a fixed CO abundance equal to the Galactic value of
[CO/\h2]$=2\times 10^{-4}$ \citep[][and see Section \ref{dis_models} for a
justification of this value]{lee96,sofia04} everywhere in the GMCs except for
$\nh2<50$\,\cmpc, where CO is not expected to survive photo-dissociation
processes \citep[e.g.][]{narayanan08a}. \sigame calculates the CO line
radiative transfer for each GMC individually and derives the CO line
emission from the entire galaxy.

\subsubsection{Individual GMCs}
For the CO radiative transfer calculations we use a slightly modified version
of the LIne Modeling Engine \cite[\lime ver.\ 1.4;][]{brinch10} - a 3D
molecular excitation and radiative transfer code.  \lime has been
modified in order to take into account the redshift dependence of the CMB
temperature, which is used as boundary condition for the radiation field during
photon transport, and we have also introduced a redshift correction in the
calculation of physical sizes as a function of distance. We use collision rates
for CO (assuming \h2 is the main collision partner) from \cite{yang2010}.  In
\lime, photons are propagated along grid lines defined by Delaunay
triangulation around a set of appropriately chosen sample points in the gas,
each of which contain information on \nh2, $T_k$, $\sigma_v$, $[{\rm CO}/\h2]$
and $\boldsymbol{v}$. \sigame constructs such a set of sample points throughout each
GMC: about 5000 points distributed randomly out to a radius of $50$\,pc, i.e.,
beyond the effective radius of typical GMCs in G1, G2, and G3 (Figure
\ref{structure_f2}) and in a density regime below the threshold density of
$50$\,\cmpc adopted for CO survival (see previous paragraph). The concentration
of sample points is set to increase towards the centre of each GMC where the
density and temperature vary more drastically.  

For each GMC, \lime generates a CO line data cube, i.e., a series of CO
intensity maps as a function of velocity. The velocity-axis consists of 50
channels, each with a spectral resolution of $1.0$\,\kms, thus covering the
velocity range $v=[-25,+25]\,{\rm km\,s^{-1}}$. The maps are $100\,{\rm pc}$ on
a side and split into 200 pixels, corresponding to a linear resolution of
$0.5\text{\,pc/pixel}$ (or an angular resolution of
$5.9\times10^{-5}\,\arcsec\text{/pixel}$ at $z=2$).  Intensities are corrected
for arrival time delay and redshifting of photons.

Figure \ref{apD4} shows the area- and velocity-integrated CO Spectral Line
Energy Distributions (SLEDs) for the same 80 GMCs used in Section \ref{Tk_GMC}
to highlight the $\Tk-\nh2$ profiles (Figure \ref{apD3}). The first thing to
note is that the CO line fluxes increase with \Mgmc, which is due to the
increase in size, i.e., surface area of the emitting gas, with cloud mass
(Section \ref{split}). Turning to the shape of the CO SLEDs, a stronger \g0
(and \cri) increases the gas temperature and thus drives the SLEDs to peak at
higher $J$-transitions.  Only the higher, $J_{\rm up}>4$, transitions are also
affected by metallicity, displaying increased flux with increased \Z.  
For GMCs with high \g0, the high metallicity
levels thus cause the CO SLED to peak at $J_{\rm up}>8$.

\subsubsection{The effects of dust}\label{dust_properties}
Dust absorbs the UV light from young O and B stars and re-emits in the far-IR,
leading to possible `IR pumping' of molecular infrared sources.  However, due
to the large vibrational level spacing of CO, the molecular gas has to be at a
temperature of at least $159$\,K, for significant IR pumping of the CO
rotational lines to take place, when assuming a maximum filling factor of 1, as
shown by \cite{carroll81}.  Most of the gas in our GMC models is at
temperatures below $100$\,K, with only a small fraction of the gas, in the very
outskirts, of the GMCs reaching $\Tk>159$\,K, as seen in Figure \ref{apD3} in
Appendix \ref{apD}. This happens only if the metallicity is low ($\Z\leq0.1$)
or in case of a combination between high FUV field ($\g0\geq4$) and moderate
metallicity. 

In principle, the high-$J$ CO lines could be subject to extinction by dust, but
this effect is significant only in extremely dust-enshrouded sources
\citep{papa10a}, and therefore unlikely to be relevant for our
simulations. 

While \lime is
capable of including dust in the radiative transfer calculations, provided that a table of dust
opacities as function of wavelength be supplied together with the input model, 
we have chosen not to include dust in the simulations presented here.

\begin{figure*}
\centering
\includegraphics[width=1\columnwidth]{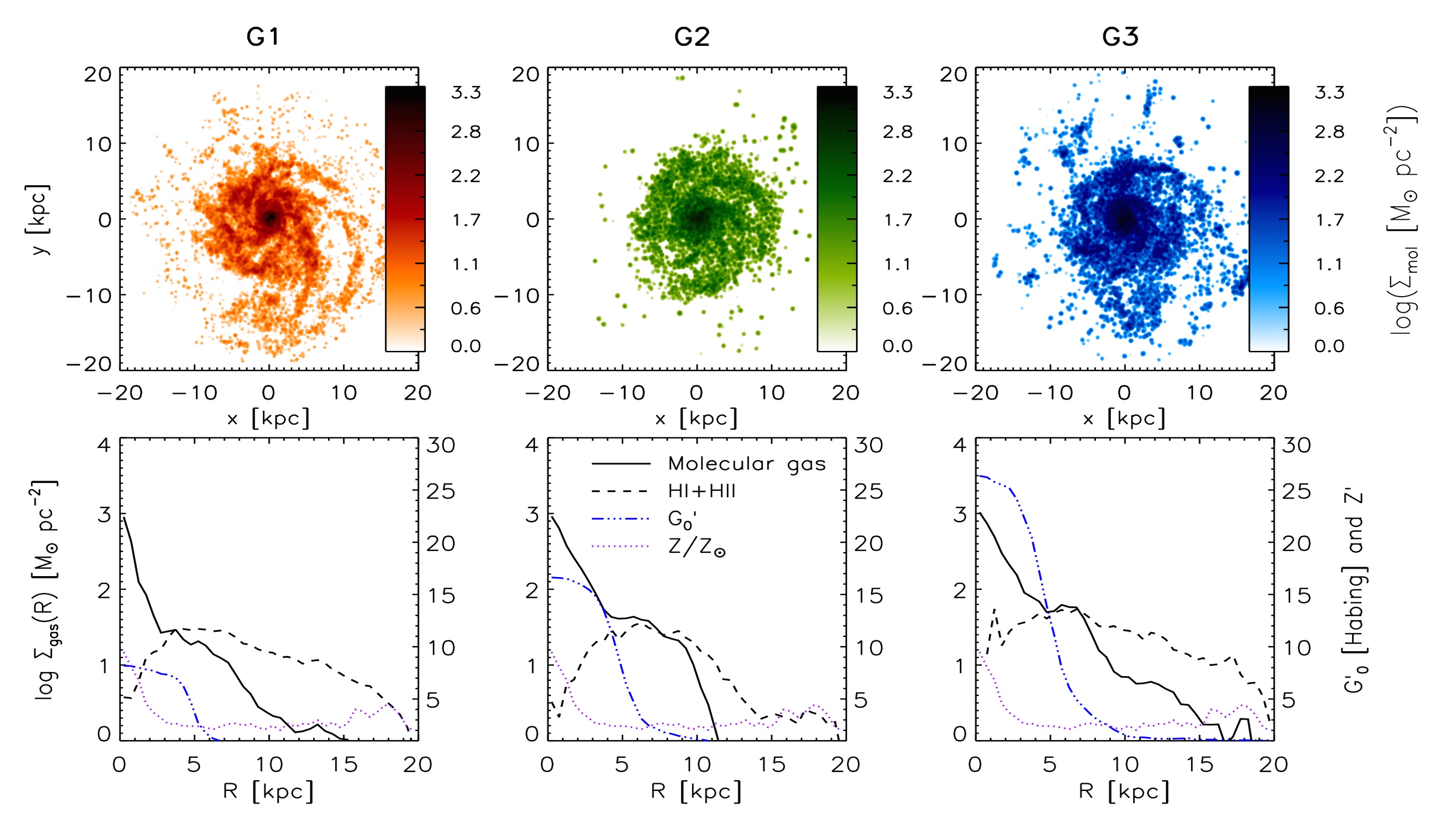}
\caption{\footnotesize{Top row: Molecular surface density maps of our model galaxies seen
face-on.  The maps have been smoothed using a circular Gaussian with full width
at half maximum (FWHM) of $3$ pixels corresponding to $0.24$\,kpc. The
molecular gas surface density maps are seen to trace the spiral arms and inner
disk as well as a few dense clumps further out. Bottom row: azimuthally
averaged radial profiles of the molecular (solid curve) and \hi+\hii (dashed
curve) gas surface densities, of the mean metallicity (dotted curve) and of \g0
(dot-dashed curve) -- determined from 50 radial bins stretching from 0 to
15\,kpc from the centre of each galaxy. The molecular gas is the result of
applying the recipes in Section \ref{HI_to_H2} to the SPH simulations presented
in Section \ref{model_galaxies}, while the $\hi+\hii$ gas is the initial SPH
gas mass minus the derived molecular gas mass (both include the contribution
from helium).  We estimate \g0 by averaging over the FUV fields impinging on
all GMCs in each radial bin.}}
\label{h2_map}
\end{figure*}

\subsubsection{CO emission maps}\label{combining}
A given GMC is assigned the CO emission line profile of the GMC model that
corresponds to the nearest grid point in the (\g0, \Mgmc, \Z) parameter space.
A spatial grid of $400\times400$\,pixels is overlayed on each galaxy viewed
face-on, and the line profiles of all GMCs within each pixel are added to a
common velocity axis, thus resulting in a position-velocity datacube.  CO
moment 0 maps are then constructed by integrating the flux within each pixel in
velocity. These maps are $40\,{\rm kpc}$ on each side and thus have a resolution of
$\sim 100\,{\rm pc/pixel}$.  By expanding the pixel size to encompass the
entire galaxy, the global CO SLED is derived. 

The approach described above assumes that the molecular line emission
from each GMC is radiatively decoupled from all other GMCs, due to the
significant velocity gradients across the galaxies ($\sim200-500$\,\kms) and
the relatively small internal line widths of the individual GMCs 
($\sim2-34$\,\kms for the GMCs in G1, G2 and G3).

\begin{figure*}
\centering
\includegraphics[width=0.95\columnwidth]{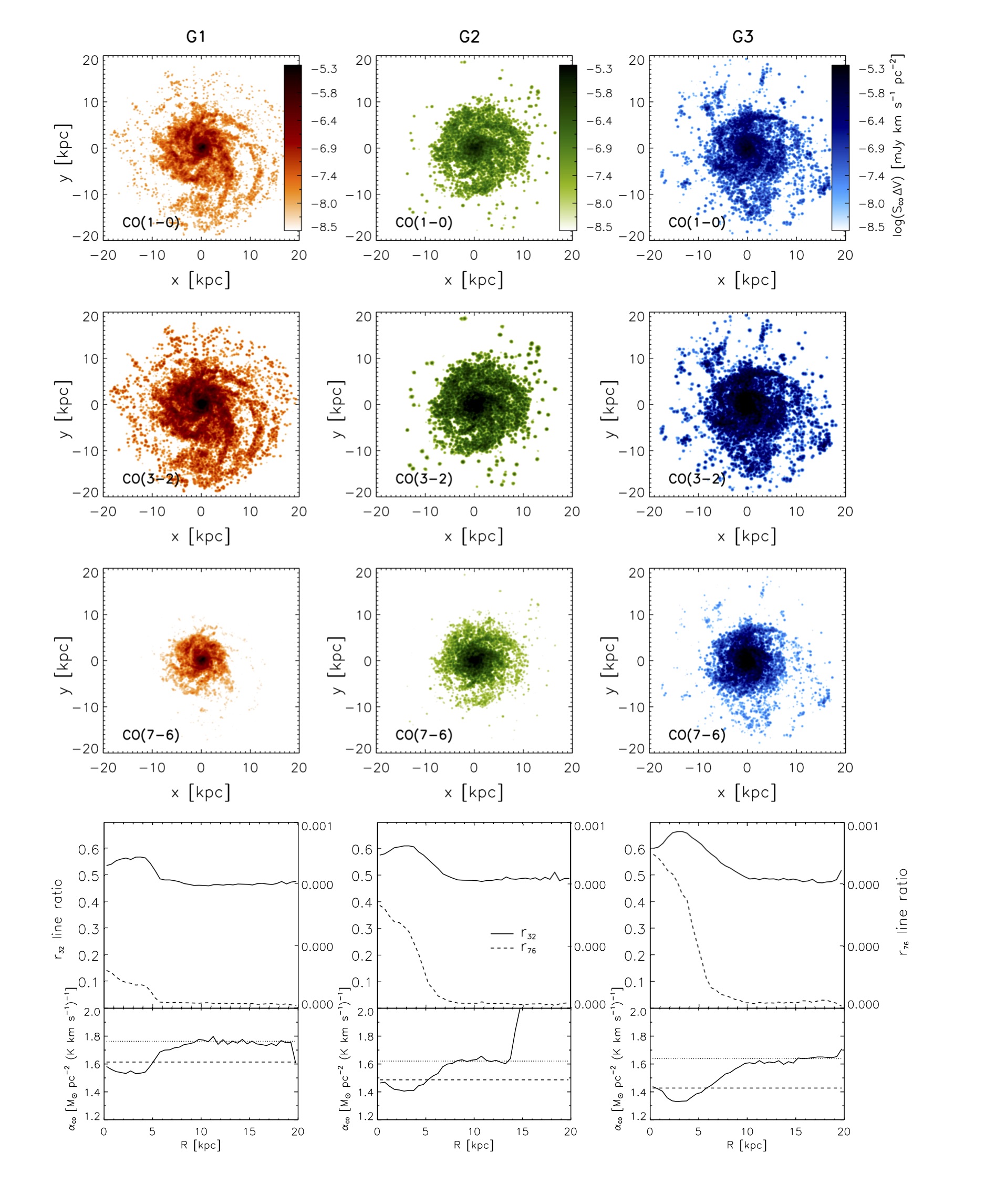}
\caption{\footnotesize{The top three rows show the moment 0 maps of the CO(1$-$0),
CO(3$-$2) and CO(7$-$6) emission from G1, G2 and G3. The CO maps have been
smoothed using a circular Gaussian with full width at half maximum (FWHM) of 3
pixels corresponding to $0.24$\,kpc, and as with the SPH and molecular gas
surface density maps, a logarithmic scale has been applied in order to better
display extended emission.  The bottom row shows the azimuthally averaged CO
3$-$2/1$-$0 and 7$-$6/1$-$0 brightness temperature line ratios (denoted
$r_{32}$ and $r_{76}$, respectively) as functions of projected radius for each
of the three galaxies.  A radial bin-size of $0.5\,{\rm kpc}$ was used. Also
shown are the azimuthally averaged radial profiles of CO-to-\h2 conversion
factor $\alpha_{\rm CO}(R)=\Sigma_{\h2}/I_{{\rm CO}(1-0)}$ in units of
\msun\,pc$^{-2}$\,(K\,km\,\ps)$^{-1}$ with a dashed line indicating the global
\aCO factor and a dotted line for the disk-averaged \aCO factor (see Section \ref{CO_mom0}).
}}
\label{CO_mom0_f_mom0}
\end{figure*}

\section{Simulating massive $z=2$ main sequence galaxies}\label{results}
In this section we examine the \h2 surface density and CO emission maps
resulting from applying \sigame to the three SPH galaxy simulations G1, G2, and
G3 at $z=2$ (Section \ref{cosmological_simulations}), and we compare with existing CO
observations of main sequence galaxies at $z\sim 1-2.5$.  
We also tested \sigame on three MW-like galaxies (see Appendix \ref{apMW}), 
finding a general agreement with the CO line observations of MW and other similar local galaxies.
As mentioned in
previous sections, our default grid will be that corresponding to a Plummer
density profile and a pressure of $\Pe/k_{\rm B}=10^4\,{\rm \cmpc\,K}$, combined
with a GMC mass spectrum of slope $\beta=1.8$, unless otherwise stated. 

\subsection{Total molecular gas content and \h2 surface density maps}\label{H2_cont}
The total molecular gas masses of G1, G2 and G3 -- obtained by summing up the
GMC masses associated with all SPH particles within each galaxy (i.e., $M_{\rm
mol} = \sum \Mgmc = \sum \fh2 m_{\rm SPH}$) -- are $7.1\e{9}$, $1.7\e{10}$, and
$2.1\e{10}$\,\msun, respectively, corresponding to about $34$, $59$ and
$45$\,\% of the original total SPH gas masses of the galaxies within $R_{\rm
cut}=20\,$kpc.\footnote{These global molecular-to-SPH gas mass fractions are
calculated as $M_{\rm mol}/M_{\rm SPH} = \sum \Mgmc / \sum m_{\rm SPH}$, where
the sums are over all SPH particles.} The global molecular gas mass fractions
(i.e., $f_{\rm mol} =M_{\rm mol}/(M_*+M_{\rm mol}$) are 11.8, 10.2 and 9.1\,\%
for G1, G2, and G3, respectively.

This is $\sim4-5\times$ below the typical molecular gas mass fraction ($\sim
40-60\,\%$) inferred from CO observations of main sequence galaxies at $z\sim
1-3$ with similar stellar masses and star formation rates as our simulations
\citep[e.g.,][]{daddi10,magnelli12,tacconi13}.

The reason for this discrepancy is the low SPH gas mass fractions of our
simulated galaxies to begin with ($f_{\rm SPH} =M_{\rm SPH}/(M_*+M_{\rm
SPH})=18-28$\,\%), which obviously restricts $f_{\rm mol}$ to lower values.
Such low gas fractions have been observed in local star forming galaxies.  For
example, assuming a MW-like \aCO factor, \cite{saintonge11} derived $f_{\rm
mol}\sim6.7\pm4.5$\,\% for a sample of 119 CO$(1-0)$ detected normal star
forming galaxies and a stack of 103 non-detections.

\smallskip

Figure \ref{h2_map} (top panels) shows the molecular gas surface density (\Sh2)
maps of G1, G2, and G3, where \Sh2 is calculated over pixels $80\,{\rm
pc}\times 80\,{\rm pc}$ in size. The pixel size was chosen in order to avoid
resolving the GMCs, which typically have sizes $\ls 40\,{\rm pc}$ (see
Figure \ref{structure_f2}).
The molecular gas is seen to extend out to radii of $\sim 10\,{\rm kpc}$ and
beyond, but generally the molecular gas concentrates within the inner regions
of each galaxy. The distribution of molecular gas broadly follows the central
disk and spiral arms where the SPH surface density ($\Sigma_{\rm SPH}$) is also
the highest (Figure \ref{model_galaxies_f1}). The correspondence is far from
one-to-one, however, as seen by the much larger extent of the SPH gas, i.e.,
regions where \h2 has not formed despite the presence of atomic and ionised
gas.  This point is further corroborated in the bottom panels of Figure
\ref{h2_map}, which show azimuthally-averaged radial surface density profiles
of the molecular gas and of the $\hi+\hii$ gas. The latter is simply the
initial SPH gas mass with the molecular gas mass subtracted, and has been
corrected (just like the molecular gas phase) for the mass contribution from
helium.

In order to compare \Shihii and \Sh2 with observations, we have set the
radial bin width to $0.5$\,kpc -- the typical radial bin size used for nearby
spirals in the work of \cite{leroy08}\footnote{The \h2 surface density maps in
Figure \ref{h2_map} are averaged over areas $80\times 80\,{\rm pc}$ in size, and
therefore give higher peak surface densities than the radial profiles which are
averaged over $\sim 0.5\,{\rm kpc}$ wide annuli.}. The radially binned \Sh2
reaches $\sim 800-1000$\,\msun\,pc$^{-2}$ in the central regions of our
simulated galaxies, which is comparable to observational estimates of \Sh2 (of
several 100\,\msun\,pc$^{-2}$) towards the centres of nearby spirals
\citep{leroy08}.  In all three galaxies, the H{\sc i}+H{\sc ii} surface density dips
within the central $\sim 1-2\,{\rm kpc}$, coinciding with a strong peak in \Sh2.
Thus, despite the marked increase in the FUV radiation field towards the centre,
the formation of \h2 driven by the increase in gas pressure is able to overcome
photodestruction of \h2 through absoption of Lyman or Werner band radiation. The central \h2 surface
densities are similar for all galaxies and is a direct consequence of very
similar SPH gas surface densities in the centre combined with molecular gas mass
fractions approaching 1.  From $R\sim 2\,{\rm kpc}$ and out to $\sim
10\,{\rm kpc}$, the $\hi+\hii$ surface density remains roughly constant with
values of $\sim 40$, $\sim 70$ and $\sim 100$\,\msun\,pc$^{-2}$ for G1, G2 and
G3, respectively. 

Radial profiles of the \hi and molecular gas surface density that are
qualitatively very similar to our simulations have been observed in several
nearby star-forming disk galaxies \citep[e.g.,][]{leroy08,bigiel08}.  In local
galaxies, however, the \hi surface density, including helium, rarely exceeds
$\sim 10$\,\msun\,pc$^{-2}$, while in our simulations we find \hi+\hii surface
densities that are $4-10\times$higher, which is due to the substantial fraction
of ionised gas in our simulated galaxies.

\subsection{CO line emission maps and resolved excitation
conditions}\label{CO_mom0} 
Moment 0 maps of the CO(1$-$0), CO(3$-$2) and
CO(7$-$6) emission from G1, G2 and G3 are shown in Figure \ref{CO_mom0_f_mom0}.
Both the CO(1$-$0) and CO(3$-$2) emission are seen to trace the \h2 gas
distribution well (Figure \ref{h2_map}), while the CO(7$-$6) emission only
trace gas in the central $\sim 7\,{\rm kpc}$ of the galaxies. 

Also shown in Figure \ref{CO_mom0_f_mom0} (bottom row) are the azimuthally
averaged CO 3$-$2/1$-$0 and 7$-$6/1$-$0 brightness temperature line ratios
(denoted $r_{\rm 32}$ and $r_{\rm 76}$, respectively) as a function of radius
for G1, G2 and G3.  The profiles show that the gas is more excited in the
central $\sim 5\,{\rm kpc}$, where typical values of $r_{\rm 32}$ and $r_{\rm
76}$ are $\sim 0.55-0.65$ and $\sim 0.02-0.08$, respectively, compared to
$r_{\rm 32}\sim 0.5$ and $r_{76}< 0.01$ further out in the disk. This radial
behaviour of the line ratios does not reflect the \h2 gas surface density,
which peaks towards the centre rather than flattens, and gradually trails off
out to $R\sim 12\,{\rm kpc}$ instead of dropping sharply at $R\sim 4-6\,{\rm
kpc}$ (Figure \ref{CO_mom0_f_mom0}). Rather, $r_{\rm 32}$ and $r_{\rm 76}$ seem
to mimick the radial behaviour of \g0 (and thus \cri), which makes sense since
\g0 and \cri are the most important factors for the internal GMC temperature
distribution (Figure \ref{apD3}). The central values for $r_{32}$ and $r_{76}$
increase when going from G1 to G3, as expected from the elevated levels of star
formation density (and of \g0 and \cri, accordingly) in their central regions.
Beyond $\sim 6\,{\rm kpc}$ the line ratios are constant ($\sim 0.5$ and $<
0.01$) and the same for all three galaxies, due to relatively similar \g0 (and
\cri) there.
 
The decrease in $r_{\rm 32}$ towards the centre has also 
been observed in nearby galaxies. 
For M\,51, a spiral galaxy with a smaller SFR than our model galaxies 
(see Appendix\,\ref{apMW}), 
\cite{vlahakis13} measured a median ratio of $r_{\rm 32}=0.54$ for pixels covering 
the central kpc region, but typical ratios of $0.2-0.4$ in the arm and inter-arm 
regions. 
In comparison, our model galaxies display larger $r_{32}$ ratios and a 
slightly less pronounced drop of $0.1$ or less when going from 
the central kpc region to the outskirts of the disk.
\cite{mao10} observed $125$ nearby galaxies of different types 
and found global $r_{32}$ values to be $0.61\pm0.16$
in normal galaxies, and $>0.89$ in starbursts and (U)LIRGs. \cite{iono09} and
\cite{papa12} found slightly lower $r_{32}$ in their samples of (U)LIRGs, with
mean values of $0.48\pm0.26$ and $0.67\pm0.62$ respectively. 
The $r_{32}$ profiles of our model galaxies reveal more highly excited gas 
in their centres than in the disks, with central values similar to the  
observed values towards (U)LIRGs by \cite{papa12} but below those reported by \cite{mao10}.
Finally, \cite{geach12} employed Large Velocity Gradient (LVG) radiative transfer models 
to examine $r_{32}$ in different environments. For quiescent clouds similar to 
very low-excitation gas clouds found in M31, 
they showed that $r_{32}$ can be as low as $\sim0.13$, 
while dense ($n(\h2)>10^4\,\cmpc$) star-forming clouds typically have $r_{32}\sim0.88$.
Most likely a result of their moderate star formation rates, the $r_{32}$ 
radial profiles of our model galaxies lie in between these more extreme cases.

\subsection{The CO-to-\h2 conversion factor}\label{section:alpha-CO}
The CO-to-\h2 conversion factor (\aCO) connects CO(1$-$0) line luminosity (in
surface brightness temperature units) with the molecular gas mass ($M_{\rm
mol}$) as follows:
\begin{equation}
\alpha_{\rm CO}=\frac{M_{\rm mol}}{L'_{\rm CO(1-0)}},
\label{alpha_eq}
\end{equation}
From the CO$(1-0)$ surface brightness and \h2 surface density maps of our
simulated galaxies, we calculate the average \aCO within radial bins from the
galaxy centres (Figure \ref{CO_mom0_f_mom0}). 
The resulting radial profiles show that \aCO is essentially constant as a function
of radius, taking on values in the range
$\sim1.3-1.8$\,\msun\,pc$^{-2}$\,(K\,km\,s$^{-1}$)$^{-1}$ across all three
galaxies, except in the outskirts of G2 where \aCO shoots up to higher values. 
The high values of \aCO in the outskirts of G2 
reflect the very low molecular gas surface densities 
(see Figure\,\ref{h2_map}) and resulting lack of CO line emission, 
rendering \aCO a meaningless quantity in this region.
In order to compare with the stellar properties in 
Table\,\ref{model_galaxies_t1}, we will use only the region within $R=15\,$kpc 
for the further analysis of CO line emission in G02. 
Small systematic changes in \aCO with radius are seen
in all three simulated galaxies as \aCO is systematically below (above) the
global \aCO value, marked with dashed lines, at $\ls 5\,{\rm kpc}$ ($\gs 5\,{\rm kpc}$).
\cite{blanc13} measured a drop in \aCO by a factor of two when going
from $R\sim7$\,kpc to the $R<2$\,kpc central region of the Sc galaxy NGC\,628, assuming a constant
gas depletion timescale when converting SFR surface densities into gas masses.
For comparison, the radial \aCO profiles of our model galaxies 
only drop by about $10\%$ from $R\gtrsim7$\,kpc 
to the $R<2$\,kpc central region.

\cite{sandstrom13} measured and examined the resolved \aCO values in a sample of 26 
nearby spiral galaxies. 
A disk-averaged \aCO value was calculated for each galaxy, by tiling them with 
pixels of spacing $37.5\arcsec$ (corresponding to pixel sizes of $0.7-3.9\,{\rm kpc}^2$)
and calculating the mean of \aCO values for pixels with 
a sufficient signal-to-noise ratio. 
It was found that, on average, the galaxies 
have a central ($R\leq1$\,kpc) \aCO value a factor of two 
below the disk-averaged \aCO value. 
We calculate disk-averaged \aCO for our model galaxies in a similar way by 
tiling the galaxies seen face-on with $1\,{\rm kpc^2}$-sized pixels 
and taking the average over pixels within the 
cut-out radii $R_{\rm cut}$ given in Table\,\ref{model_galaxies_t1}. 
The resulting values are shown with dotted lines in Figure 9 and 
are no more than $\sim1.2\times$ the central ($R\leq1\,$kpc) \aCO. 
Our simulated galaxies, therefore do not quite reproduce the drop in 
\aCO typically observed when going from the disk to the central regions 
of local, spiral galaxies. 
Although, we note that for three galaxies from the \cite{sandstrom13} sample 
(NGC\,3938, NGC\,3077 and NGC\,4536) \aCO changes by only $10-16\,\%$  
from the disk to the centre, which is in line with our simulations. 

The \aCO factor is expected to depend on \Z, as higher metallicity means higher C and O
abundances as well as more dust that helps shield CO from photodestruction by
FUV light, thereby leading to a possibly lowering of \aCO. A comparison of the
\aCO radial profiles with those of \g0 and \Z in the bottom panel of Figure
\ref{h2_map}, suggests that the transition in \aCO is caused by a change in \g0
rather than a change in \Z, since \aCO and \g0 generally start to drop at
around $R\sim6$\,kpc while \Z already drops drastically at 1\,kpc from the
centre.  Our modeling therefore implies that \aCO is controlled by \g0 rather
than \Z in normal star-forming galaxies at $z\sim2$, in agreement with the
observations by \cite{sandstrom13} who do not find a strong correlation with
\Z. 

From the total molecular gas masses and CO$(1-0)$ luminosities of G1, G2, and
G3, we derive global \aCO factors of $\aCO=1.6$, $1.5$ and
$1.4$\,\msun\,pc$^{-2}$\,(K\,km\,s$^{-1}$)$^{-1}$, respectively.  These values
are lower (by a factor $\sim 3$) than the inner disk MW value \newline ($\alpha_{\rm CO,
MW} \simeq 4.3\pm 0.1\,{\rm \msun\,pc^{-2}\,(K\,km\,s^{-1})^{-1}}$), 
and closer to the typical
mergers/starburst \aCO-values ($\sim 0.2\times \alpha_{\rm CO, MW}$) inferred
from CO dynamical studies of local ULIRGs
\citep[e.g.,][]{solomon97,downes98,bryant99} and $z\sim 2$ SMGs
\citep[e.g.,][]{tacconi08}. Our \aCO-values are also below those inferred from
dynamical modeling of $z\sim1.5$ BzKs \citep[$\alpha_{\rm CO}=3.6\pm 0.8\,{\rm
\msun\,pc^{-2}\,(K\,km\,s^{-1})^{-1}}$; ][]{daddi10}.

The \aCO factors of the $z\sim1-1.3$ star-forming galaxies
studied by \cite{magnelli12} occupy the same region of the $M_*$--\,SFR plan as
our model galaxies, but have \aCO factors at least a factor $\sim6$ higher.
Their \aCO are $\sim5-20\,{\rm \msun\,pc^{-2}\,(K\,km\,s^{-1})^{-1}}$ when
converting dust masses to gas masses using a metallicity-dependent gas-to-dust
ratio, and \cite{magnelli12} further find that the same galaxies have relatively
low dust temperatures of $\lesssim28$\,K compared to the galaxies further above
the main sequence of \cite{rodighiero10}.  The fact that the main-sequence
galaxies of \cite{magnelli12} are all of near-solar metallicity as ours, cf.
Table \ref{model_galaxies_t1}, suggests that other factors, such as dust
temperature caused by strong FUV radition, can be more important than
metallicity in regulating \aCO, in line with our study of the \aCO radial
profiles.
 
\begin{figure*}
\centering
\includegraphics[width=0.95\columnwidth]{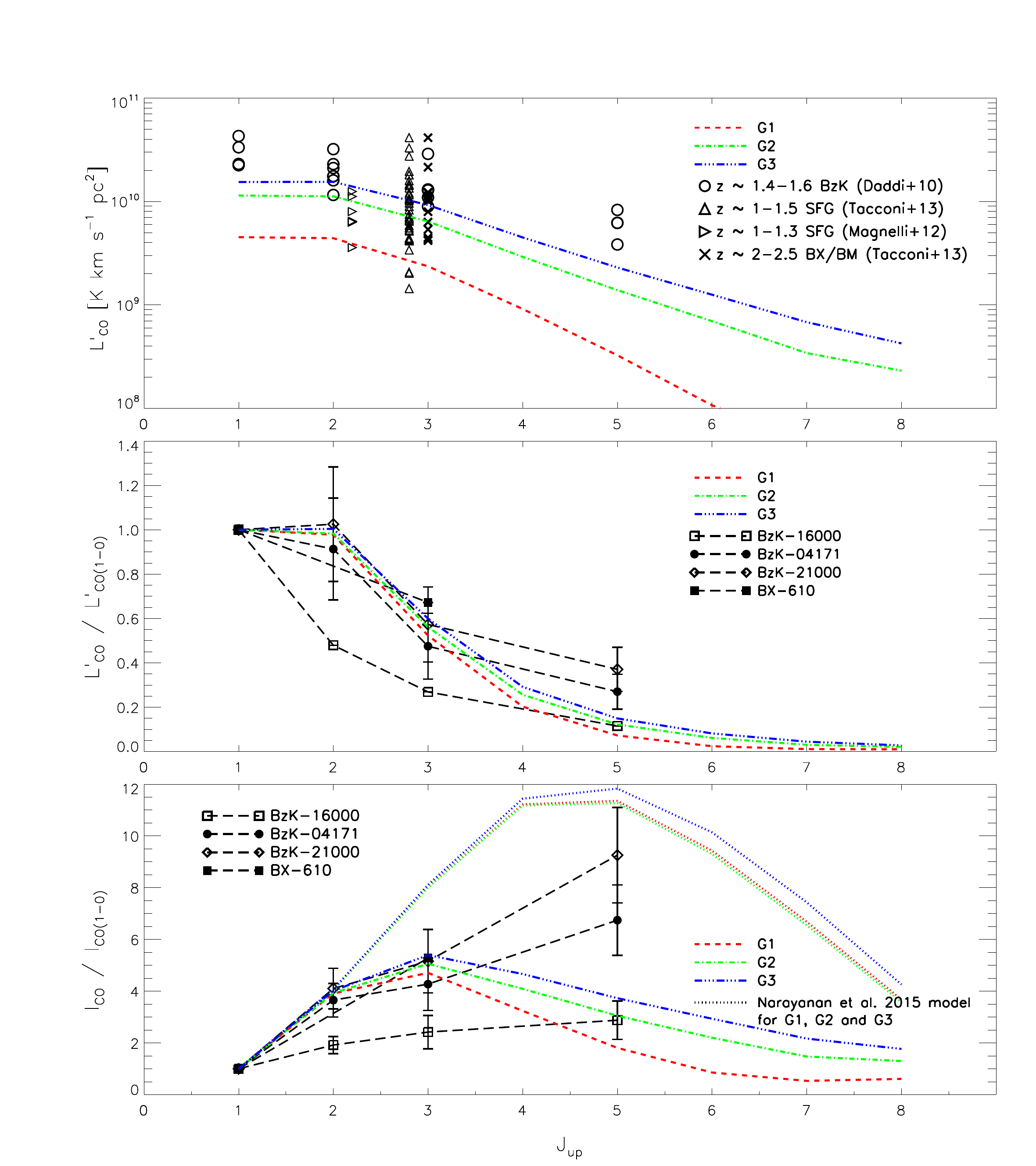}
\caption{\footnotesize{Global CO SLEDs of our three model galaxies G1, G2 and G3 shown as 
red (dashed), green (dash-dot) and blue (dash-dot-dot-dot) curves, respectively.  
The SLEDs are given as absolute line
luminosities in units of K\,\kms\,pc$^{-2}$ (top panel), as brightness
temperature ratios normalised to the CO(1$-$0) transition (middle panel), and as
velocity-integrated intensity ratios normalised to CO(1$-$0) (bottom panel).
The model CO SLEDs are compared with observations of $z\sim1.4-1.6$ BzK galaxies
\citep[open circles;][]{dannerbauer09,daddi10,daddi15,aravena10,aravena14}, $z\sim1-1.5$
star-forming galaxies \citep[CO(3$-$2); empty triangles;][]{tacconi13},
$z\sim1-1.3$ star-forming galaxies \citep[CO(1$-$0) and CO(2$-$1); right-facing
triangles;][]{magnelli12}, and $z\sim2-2.5$ BX/BM galaxies \citep[CO(3$-$2);
crosses;][]{tacconi13}.  Four BzK galaxies (BzK$-$16000, BzK$-$4171, BzK$-$21000 
and BzK$-$610) have been observed in CO$(1-0)$ and at least one additional
transition to date, and are highlighted in the bottom two panels by connecting 
dashed lines and individual symbols. Also shown in the bottom panel, 
with dotted lines, are the line ratio predictions of
\protect\cite{narayanan14} (D14), calculated for the \SFRsd of our galaxies, 
G1 to G3 from bottom to top (see Section \ref{dis_models}).}}
\label{CO_sled_f_sled}
\end{figure*}

\subsection{Global CO line luminosities and spectral line energy distributions} \label{CO_sled}
The global CO SLEDs of G1, G2 and G3 are shown in Figure \ref{CO_sled_f_sled}
in three different incarnations: 1) total CO line luminosities ($L'_{\rm
CO_{J,J-1}}$), 2) brightness temperature ratios ($T_{\rm B,CO_{J,J-1}}/T_{\rm
B,CO_{1,0}}=L'_{\rm CO_{J,J-1}}/L'_{\rm CO_{1,0}}$) and 3) line intensity
ratios ($I_{\rm CO_{J,J-1}}/I_{\rm CO_{1,0}} = L'_{\rm CO_{J,J-1}}/L'_{\rm
CO_{1,0}} \times \left ( \nu_{\rm CO_{J,J-1}}/\nu_{\rm CO_{1,0}}\right )^2$).
We have also compiled relevant CO observations of normal star-forming galaxies
at $z \sim 1-3$ in order to facilitate a comparison with our simulated CO
SLEDs. In addition to the $z\sim 1-1.5$ BzK and $z\sim 2-2.5$ BX/BM samples by
\cite{daddi10} and \cite{tacconi13}, respectively, this includes 7 star-forming
galaxies (SFGs) at $z\sim1.2$ \citep{magnelli12} and 39 SFGs at $z\sim1-1.5$
\citep{tacconi13}. The galaxies in the sample from \citet{magnelli12} are all
detected in CO(2$-$1) as well as in {\it Herschel}/PACS bands, and have
$\log({M_*/\msun})=10.36-11.31$ and SFR\,$\sim29-74$\,\sfru. The $z\sim 1-1.5$
SFG sample from \citet{tacconi13} comes from the Extended Growth Strip
International Survey (EGS), is covered by the CANDELS and 3D-HST programs
($J-H$ bands and H$\alpha$ respectively), and has
$\log({M_*/\msun})=10.40-11.23$ and SFR\,$\sim28-630$\,\sfru.

The $z\sim2-2.5$ BX/BM galaxies of \cite{tacconi13} are not only closest in
redshift to our simulated galaxies but also occupy the same region of the
SFR$-M_*$ plane (see Figure\,\ref{model_galaxies_f_M_SFR}). We find that G2 and
G3, the two most massive ($M_*>10^{11}$\,\msun) and star-forming (SFR
$\gtrsim80$\,\sfru) galaxies of our simulations, have CO(3$-$2) luminosities of
$6.4\e{9}$ and $9.2\e{9}\,{\rm K\,km\,\ps\,pc^2}$, respectively, in the range of
CO(3$-$2) luminosities of the BX/BM galaxies.  G1 is just below this range and
about a factor of 4 below the average BX/BM luminosity
($\sim1.2\pm0.9\times10^{10}\,{\rm K\,km\,\ps\,pc^2}$).  All three simulated
galaxies have CO(2$-$1) and CO(3$-$2) luminosities consistent with those of the
$z\simeq1-1.5$ SFGs observed by \citet{magnelli12} and \citet{tacconi13}, which
span a range in CO(2$-$1) and CO(3$-$2) luminosity of $(3.6-12.5)\times10^9$
and $(1.4-41.3)\times10^9\,{\rm K\,km\,s^{-1}\,pc^2}$, respectively.  Finally,
we see that the $z\sim 1.5$ BzK galaxies in general have higher CO line
luminosities across all observed CO transitions (up to $J_{\rm up} = 5$),
although at the $J=2-1$ and $J=3-2$ transitions there is overlap with G3.
As mentioned in Section \ref{H2_cont}, the molecular gas mass fractions of our
galaxies is a factor $\sim 4-5$ below the mean of the observed galaxies at
$z\sim1-2.5$ with which we compare, and we propose this as the main reason for
the comparatively low CO luminosities. 

Differences in the global CO excitation conditions between G1, G2 and G3 are
best seen in the CO(1$-$0) normalised luminosity (and intensity) ratios, i.e.,
middle (and bottom) panel in Figure \ref{CO_sled_f_sled}. The CO SLEDs of all
three galaxies follow each other quite closely up to the $J=3-2$ transition
where the SLEDs all peak; at $3<J_{\rm up}<7$ the SLEDs gradually diverge.
While the metallicity distributions in G1, G2 and G3 are similar (see Figure \ref{apD1}), 
the rise in \g0 presents a likely cause to the increasing high-$J$ flux 
when going from G1 to G3, as higher \g0 leads to more flux primarily in the $J>4$ 
transitions (see Figure \ref{apD4}).  

Our simulated galaxies are seen to have CO 2$-$1/1$-$0, 3$-$2/1$-$0, and
5$-$4/1$-$0 brightness temperature ratios of $r_{\rm 21}\simeq 1$, $r_{\rm
32}\simeq 0.6$, and $r_{\rm 54}\simeq 0.15$, respectively. The first two ratios
compare extremely well with the line ratios measured for BzK$-$4171 and
BzK$-$21000, i.e., $r_{\rm 21}\simeq 0.9-1$, and $r_{32}\simeq 0.5-0.6$
\citep{dannerbauer09,aravena10,daddi10,daddi15}, and suggest that our
simulations are able to emulate the typical gas excitation conditions
responsible for the excitation of the low-$J$ lines in normal $z\sim 1-3$ SFGs.
In contrast, $r_{\rm 54}=0.3-0.4$ observed in BzK$-$4171 and BzK$-$21000
\citep{daddi15}, is nearly $2\times$\,higher than our model predictions.
\cite{daddi15} argue that this is evidence for a component of denser and
possibly warmer molecular gas, not probed by the low-$J$ lines.  In this
picture, we would expect CO(4$-3$) to probe both the cold, low-excitation gas as
well as the dense and possibly warm star-forming gas traced by the CO(5$-$4)
line, and we would expect CO(6$-$5) to be arising purely from this more highly
excited phase and thus departing even further from our models. 

However, significant scatter in the CO line ratios of main-sequence galaxies is
to be expected, as demonstrated by the significantly lower line ratios observed
towards BzK$-$16000: $r_{\rm 21}\simeq 0.4$, $r_{\rm 32}\simeq 0.3$, and $r_{\rm
54}\simeq 0.1$ \citep{aravena10,daddi10,daddi15} and, in fact, the average
$r_{\rm 54}$ for all three BzK galaxies above \citep[$\simeq 0.2$;][]{daddi15} is
consistent with our models.  It may be that G2 and G3, and perhaps even G1, are
more consistent with the CO SLEDs of the bulk $z\sim 1-3$ main-sequence
galaxies.  We stress, that to date, no $z\sim 1-3$ main sequence galaxies have
been observed in the $J=4-3$ nor the $6-5$ transitions, and observations of
these lines, along with low- and high-$J$ lines in many more BzK and
main-sequence $z\simeq 1-3$ galaxies are needed in order to fully delineate the
global CO SLEDs in a statistically robust way. 

\section{Testing different ISM models} \label{dif_ISM}
In this section we investigate the effects on our simulation results when
adopting i) a more top-heavy GMC mass spectrum with a slope of $\beta=1.5$, ii)
a steeper GMC density profile, i.e., a Plummer profile exponent of $-7/2$, and
iii) a GMC model grid that includes \Pe as a fourth parameter. In order to
carry out option iii), the GMC model grid was produced for $\Pe/k_{\rm B} =
10^{4}\,{\rm cm^{-3}\,K}$ (default), $10^{5.5}\,{\rm cm^{-3}\,K}$ and
$10^{6.5}\,{\rm cm^{-3}\,K}$, allowing for a look-up table of CO SLEDs in (\g0,
$m_{\rm GMC}$, \Z, \Pe) parameter space.  

\begin{figure}
\centering
\includegraphics[width=0.6\columnwidth]{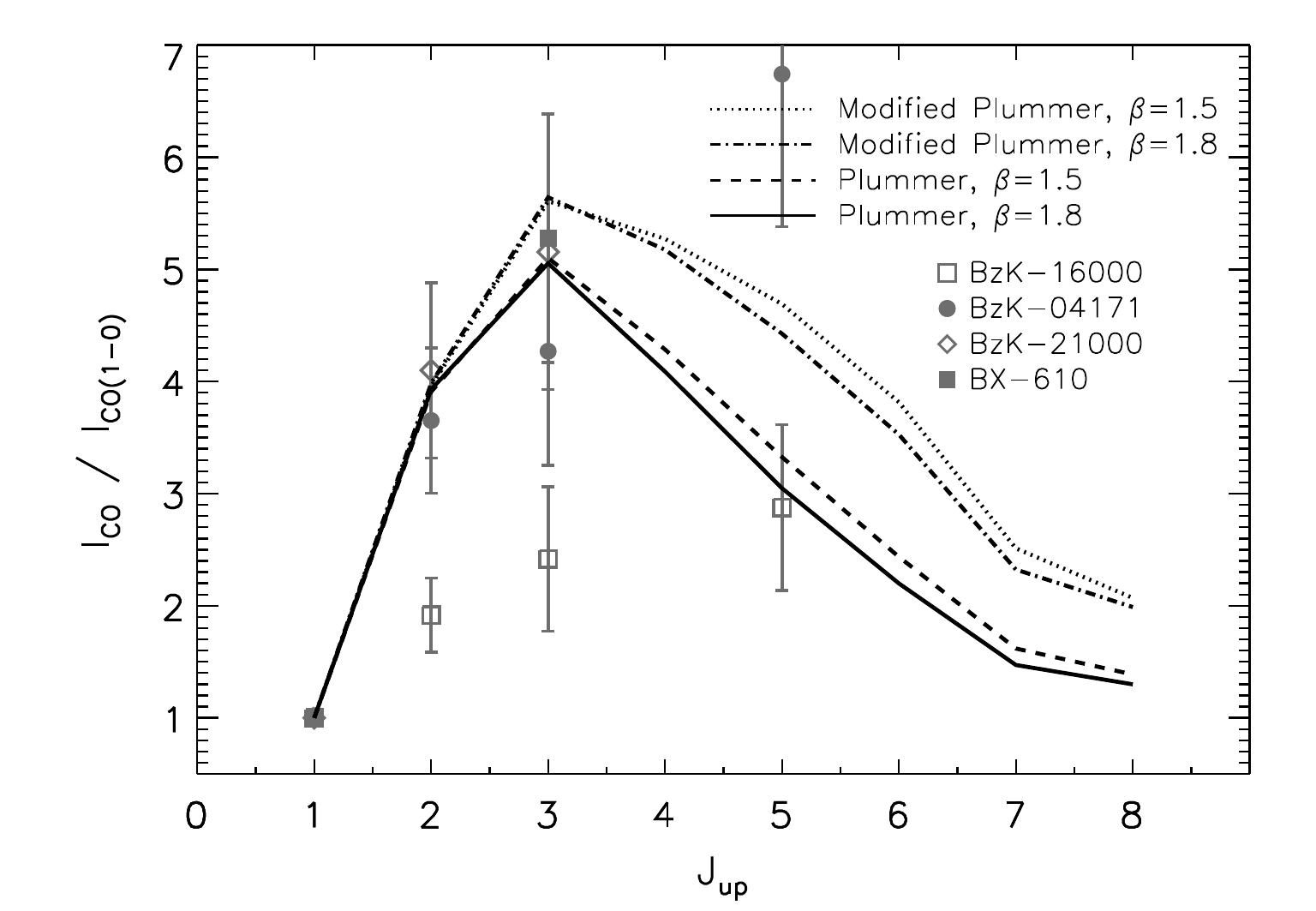}\\
\includegraphics[width=0.6\columnwidth]{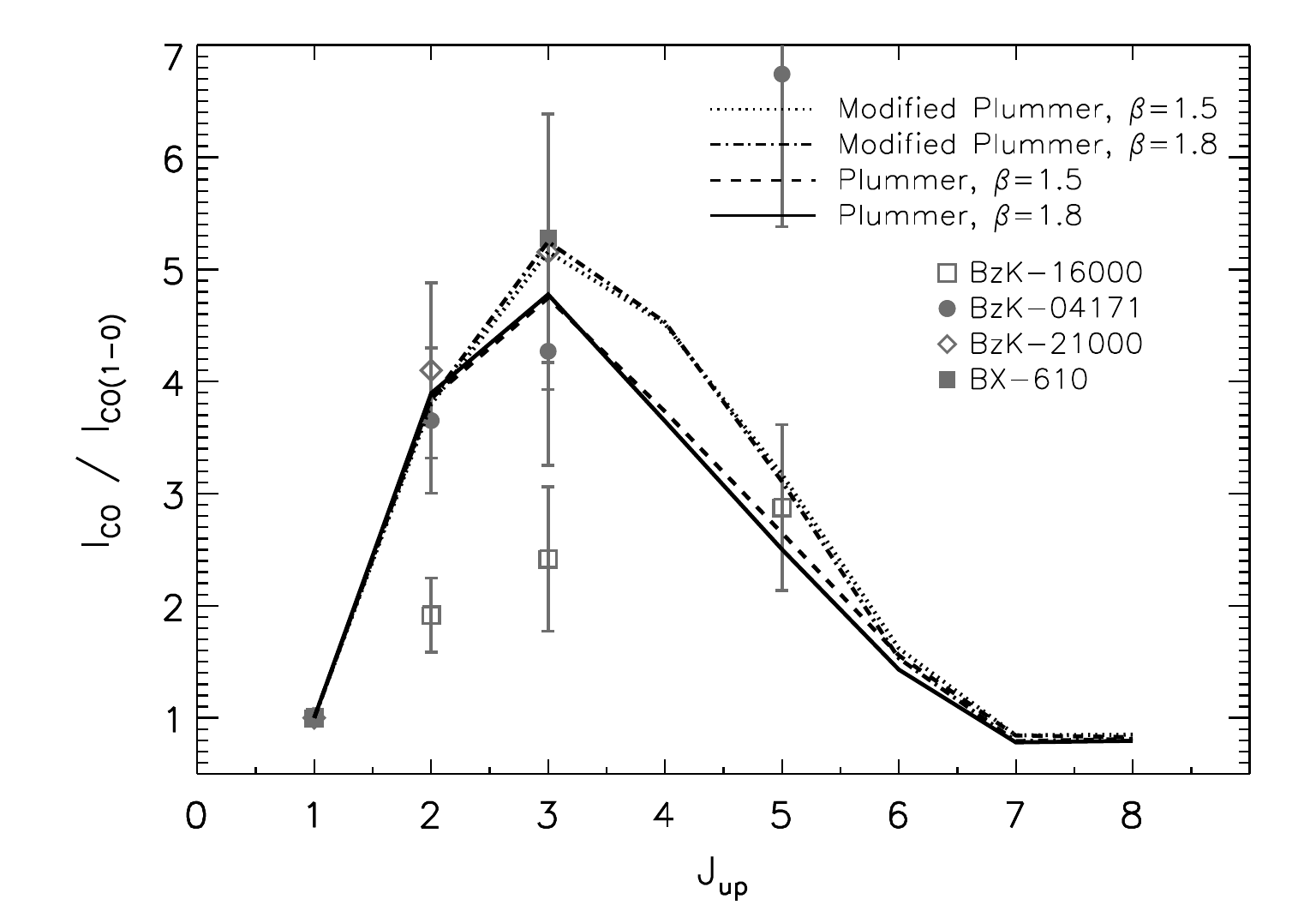}
\caption{\footnotesize{Global CO SLEDs of our model galaxy G2 for different choices of ISM
prescriptions. {\it Top}: For a pressure fixed of $10^4\,{\rm \cmpc\,K}$.  {\it
Bottom}: Using the \Pe as a fourth parameter in the GMC model grid (see Section
\ref{gmcgrid}).}}
\label{CO_sled_ism}
\end{figure}

We examine the effects that options i)-iii) have on the global CO SLED of
galaxy G2 (Figure \ref{CO_sled_ism}) and how combinations of option i) and ii)
change the values of the global CO-to-H$_2$ conversion factor for all three
simulated galaxies (Table \ref{aCOt}).  Also, we show the impact that changes
ii) and iii) have on the CO SLEDs of the individual GMC grid models in Figures
\ref{apD5} and \ref{apD6}. 

Changing the GMC mass spectrum from $\beta=1.8$ to $1.5$ leaves the CO SLED
virtually unchanged with only a marginal increase in the line flux ratios for
$J_{\rm up}\geq4$. This holds true regardless of what has been assumed for ii)
and iii) (compare solid vs.\ dashed and dot-dashed vs.\ dotted curves in Figure
\ref{CO_sled_ism}). Also, changing $\beta$ from 1.8 to 1.5 does not lead to any
significant change in \aCO (Table \ref{aCOt}).

Adopting the modified Plummer density profile with an exponent of $-7/2$
instead of $-5/2$ results in significantly higher global line ratios for
$J_{\rm up} \ge 3$ (see dotted vs.\ dashed curve and dot-dashed vs.\ solid
curves in Figure \ref{CO_sled_ism}). This is due to the higher central
densities achieved for the modified Plummer profile which, as shown in Figure
\ref{apD5}, leads to an increase in the excitation of the higher $J$ lines
relative to that of the low-$J$ lines. Significantly higher \aCO values
($\simeq 3.6\,{\rm \msun\,pc^{-2}\,(K\,km\,s^{-1})^{-1}}$) are obtained when
adopting a modified Plummer profile (Table \ref{aCOt}). These are in excellent
agreement with the \aCO values inferred for $z\simeq 1.5-2$ BzK galaxies by
\citet{daddi10}. In our simulations, the higher \aCO values are due the
steeper density profile which leads to smaller GMC sizes and therefore smaller
CO$(1-0)$ fluxes (Figure\,\ref{apD5}) and, ultimately, a lowering of the total
CO$(1-0)$ luminosity of the galaxy.

Including external pressures of $\Pe/k_{\rm B}=10^{5.5}\,{\rm cm^{-3}\,K}$ and
$10^{6.5}\,{\rm cm^{-3}\,K}$ in the GMC model grid (see Section \ref{gmcgrid}),
means that these pressure values are assigned to GMCs with $\Pe/k_{\rm B}$ in
the ranges $10^{4.75}-10^6$ and $> 10^{6}\,{\rm cm^{-3}\,K}$, respectively.
The higher pressures now experienced by these subsets of GMCs results in
smaller radii since $R_{\rm GMC} \propto \Pe^{-1/4}$ for a given mass (see eq.\
\ref{structure_e1}). Their velocity dispersions and densities increase since
$\sigma_v \propto P_{\rm ext}^{1/8}$ and $n_{\rm H_2}(R=0) \propto P_{\rm
ext}^{3/4}$ for a given mass (see eqs.\ \ref{structure_e2} and
\ref{equation:plummer-profile}). The net effect this has on the global CO SLED
is depicted in the bottom panel of Figure \ref{CO_sled_ism}.  Line ratios with
$J_{\rm up}=3$ are lowered, and increasingly so for higher $J$, resulting in a
steeper decline at high ($J_{\rm up}\geq4$) transitions. 

\begin{center}
\begin{table}
\centering
\caption{The global \h2-to-CO conversion factors (in units of ${\rm
\msun\,pc^{-2}\,(K\,km\,s^{-1})^{-1}}$), averaged over G1, G2, and G3, for combinations
of assumptions i) and ii). The pressure has been kept fixed at $\Pe/k_{\rm
B}=10^4\,{\rm\cmpc\,K}$.}
\begin{tabular}{l|ll} 
\hline
\hline
 	 				&	$\beta=1.8$     	& 	$\beta=1.5$      	\\\hline 
Plummer profile				&	$1.5\pm0.1$		& 	$1.5\pm0.1$ 		\\ 
Modified Plummer profile		&	$3.6\pm0.4$		& 	$3.5\pm0.5$     	\\ 
\hline
\end{tabular}
\\
\label{aCOt}
\end{table}
\end{center}

\section{Comparison with other models}
\label{dis_models}
The sub-grid physics implemented by \sigame assumes that the scaling-laws and
thermal balance equations that have been established for GMCs in
our own Galaxy can be applied to the ISM conditions in high-$z$ galaxies.
In this respect, \sigame does not differ from most other numerical simulations
of the molecular line emission from galaxies
\citep[e.g.,][]{narayanan06,pelupessy06,greve08,christensen12,lagos12,munoz13,narayanan14,lagos12,popping14},
although there are of course differences in the range and level of detail of
the various sub-grid physics implementations.  Here we highlight and discuss
some of these differences between \sigame and other simulations.

First, however, we compare the \sigame CO SLEDs of G1, G2, and G3 with those
predicted by other CO emission simulations of similar main-sequence galaxies.
A direct comparison can be made with the models presented by
\cite{narayanan14}, where the global CO SLED is parametrised as a function of
the luminosity weighted SFR surface density ($\Sigma_{\rm SFR}$). 
We approximate a luminosity weighted $\Sigma_{\rm SFR}$ by calculating a mean of 
$\Sigma_{\rm SFR}$ for $1\,{\rm kpc}^2$ pixels across the galaxies seen face-on, 
weighted by the total SFR within the pixels. 
With this procedure, G1, G2 and G3 have $\Sigma_{\rm SFR}$ of 
$3.0$, $2.9$ and $3.8\,\sfru\,{\rm kpc}^2$, respectively, and the resulting 
CO SLEDs inferred from the \cite{narayanan14} parametrisation
are shown as dotted lines in the bottom panel of Figure \ref{CO_sled_f_sled}.
These are seen to peak at $J_{\rm up}=5$ and not $3$ as the \sigame CO SLEDs do.
Also, the line flux ratios are significantly higher for all transitions above 
$J_{\rm up}=2$. 
Both set of models agree on the $r_{21}$ ratio, but follow distinct CO SLED 
shapes for the remaining transitions. 
The models also seem to suggest one dominating ISM 
phase, whereas the observed CO SLED of the BzK galaxies 
indicate a more complex multi-phased ISM.

Combining a semi-analytical galaxy formation model with a photon-dominated
region code, \cite{lagos12} made predictions of the CO SLEDs of $z=2$ galaxies
as a function of their IR luminosities. They found that for galaxies with
infrared luminosities $L_{\rm IR}\sim10^{11.6}-10^{12.2}\,\lsun$ -- which is
the range found for G1, G2, and G3 if one adopts $L_{\rm
IR}/L_{\odot}=\rm{SFR}/[\msun\,\rm{yr}^{-1}]\times10^{10}$
(\citealt{kennicutt98} for a Chabrier IMF) -- the CO SLEDs (when
converted to velocity integrated flux ratios in order to compare with the
bottom panel of Figure \ref{CO_sled_f_sled}) peak at $J=3-2$,  
in agreement with our model galaxies. 
However, the line ratios of the CO SLEDs of \cite{lagos12} 
are consistently above ours up to $J_{\rm up}=7$, 
but cross over our models at $J_{\rm up}=7$ and go 
below at $J_{\rm up}=8$. 
In terms of CO line luminosity, our model G1 agrees best with the models of 
\cite{lagos12}, lying within the 10-90 percentile ranges ($80\%$ of their galaxy distribution) 
for the CO luminosities of their sample 
for $2<J_{\rm up}<8$, but slightly below for the first two transitions 
and for $J_{\rm up}>7$. G2 and G3, lying roughly a factor 2-4 above G1, 
are outside the 10-90 percentile ranges of the models from \cite{lagos12}.

\cite{popping14}, in their semi-analytical study of
MS galaxies at $z=0$, 1.2 and 2, found that 
for galaxies with far-infrared (FIR)
luminosities of $L_{\rm FIR}=10^{11}-10^{12}\,\lsun$, the CO SLED
peaks at CO(3$-$2) or CO(4$-$3) at $z=0$, but at CO(6$-$5) at $z=2$, 
as a result of denser and warmer gas in their model galaxies. 
The CO luminosities of our galaxies G2 and G3 are in broad agreement with 
the corresponding models by \cite{popping14} at CO(4$-$3) and (5$-$4) for similar SFRs. 
But at $J_{\rm up}<4$ our galaxies are above 
while at $J_{\rm up}>5$, our galaxies are below those of \cite{popping14} up until 
$J_{\rm up}=9$ where the SLEDs cross again.

\bigskip

\noindent$\bullet$ {\it Implementation of \g0 and \cri}\\
\sigame stands out from most other simulations to date in the way the FUV
radiation field and the CR flux that impinge on the molecular clouds are
modelled and implemented.  Most simulations adopt a fixed, galaxy-wide value of
\g0 and \cri, scaled by the total SFR or average gas surface density across the
galaxy \citep[e.g.,][]{lagos12,narayanan14}.  \sigame refines this scheme by
determining a spatially varying \g0 (and \cri) set by the local SFRD, as
described in Section \ref{HI_to_H2}. By doing so, we ensure that the molecular
gas in our simulations is calorimetrically coupled to the star formation in
their vicinity.  If we adopt the method of \cite{narayanan14} and calculate a
global value for \g0 by calibrating to the MW value, using SFR$_{\rm
MW}=2$\,\sfru and $G_{\rm 0,MW}=0.6$\,Habing, our galaxies would have \g0
ranging from about $12$ to $42$, whereas, with the SFR surface density scaling
of \cite{lagos12}, global \g0 values would lie between about $6$ and $9$.  For
comparison, the locally determined \g0 in our model galaxies spans a larger
range from $0.3$ to $27$ (see Figure \ref{apD1}).  \cite{narayanan14} determine
the global value of \cri as $2\e{-17}\Z\ps$, corresponding to values of
$3.7-6.2\e{-17}\ps$ in our model galaxies, when using the mass-weighted mean of
\Z in each galaxy.  Typical values adopted in studies of ISM conditions are around
$(1-2)\e{-17}$\ps \citep{wolfire10,glover12}, but again, the local values of
\cri in our galaxies span a larger range, from $1.3\e{-17}$ to
$1.3\e{-15}$\,\ps.
Adopting the method of \cite{narayanan14} for \g0 and \cri in our galaxies, leads to higher 
CO luminosities at $J_{\rm up}>2$, but very similar CO$(1-0)$ luminosities which together 
with slightly reduced molecular gas masses result in smaller \aCO factors by about $7-8$\,\%.
\bigskip

\begin{figure} 
\hspace{0.5cm}
\includegraphics[width=0.9\columnwidth]{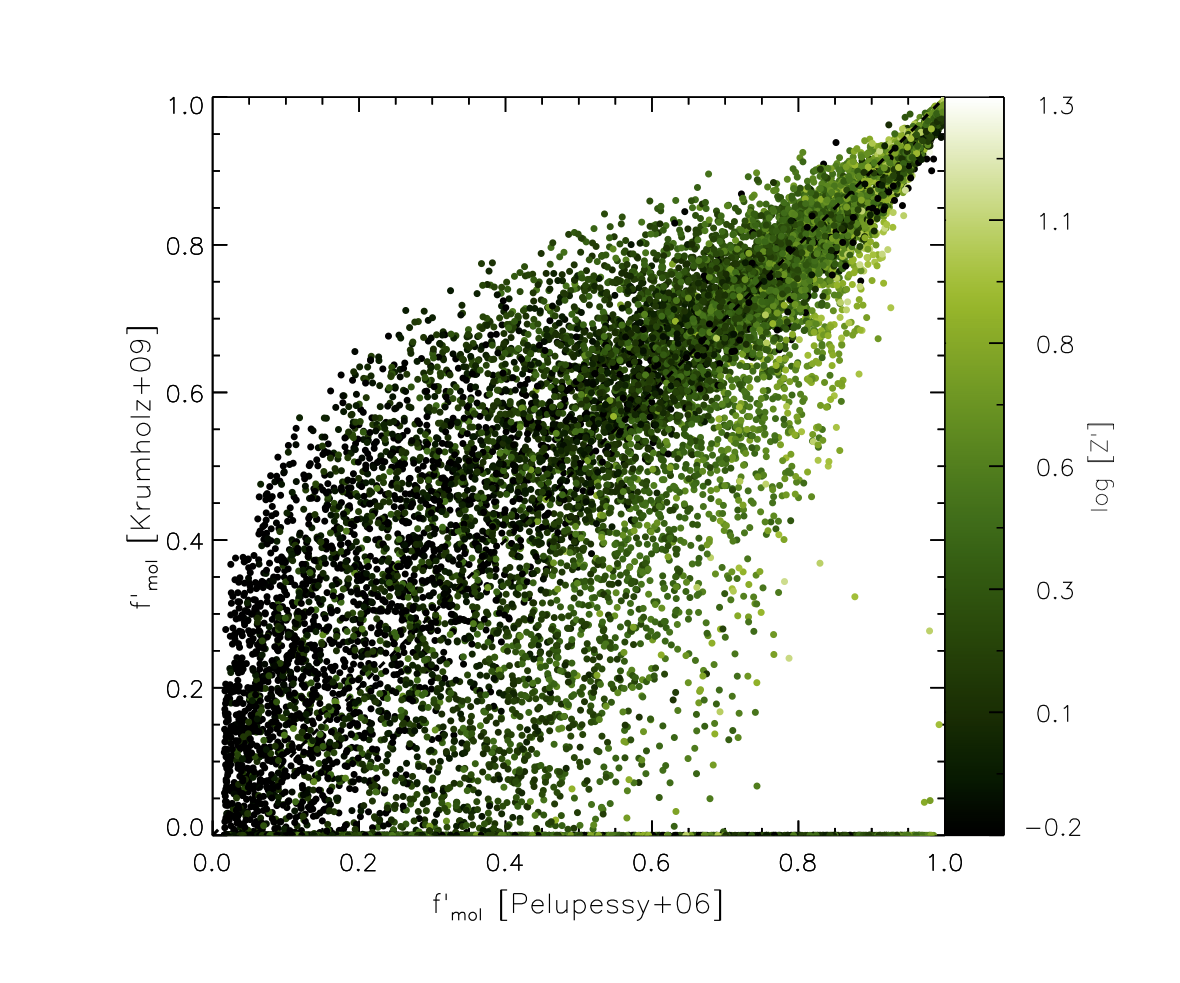}
\caption{\footnotesize{Comparison of two methods for calculating the molecular gas mass
fraction, \fh2, of SPH particles in G1: That of \protect\cite{pelupessy06} with
external cloud pressure derived from hydrostatic mid-plane equilibrium as done
in this work (abscissa) and that of \protect\cite{krumholz09b} (ordinate) used
by e.g., \protect\cite{narayanan14}, color-coded by metallicity.}}
\label{h2_models}
\end{figure}

\noindent$\bullet$ {\it The molecular gas mass fraction}\\ In \sigame the
molecular gas mass fraction (\fh2) is calculated following the work by
\cite{pelupessy06} (P06), in which \fh2 depends on temperature, metallicity,
local FUV field, and boundary pressure on the gas cloud in question.  Other
methods exist, such as those of \cite{blitz06} and \cite{krumholz09b} (K09).
The K09 method, which was adopted by e.g. \cite{narayanan14}, 
establishes \fh2 from local cloud metallicity, dissociating radiation field and 
column density.
In Figure \ref{h2_models} we
compare the K09 method to that of
P06.  The gas surface density that enters in the K09 method was estimated
within 1 kpc of each SPH particle, as used in our calculation of \Pe (see
Section \ref{HI_to_H2}). 

The two methods agree for molecular gas fractions $\gs 80\%$, which are seen to
be dominated by SPH particles with high metallicities (i.e., $\log \Z \gs
0.3$). At lower molecular gas fractions the agreement between the two methods
exhibit an increasing scatter. Also, systematic, metallicity dependent
differences are seen between the two methods.  For high metallicity gas, the
P06 method gives systematically higher molecular gas fractions than the K09
method. At low metallicities (i.e., $\log \Z \ls 0.2$) the K09 method tends
to result in higher molecular gas fractions than P06. The above is the result
of the K09 method having a much weaker dependance on \Z \citet{krumholz09b}
than the P06 method (see Section \ref{HI_to_H2}).  Despite these differences,
we find that using the K09 instead of the P06 method does not change the total
molecular gas mass significantly: $f_{\rm mol}$ for G1, G2 and G3 goes from
12.2, 10.0, 9.2\,\% to 11.4, 10.7 and 10.3\,\%, respectively. 

A common feature of the above \hi $\longrightarrow \h2$ prescriptions is that
they assume an instantaneous \hi to \h2 conversion.  However, as pointed out by
\cite{pelupessy06}, the typical timescales for \h2 cloud formation are of order
$\sim10^7$\,yr, which are comparable to the time-scales of a number of
processes, such as cloud-cloud collisions and star formation, that can
potentially alter or even fully disrupt typical molecular clouds and drive \h2
formation/destruction in both directions.  Ideally, therefore, a comprehensive
modeling of the \hi $\longrightarrow \h2$ transition should be time-dependent.
Relying on the findings of \cite{narayanan11} and \cite{krumholz11}, the static
solution for \h2 formation is a valid approximation for $\Z>0.01$, which is the
case for the three model galaxies studied here.

\bigskip

\noindent$\bullet$ {\it CO abundance}\\
\sigame assumes a constant CO abundance relative to H$_2$, which for the
simulations presented in this work was set to the Galactic CO abundance, i.e.,
$[{\rm CO}/\h2]=2\times10^{-4}$.  In reality, \h2 gas can self-shield better
than the CO gas, creating an outer region or envelope of `CO-dark' gas
\citep{bolatto08}.  Observations in the MW by \cite{pineda13} indicate that the
amount of dark gas grows with decreasing metallicity and the resulting absence
of CO molecules.  By modeling a dynamically evolving ISM with cooling physics
and chemistry incorporated on small scales, \cite{smith14} showed that in a
typical MW-like disk of gas, the CO-dark gas mass fraction, $f_{\rm DG}=$, defined
as having integrated CO intensity $W_{\rm CO}<0.1$\,K\,km\,\ps, 
is about $42$\,\%, but up to $62$\,\% in a radiation field ten times that of
the solar neighbourhood.  While \cite{smith14} kept metallicity fixed to 
solar, \cite{wolfire10} found $f_{\rm DG}$ values of about
$0.5-0.7$ for the low metallicity cases with $\Z=0.5$ in $10^6$\,\msun clouds 
immersed in the local interstellar radiation field. 
Recently, Bisbas et al.\ (2015) argued that CR
ionization rates of $10-50\times\zeta_{\rm{CR,MW}}$ can effectively destroy
most CO in GMCs.  

Thus, it is expected that $[{\rm CO}/\h2]$ is lower in regions of low metallicity
and/or intense FUV and CR fields, effectively leading to underestimates of the \aCO
conversion factor when not accounted for in models. Since in this work we have
restricted our simulations to main-sequence galaxies with solar or higher than
solar metallicities, adopting a constant Galactic $[{\rm CO}/\h2]$ seems a
reasonable choice, but see \cite{lagos12}, \cite{narayanan12} and
\cite{narayanan14} for alternative approaches.

\bigskip

\noindent$\bullet$ {\it GMC density and thermal structure}\\
When solving for the temperature structure of each GMC, \sigame includes
only the most dominant atomic and molecular species in terms of heating and
cooling efficiencies. 
\sigame also ignores heating via X-ray irradiation and turbulence. 
In particular, it has been shown with models that 
turbulent heating can completely control the mean temperature of 
molecular clouds, exceeding the heating from cosmic rays \citep{pan09}. 
Our comparison of observed CO SLEDs at $z=2$ with those of \sigame, 
suggest that our simulations are missing a warm gas component, 
which could be explained, at least in part, by the exclusion of turbulent heating.
We have, however, checked our \Tk$-$\nh2 curves in Figure \ref{apD3} against those of
\cite{glover12}, who performed time-resolved, high-resolution ($\delta m \simeq
0.05-0.5\,\msun$) SPH simulations of individual FUV irradiated molecular clouds
using a chemical network of 32 species (see their Figure 2). Overall, there is
great similarity in the \Tk vs.\ \nh2 behaviour of the two sets of
simulations. This includes the same main trend of decreasing temperature with
increasing hydrogen density, as well as the local increase in \Tk at $\sim
10^3-10^4$\,\cmpc and the subsequent decrease to $\Tk<10$\,K at
$\nh2>10^{5.5}$\,\cmpc.  Thus, our GMCs models, despite their simplified density
profiles and chemistry, seem to agree well with much more detailed simulations.

\bigskip

\section{Summary}\label{CO:con}
In this chapter we have presented \sigame, a code that simulates the
molecular line emission of galaxies via a detailed post-processing of the
outputs from cosmological SPH simulations. 
A sequence of sub-grid prescriptions are applied to a simulation snapshot 
in order to derive the molecular gas density and temperature from the 
SPH particle information, which includes SFR, gas density, 
temperature and metallicity.

A key aspect of \sigame is the {\it localised} coupling between the star
formation and the energetics of the ISM, where the strength of the local  FUV
radiation field and CR ionization rate that impinge and heat a cloud scales with
the local star formation rate density.
The radial temperature profile of each GMC is calculated by balancing the
heating rate with cooling from \h2, CO, \oi, and \cii lines in addition to
gas-dust interactions. In these thermal balance calculations, \sigame takes
into account the local enrichment of the gas, which is a critical parameter
for the cooling rates.  The CO emission line spectrum from a grid of GMC models
is calculated using the 3D radiative transfer code \lime. 

We used \sigame to create line
emission velocity-cubes of the full CO rotational ladder for three cosmological
N-body/SPH simulations of massive ($M_* \gtrsim 10^{10.5}\,\msun$)
main-sequence galaxies at $z=2$. 

Molecular gas is produced more efficiently towards the centre of each galaxy, 
and while \hi surface gas densities (including helium) do not exceed $\sim100$\,\msun\,pc$^{-2}$ anywhere in the disk, 
central molecular gas surface densities reach $\sim1000$\,\msun\,pc$^{-2}$ 
on spatial scales of $100\,{\rm pc}\times 100\,{\rm pc}$, in good agreement with observations made at 
similar spatial resolution. This strong increase in molecular surface density 
is brought on by a similar increase in total gas surface density, 
overcoming the increase in photo-dissociating FUV field towards the centre of each galaxy.

Turning to the CO emission, the velocity-integrated moment 0 maps reveal distinct 
differences in the various transitions as molecular gas tracers. 
The morphology of molecular gas in our model galaxies is well reproduced in CO(1-0), 
but going to higher transitions, the region of CO emitting gas shrinks towards the galaxy centres.
The global CO SLEDs of our simulated galaxies all peak at $J=3-2$. 
Recent CO$(5-4)$ observations of $z\sim1.5$ BzK galaxies seem to suggest that 
these galaxies actually peak at higher $J$. 
The CO$(3-2)$ line luminosities of our model galaxies 
are within the range of 
corresponding observed samples at redshifts $z\sim1-2.5$, 
however on the low side. 
In particular, the model galaxies are below or at the CO luminosities 
of BzK-selected galaxies of comparable mass and SFR but at $z\sim1.5$. 
The low luminosities are most likely a consequence of molecular 
gas mass fractions in our galaxies being 
about $\sim4-5$ times below the observed values in the star-forming galaxies 
at $z=1-2.5$ used to compare with. 

Combining the derived \h2 gas masses with the CO$(1-0)$ line emission found, 
we investigate local variations in the CO-\h2 conversion factor \aCO. 
The radial \aCO profiles all show a decrease towards the galaxy centres, 
dropping by a factor of $\sim1.2$ in the central $R\leq2$\,kpc region compared to the disk average, 
the main driver being the FUV field rather than a gradient in density or metallicity. 
Global \aCO factors range from $1.4$ to $1.6$\,\msun\,pc$^{-2}$\,(K\,km\,s$^{-1}$)$^{-1}$ 
or about $0.3$ times the MW value, but closer to values for $z\sim1.5$ normal star-forming galaxies 
identified with the BzK colour criteria. 
Changing the GMC properties from what is observed in the MW and local galaxies to a 
steeper GMC density profiles and/or a shallower GMC mass spectra, 
resulted in elevated \aCO values of up to $0.9$ times the MW value.

The CO luminosity ratios of CO 3$-$2/1$-$0 and 7$-$6/1$-$0 ($r_{32}$ and $r_{76}$ respectively) 
drop off in radius about where the FUV radiation drops in intensity, and are thus likely controlled by 
the FUV field as is \aCO. The global ratios of $r_{21}\simeq 1$ and 
$r_{32}\simeq 0.6$ agree very well with observations of BzK galaxies, 
while the $r_{54}$ of about $0.15$ 
is low compared to recent observations in BzK$-$4171 and BzK$-$21000. 
The low flux from high CO transitions in our models compared to observations could be 
explained, at least in part, by our omission of turbulent heating in \sigame.
However, more observations of $J_{\rm up}>3$ lines towards high-$z$ main-sequence galaxies,
such as the BzKs, are still needed in order to determine the
turn-over in their CO SLEDs and better constrain the gas excitation. 

Finally, we note that \sigame in principle is able to simulate the
emission from a broad range of molecular and atomic lines in the far-IR/mm
wavelength regime provided measured collision rates exist, such as those found in the
LAMDA database\footnote{\url{http://www.strw.leidenuniv.nl/~moldata/}, \cite{schoier2005}}. 
For more on future applications of \sigame, 
see the Outlook chapter after the following chapter.

\clearpage

\begingroup
\section{References}
\def\chapter*#1{}
\bibliographystyle{apj} 
	\setlength{\bibsep}{1pt}
	\setstretch{1}
\bibliography{bibsCO}
\endgroup

%% file: CII.tex
\chapter{\cii line emission from galaxies} \label{gas2}

\section{Probing the neutral and ionized gas}\label{1:CII}

\begin{wrapfigure}[15]{l}{6cm}
\centering
\includegraphics[width=6cm]{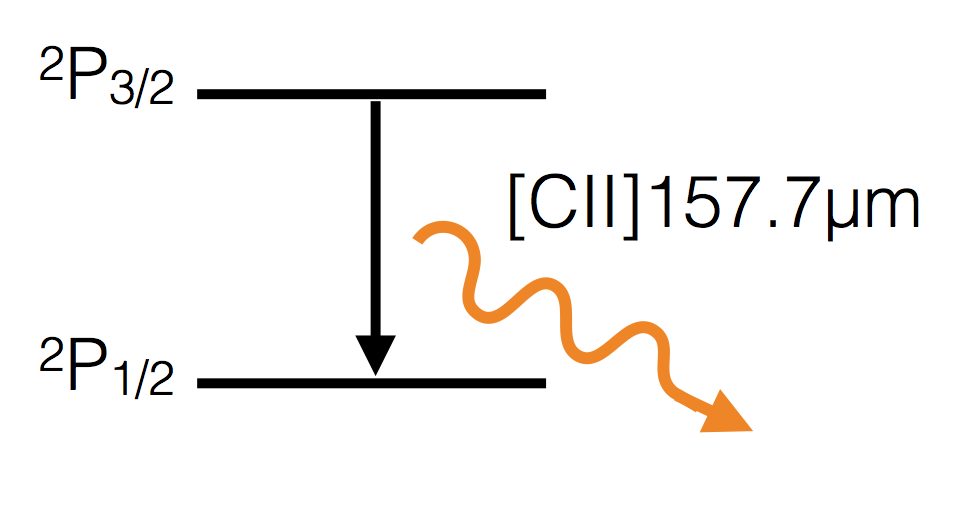}
\caption{\footnotesize{The two level system of ionized carbon leading to the line emission at $\sim158$\,$\mu$m. 
The excitation happens via collisions with either electrons, atoms or molecules.}}
\label{f:CIIillu}
\end{wrapfigure}

Being the fourth most common element after hydrogen in the Milky Way (MW), 
it comes as no surprise that carbon is nearly ubiquitous thoughout the various phases of the 
ISM in star-forming galaxies \citep[e.g.][]{dartois07,esteban14,james14}. 
Much of that carbon is ionized by the ultraviolet (UV) radiation permeating the ISM, due 
to the low ionization potential of carbon (11.3\,eV cf. 13.6\,eV for hydrogen). 
As a result, singly ionized carbon (C{\sc ii}) is found in regions of ionized 
as well as neutral gas. 
In both regions, the $^2$P$_{3/2}$-$^2$P$_{1/2}$ fine structure transition of 
C{\sc ii} can be collisionally excited by electrons ($e^-$), atoms (\hi) or molecules (\h2), 
depending on the gas phase, resulting 
in an emission line at $157.714\,{\rm \mu m}$ or $1900.5369\,{\rm GHz}$ 
(hereafter \cii). 
The critical densities of \cii are only $16\,$cm$^{-3}$, $2400\,$cm$^{-3}$ and
$4800\,$cm$^{-3}$ for collisions with $e^-$, \hi and \h2 respectively 
at a temperature of 500\,K \citep{goldsmith12}. 
Together, these facts are what makes \cii one of the strongest cooling
lines of the ISM, with a line luminosity equivalent to $\sim 0.1-1\%$ of
the far-infrared (FIR) luminosity of galaxies
\citep[e.g.][]{stacey91,brauher08}.

\section{Observations of \cii emission in galaxies at high and low redshift}
Due to high atmospheric opacity at these frequencies,
observations of \cii\ in the local Universe must be done at high
altitudes or in space. 
Indeed, the very first detections of \cii\ towards
Galactic objects \citep{russell80,stacey83,kurtz83} 
and other galaxies \citep{crawford85,stacey91,madden92} were done with
airborne observatories such as the NASA Lear Jet and the Kuiper Airborne
Observatory. 
The advent of the Infrared Space Observatory (ISO) allowed for the first
systematic \cii\ surveys of local galaxies \citep[e.g.][]{malhotra97,luhman98,luhman03}.
Detections of \cii\ at high-$z$ have also become feasible over recent years 
with ground-based facilities \citep{hailey-dunsheath10,stacey10}, as well as the
{\it Herschel} Space Observatory \citep[see the recent review by ][]{casey14}.

\subsection{The \cii deficit}
\begin{figure}[!htbp] 
\centering
\includegraphics[width=0.6\columnwidth]{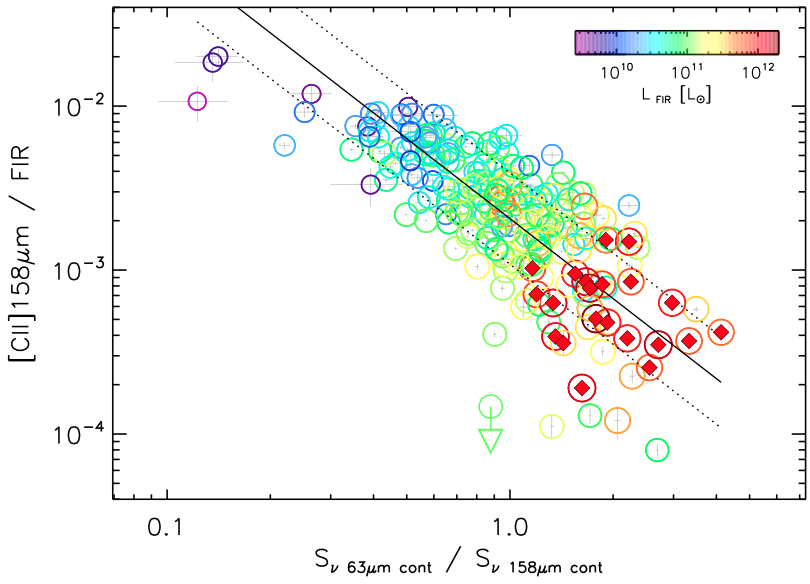}
\caption{\footnotesize{The \cii deficit as investigated and plotted by \cite{diaz-santos13}
for local luminous infrared galaxies (LIRGs). 
Deviation from the global \cii-FIR relation increases 
with dust temperature corresponding to increasing 60/100\,$\mu$m color.}}
\label{f:CIIdef}
\end{figure}

Early observations and modeling suggested that \cii is predominantly associated 
with Photodissociation Regions (PDRs) in the outskirts of molecular clouds exposed 
to intense far-ultraviolet (FUV) ratiation, hence predicting a strong 
correlation between \cii luminosity, \Lcii, 
and that of reprocessed FUV light as measured in far-infrared (FIR), \Lfir 
\citep{tielens85,crawford85}. 
However, a deficit in \Lcii relative to \Lfir was soon observed towards 
local ultraluminous infrared galaxies (ULIRGs) and high-$z$ galaxies 
dominated by an AGN
\citep[e.g.][]{malhotra97,luhman98,malhotra01,luhman03,diaz-santos13,farrah13}. 
Some indications suggest that this `\cii\ deficit' persists at high-$z$, with
the ratio extending to other FIR lines \citep[e.g.][]{gracia-carpio11},
though the existence of a high-$z$ deficit is debated
\citep{hailey-dunsheath10,wagg10,debreuck11,ferkinhoff11,swinbank12}.
\cite{magdis14} find that intermediate $z\sim0.3$ ULIRGs actually fall on the 
relation for local normal galaxies measured by \cite{malhotra01}, suggesting 
that high-$z$ ULIRGs are in fact scaled-up versions of local star-forming galaxies, 
rather than the disturbed systems resulting from mergers that are typically 
associated with local $z<0.2$ ULIRGs.
The \cii\ deficit increases with higher dust temperatures 
on both global and resolved scales in local galaxies \citep[e.g][]{croxall12,diaz-santos13,herrera-camus15}, 
shown with an example in Fig.\,\ref{f:CIIdef}, 
suggesting that dust plays a crucial role in 
controlling the \Lcii/\Lfir\ ratio. 

Various scenarios have been put forward as causing the \cii deficit 
\citep[see a summary of these in e.g.][]{malhotra01,herrera-camus15}, 
but a consesus has not been achieved with observations and models so far. 
Favoured scenarios causing the \cii deficit are ones in which the efficiency of the 
FUV light to heat the gas is decreased or obscuring dust absorbs the FUV 
radiation before it reaches the potential sources of \cii emission. 
Such scenarios include i) large \hii regions created by high ionization parameters 
where most of the FUV radiation is absorbed by dust rather than goes to heating the gas 
via photo-ionization \citep[e.g.][]{tielens85,abel09,gracia-carpio11,croxall12}, 
ii) high levels of grain charging, primarily in PDRs, 
which decrease the efficiency of photoelectric heating of the gas by FUV 
radiation \citep[e.g.][]{malhotra01,croxall12,ibar15}, 
and iii) dust-bounded clouds or PDRs in ULIRGs, where high dust masses in the \hii\ 
regions absorb most FUV light before it reaches regions with more ionized carbon 
\citep{luhman03,farrah13,abel09}. 

\subsection{Contributing gas phases to the \cii emission}
Knowing the relative contributions to \Lcii from different gas phases in the ISM would help 
sorting out in the possible causes for the \cii deficit. 
But observations of \cii emission alone, will not readily give away information 
on the different ISM phases contributing, 
unless combined with some other gas tracer such as 
\nii (122\,$\mu$m) \citep{malhotra01} or 
\hi (21\,cm) as well as CO line emission as done by \cite{pineda13} for the MW, 
in order to separate contributions from ionized, neutral and molecular gas. 
Updated in \cite{pineda14}, the ISM phases considered are found to contribute with rougly amounts to the 
total \Lcii with 30\pct from dense PDRs, 25\pct from cold \hi, 25\pct from CO-dark \h2 gas 
and 20\pct from ionized gas. 
Observing 60 normal, star-forming galaxies in \cii and \nii, \cite{malhotra01} 
estimated that a rough mean of 50\pct of \Lcii comes from ionized gas and the rest from PDRs.
Even larger PDR contributions were found in the giant \hii region N11 in the Large Magellanic Cloud (LMC), 
by \cite{lebouteiller12} who conclude, 
from comparison of \cii with \nii, that 95\pct of \Lcii arises in diffuse PDRs, whereas dense PDRs are better traced by \oil, 
also considered the second most important cooling line in neutral gas. 

\subsection{\cii as a star formation rate tracer} \label{cii:intro:sfr}
Since \cii is sensitive to the local FUV field which itself probes OB 
star formation activity, it was early on suggested that 
\Lcii correlates with SFR of a galaxy, and 
observations soon revealed a \Lcii-SFR correlation for nearby 
galaxies \citep{stacey91,leech99,boselli02}. 
More comprehensive studies were performed by
\citet{delooze11,delooze14} and \citet{farrah13} for
modest star-forming galaxies and ULIRGs in the local Universe. 
The scatter in the \Lcii-SFR relation is substantial though, 
apparently irregardless of the overall galaxy 
classification \citep{sargsyan12}, 
and increasing towards low metallicity, warm dust temperatures and 
large filling factors of ionized, diffuse gas \citep{delooze14}.

With the Photodetector Array Camera \& Spectrometer (PACS) on board 
{\it Herschel}, resolved observations of \cii\ in local galaxies 
became possiple \citep[e.g.][]{delooze14,herrera-camus15,kapala15}, 
and the advent of the Atacama Large Millimeter Array (ALMA) promises to
make such detections and observations relatively routine, even at high-$z$ 
\citep[e.g.][]{wang13}. 
The \Lcii-SFR relation has now also observed on kpc-scales in local galaxies 
as a kind `\cii KS relation' between surface density of SFR, 
\SFRsd, and that of \cii\ luminosity, \CIIsd\ 
\citep[e.g.][]{delooze14,herrera-camus15,kapala15}. 
\cite{kapala15} concluded that \cii\ traces SFR in the spiral arms 
of the Andromeda Galaxy (M31) similarly 
to what is seen in larger samples of more distant galaxies, 
although with a significant contribution to \cii\ from outside star-forming regions. 
and a shallower slope of the \CIIsd-\SFRsd relation on $\sim50$\,pc 
scales than on kpc scales.
For the MW, \cite{pineda14} find that only the combined emission of all gas phases, 
leads to a slope of the \CIIsd-\SFRsd\ relation in agreement with extragalactic observations.
But the \CIIsd-\SFRsd\ relation for local galaxies 
also suffers from a great deal of scatter. 
\cite{delooze14} find, when observing $32$ local dwarf galaxies on kpc-scales, 
that this scatter is most likely due to internal ISM conditions, 
rather than large variations within individual galaxies. 
Analyzing $46$ nearby (mostly spiral) galaxies from the 
{\it Herschel} KINGFISH sample, \cite{herrera-camus15} succeeded in reducing the scatter on 
the \CIIsd-\SFRsd ~relation at warm IR colors, by 
deriving a set of IR color adjustments that can be applied to 
normal, star-forming galaxies in the absence of strong AGNs. 
But a firm physical reason for the scatter is still missing.

Metallicity is a potentially important factor for the \Lcii-SFR, as 
an increase in metallicity translates into a higher mass fraction of carbon 
and dust, both of which affect the \Lcii-SFR relation.
Similar to \cite{delooze14}, \cite{herrera-camus15} found an increased scatter 
around the \Lcii-SFR relation at low metallicities, 
implying that at low metallicities, a non-negligible fraction of the neutral gas cooling 
takes place via the [\oi]\,63\,$\mu$m cooling line instead the \cii\ line. 
In M31, \cite{kapala15} see an increasing trend in \Lcii/\Ltir 
~with radius as expected if \Lcii ~itself dependeded strongly on metallicity, since M31 
excibits a clearly decreasing metallicity with radius \citep{sanders12}. 
However, \cite{kapala15} cannot rule out other factors such as stellar density and radiation field strength. 
\cite{kramer13} found a sharp increase in \Lcii/\Lfir ~at $R\sim4.5$\,kpc in M33, 
but dismiss metallicity as the sole cause, due to the rather shallow metallicity 
gradient in M33 \citep{magrini10}. 
The effect of low metallicity together with the disruption of molecular clouds 
is also becoming an important subject for studies of 
the Epoch of Reionization (EoR) at $z\sim6$, 
as a possible explanation for several non-detections of 
normal star-forming galaxies. Examples include the
lensed Ly$\alpha$ emitter at $z=6.56$ and the Himiko galaxy at
$z\sim6.5$ forming stars at a rate of $\approx10$ and $\sim100$~\sfru,
respectively \citep{kanekar13,ouchi13}, and more recently 3 Lyman Break Galaxies 
at $z\sim7$ of SFR~$\sim10$\,\sfru \citep{maiolino15}.

\section{Modeling \cii emission}
How the physical conditions of the ISM control 
the \cii\ emission can be investigated by applying sub-grid treatment to 
the gas in semi-analytical \citep{popping14,munoz14}
or fully hydrodynamical \citep{nagamine06,vallini13}
simulations of galaxy evolution. 
In their models of \cii emission in high-$z$ galaxies \cite{nagamine06} found that 
\cii emission depends significantly on the amount of neutral gas. 
\cite{vallini13} improved on the method of \cite{nagamine06}, by implementing radiative transfer 
of the UV field and higher resolution simulations of a single $z=6.6$ Ly$\alpha$ emitter, 
though only at two fixed metallicities ($Z=\Zsun$ and $Z=0.02\Zsun$). 
\cite{popping14} managed to reproduce the \cii deficit for all FIR luminosities at $z>1$ as well as 
the \Lcii-FIR correlation at $z=0$, however with the simplification of 
averaging galaxy properties across annuli in the galactic disks. 
\cite{munoz14} modeled $z>6$ Ly$\alpha$ emitters, 
slightly overpredicting the \cii\ while underpredicting the CO line emission, 
when compared to the few observations available, 
but demonstrating that line modeling during the Epoch of Reionization is possible.
In addition, other works have concentrated on single clouds employing either 
radiative transfer \citep{abel09} or an escape probability approach \citep{goldsmith12} 
to derive the \cii\ strength. 

\chapter{Understanding the \Lcii-SFR relation with simulations (Paper II)} \label{paper2}

\section{Aim of this project}
As described in the previous chapter, simulations hold promise in aiding interpretation 
of the \cii emission. 
In this chapter, an adapted version of our code \sigame is presented that is
capable of incorporating \cii\ emission into smoothed particle hydrodynamics
(SPH) simulations of galaxies. We will consider a multi-phased ISM consisting of
molecular clouds, whose surface layers are stratified by FUV-radiation from
localized star formation, embedded within a neutral medium of atomic gas.  In
addition, we will include the diffuse ionized gas intrinsic to the SPH simulations
as a third ISM phase. The temperatures of the molecular and atomic gas are
calculated from thermal balance equations sensitive to the local FUV-radiation
and CR ionization rate.  \sigame will be applied to GADGET-3 cosmological SPH
simulations of seven star-forming galaxies on the main-sequence (MS) at $z=2$ in
order to simulate the \cii\ emission from normal star-forming galaxies at this
epoch, examine the relative contributions to the emission from the molecular,
atomic and ionized ISM phases, and the relationship to the star formation
activity in the galaxies.  Throughout, we adopt a flat cosmology with
$\Omega_{\rm M} = 0.27, \Omega_{\Lambda} = 0.73$, and $h = 0.71$
\citep{spergel2003}.

\section{Methodology overview}\label{section:methodology}
This section gives an overview of our multi-phased ISM model. 
To a large extent the model follows the steps laid out in Chapter\,\ref{paper1}, 
in that it is applied
at the post-processing stage of an SPH simulation. 
As input, the ISM model requires: 
the position ($[x,y,z]$), velocity ($[v_x,v_y,v_z]$), smoothing length ($h$), gas
mass (\msph), hydrogen density (\nH), gas kinetic temperature (\Tk), electron
fraction ($\xe = n_{e}/n_{\rm H}$), SFR, metallicity ($Z$)
as well as the relative abundances of carbon ([C/H]) and oxygen ([O/H]). The key
steps involved in the post-processing are illustrated in
Fig.\,\ref{figure:cartoon} and briefly listed below (with details given in
subsequent sections):

\begin{figure*}[t]
\begin{center}
\includegraphics[width=1.\columnwidth]{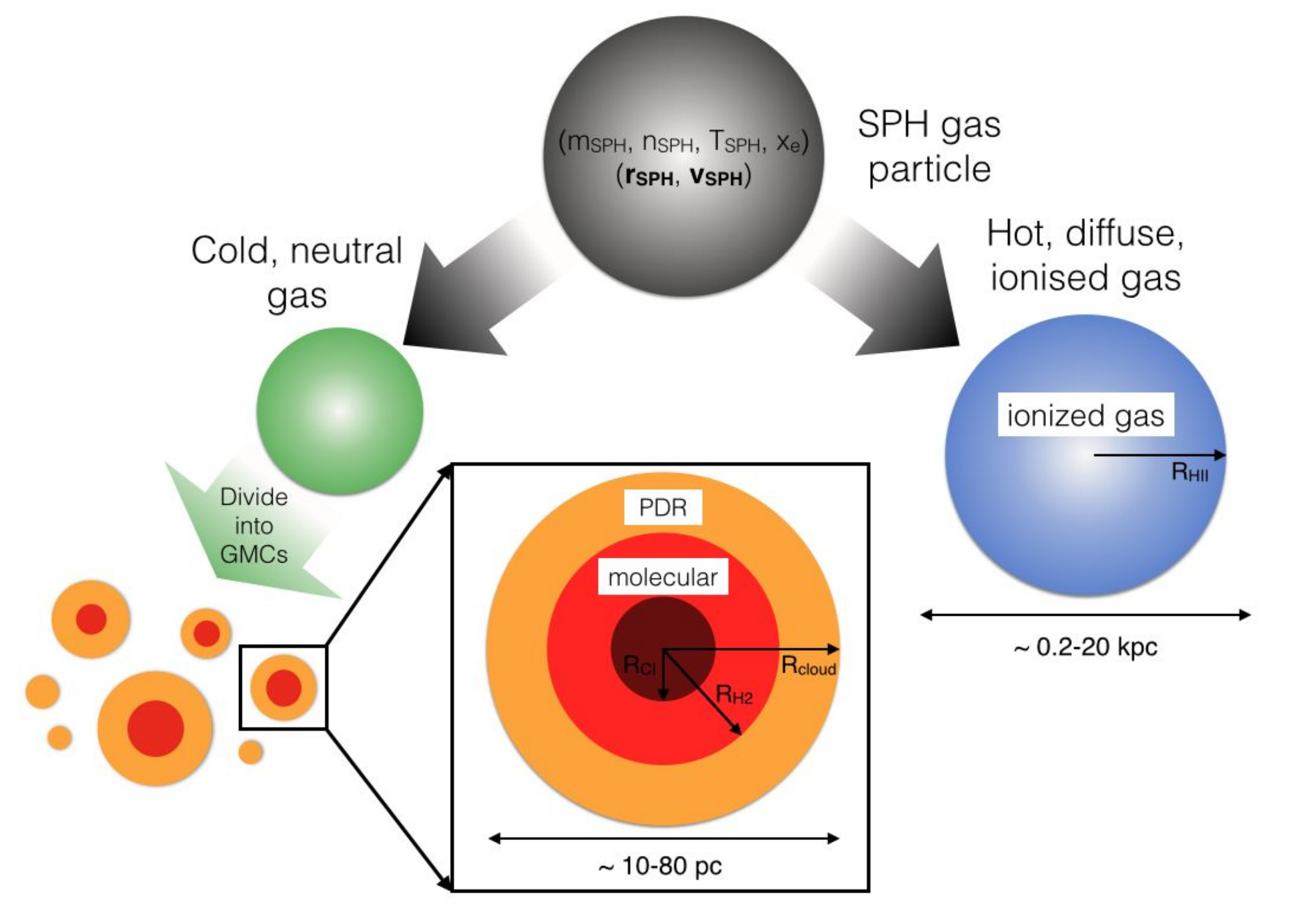}
\end{center}
\caption{\footnotesize{Schematic illustrating the sub-grid procedures applied to the SPH
simulation in post-processing. Each SPH particle is a hybrid of neutral and
ionized gas. The neutral gas associated with each SPH gas particle is divided
into GMCs with masses and sizes following the Galactic mass-spectrum and
mass-size relation for GMCs.  Each GMC has an onion-layer structure, set by the
stratification of the impinging FUV-field, which consists of outer layer of PDR/atomic
gas of H{\sc i} and C{\sc ii}, and an inner molecular region where
carbon is found in its single ionized state, and in neutral form further in
(Section \ref{split1}).  The ionized gas associated with each SPH particle is
assumed to reside in spherical clouds with radii and temperatures given by the
SPH smoothing length and gas temperature (Section \ref{split2}).}}
\label{figure:cartoon}
\end{figure*}

The key steps involved in this process are illustrated in Fig.\,\ref{figure:cartoon} and briefly listed below
(with full details given in subsequent sections):

\begin{enumerate}
\item The SPH gas is separated into its neutral and ionized constituents as
dictated by the electron fraction provided by the \gadget simulations.

\item The neutral gas is divided into giant molecular clouds (GMCs) according to
the observed mass function of GMCs in the Milky Way (MW) and nearby quiescent
galaxies. The GMCs are modeled as logotropic spheres with their sizes and
internal velocity dispersions derived according to pressure-normalized scaling
relations.

\item Each GMC is assumed to consist of three spherically symmetric regions: (1)
a FUV-shielded molecular core region where all carbon is locked up in \ci and
CO; (2) an outer molecular region where both \ci\ and C{\sc ii} can exist; and
(3) a largely neutral atomic layer of \hi, \hii ~and C{\sc ii}.  The last region
mimics the FUV-stratified PDRs observed at the surfaces of molecular clouds
\citep{hollenbach99}. 
This layer can contain both atomic and ionized gas, but 
we shall refer to it simply as the PDR gas.
The relative extent of these regions within each cloud,
and thus the densities at which they occur, ultimately depends on the strength
of the impinging FUV-field and CR ionization rate. The latter are set to scale
with the local SFR volume density and, by requiring thermal balance with the
cooling from line emission (from C{\sc ii}, \oi, and \h2), determine the
temperatures of the molecular and atomic gas phases.

\item The remaining ionized gas of the SPH simulation is divided into \hii\ clouds 
of radius equal to the smoothing lengths, temperature equal to that of the SPH 
simulation and constant density. 

\item The \cii\ emission from the molecular, PDR, and diffuse ionized gas
is calculated separately and summed to arrive at the total \cii\ emission from
the galaxy. In doing so it is assumed that there is no radiative coupling
between the clouds in the galaxy.
\end{enumerate}
The SPH simulations used in this study, and the galaxies extracted from them,
are described in the following section.

%
\begin{table*}
\centering
\caption{Global properties of the seven simulated galaxies used for this work at $z=2$.} 
\begin{tabular}{lccccccc}
\hline
                                		& G1        & G2        & G3       	&	G4		&	G5		&	G6     &	G7		  \\	
\toprule                                                                                                       
\Mstar [10$^{10}$\,\msun]			& 0.36		&	0.78    &	0.95	& 	1.80	&	4.03	&	5.52   &	6.57	 \\
\Mgas [10$^{10}$\,\msun]			& 0.42		&   0.68	&	1.26	&  	1.43	&	2.59	&	2.16   &	1.75	 \\
\Mneu [10$^{10}$\,\msun]			& 0.09		&   0.13	&	0.57	&  	0.20	&	0.29	&	0.29   &	0.39	 \\
$M_{\rm ionized}$ [10$^{10}$\,\msun]	        & 0.33		&   0.55	&	0.69	&  	1.23	&	1.30	&	1.87   &	1.36	 \\
SFR [\sfru]				        & 4.9		&   10.0 	&	8.8	&	25.1	&	19.9	&	59.0   &	37.5	 \\
\SFRsd [\sfru kpc$^{-2}$]	    		& 0.016		&   0.032	&	0.028	&	0.080	&	0.063	&	0.188  &	0.119	 \\	
\Z	    					& 0.43		&   0.85	&	0.64	&	1.00	&	1.00	&	1.67   &	1.72	 \\	
\hline                                                       	                    
\hline                                                                           
\label{table:prop}
\end{tabular}

All quantities have been calculated at $z=2$ using a fixed cut-out radius of
$R_{\rm cut}=10\,{\rm kpc}$, which is the radius at which the accumulative stellar
mass function of each galaxy flattens. \Mgas is the total gas mass, and \Mneu
and $M_{\rm ionized}$ the gas masses in neutral and ionized form, respectively
(see Section \ref{modelISM}). The metallicity ($\Z=Z/Z_{\odot}$) is the mean of
all SPH gas particles within $R_{\rm cut}$. \label{table:global-properties}
\end{table*}

\section{SPH Simulations}\label{sph}
Our simulations are evolved with an updated version of the public \gadget
cosmological SPH code (\citealt{Springel05} and S. Huang et al. 2015 in preparation).  It
includes cooling processes using the primordial abundances as described in
\citet{Katz96}, with additional cooling from metal lines assuming
photo-ionization equilibrium from \citet{wiersma09}.  We use the more recent
`pressure-entropy' formulation of SPH which resolves mixing issues when
compared with standard `density-entropy' SPH algorithms \citep[see][for further
details]{Saitoh13,Hopkins13DISPH}.  Our code additionally implements the
time-step limiter of \cite{Saitoh09}, \cite{Durier12} which improves the accuracy of
the time integration scheme in situations where there are sudden changes to a
particle's internal energy.  To prevent artificial fragmentation
\citep{Schaye08,Robertson08}, we prohibit gas particles from cooling below
their effective Jeans temperature which ensures that we are always resolving at
least one Jeans mass within a particle's smoothing length.  This is very
similar to adding pressure to the ISM as in
\cite{Springel03}, \cite{Schaye08}, except instead of directly pressurizing the gas we
prevent it from cooling and fragmenting below the Jeans scale.

We stochastically form stars within the simulation from molecular gas following
a \citet{Schmidt59} law with an efficiency of 1\% per local free-fall time
\citep{Krumholz07,Lada10}.  The molecular content of each gas particle is
calculated via the equilibrium analytic model of
\citet{Krumholz08,Krumholz09,mckee10}.  This model allows us to regulate star
formation by the local abundance of H$_2$ rather than the total gas density,
which confines star formation to the densest peaks of the ISM.
Further implementation details can be found in \citet{thompson14}.  Galactic
outflows are implemented using the hybrid energy/momentum-driven wind (ezw)
model fully described in \citet{Dave13,Ford15}.  We also account for metal
enrichment from Type II supernovae (SNe), Type Ia SNe, and AGB stars as
described in \citet{Oppenheimer08}.

\subsection{SPH simulations of $z=2$ MS galaxies} \label{tcase}
We use the cosmological zoom-in simulations presented in \citet{thompson15}, and
briefly summarized here.  Initial conditions were generated using the {\small
MUSIC} code \citep{MUSIC} assuming cosmological parameters consistent with
constraints from the {\it Planck} \citep{Planck14} results, namely
$\Omega_{m}=0.3, \Omega_{\Lambda}=0.7,H_0=70,\sigma_8=0.8,n_s=0.96$. Six target
halos were selected at $z=2$ from a low-resolution $N$-body simulation consisting
of $256^3$ dark-matter particles in a $(16h^{-1}{\rm Mpc})^3$ volume with an
effective co-moving spatial resolution of $\epsilon=1.25\,h^{-1}$ kpc. Each
target halo is populated with higher resolution particles at $z=249$, with the
size of each high resolution region chosen to be $2.5$ times the maximum radius
of the original low-resolution halo. The majority of halos in our sample are
initialized with a single additional level of refinement
($\epsilon=0.625\,h^{-1}$kpc), while the two smallest halos are initialized with
two additional levels of refinement ($\epsilon=0.3125\,h^{-1}$kpc).

The six halos produce seven star-forming galaxies at $z=2$ that are free from
all low-resolution particles within the virial radius of their parent halo.
Their stellar masses (\Mstar) range from $3.6\times10^9$ to $6.6\times
10^{10}\,\msun$ and their SFRs from $5$ to $60$\,\sfru (Table
\ref{table:global-properties}). We hereafter label the galaxies G1, ..., G7 in
order of increasing \Mstar. Other relevant global properties directly inferred
from the SPH simulations, such as total gas mass (\Mgas), neutral and ionized
gas masses (\Mneu and $M_{\rm ionized}$, respectively), average SFR surface
density ($\Sigma_{\rm SFR}$), and average metallicity (\Z), can also be found in
Table \ref{table:global-properties}.
\begin{figure}[t]
\begin{center}
\includegraphics[width=\columnwidth]{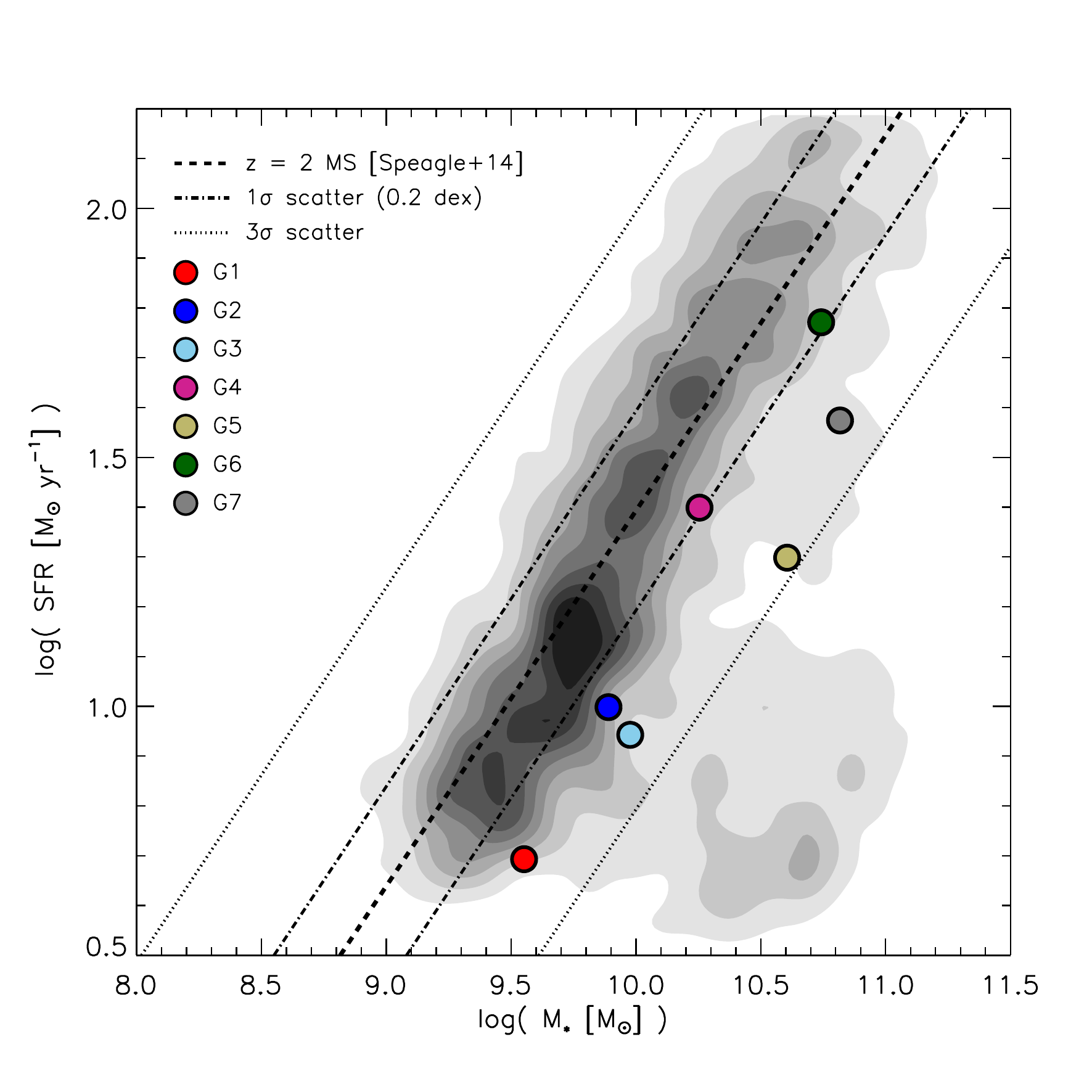}
\end{center}
\caption{\footnotesize{The SFR--M$_{\ast}$ relation at $z\simeq 2$ as determined by
\citet{speagle14} (dashed line) with the location of our seven simulated
galaxies highlighted (filled circles). Dotted-dashed and dotted lines indicate the
$1\sigma$ and $3\sigma$ scatter around the relation of \cite{speagle14}. For
comparison we also show the locus defined by $3754$ $1.4<z<2.5$ galaxies from
the NEWFIRM Medium-Band Survey (gray filled contours), with masses and SFRs
calculated using a Kroupa IMF \citep{whitaker11}.}}
\label{M_SFR}
\end{figure}

Fig.\,\ref{M_SFR} shows the locations of G1, ..., G7 in the SFR$-$\Mstar
diagram.  The galaxies are consistent with observational determinations of the
$z\sim2$ MS of star-forming galaxies
\citep{whitaker11,speagle14}.

\section{Modeling the ISM} \label{modelISM}
\begin{figure}[htbp] 
\hspace*{1.5cm}
\includegraphics[width=0.9\columnwidth]{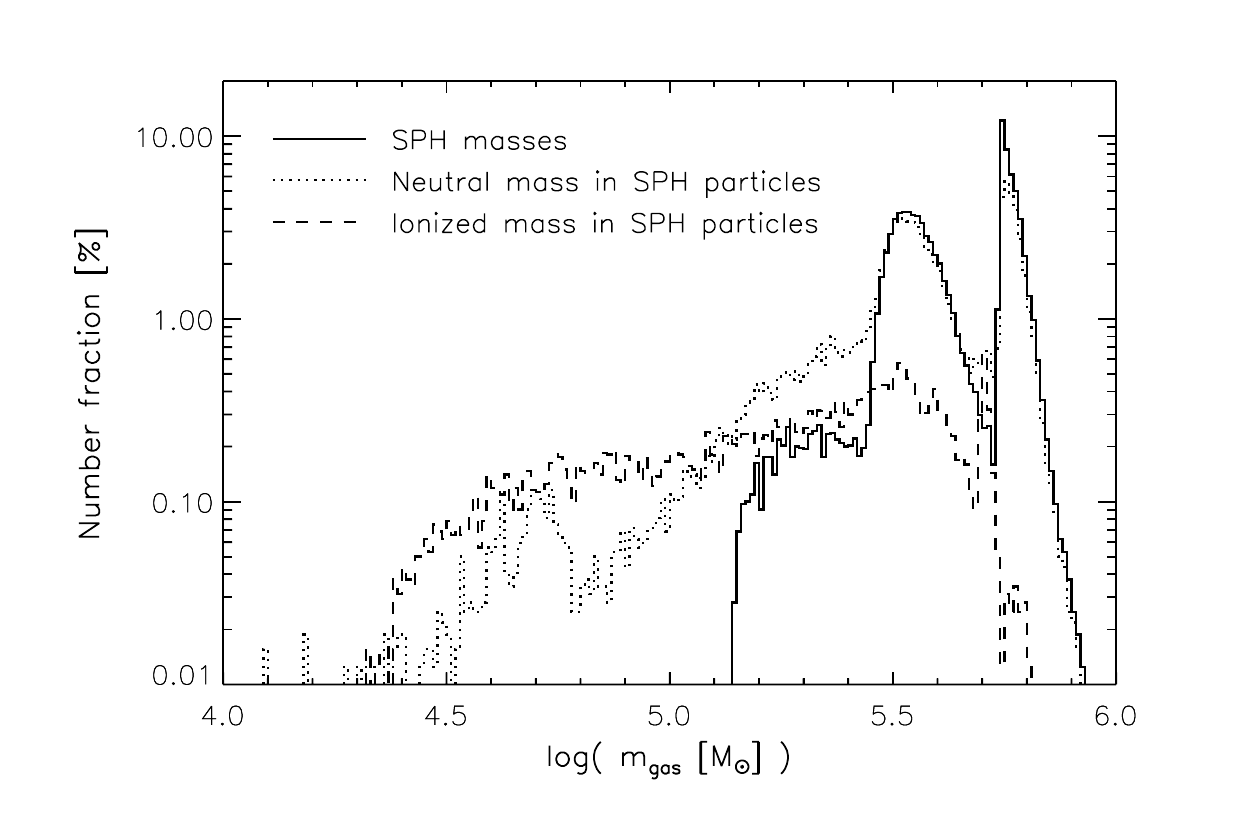}
\caption{\footnotesize{The distribution of SPH gas particle masses (solid histogram) in G4.
The distribution peaks at two characteristic masses ($\sim 10^{5.5}\,\msun$ and
$\sim 10^{5.8}\,\msun$), where the lower mass peak represents gas particles left
over from the first generation of stars in the simulation. Splitting the gas
into its neutral and ionized components according to eqs.\ \ref{eq:mwcnm} and
\ref{eq:mhii} results in the mass distribution given by the dotted and dashed
histograms, respectively.}}
\label{m_sph}
\end{figure}

As illustrated in Fig.\,\ref{figure:cartoon}, the first step in modeling the
ISM is to split each SPH particle into an ionized and a neutral gas component.
This is done using the electron fraction, $\xe$, associated with each SPH
particle, i.e.,: 
\begin{align}
	m_{\rm neutral}			&=	(1-\xe) \msph	\label{eq:mwcnm} \\
	m_{\rm ionized}			&=	\xe \msph		\label{eq:mhii}.
\end{align}
The electron fraction from \gadget gives the density of electrons relative to
that of hydrogen, $\ne/\nH$, and can therefore reach values of $\sim1.16$ in the
case where helium is also ionized. As a result we re-normalized the distribution
of \xe ~values to a maximal \xe ~of 1 so as to not exceed the total gas mass in
the simulation. Fig.\ \ref{m_sph} shows the distribution of SPH gas particles
masses in G4 -- chosen for its position near the center of the stellar and gas
mass ranges of G1, ..., G7 -- along with the mass distributions of the neutral
and ionized gas components obtained from eqs.\ \ref{eq:mwcnm} and \ref{eq:mhii}.
The ionized gas is seen to have a relatively flat distribution spanning the mass
range $\sim 10^{4.3} - 10^{5.8}\,\msun$. The neutral gas, however, peaks at two
characteristic masses ($\sim 10^{5.5}\,\msun$ and $\sim 10^{5.8}\,\msun$), where
the lower mass peak represents gas particles left over from the first generation
of stars in the simulation.

\subsection{GMCs}\label{split1}
\subsubsection{Masses, sizes and velocity dispersion}\label{split11}
The neutral gas mass, \mneu, associated with a given SPH particle is divided
into GMCs by randomly sampling the GMC mass spectrum as observed in the
Galactic disk and Local Group galaxies: $\frac{dN}{dm_{\rm GMC}} \propto m_{\rm
GMC}^{-\beta}$ with $\beta=1.8$ \citep{blitz07}.  Similar to \citet{narayanan08,narayanan08a}, 
a lower and upper cut in mass of $10^4\,\msun$
and $10^6\,\msun$, respectively, are enforced in order to ensure the GMC masses
stay within the range observed by \cite{blitz07}. Up to 40 GMCs are created per
SPH particle but typically most ($>90\%$) of the SPH particles are split into
four GMCs or less.  While \mneu~never exceeds the upper GMC mass limit
($10^6\,\msun$), there are instances where \mneu~is below the lower GMC mass
limit ($10^4\,\msun$).  In those cases we simply discard the gas, i.e., remove it
from any further sub-grid processing. For the highest resolution simulations in
our sample (G1 and G2), the discarded neutral gas amounts to $\sim6\,\%$ of the
total neutral gas mass, and $\ls 0.02\,\%$ in the remaining five galaxies.  We
shall therefore assume that it does not affect our results significantly.

The GMCs are randomly distributed within $0.2\times$ the smoothing length of the
original SPH particle, but with the radial displacement scaling inversely with
GMC mass in order to retain the original gas mass distribution as closely as
possible. To preserve the overall gas kinematics as best possible, all GMCs
associated with a given SPH particle are given the same velocity as that of the
SPH particle.

GMC sizes are obtained from the pressure-normalized scaling relations for
virialized molecular clouds which relate cloud radius (\rcl) with mass ($m_{\rm
GMC}$) and external pressure (\Pe):
\begin{align}
 	\frac{\rcl}{\text{pc}} =\left( \frac{\Pe/k_{\text{B}}}{10^4\,\cmpc\,\text{K}} \right)^{-1/4} \left (\frac{\mgmc}{290\,\text{\msun}}\right )^{1/2}.	
	\label{Pe_size}
\end{align}
We assume  $\Pe = P_{\rm tot}/(1+\alpha_0+\beta_0)$ for relative cosmic and
magnetic pressure contributions of $\alpha_0=0.4$ and $\beta_0=0.25$
\citep{elmegreen89b}. For the total pressure ($P_{\rm tot}$) we adopt the
external hydrostatic pressure at mid-plane for a rotating disk of gas and stars,
i.e.,:
\begin{equation}
	P_{\rm tot}\approx\frac{\pi}{2}\rm{G}\gassd \left[ \gassd+\left( \frac{\sigma_{\rm gas,\perp}}{\sigma_{\rm *,\perp}} \right) \starssd\right], 
\end{equation}
where \gassd\ and \starssd\ are the local surface densities of gas and stars,
respectively, and $\sigma_{\rm gas,\perp}$ and $\sigma_{\rm *,\perp}$ their
local velocity dispersions measured perpendicular to the mid-plane (see e.g.,
\cite{elmegreen89b,swinbank11}). For each SPH particle, all of these quantities
are calculated directly from the simulation output (using a radius of $R=1$\,kpc
from each SPH particle), and it is assumed that the resulting $\Pe$ is the
external pressure experienced by all of the GMCs generated by the SPH particle.
We find that GMCs in our simulated galaxies are subjected to a wide range of
external pressures ($\Pe/k_{\rm B} \sim10^2-10^7\,{\rm cm^{-3}\,K}$). For
comparison, the range of pressures experienced by clouds in our Galaxy and in
Local Group galaxies is $\Pe/k_{\rm B} \sim10^3-10^7\,{\rm cm^{-3}\,K}$ with an
average of $P_{\rm ext}/k_{\rm B} \sim10^4$\,\cmpc\,K in Galactic clouds
\citep{elmegreen89b,blitz07}. This results in GMC sizes in our simulations
ranging from $\rcl=1 - 300\,{\rm pc}$. 

The internal velocity dispersion ($\sigma_v$) of the GMCs is inferred from the virial
theorem, which provides us with a pressure-normalized $\sv-R_{\rm GMC}$
relation:
\begin{equation}
	\sigma_v = 1.2\,\kms \left( \frac{\Pe/k_{\rm B}}{10^4\,\cmpc\,{\rm K}}\right)^{1/4}\left ( \frac{\rcl}{{\rm pc}}\right )^{1/2},
	\label{sigma_v_Pe}
\end{equation}
where the normalization of $1.2\,{\rm km\,s^{-1}}$ comes from studies of Galactic GMCs
\citep[][]{larson81,elmegreen89b,swinbank11}.

\subsubsection{GMC density and temperature structure} \label{split12}
\begin{figure}[htbp] 
\hspace*{0.5cm}
\includegraphics[width=0.8\columnwidth]{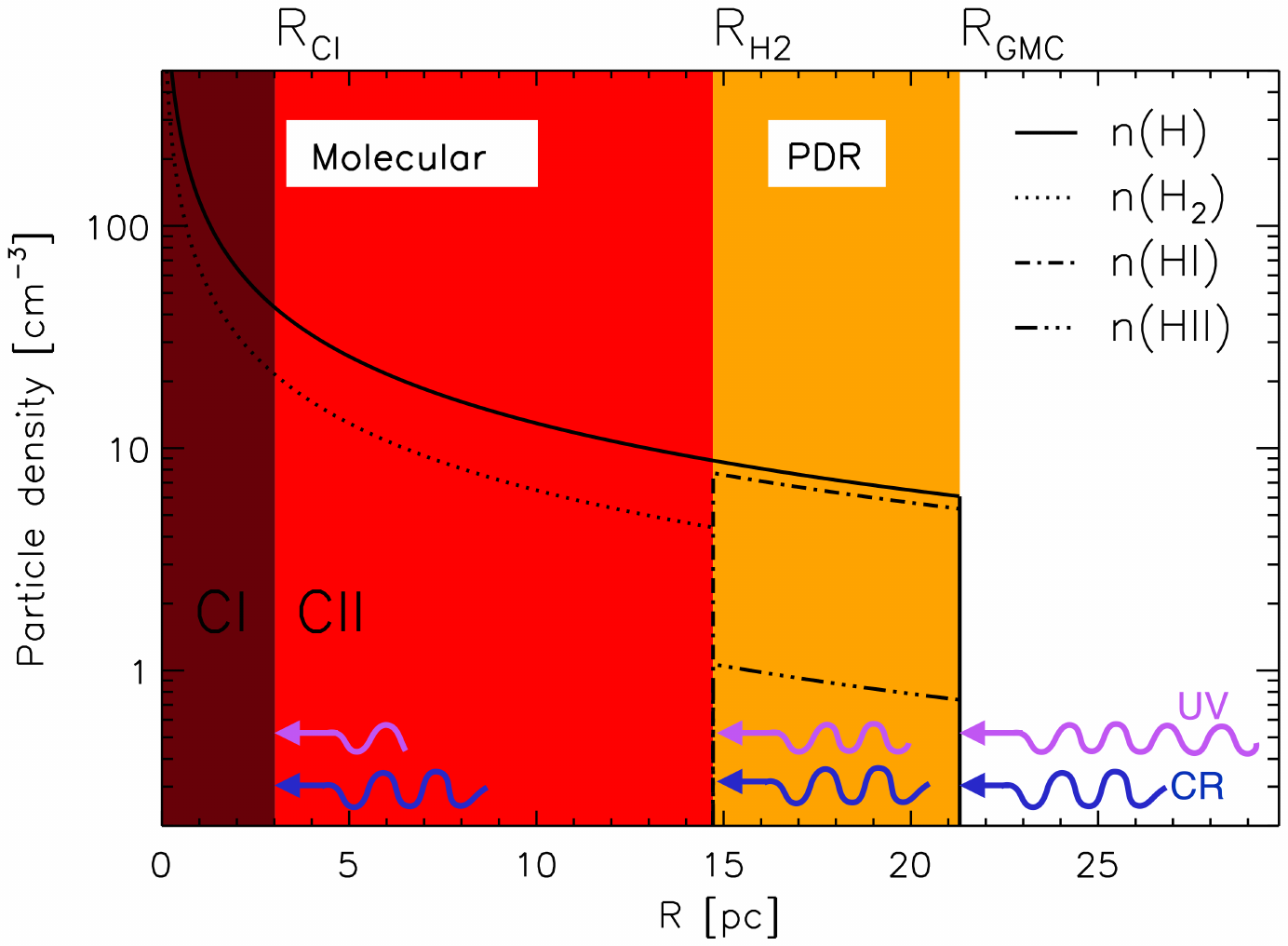}
\caption{\footnotesize{Example H number density profile (solid line) for a GMC of mass $m_{\rm
GMC}=1.3\times 10^{4}$\,\msun ~and radius $R_{\rm GMC}=21$\,pc. Also shown are
the density profiles of H$_2$ ($=0.5 n_{\rm H}$), H{\sc i} and H{\sc ii}. Note,
the transition from molecular to atomic H is assumed to happen instantaneously
at $R_{\rm H_2}$. The total C abundance follows that of H but scaled with the
[C/H] abundance (as provided by the `parent' SPH particle, see Section
\ref{section:methodology}).  C{\sc ii} can exist throughout the cloud except
for the very inner region ($R < R_{\rm CI}$; indicated in brown), with the C{\sc
i} to C{\sc ii} transition happening instantaneously at $R_{\rm CI}$. \cii emission
from the layer $R_{\rm CI}\le R \le R_{\rm H_2}$ (indicated in red) is referred
to as `molecular' emission, while \cii emission from $R_{\rm H_2}< R \le R_{\rm
GMC}$ (indicated in orange) is referred to as `PDR' emission.  The
relative thickness of these layers is set by the impinging FUV radiation field
and cosmic rays (illustrated as purple and blue arrows, respectively), with the former
undergoing attenuation further into the cloud.}}
\label{onion}
\end{figure}

We assume a truncated logotropic profile for the total hydrogen number density
of the GMCs, i.e.,: 
\begin{align}
	\nH(R)			=	\next\left( \frac{\rcl}{R} \right), 	
	\label{eq:nH}
\end{align}
where $\nH(R>\rcl)=0$. For such a density profile it can be shown that the
external density, \next, is 2/3 of the average density:
\begin{align}
	\next			=	2/3\nav=2/3\frac{\mgmc}{4/3\pi\mH \rcl^3}.
	\label{eq:next}
\end{align}
While the total hydrogen density follows a logotropic profile, the transition from
H$_2$$\longrightarrow$H{\sc i}/H{\sc ii} is assumed to be sharp. Similarly for
the transition from C{\sc i}$\longrightarrow$C{\sc ii}. This is illustrated in
Fig.\,\ref{onion}, which shows an example density profile of a GMC from our simulations.
From the center of the GMC and out to \rh2, hydrogen is in molecular form.
Beyond \rh2, hydrogen is found as \hi and \hii out to \rcl. 

The size of the molecular region, when adopting the logotropic density profile,
is related to the total molecular gas mass fraction (\fmol) of each
GMC:
\begin{align}
	\fmol	&=	\frac{\mmol}{\mgmc}
			=	\frac{\int_0^{R_{\rm H_2}} \rho_{\rm ext}\frac{R_{\rm GMC}}{R} 4\pi R^2 dR}
			{\int_0^{R_{\rm GMC}}\rho_{\rm ext}\frac{R_{\rm GMC}}{R} 4\pi R^2 dR}\\
			&=	\frac{\int_0^{R_{\rm H_2}} R \,dR}{\int_0^{R_{\rm GMC}} R \,dR}
			=	\left( \frac{R_{\rm H_2}}{R_{\rm GMC}} \right)^2\\
	&\Rightarrow \rrh2	=	\sqrt{\fmol}, \label{eq:fmol}
\end{align}
where \rrh2 is the fractional radius, $\rrh2=\rh2/\rcl$.
We find \fmol by assuming $\hi\leftrightarrow\h2$ equilibrium and using
the analytical steady-state approach of \cite{pelupessy06} for inferring
\fmol for a logotropic cloud
subjected to a radially incident FUV radiation field \citep[see also][]{olsen15}.  
In this framework, the value of \fmol\ (and thereby $\rrh2$)
depends on:

\begin{enumerate}
\item The cloud boundary pressure (\Pe), which is calculated as
explained in Section\,\ref{split11}.

\item The metallicity (\Z) of the GMC, which is inherited from
its parent SPH particle and assumed constant throughout the cloud.

\item The kinetic temperature of the gas at the GMC surface
($T_k(R_{\rm GMC})$). This quantity is calculated in an iterative process
together with \fmol\ by solving the following thermal balance equation:
\begin{equation}
 	\Hpe+\Hcrhi = \Ccii+\Coi \label{Tk_rcl}, 
\end{equation}
where \Hpe is the heating rate associated with the photoelectric ejection of
electrons from dust grains by the FUV field and \Hcrhi is the heating rate by
cosmic rays in atomic gas. The main cooling agents are assumed to be due to \cii
and [O{\sc i}]($63\,{\rm \mu m}$ and $145\,{\rm \mu m}$) line emission (i.e.,
\Ccii and \Coi, respectively). \Hpe, \Hcrhi, and \Ccii all depend on the
electron fraction at $R_{\rm GMC}$, which is determined by the degree of H{\sc i}
ionization by the local FUV radiation field and CR ionization rate 
(see below for how these quantities are derived). For analytical expressions for
the heating rates, we refer to \cite{olsen15}. For $\Lambda_{\rm
O\textsc{i}}$ we use the expressions given in \citet{rollig06}. The
calculation of \Ccii at the GMC surface is detailed in Section \ref{cii_em} and
Appendix \ref{apD}.

\item The strength of the local FUV radiation field (\g0) and the
CR ionization rate (\cri) impinging on the GMCs.  These quantities do not come
out from the simulation directly, and instead they are calculated by scaling the
Galactic FUV field ($G_{\rm 0,MW}$) and CR ionization rate ($\zeta_{\rm CR,MW}$)
with the local SFR volume density in the simulations, i.e., $G_{\rm 0} \propto
G_{\rm 0,MW} \left ({\rm SFRD}_{\rm local}/{\rm SFRD}_{\rm MW}\right )$ and
$\zeta_{\rm CR} \propto \zeta_{\rm CR,MW} \left ({\rm SFRD}_{\rm local}/{\rm
SFRD}_{\rm MW}\right )$, where ${\rm SFRD_{\rm local}}$ is estimated for each
SPH particle as the volume averaged SFR within a $2\,{\rm kpc}$ radius.  We have
adopted Milky Way values of $G_{\rm 0,MW} = 0.6\,{\rm Habing}$ \citep{seon11}
and $\zeta_{\rm CR, MW} = 3\times 10^{-17}\,{\rm s^{-1}}$ \citep{webber98}.  For
${\rm SFRD}_{\rm MW}$ we adopt $0.0024\,{\rm \msun\,yr^{-1}\,kpc^{-3}}$,
inferred from the average Galactic SFR ($0.3\,{\rm \msun\,yr^{-1}}$) within a
disk $10\,{\rm kpc}$ in radius and $0.2\,{\rm kpc}$ in height
\citep{heiderman10,bovy12}.

\item The electron fraction at the cloud boundary. This fraction is
not the previously introduced \xe, which was inherent to the SPH simulations and
used to split the SPH gas into a neutral and ionized gas phase. Instead, it is
the electron fraction given by the degree of ionization of H{\sc i} caused by
the \g0 and \cri impinging on the cloud.  This fraction (and thus the H{\sc
i}:H{\sc ii} ratio) is calculated with \texttt{CLOUDY} v13.03 \citep{ferland13}
given the hydrogen density and temperature at the cloud boundary and assuming an
unattenuated \g0 and \cri at $R_{\rm GMC}$.

\end{enumerate}

\bigskip

The \cii emitting region in each of our GMCs is defined as the layer between the
surface of the cloud and the depth at which the abundances of C and C$^+$ are
equal. Hence, if the latter occurs at a radius $\rci$ from the cloud center, the
thickness of the layer is $\rcl - \rci$ (Fig.\ \ref{onion}). At radii $< \rci$,
all carbon atoms are for simplicity assumed to be in neutral form. In order to
determine the fractional radius \rrci ($=R_{\rm CI}/R_{\rm GMC}$) we follow the
work of \cite{rollig06} \citep[but see also][]{pelupessy09}, who considers the
following dominant reaction channels for the formation and destruction of C$^+$:
\begin{align}
    \rm{C} +\gamma             &\rightarrow    \rm{C}^+ + {\it e^-}  \label{CIIform1} \\
    \rm{C}^+ +{\it e^-}           &\rightarrow    \rm{C} + \gamma  \label{CIIform2} \\
    \rm{C}^+ +\rm{H}_2      &\rightarrow    \rm{CH}^+_2 + \gamma.   \label{CIIform3}
\end{align}
In this case, \rrci can be found by solving the following equation:
\begin{align}
    5.13\times10^{-10}&{\rm s^{-1}}\g0\int_{1}^{\infty}\frac{e^{-\mu \xi_{\rm FUV}A_{\rm V}(\rrci)}}{\mu^2}d\mu \label{eq:R06} \\
    =&~n_{\rm H}(r_{\rm C{\textsc i}})[a_{\rm C}\Xc+0.5k_{\rm C}], \nonumber
\end{align}
where the left-hand side is the C$^+$ formation rate due to photo-ionization by
the attenuated FUV field at \rrci (eq.\ \ref{CIIform1}), and the right-hand side
is the destruction rate of C$^+$ due to recombination and radiative association
(eqs.\ \ref{CIIform2} and \ref{CIIform3}). The constants $a_{\rm
C}=3\e{-11}$\,\cmpc\ps ~and $k_{\rm C}=8\e{-16}$\,\cmpc\ps ~are the
recombination and radiative association rate coefficients.  Note, we have
accounted for an isotropic FUV field since $\mu = \cos \theta$, where $\theta$
is the angle between the Poynting vector and the normal direction.  $A_{\rm
V}(\rrci)$ is the visual extinction corresponding to the $\rcl - \rci$ layer,
and is given by $A_{\rm V}(\rrci) = 0.724\sigma_{\rm dust}\Z \langle n_{\rm
H}\rangle\rcl \ln\left( \rrci^{-1} \right)$, where $\sigma_{\rm dust} =
4.9\e{-22}\,\cms$ is the FUV dust absorption cross section
\citep{pelupessy09,mezger82}. $\xi_{\rm FUV}$ accounts for the difference in
opacity between visual and FUV light and is set to $3.02$.  \Xc~is the carbon
abundance relative to H and is calculated by adopting the carbon mass fractions
of the parent SPH particle (self-consistently calculated as part of the overall
SPH simulation) and assuming it to be constant throughout the GMC.  
%
%
\begin{figure}[htbp] 
\hspace*{1cm}
\includegraphics[width=0.8\columnwidth]{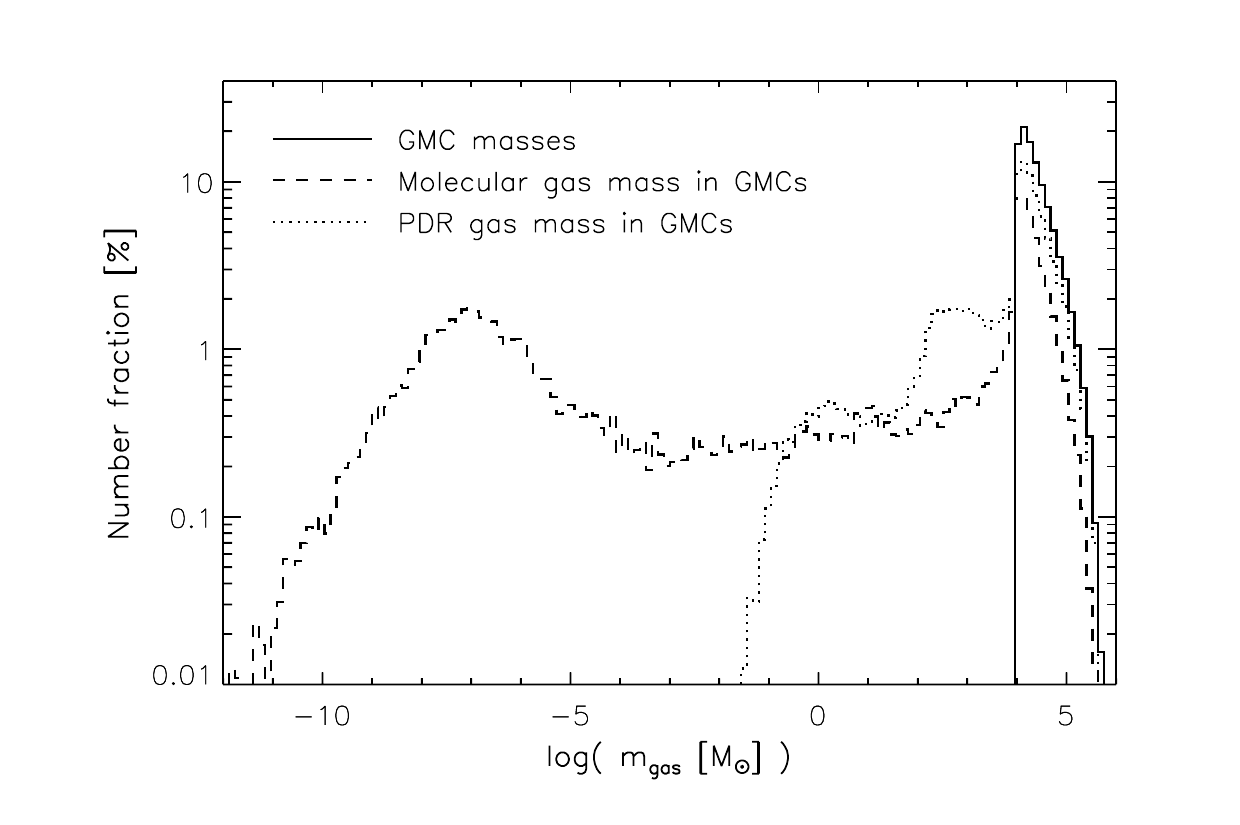}\\
\hspace*{1cm}
\includegraphics[width=0.8\columnwidth]{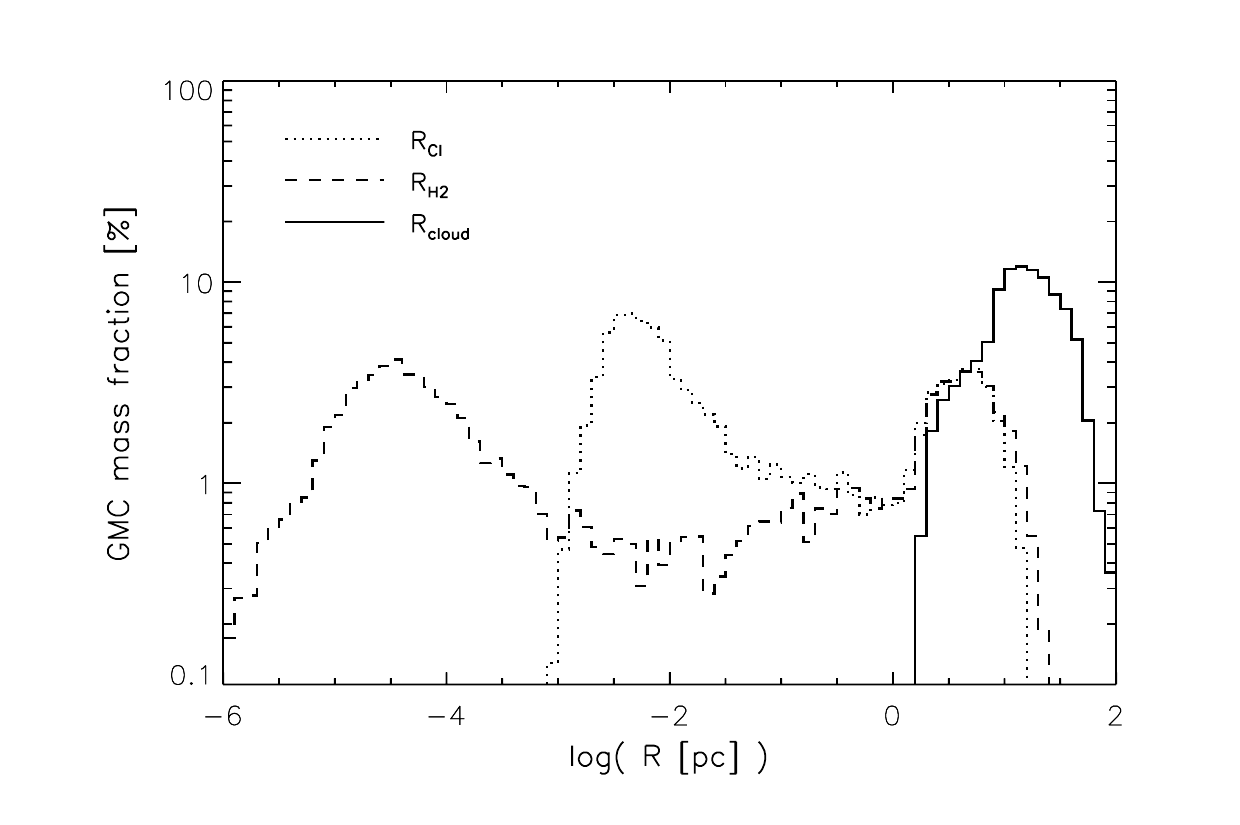}
\caption{\footnotesize{{\bf Top:} the number distribution of GMC masses (solid histogram)
together with the number distributions of the molecular (dashed histogram) and
PDR (dotted histogram) gas component.  {\bf Bottom:} GMC-mass-weighted
distributions of \rci (dotted), \rh2 (dashed) and \rcl for GMCs in G4.}}
\label{figure:R-T-distributions-GMCs1}
\end{figure}

\begin{figure}[htbp] 
\hspace*{1cm}
\includegraphics[width=0.8\columnwidth]{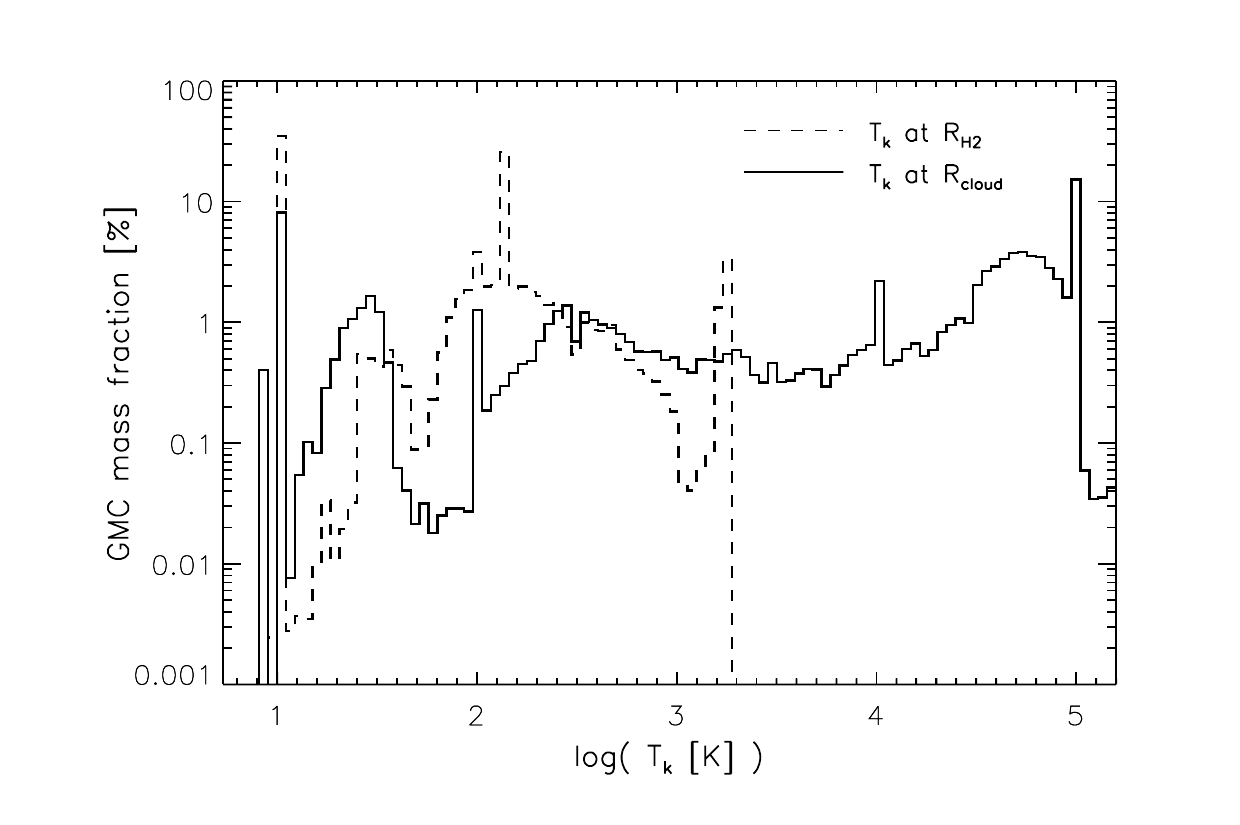}\\
\caption{\footnotesize{GMC-mass-weighted distributions of the temperature at 
\rh2 (dashed) and at the cloud surfaces (solid) for GMCs in G4.}}
\label{figure:R-T-distributions-GMCs2}
\end{figure}

\bigskip

The temperature of the \cii-emitting molecular gas (i.e., from the gas layer
between \rci and \rh2) is assumed to be constant and equal to the temperature at
\rh2. The latter is given by the thermal balance:
\begin{equation}
   	\Hpe+\Hcrh2 =\Ch2+\Ccii+\Coi,
	\label{Tk_rH2}
\end{equation}
where \Hcrh2 ~is the CR heating in molecular gas, and \Ch2 is the
cooling due to the S(0) and S(1) rotational lines of H$_2$ (\citet{papa14}; see
also \citealt{olsen15}). \Hcrh2 depends on \xe, which is calculated with
\texttt{CLOUDY} for the local CR ionization rate and the attenuated FUV field at
\rrh2, i.e.\ $\g0 e^{-\xi_{\rm FUV}A_{\rm V}(\rrh2)}$. $A_{\rm V}(\rrh2)$
corresponds the extinction through the outer atomic transition layer of each
GMC. 

The temperature of the PDR gas (i.e., from the gas between \rh2 and \rcl)
is assumed to be constant and equal to the temperature at \rcl.

\bigskip

For each GMC we solve in an iterative manner simultaneously for \rh2 (eq.\
\ref{eq:fmol}) and \rci (eq.\ \ref{eq:R06}), as well as for the gas temperature
at \rcl and at \rh2 (eqs.\ \ref{Tk_rcl} and \ref{Tk_rH2}, respectively)

The resulting distributions of \rh2 and \rci for the GMC population in G4 are
shown in Fig.\,\ref{figure:R-T-distributions-GMCs1} (bottom panel), along with
the distribution of $R_{\rm GMC}$ (obtained from eq.\ \ref{Pe_size}).  In GMCs
in general, we expect $R_{\rm C\textsc{i}} < R_{\rm H_2}$, due to efficient
H$_2$ selfshielding.  However, as Fig.\ \ref{figure:R-T-distributions-GMCs1}
shows some of the GMCs in our simulations have very small \rh2 (due to them
having virtually zero molecular gas fractions), and in those cases \rci can be
equal to or even exceed \rh2. The latter implies that the \cii emission is only
coming from the PDR phase, with no contribution from the molecular gas.

The distributions of the kinetic temperature of the gas at \rh2 and \rcl for the
GMC population in G4 are shown in Fig.\,\ref{figure:R-T-distributions-GMCs2}. 
The temperatures at the cloud surfaces range from $\sim 8\,{\rm
K}$ to $\sim10^{5.5}\,{\rm K}$, and the temperatures at \rh2 lies between $\sim
8$ and $1800\,{\rm K}$.  

With \rh2 and \rci determined for each GMC we can calculate the gas masses
associated with the molecular and PDR gas phase, respectively. The
resulting mass distributions are shown in the top panel of
Fig.\,\ref{figure:R-T-distributions-GMCs1}, along with the distribution of total
GMC masses ($m_{\rm GMC}$) as determined by the adopted GMC mass spectrum
(Section \ref{split11}). We see that most of the molecular and PDR gas
masses follow the total GMC mass spectrum. There is, however, a fraction of GMCs
with extremely small molecular gas masses (corresponding to $\fmol\sim 0$).  On
the other hand, there are some GMCs with small PDR gas masses, i.e.,
clouds that are so shielded from FUV radiation that they are almost entirely
molecular.

\subsection{The ionized gas} \label{split2}
\begin{figure}[htbp] 
\hspace*{1cm}
\includegraphics[width=0.8\columnwidth]{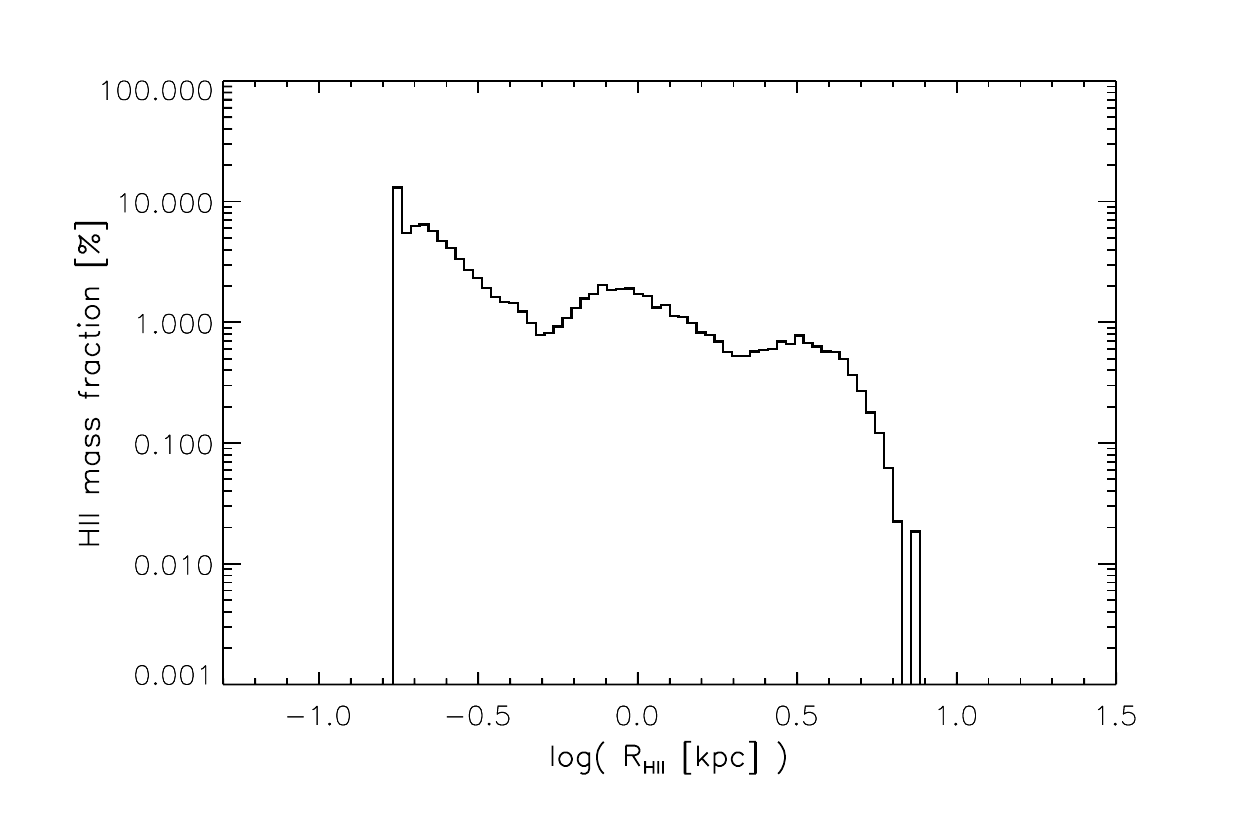}\\
\hspace*{1cm}
\includegraphics[width=0.8\columnwidth]{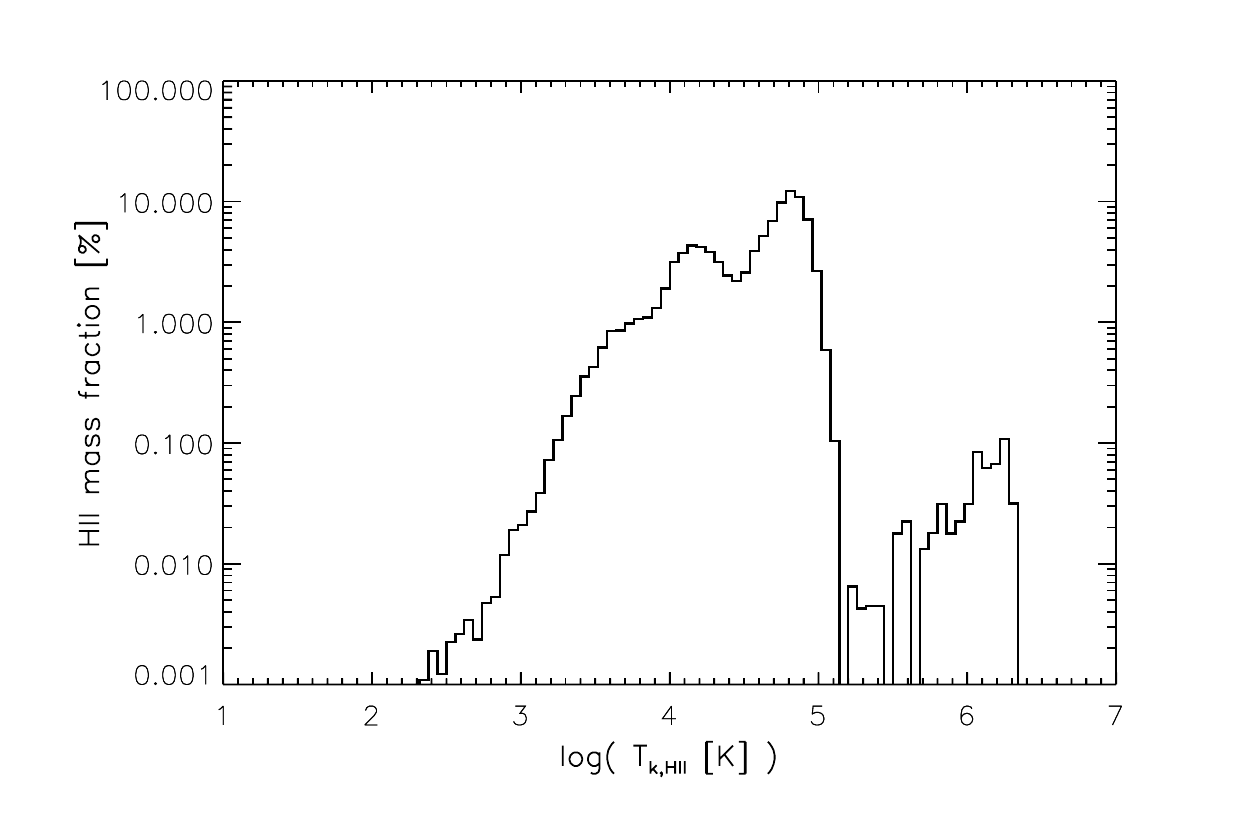}
\caption{\footnotesize{Mass-weighted histograms of the size (top) and temperature (bottom)
distributions of the ionized clouds in G4.}}
\label{figure:R-T-distributions-HII}
\end{figure}

The ionized gas in our simulations (see eq.\ \ref{eq:mhii}), is assumed to be
distributed in spherical clouds of uniform densities and with radii ($R_{\rm
H\textsc{ii}}$) equal to the smoothing lengths of the original SPH particles.
These ionized regions in our simulations are furthermore assumed to be
isothermal with the temperatures equal to that of the SPH gas.  Fig.\
\ref{figure:R-T-distributions-HII} shows the size (top) and temperature (bottom)
distribution for the ionized clouds in G4. Cloud sizes range from $\sim
0.1$ to $\sim 10\,{\rm kpc}$, with more than $60\,\%$ of the ionized gas
mass residing in clouds of size $\ls 1000\,{\rm pc}$. For comparison, the range
of observed sizes of \hii clouds in nearby galaxies is $10-1000\,{\rm pc}$
\citep{oey97,hodge99}.  The temperatures range from $\sim 10^2\,{\rm K}$ to
$\sim 10^6\,{\rm K}$ with the bulk of the ionized gas having temperatures $\sim
10^{4-5}\,{\rm K}$.

\section{The \cii\ line emission}\label{cii_em}
The \cii\ luminosity of a region of gas is the volume-integral of the effective \cii\
cooling rate per volume, i.e.,:
\begin{align}
	\Lcii			=	\int_{\rm \Delta V} \Ccii \,dV.
	\label{eq:Lcii}
\end{align}
Since we have adopted spherical symmetry in our sub-grid treatment of the clouds
(both neutral and ionized), we have: 
\begin{align}
	\Lcii 	&= 4\pi\int_{R_1}^{R_2} \Ccii R^2 dR,
	\label{eq:CII-integral}
\end{align}
where $R_1 = R_{\rm C\textsc{i}}$ and $R_2 = R_{\rm H_2}$ for the molecular
phase; $R_1 = R_{\rm H_2}$ and $R_2 = R_{\rm GMC}$ for the PDR region, and $R_1
= 0$ and $R_2 = R_{\rm H\textsc{ii}}$ for the ionized gas.

The effective \cii cooling rate is: 
\begin{align}
	\Ccii			=	A_{\rm ul}\beta f_{\rm u}\ncii h \nu,
	\label{eq:ccii}
\end{align}
where $A_{\rm ul}$ ($=$2.3\e{-6}$\,\ps$)\footnote{Einstein coefficient for
spontaneous emission taken from the LAMDA database:
\url{http://www.strw.leidenuniv.nl/~moldata/}, \cite{schoier05}} is the Einstein
coefficient for spontaneous decay. We ignore the effects of any background
radiation field from dust and the cosmic microwave background (CMB). $\beta$ is the \cii photon escape
probability for a spherical geometry (i.e., $\beta = (1 - exp(-\tau))/\tau$,
where $\tau$ is the \cii optical depth). $f_u$ is the fraction of singly ionized
carbon in the upper $^2P_{3/2}$ level and is determined by radiative processes
and collisional (de)excitation (see Appendix \ref{apD}).  The latter can occur
via collisions with $e^-$, H{\sc i}, or H$_2$, depending on the state of the
gas. In our simulations the collisional partner is H$_2$ in the molecular phase,
H{\sc i} and $e^-$ in the PDR regions, and $e^-$ in the ionized gas.  Analytical
expressions for the corresponding collision rate coefficients as a function of
temperature are given in Appendix \ref{apD}.  \ncii\ is the number density
of singly ionized carbon and is given by $\ncii = X_{\rm C} f_{\rm C\textsc{ii}}
n_{\rm H}$, where $f_{\rm C\textsc{ii}}$ is the fraction of carbon atoms in the
singly ionized state.  For the latter we used tabulated fractions from
\texttt{CLOUDY} v13.03 over a wide range in temperature, hydrogen density, FUV
field strength, and CR ionization rate.

We calculate the integral in eq.\ \ref{eq:CII-integral} numerically by splitting
the $R_2 - R_1$ region up into 100 radial bins. In each bin, $n_{\rm H}$ is set
to be constant and -- in the case of the molecular and PDR regions -- given by
the logotropic density profile at the radius of the given bin
(Fig.\,\ref{onion}). For the ionized clouds, $n_{\rm H}$ is constant throughout
(Section \ref{split2}). For the PDR regions (i.e., from $R_{\rm H_2}$ to $R_{\rm
GMC}$) we assume that the temperature, electron fraction and \g0 are kept fixed
to the outer boundary value at $R_{\rm GMC}$ (i.e., no attenuation of the FUV
field). This implies that the \cii luminosity from the PDR gas is an upper
limit. Similar for the \cii emission from the molecular region (i.e., from
$R_{\rm C\textsc{i}}$ to $R_{\rm H_2}$), where we assume that the temperature
throughout this region is fixed to its values at $R_{\rm H_2}$. Also, throughout
this region we adopt the attenuated FUV field at $R_{\rm H_2}$.

\section{Results and discussion}
Having divided the ISM in our galaxies into molecular, atomic and ionized gas
phases, and having devised a methodology for calculating their \cii\ emission,
we are now in a position to quantify the relative contributions from the
aforementioned gas phases to the total \cii emission, and examine their
relationship to the on-going star formation.
\begin{figure*}[t!] 
  \begin{center}
  \includegraphics[width=0.8\textwidth]{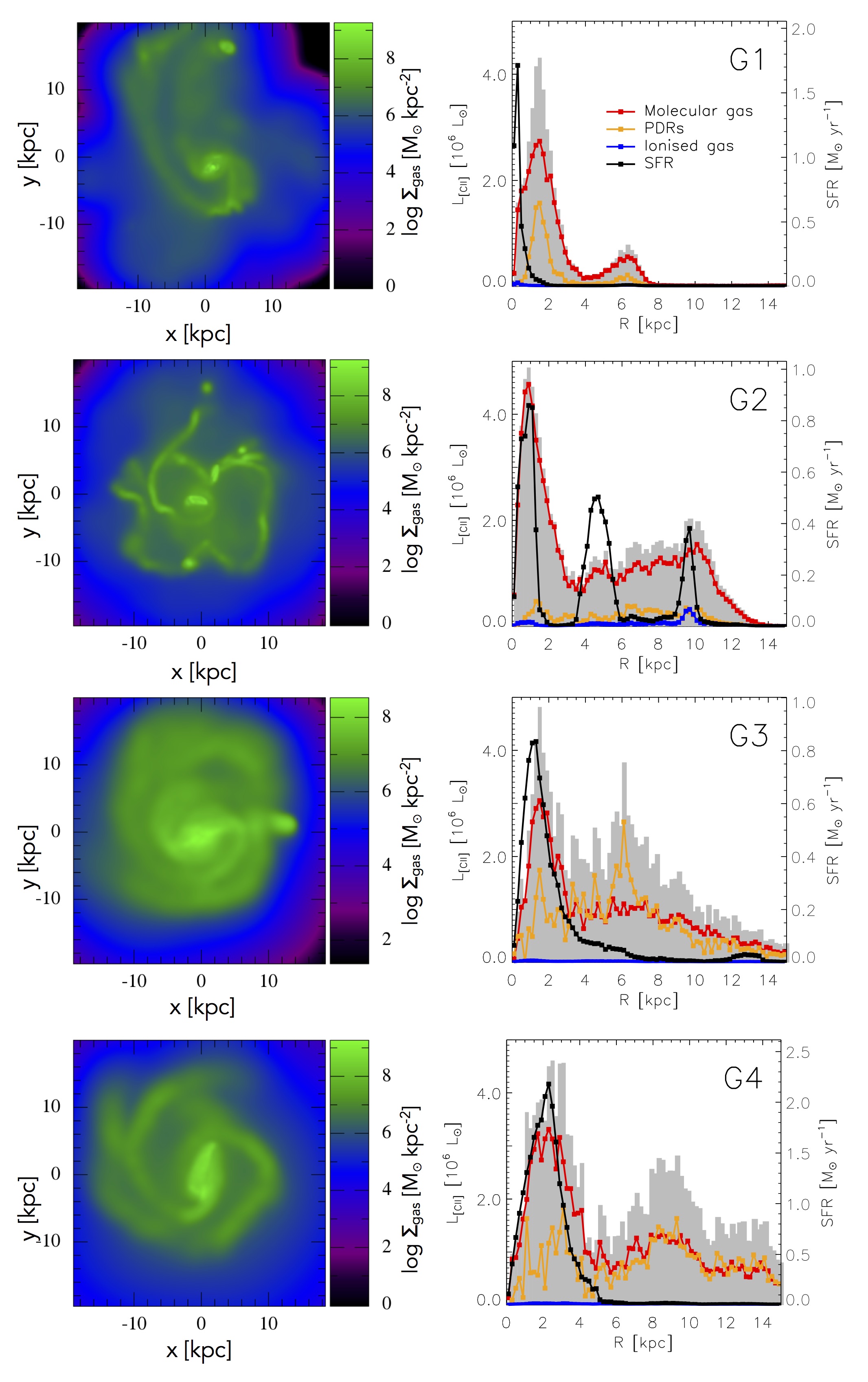}
  \caption{\footnotesize{Gas and SFR surface density maps (left and middle columns,
  respectively) of our SPH simulated galaxies viewed face-on (G1, ..., G7 from
  top to bottom; see also \citet{thompson15}).  The horizontal white bars
  correspond to a physical scale of $5\,{\rm kpc}$. The right-hand column shows
  the radial profiles of the total \cii luminosity (gray histogram) and the
  contributions from molecular gas (red curve), PDR (orange curve) and
  ionized gas (blue curve). The SFR radial profiles are also shown (black
  curve).  The radial profiles were determined by summing up the \cii luminosity
  and SFR within concentric rings with fixed width of $0.2\,{\rm kpc}$.}}
 \label{fig:maps1}
 \end{center}
\end{figure*}
\setcounter{figure}{6}
\begin{figure*}[t!]
\begin{center}  
  \includegraphics[width=0.8\textwidth]{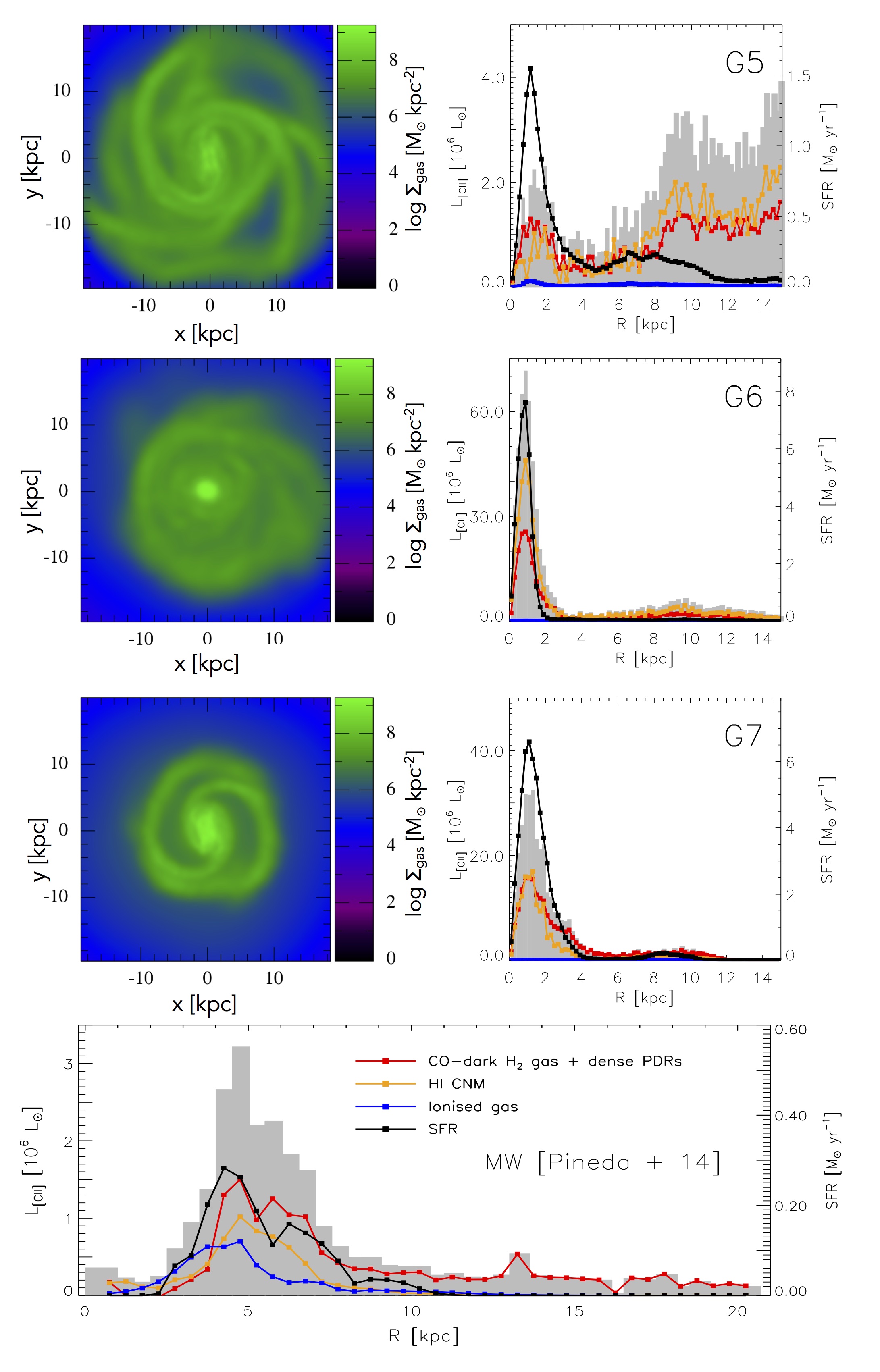}
  \caption{\footnotesize{Continued. In the bottom panel we show for reference the radial \Lcii
  profile of the Galaxy (gray histogram) along with the contributions from
  CO-dark H$_2$ gas and dense PDRs (red curve), cold H\,{\sc i} (orange curve),
  and ionized gas (blue curve) \citep{pineda14}. Also shown is the radial SFR
  profile (black curve) inferred from the $1.4\,$GHz intensity distribution.
  Fixed radial bin widths of $0.5\,{\rm kpc}$ have been adopted (see
  \citet{pineda14} for details).}}
  \label{fig:maps2}
\end{center}
\end{figure*}

\subsection{Radial \cii luminosity profiles}\label{subsection:CII-profiles}
First, however, to get a sense of the distribution of gas and star formation in
our simulated galaxies, we show in Fig.\ \ref{fig:maps1} surface density maps of
the total SPH gas (left column) and star formation rate (middle column) when
viewed face-on. The maps reveal spiral galaxy morphologies, albeit with some
variety: some (G1, G2 and G3) show perturbed spiral arms due to on-going mergers
with satellite galaxies; others (G4, G5, G6 and G7) have seemingly undisturbed,
grand-design spiral arms; a central bar-like structure is also seen in some (G2,
G4, G5 and G7). Overall, the star formation is seen to be much more centrally
concentrated than the SPH gas. This is especially true for G1 and G2, which have
very centrally peaked star formation. The radial SFR and \cii luminosity
profiles of G1, ..., G7 -- derived by summing up the SFR and the \cii luminosity
within concentric rings (of fixed width: $0.2\,{\rm kpc}$) -- are also shown in
Fig.\,\ref{fig:maps1}.  We have inferred the radial \cii luminosity distribution
for the full ISM as well as for the individual gas phases.

\clearpage

The radial SFR profiles of our model galaxies typically peak at $R\sim0.5\,{\rm
kpc}$ (in G1 at $R\ls 0.5\,{\rm kpc}$) and then tail off with radius.  In some
cases, local peaks in the star formation activity occur at galactocentric
distances $\gs 2\,{\rm kpc}$, corresponding to the locations of either satellite
galaxies (G2 and G3) or spiral arms (G5 and G7). 

The total \cii luminosity profiles (gray histograms) also peak at
$R\sim0.5\,{\rm kpc}$, and within the central $R\ls 1.5\,{\rm kpc}$ there is in
general a good correspondence between the total \cii emission and the star
formation activity. This correspondence is driven by the molecular gas phase
which dominates the \cii emission in the central regions. The molecular \cii
emission is seen to correlate strongly with the star formation out to radii
$\sim 5\,{\rm kpc}$ and beyond (e.g., G3 and G4). There are cases, however,
where localized enhancements in the SFR are not matched by
increased \cii emission from the molecular gas (e.g.\ G2, and G3).  These SFR
enhancements are reflected in the \cii emission profile of the ionized gas,
which at galactocentric distances $\gs 2\,{\rm kpc}$ follow the SFR closely
(despite contributing only a small fraction to the total \cii emission
budget, see below). In contrast, the \cii emission from the PDR gas,
which dominates the total \cii luminosity at $2.0\,{\rm kpc} \ls R \ls 10\,{\rm
kpc}$, does not appear to be a sensitive tracer of the SFR. At $R\ls 1.5\,{\rm
kpc}$ where the SFR peaks, the \cii emission from this phase is seen to drop. At
larger radii, the PDR \cii emission declines but at a more gradual rate
than the star formation.

\bigskip

In the bottom right panel of Fig.\,\ref{fig:maps1} we show the Galactic SFR and
\cii luminosity radial profiles from \cite{pineda14}, who observed the \cii
emission from (1) CO-dark H$_2$ gas and dense PDRs, (2) cold neutral \hi gas, and
(3) hot ionized gas in our own Galaxy. In order to facilitate an approximate
comparison with our simulations we identify these three Galactic ISM phases with
the molecular, atomic, and ionized gas in our simulations. It is important to
keep in mind, however, that the ISM in our simulations has a higher pressure, is
kinematically more violent, and is more actively forming stars than is the 
case in our Galaxy. 

The SFR and \cii profiles in our Galaxy peak at larger radii ($R\sim4-5\,$kpc)
than in our simulated galaxies. This is not surprising given the low content of
star formation and gas in the Galactic bulge, and the fact that the bulk of star
formation in our Galaxy takes place in the disk. In contrast, gas is still being
funneled toward the central regions of our simulated galaxies where it is converted into stars.
Thus the SFR level in our simulated galaxies is much higher (by $\gs 10\times$)
and more centrally concentrated than in our Galaxy, where stars form at a rate
of $\ls 0.1-1\,{\rm \msun\,yr^{-1}}$ across the disk. Remarkably,
significant levels of \cii emission extend out to $R\sim 20\,{\rm kpc}$ in our
Galaxy, well beyond the point where star formation has ceased. Here, the
emission is completely dominated by CO-dark H$_2$ gas and dense PDR regions.
This is similar to the picture seen in our simulated galaxies where the \cii
emission at large radii is dominated by a neutral gas phase -- designated PDR
gas in our simulations, dubbed CO-dark H$_2$ $+$ dense PDR gas in
\cite{pineda14} -- that is largely uncoupled from star formation.  The radial
\cii profile of the ionized gas in the Galaxy roughly follows the SFR profile,
as is the case in our simulations. 

\bigskip

\begin{figure}[h] 
\hspace{1.5cm}
\includegraphics[width=0.8\columnwidth]{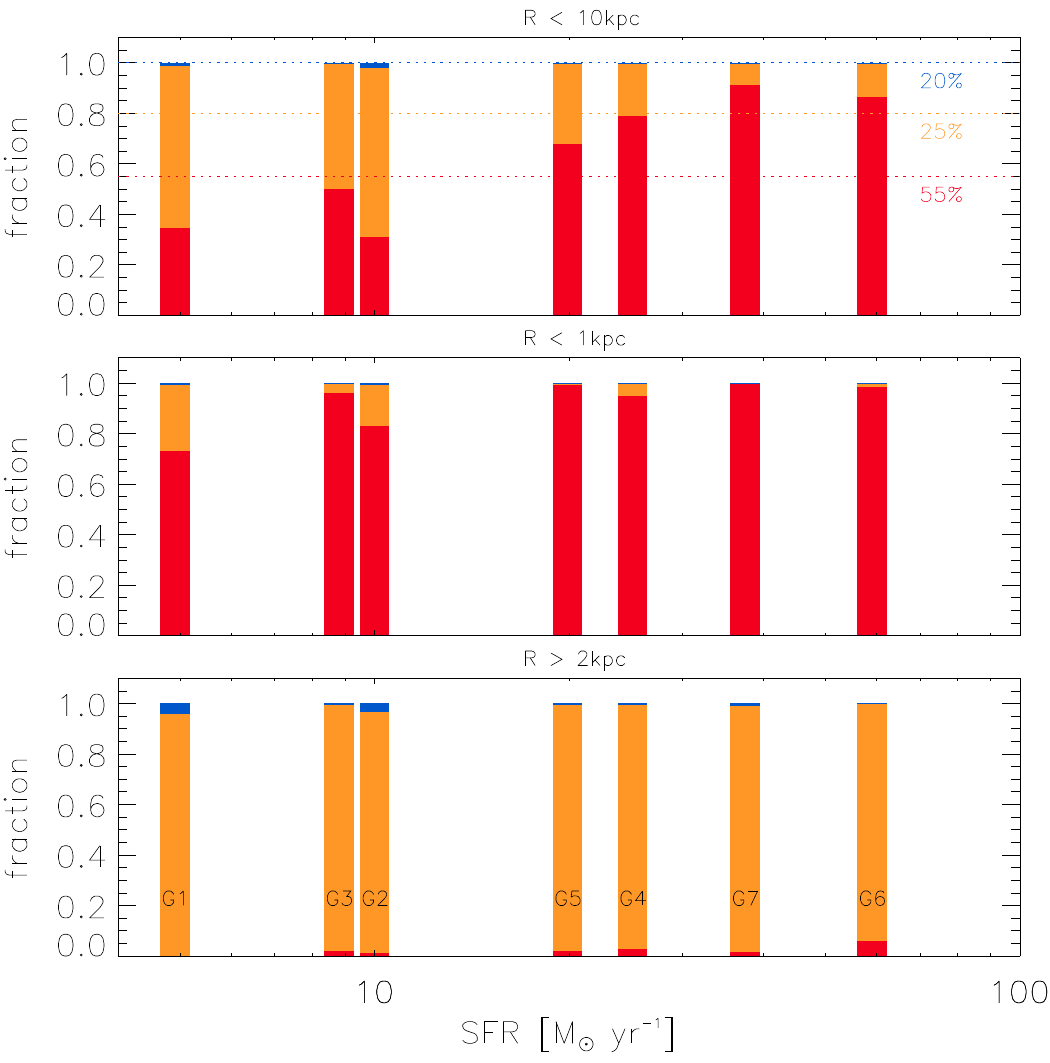}
\caption{\footnotesize{Contributions to the total \cii luminosity from the molecular (red),
PDR (yellow), and ionized (blue) gas phases for each of our simulated galaxies,
ranked according to their total SFRs. The three panels show the relative
contributions within $R < 10\,{\rm kpc}$ (top), for $R < 1\,{\rm kpc}$ (middle),
and for $R > 2\,{\rm kpc}$ (bottom). The horizontal dashed lines and the
percentages given indicate the relative contributions to the total \cii
luminosity of our own Galaxy from CO-dark H$_2$ + dense PDR gas (red; 55\%),
cold atomic gas (yellow; 25\%), and ionized gas (blue; 20\%) \citep{pineda14}.
The total SFR of the Galaxy is $1.9\,{\rm \msun\,yr^{-1}}$ \citep{chomiuk11}. }}
\label{figure:CII-percentages}
\end{figure}

Fig.\,\ref{figure:CII-percentages} displays the fractional \cii luminosity from
the different ISM phases: top panel for the entire disk ($R<10\,{\rm kpc}$),
middle panel for the central region ($R\le 1\,{\rm kpc}$), and bottom panel for
the outer disk ($R>2\,{\rm kpc}$).  Within $R\le 10\,{\rm kpc}$, the molecular
gas can constitute from $\sim31\%$ (G2) to $\sim 91\,\%$ (G7) of the total \cii
luminosity; for the PDR gas the range is $\sim 9\%$ (G7) to $\sim 67\%$ (G2).
Fig.\ \ref{figure:CII-percentages} shows that the contribution from the
molecular gas to the total \cii emission increases with the overall SFR of the
galaxy. A reverse trend is seen for the PDR gas.  As expected, the total
\cii emission from the central regions ($R \le 1\,{\rm kpc}$) is dominated by
the molecular gas ($\gs 70\%$) while further our ($R > 2\,{\rm kpc}$) the
PDR gas phase dominates ($\gs 90\%$)  (see Fig.\
\ref{figure:CII-percentages}, bottom two panels).  For reference, we note that
in our own Galaxy about $55\,\%$ of the total Galactic \cii luminosity (within
$R\ls 20\,{\rm kpc}$) is from molecular gas and dense PDRs, $25\,\%$ from cold
H{\sc i}, and $20\,\%$ from the ionized gas \citep{pineda14}. Thus, the ionized
phase is a more important contributor to the overall \cii budget in our Galaxy
than in the simulated galaxies presented here where the contribution from the
ionized gas is $\ls 3\%$ in all cases.

\clearpage

\subsection{The integrated $\Lcii-{\rm SFR}$ relation}\label{subsection:integrated-CII-SFR-relation}
\begin{figure*}[htbp] 
\hspace{0.5cm}
\includegraphics[width=0.9\textwidth]{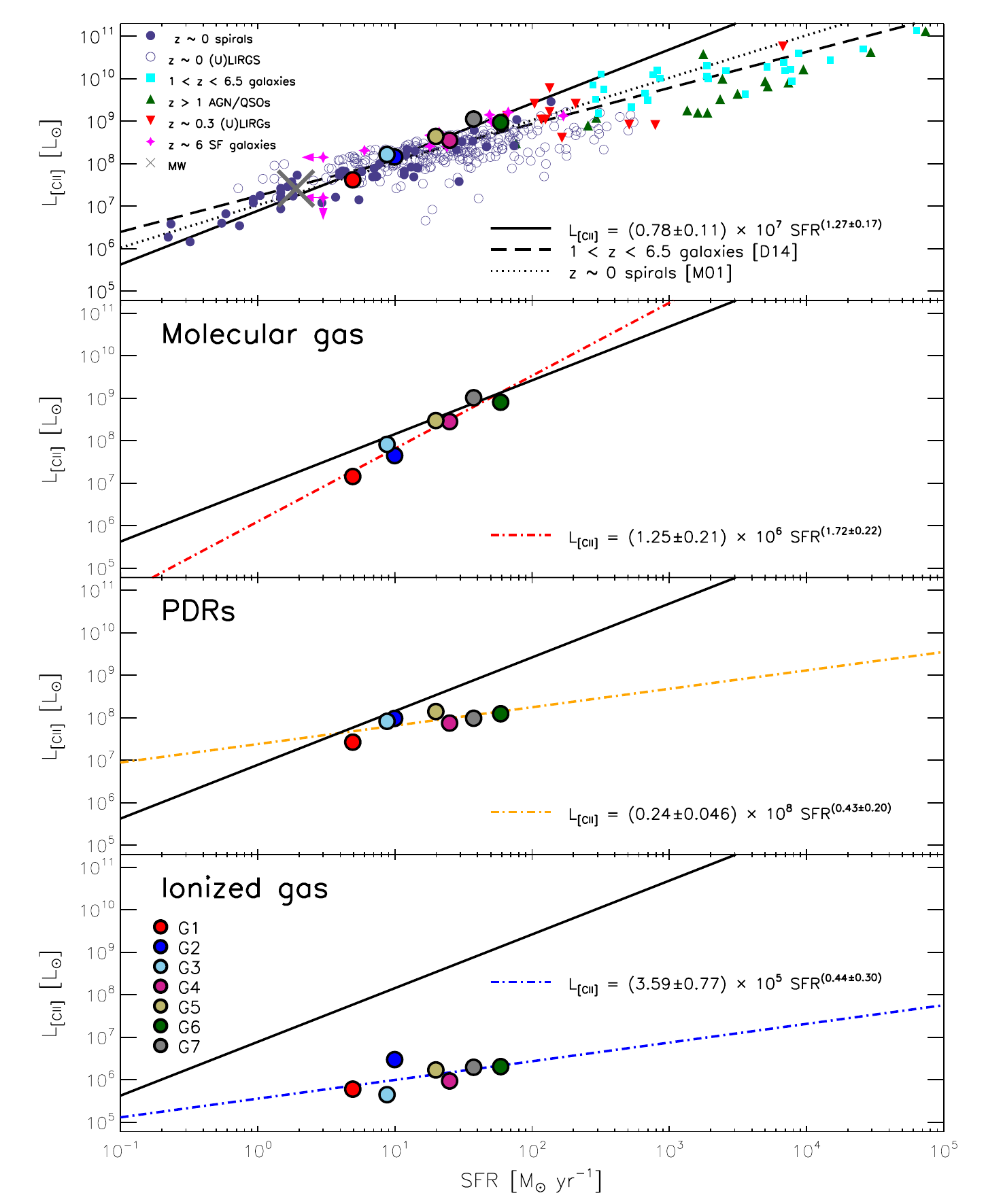}
\caption{\footnotesize{\Lcii vs.\ SFR for our simulated galaxies (big filled circles).  The
same color-coding as in Fig.\,\ref{M_SFR} is used and listed again for
convenience in the bottom panel. From top to bottom panel we show the \cii
luminosity from the full ISM (gray), the molecular gas (red), PDRs (orange), and
ionized gas (blue).  For comparison, we show individual \cii observations of 54
$z\sim 0$ spirals \citep[purple filled circles;][]{malhotra01}, with the best
fit to this sample given by the dotted line (see main text for details).  Also
shown are  240 $z\sim 0$ (U)LIRGs \citep[purple open
circles;][]{diaz-santos13,farrah13}, and 12 $z\sim0.3$ (U)LIRGs \citep[red
triangles;][]{magdis14}.  The high-$z$ comparison samples include 25 $1<z<6$
star-forming galaxies (cyan squares; D14) and the dashed line indicate the best fit
to this sample.  Also shown are 16 $z\sim4-7$ quasars \citep[green
triangles;][]{iono06,walter09,wagg10,gallerani12,wang13,venemans12,carilli13,willott13},
and 10 normal star-forming $z\sim5-6$ galaxies (magenta stars) observed by
\cite{capak15}. The Galaxy is shown for reference with a gray cross \citep{pineda14}.}}
\label{cii_sfr}
\end{figure*}

Fig.\,\ref{cii_sfr} shows the integrated ${\rm \Lcii-SFR}$ relations for our
simulated galaxies: top panel for the full ISM and, in separate panels below,
for each of the three ISM phases considered in our simulations.

When considering the entire ISM a tight correlation between \Lcii and SFR
emerges, which is well fit in log-log space by a straight line with slope
$1.27\pm 0.17$ (solid line in the top panel). This relation is largely set by
the molecular and PDR gas phases. The molecular gas phase, itself
exhibiting a strong correlation between \cii\ and SFR with slope $1.72\pm
0.22$, drives the slope of the total correlation. The PDR gas on the
other hand, shows a weaker \cii$-$SFR correlation with a slope of $0.43\pm
0.20$ but contributes significantly to the normalization of the total ${\rm
\cii-SFR}$ relation, especially at the low SFR end (see Fig.\ \ref{cii_sfr}).
The \cii emission from the ionized gas also shows a weak dependency on SFR
(slope $0.44\pm 0.30$) but does not contribute significantly to the total \cii\
emission, its normalization factor being $\gs 10\times$ below that of the
molecular and PDR gas (Fig.\ \ref{cii_sfr}, bottom panel). 

\smallskip

In the top panel of Fig.\ \ref{cii_sfr} we compare the ${\rm \Lcii-SFR}$
relation obtained from our simulated galaxies with samples of \cii\-detected
galaxies in the redshift range $z\sim0-6.5$ compiled from the literature.  Our
simulated galaxies are seen to match the observed ${\rm \Lcii-SFR}$ relation
both in terms of the slope of the relation and its overall normalization. A
power-law fit to the simulated galaxies (shown as solid line in Fig.\
\ref{cii_sfr}) yields a near-linear slope ($1.27\pm 0.17$).  Normal star-forming
galaxies at $z\sim 0$, with similar levels of SFR ($\sim 0.2 - 100\,{\rm
\msun\,yr^{-1}}$; \citealt{malhotra01}) as our simulated galaxies, are
consistent with a linear correlation given by $L_{\rm [C\,\textsc{ii}]} =
1.05\times 10^7{\rm SFR}$ (dotted line in Fig.\ \ref{cii_sfr}) with a scatter
of $0.3\,{\rm dex}$ \citep{magdis14}\footnote{This expression is inferred
from a power-law fit by \citet{magdis14} to the \cii and FIR ($42.5-122.5\,{\rm
\mu m}$) luminosities (in units of \lsun) of the \cite{malhotra01} sample:
$L_{\rm [C\,\textsc{ii}]} = 10^{-2.51\pm 0.39} L_{\rm FIR}$, where we have made
use of the conversion ${\rm SFR} = L_{\rm FIR} / 3.4\times 10^9$.}. 
The scatter of our simulated galaxies around their best-fit relation is
$0.14\,{\rm dex}$, i.e., significantly lower. We attribute this to the fact
that our simulated galaxies constitute a fairly homogeneous (and small) sample
spanning a rather small range in SFR, $L_{\rm [C\textsc{ii}]}$, and \Z, unlike
the observed samples with which we are comparing.

\bigskip

A direct comparison with \cii-detected galaxies at high redshifts is
complicated by the fact that the latter typically have significantly larger SFRs
than our model galaxies. Furthermore, high-$z$ samples are often heterogeneous
\citep[e.g., significant AGN contribution, cf.][]{gullberg15}.  One exception,
however, is the recent \cii-detected sample of star-forming galaxies at $z\simeq 5-6$
presented by \citet{capak15} (shown as magenta stars in Fig.\ \ref{cii_sfr}),
which span the same range in SFR as our simulations. These galaxies are seen to
be in excellent agreement with the ${\rm \Lcii-SFR}$ relation defined by our
simulations, both in terms of slope and normalization.  

Extrapolating our best-fit relation to SFRs $\gs
300\,{\rm \msun\,yr^{-1}}$ in order to compare with other high-$z$
samples, the relation is seen to overshoot the data. A power-law fit
to the $z > 1$ star-forming galaxies with SFR\,$\gtrsim300$\,\sfru
compiled by D14 yields: $L_{\rm [C\,\textsc{ii}]} = 1.7\times 10^7{\rm
  SFR}^{0.85}$ (D14; shown as the dashed line in Fig.\ \ref{cii_sfr}),
i.e., formally, a shallower relation than that of our simulated
galaxies and that of the $z\sim 0$ sample (albeit less so).  Finally,
we stress that the \cii-detected galaxies at $z > 1$ with SFRs $\gs
300\,{\rm \msun\,yr^{-1}}$ likely derive from rather complex
environments \citep[e.g.][]{narayanan15}, which may not correspond to
the relatively quiescent MS star-forming galaxies modeled
here.

\subsection{The resolved $\CIIsd-\SFRsd$ relation}
 
\begin{figure*}[htbp] 
\hspace{0.5cm}
\includegraphics[width=0.9\textwidth]{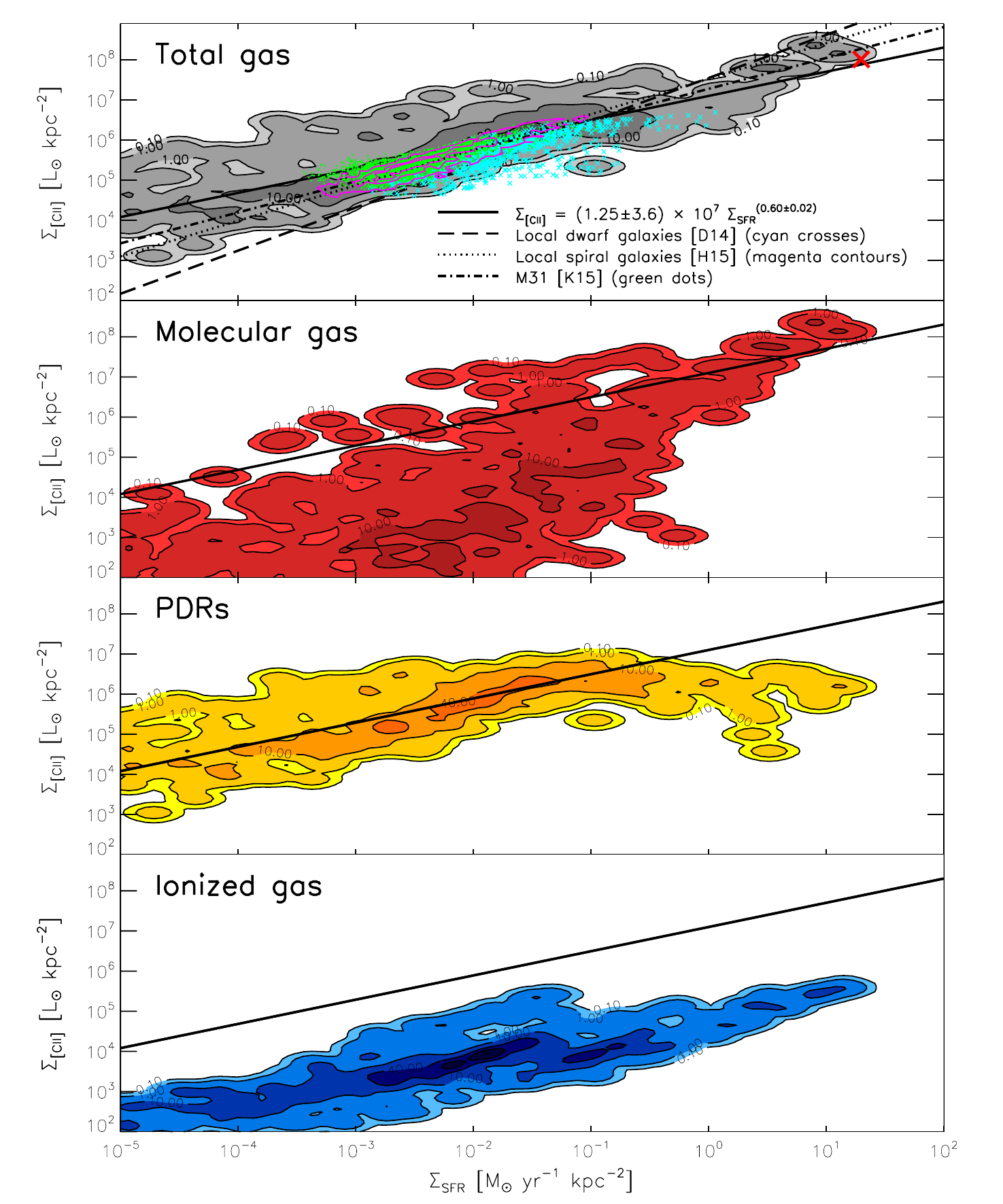} 
\caption{\footnotesize{\cii\ luminosity surface density (\CIIsd) vs.\ SFR surface density
(\SFRsd) for the full ISM in our simulated galaxies (top panel), and for each of
the three ISM phases (bottom three panels). \CIIsd and \SFRsd are determined
over $1\,{\rm kpc}\times 1\,{\rm kpc}$ regions within all 7 galaxies. The filled
colored contours indicate the number of such regions with this combination of
\SFRsd and \CIIsd as a percentage (at 0.1\%, 1\%, 10\%, 40\% and 70\%) of the maximum
number of regions. The solid line shown in all panels is the best-fit power-law
to the total gas \CIIsd vs.\ \SFRsd (see legend). For comparison we show
observed \CIIsd vs.\ \SFRsd for individual $20\,{\rm pc}$ regions in M31
(green dots; K15), $1\,{\rm kpc}$ regions in nearby (mostly spiral) galaxies
(magenta contours, 
indicating regions containing 25\%, 45\%, and 95\% of the total
number of data points; H15), and $1\,{\rm kpc}$ regions in local dwarf galaxies
(cyan crosses; D14).  
The best-fit power-laws to these three data-sets are shown
as dashed-dotted, dotted, and dashed lines, respectively. In the case of M31 the
power-law is fitted to $50\,{\rm pc}$ regions (see K15). We also show the
galaxy-averaged \CIIsd and \SFRsd of ALESS\,73.1 (red cross), a $z=4.76$
submillimeter-selected galaxy which was marginally resolved in \cii and in the
FIR continuum with ALMA, revealing a source size of $R\sim 2\,{\rm kpc}$,
$L_{\rm [CII]} = 5.15\times 10^9\,{\rm \lsun}$, and ${\rm SFR}\sim 1000\,{\rm
\msun\,yr^{-1}}$ \citep{debreuck14}.
}}
\label{cii_sfr_res}
\end{figure*}

In Fig.\,\ref{cii_sfr_res} we show the combined ${\rm \CIIsd-\SFRsd}$
relation of all seven simulated galaxies in their face-on
configuration. The relation is shown for the entire ISM (top panel)
and for each of the separate gas phases (bottom three panels). Surface
densities were determined within 1\,kpc $\times$ 1\,kpc
regions. Contours reflect the number of regions at a given
(\SFRsd,\CIIsd)-combination and are given as percentages of the peak
number of regions.

Given the variations in the local star formation conditions within our model
galaxies, we can explore the relationship between \cii\ and star formation over
a much wider range of star formation intensities than is possible with the
integrated quantities. A ${\rm \CIIsd - \SFRsd}$ correlation spanning more than
five decades in \SFRsd\ is seen for the entire ISM as well as for the individual
ISM phases. Similar to the integrated \cii\ vs.\ SFR relations in the previous
section, the molecular and PDR phases dominate the resolved \cii\ emission
budget at all SFR surface densities, and with the molecular
relation being steepest and exhibiting the largest degree of scatter.  The
resolved \cii emission from the ionized phase, while clearly correlated with
\SFRsd, is largely negligible at all star formation densities.

We compare the ${\rm \CIIsd-\SFRsd}$ relation for our simulated galaxies
to that of three resolved surveys of nearby galaxies: (1) the $\sim
50\,$pc-scale relation derived from five $3'\times 3'$ fields toward
M\,31 (K15; shown as green points in Fig.\ \ref{cii_sfr_res}), (2) the
kpc-scale relation obtained for 48 local dwarf galaxies covering a wide
range in metallicities ($Z/Z_{\rm \odot} = 0.02-1$, D14; cyan crosses),
and (3) the kpc-scale relations for local, mostly spiral, galaxies (H15;
magenta contours). The data from these surveys have been converted to
the Chabrier IMF assumed by our simulations by multiplying \SFRsd with a
factor 0.94 when a Kroupa IMF was adopted (D14; K15) or 0.92 when a
truncated Salpeter IMF was used (H15), following \cite{calzetti07} and
\cite{speagle14}.  Regarding the observations by H15, we adopt the raw
\CIIsd measurements rather than the IR-color-corrected ones. From Fig.\
\ref{cii_sfr_res} we see that over the range in \SFRsd
($\sim0.001-1\,\sfru$\,kpc$^{-2}$) spanned by these three surveys, the
${\rm \CIIsd - \SFRsd}$ relation defined by our simulations is in
excellent agreement with the observations. Also, the majority of our
simulated $1\,{\rm kpc}\times 1\,{\rm kpc}$ regions fall within the
observed \SFRsd and \CIIsd ranges. A small fraction (a few procent) of
regions in our simulations exhibit an excess in \CIIsd for a given
\SFRsd relative to the observed relations, but the majority coincide
with the observations.  We note that the observed relations exhibit
significant scatter ($\sim 0.2-0.3\,{\rm dex}$; D14, H15, K15) as well
as small systematic offsets relative to each other (in particular in the
case of D14).  The ${\rm \CIIsd-\SFRsd}$ relations observed in the five
fields in M\,31 by K15 yield best-fit power-law slopes in the range
$\sim 0.67-1.03$, with an average of $0.77$ (shown as dashed-dotted line
in Fig.\ \ref{cii_sfr_res}). Similar power-law fits to the samples of
D14 and H15 yield slopes of $\sim 1.07$ and $\sim 0.88$, respectively
(shown as dashed and dotted lines in Fig.\ \ref{cii_sfr_res}). In
comparison, a power-law fit to our simulations -- across the full
\SFRsd-range -- results in a slope of $\sim 0.60$, i.e., on the low side
of the observed range.  However, if instead we fit only to simulated
regions within the $\SFRsd =0.001-0.1\,\sfru$\,kpc$^{-2}$ range, thereby
matching the range with the most observations, we find a slope of $\sim
0.75$, i.e., well within the observed range.

Beyond the above \SFRsd-range a comparison with the observations has to rely on
extrapolations of the simple power-law fits to the observed ${\rm \CIIsd -
\SFRsd}$ relations.  At low \SFRsd\ ($\ls 0.001\,{\rm \sfru\,kpc^{-2}}$), the
simulations broadly follow the extrapolations of the observed relations, except
for D14 where the systematic offset noted at higher \SFRsd is compounded at the
lower \SFRsd values owing to the relatively steep slope of the fit to the D14
data.  At these low \SFRsd levels a larger (but still minor, overall) fraction
of the simulated regions display excess \cii\ emission levels ($\gs10\times$)
compared to the survey-based power-law fits.  This excess \cii\ emission is
driven by the PDR gas in our simulations (second panel in
Fig.\,\ref{cii_sfr_res}), which dominates the \cii emission at these \SFRsd
levels.  Interestingly, in at least two of the five fields in M\,31 studied by
K15, a similar \cii-excess relative to the power-law fit is observed at
$\SFRsd\ls 0.001\,$\sfru\,kpc$^{-2}$ (see Fig.\ 7 in K15). K15 argues that the
excess is due to a contribution from diffuse, ionized gas (\hii), although they
cannot discount the possibility that the larger dispersion is at least partly
due to being close to the sensitivity limit of their survey at such low \CIIsd.
Our simulations, however, clearly show no significant contribution from the
diffuse \hii gas phase at low \SFRsd.

At high \SFRsd ($\gs 1\,{\rm \msun\,yr^{-1}\,kpc^{-2}}$), the scatter in the
${\rm \CIIsd - \SFRsd}$ relation from our simulations decreases significantly,
and is seen to broadly match the extrapolated locally observed relations.
Furthermore, the simulation contours are seen to agree with the few galaxies
observed to date to have \SFRsd$\gs1$\,\sfru\,kpc$^{-2}$: a few sources from the
D14 sample (there are two D14 galaxies with \SFRsd$\gs 1$\,\sfru\,kpc$^{-2}$)
and ALESS\,73.1, a $z=4.76$ sub-millimeter selected galaxy, marginally resolved
in \cii and with a disk-averaged \SFRsd of $\sim 80\,{\rm
\msun\,yr^{-1}\,kpc^{-2}}$ (\citealt{debreuck14}; indicated by a red cross in
Fig.\ \ref{cii_sfr_res}).

The aforementioned studies of local galaxies typically measure both obscured
(e.g.\ from $24\,\mu$m) and un-obscured (from either FUV or H$\alpha$) SFRs in
order to estimate the total \SFRsd and, while intrinsic uncertainties are
inherent in the empirical $24\,\mu$m/FUV/H$\alpha\rightarrow\,$SFR calibrations,
the \SFRsd values should be directly comparable to those from our simulations.
Nonetheless, some caution is called for when comparing our simulations to
resolved observations, primarily regarding the determination of \SFRsd on
$\lesssim\,$kpc-scales.  In particular, as pointed out by D14, the emission from
old stellar populations in diffuse regions with no ongoing star formation can
lead to overestimates of the SFR when using the above empirical SFR
calibrations. Both K15 and H15 account for this by subtracting the expected
cirrus emission from old stars in $24\,\mu$m (using the method presented in
\citealt{leroy12}), although on scales $\ll 1\,{\rm kpc}$ this correction for
$24\,\mu$m cirrus emission becomes particularly challenging as it was calibrated
on $\gs 1\,$kpc scales.  K15 further points to the problem arising from the
possibility of photons leaking between neighboring stellar populations, meaning
that average estimates of the SFR on scales $<50$\,pc might not represent the
true underlying SFR.

\subsection{Physical underpinnings of the \Lcii-SFR relation}

\begin{figure}[htbp]
\begin{center} 
\includegraphics[width=0.8\columnwidth]{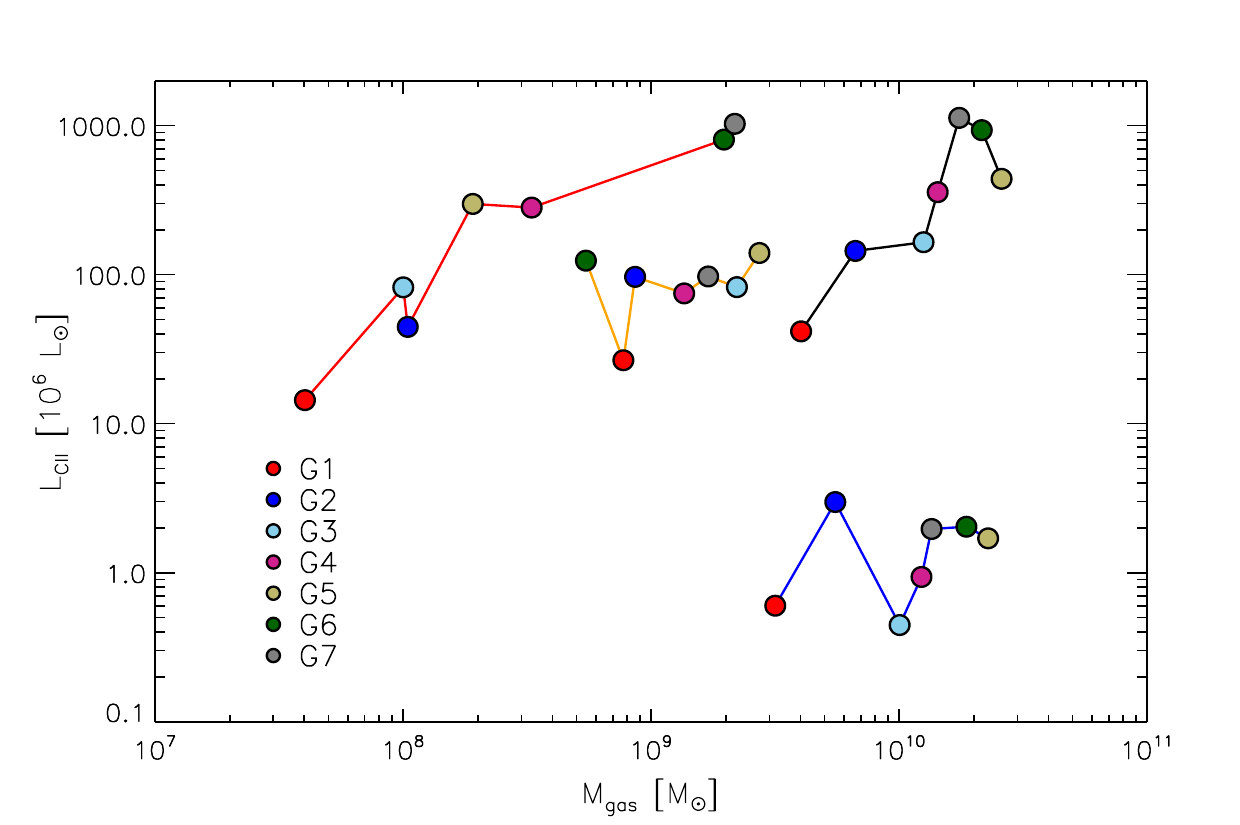}
\end{center} 
\caption{\footnotesize{\cii luminosity versus gas mass of the total ISM (black line), the
molecular (red line), the PDR (orange line), and the \hii (blue line) gas
phase for each of our seven simulated galaxies. The \cii\ emission from the
molecular gas is seen to increase with the amount of molecular gas available,
whereas this is not the case for gas associated with PDR and \hii regions.}}
\label{m_cii}
\end{figure}

The SFRs of galaxies at both low and high redshifts is
observed to strongly correlate with their molecular gas content (typically
traced by CO or dust emission)
\citep[e.g.][]{kennicutt98,daddi10a,genzel10,narayanan12a}.  This ${\rm
SFR}-{\rm H_2}$ dependency is incorporated into our SPH simulations (see
Section \ref{sph}), and it is therefore natural to ask whether the integrated
${\rm \cii-SFR}$ relation examined in Section
\ref{subsection:integrated-CII-SFR-relation} might simply be a case of galaxies
with higher SFRs having larger (molecular) gas masses with
higher associated \cii luminosities. To investigate this, we show in
Fig.\,\ref{m_cii} \cii luminosity vs.\ gas mass for the full ISM in our
simulated galaxies and for the individual gas phases separately.
For the molecular gas phase we see a significant luminosity$-$mass scaling (red
curve), which would explain the strong $L_{\rm [C\textsc{ii}]}-{\rm SFR}$
relation for this phase and, at least in part, the $L_{\rm [C\textsc{ii}]}-{\rm
SFR}$ relation for the full ISM. The PDR gas and the ionized gas
do not exhibit similar luminosity$-$mass scaling relations. Since, on average,
significantly more mass resides in each of these two phases than in the
molecular phase, this has the effect of weakening the correlation between $L_{\rm
[C\textsc{ii}]}$ and SFR when considering the full ISM.

\begin{figure}[htbp] 
\begin{center} 
\includegraphics[width=0.8\columnwidth]{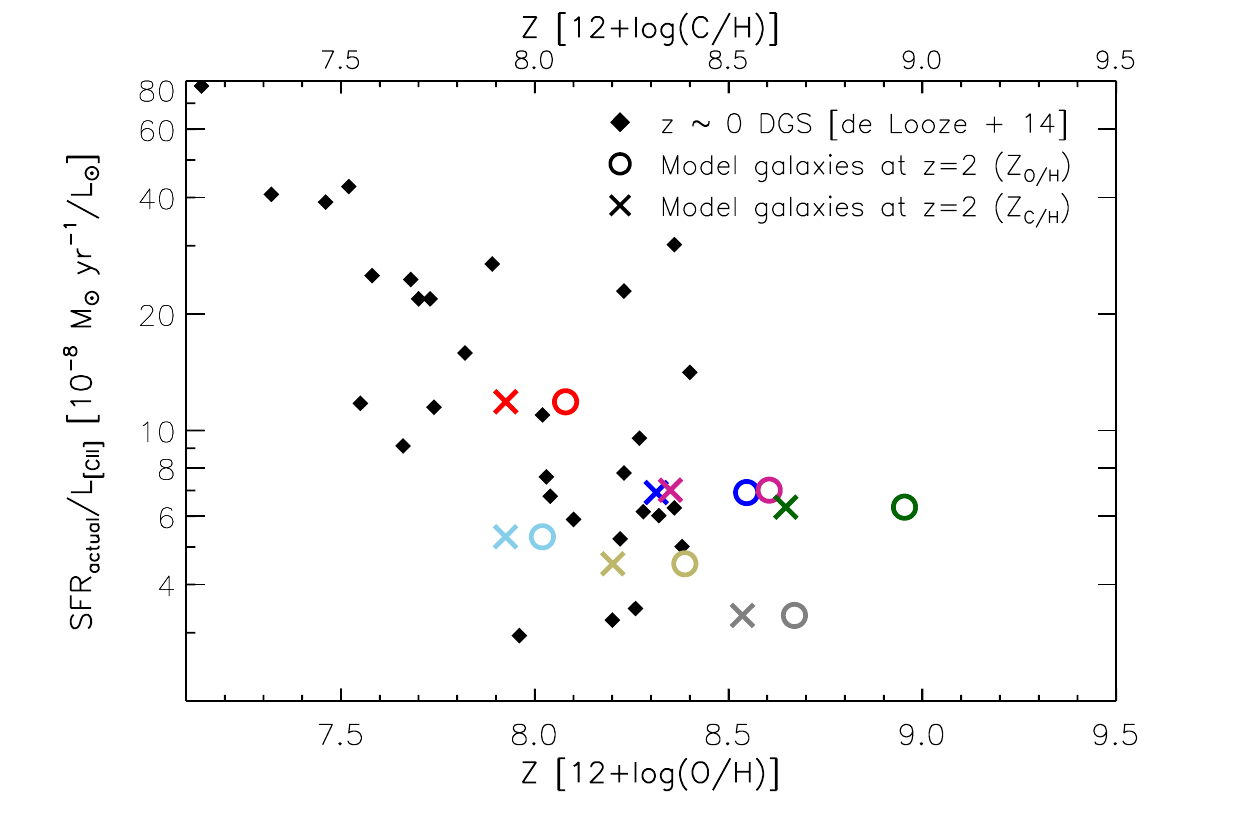}
\end{center}
\caption{\footnotesize{${\rm SFR}/\Lcii$ as a function of metallicity for our simulated
galaxies.  The metallicity is parametrized both relative to the oxygen (circles)
and the carbon (crosses) abundance. The galaxies are color coded in the usual
manner.  For comparison, we show ${\rm SFR}/\Lcii$ vs.\ $Z_{\rm [O/H]}$ for the
local dwarf galaxy sample of D14, where SFRs are from combined UV and $24\,\mu$m
measurements.}}
\label{cii_zm}
\end{figure}

The abundances of metals in the ISM is expected to affect the \cii\ emission.
In their study of low-metallicity dwarf galaxies, D14 compared total (IR$+$UV)
SFRs to SFRs derived using the best-fit $\Lcii-{\rm SFR}$ relation to their
entire sample.  They found an increase in the SFR$_{\rm{IR+UV}}$/SFR$_{{\rm
C\textsc{ii}}}$ fraction, corresponding to weaker \cii emission, toward lower
metallicities.  In Fig.\ \ref{cii_zm} we show ${\rm SFR/\Lcii}$ as a function of
$Z_{\rm [O/H]}$ ($=12+\log {\rm [O/H]}$) for our simulated galaxies as well as
for the dwarf galaxy sample of D14.  As D14 points out, $Z_{\rm [O/H]}$ does not
take into account potential deficits of carbon relative to the [O/H] abundance,
potentially obscuring any potential correlation between \cii luminosity and
actual carbon abundance. In Fig.\ \ref{cii_zm} we therefore also plot ${\rm
SFR/\Lcii}$ as a function of  $Z_{\rm [C/H]}$ ($=12+\log {\rm [C/H]}$) for our
simulated galaxies. Arguably, our simulated galaxies exhibit a weak trend in
${\rm SFR/\Lcii}$ with metallicity (in particular for [C/H]). Certainly, there
is a very good overall agreement with the findings of D14. The lack of a strong
trend within our simulation sample is not suprising, however, since our galaxies
span a limited range in metallicity ($Z_{\rm [O/H]} = 8.03-8.96$) and, as we saw
in Section \ref{subsection:integrated-CII-SFR-relation}, form a tight $L_{\rm
CII}-{\rm SFR}$ relation.  

\begin{figure}[htbp] 
\begin{center} 
\includegraphics[width=0.8\columnwidth]{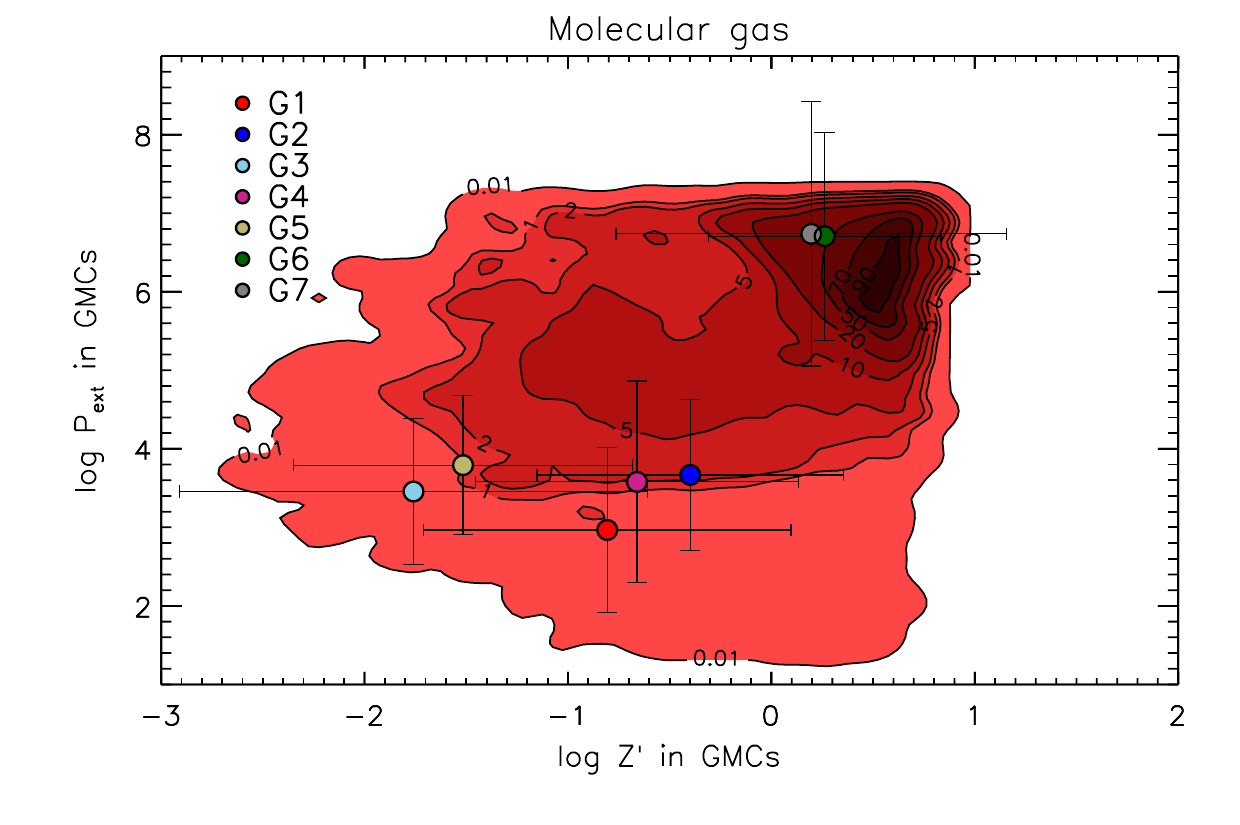}\\
\includegraphics[width=0.8\columnwidth]{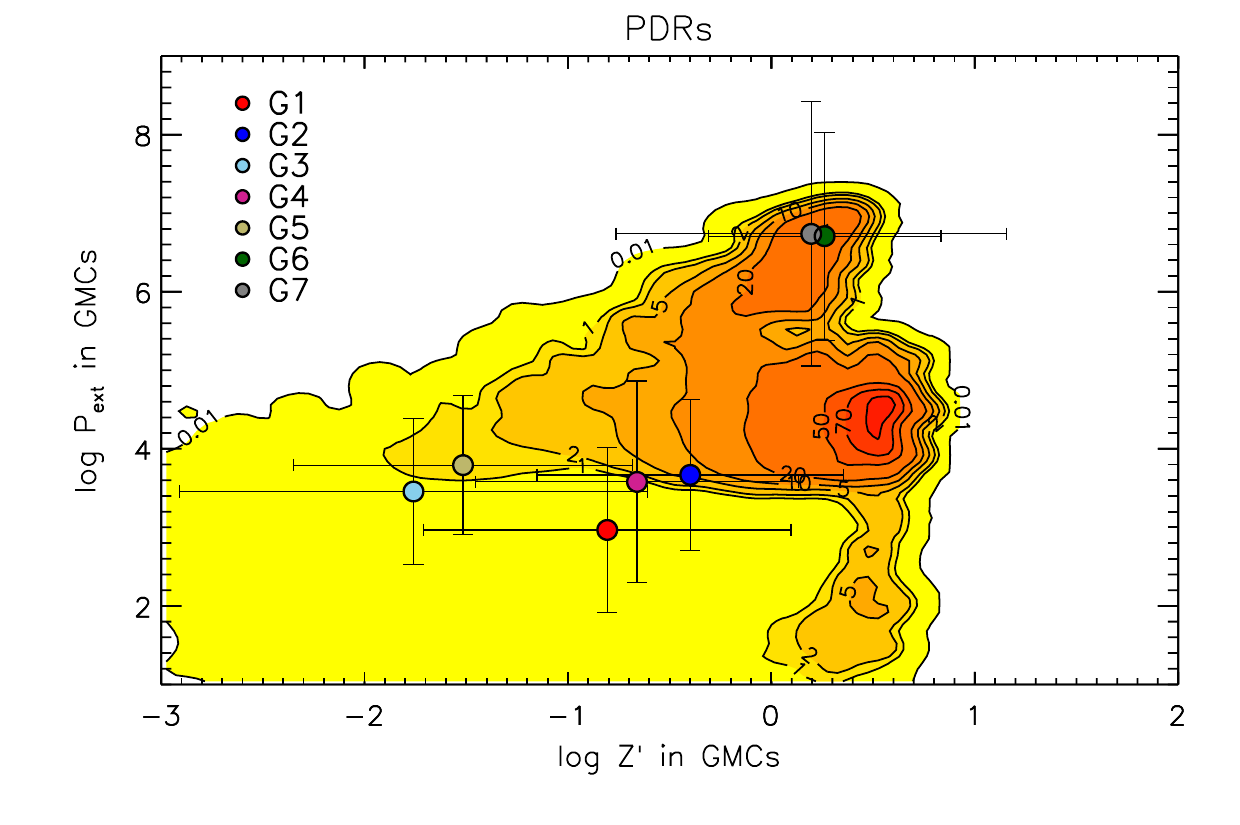}
\end{center} 
\caption{\footnotesize{Contours of (median) $\epsilon_{\rm [C\textsc{ii}]}$ (i.e.\ \cii
luminosity per gas mass) as a function of \Zn\ and $\Pe/k_{\rm B}$ for the
molecular (top) and PDR (bottom) gas phases in our model
galaxies. In each panel the contours indicate 0.01\%, 1\%, 2\%, 5\%, 10\%, 20\%, 50\%, 70\%, and
90\% percentages of the maximum efficiency. For reference, the maximum (median)
$\epsilon_{\rm [C\textsc{ii}]}$ for the molecular and PDR gas phases
are: $3.7$ and $6.6\,{\rm \lsun/\msun}$, respectively.  Median values
of \Zn\ and $\Pe/k_{\rm B}$ and for the GMC and ionized cloud population in each
model galaxies are indicated as colored circles (error bars are the $1\sigma$
dispersion of the distributions).}}
\label{Pe_Z_grid1}
\end{figure}

\begin{figure}[htbp] 
\begin{center} 
\includegraphics[width=0.8\columnwidth]{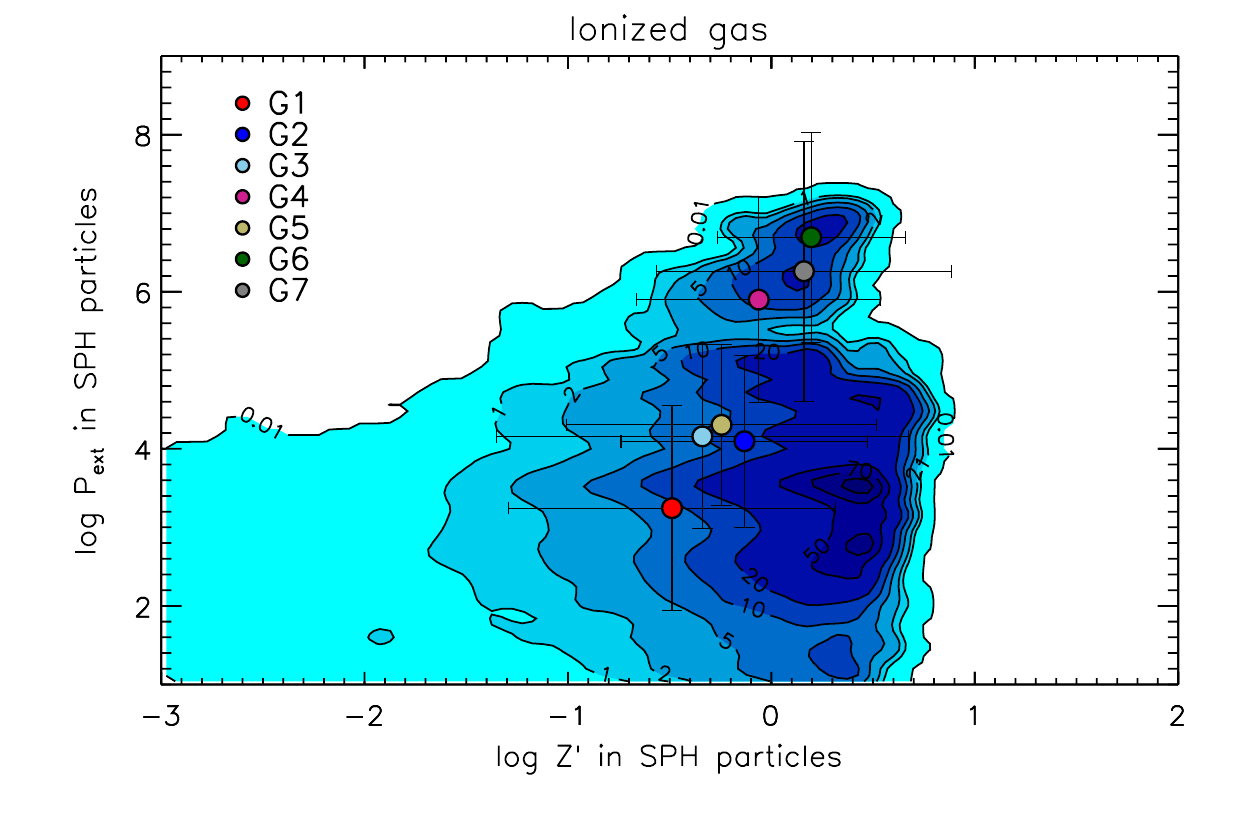}
\end{center} 
\caption{\footnotesize{Contours of (median) $\epsilon_{\rm [C\textsc{ii}]}$ (i.e.\ \cii
luminosity per gas mass) as a function of \Zn\ and $\Pe/k_{\rm B}$ for the 
\hii\ gas phases in our model
galaxies. The contours and galaxy positions are defined as in Fig.\,\ref{Pe_Z_grid1}. 
The maximum (median) ionized gas phase
is: $0.02\,{\rm \lsun/\msun}$.}}
\label{Pe_Z_grid2}
\end{figure}

To better understand how the metallicity and also the ISM pressure affects the
\cii emission, we define a `\cii\ emission efficiency' ($\epsilon_{\rm
[C\textsc{ii}]}$) of a given gas phase as the \cii luminosity of the gas phase
divided by its mass, and examine how it depends on \Zn\ and $\Pe/k_{\rm B}$.
Specifically, we create a grid of (\Zn, $\Pe/k_{\rm B}$) values, and in each
grid point the median $\epsilon_{\rm [C\textsc{ii}]}$ of all clouds in our
simulated galaxies is calculated.  The resulting $\epsilon_{\rm [C\textsc{ii}]}$
contours as a function of \Zn\ and $\Pe/k_{\rm B}$ are shown in Fig.\
\ref{Pe_Z_grid1} for the molecular (top) and PDR (middle), and Fig.\
\ref{Pe_Z_grid2} for the \hii\ gas phases. 
For comparison, we also show the median \Zn- and $\Pe/k_{\rm B}$-values
for each of our galaxies.

The molecular phase in our simulations is seen to radiate most efficiently in
\cii at relatively high metallicities ($\log \Z\sim0.5$) and cloud external
pressures ($\Pe/k_{\rm B} \sim10^6\,{\rm cm^{-3}\, K}$).  The PDR and ionized
phases, however, have their maximal $\epsilon_{\rm [C\textsc{ii}]}$ at $\log
\Z\gs0$ and $\Pe\sim10^4\,{\rm cm^{-3}\,K}$, i.e.\ at $\sim 100\times$ lower
pressures than the molecular phase, and can maintain a significant \cii
efficiency ($\gs 20\%$) even at $\Pe/k_{\rm B}\sim 10^2\,{\rm cm^{-3}\,K}$
provided $\log \Zn\gs 0$. This is consistent with our finding in Section
\ref{subsection:CII-profiles}, that the molecular phase dominates the \cii
emission in the central, high-pressure regions, while the PDR gas dominates
further out in the galaxy where the ISM pressure is less extreme.  In all three
panels, G6 and G7 lie in regions of high $\epsilon_{\rm [C\textsc{ii}]}$ ($\gs
50\%$ for the molecular and PDR gas). Thus,
the reason why these two galaxies have the highest total \cii luminosities is in
part due to the fact that they have the highest molecular gas masses (by nearly
an order of magnitude, see Fig.\ \ref{m_cii}), and in part due to the bulk of
their cloud population (GMCs or ionized clouds) having sufficiently high
metallicities (on average $\gs 5\times$ higher than G2) and experiencing external
pressures ($\sim 100 \times$ higher on average than the remaining galaxies)
which drives their \cii efficiencies up.

\section{Comparing with other \cii\ simulations}
A direct quantitative comparison with other \cii\ emission simulation studies in
the literature \citep{nagamine06,vallini13,vallini15,popping14b,munoz14} is complicated by
the fact that there is not always overlap in the masses and/or redshifts of the
galaxies simulated by the aforementioned studies and our model galaxies.
Furthermore, there are fundamental differences in the simulation approach, with
some adopting semi-analytical models \citep{popping14b,munoz14} and others
adopting SPH simulations \citep{nagamine06,vallini13,vallini15}. Also, differences in the
numerical resolution of both types of simulations, and in the specifics of the
sub-grid physics implemented, can lead to diverging results and make comparisons
difficult.

The simulations presented here combine cosmological SPH galaxy simulations with
a sub-grid treatment of a multi-phase ISM that is locally heated by FUV
radiation and CRs in a manner that depends on the local SFR 
density within the galaxies.  We have applied our method to a high
resolution ($\delta m_{\rm SPH}\simeq 1.7\times 10^4\,{\rm \msun}$) cosmological
SPH simulation of $z=2$ star-forming galaxies (i.e., baryonic mass resolution
$\delta m_{\rm SPH}\simeq 1.7\times 10^4\,{\rm \msun}$ and gravitational
softening length $\simeq 0.6\,h^{-1}\,{\rm kpc}$, see Section \ref{sph}).  A
novel feature of our simulations is the inclusion of molecular, neutral and
ionized gas as contributors to the \cii emission.  Another unique feature of our
model is the inclusion of CRs as a route to produce C$^+$ deep inside the GMCs
where UV photons cannot penetrate (see Section \ref{cii_em}). 

Our simulations probably come closest -- both in terms of methodology and
galaxies simulated -- to those of \cite{nagamine06} who 
employed cosmological simulations with \gadgettwo (the precursor for \gadget used here) to predict
total \cii luminosities
from dark matter halos associated with $z\simeq 3$ LBGs 
with $M_{\rm *} \sim 10^{10}\,{\rm \msun}$ and ${\rm SFR}
\gs 30\,{\rm \msun\,yr^{-1}}$ \citep[see also][]{nagamine04}.  Their simulations
employed gas mass resolutions in the range $\delta m_{\rm SPH}\simeq 3\times
10^{5-8}\,{\rm \msun}$ and gravitational softening lengths of typically
$>1h^{-1}\,{\rm kpc}$.  In their simulations the ISM consist of a CNM 
($T\sim 80\,{\rm K}$ and $n\sim 10\,{\rm cm^{-3}}$) and a warm neutral
medium ($T\sim 8000\,{\rm K}$ and $n\sim 0.1\,{\rm cm^{-3}}$) in
pressure-equilibrium, and the assumption is made that the \cii emission only
originates from the former phase. The thermal balance calculation of the ISM
includes heating by grain photo-electric effect, CRs, X-rays, and
photo-ionization of C\,{\sc i} -- all of which are assumed to scale on the local
SFR surface density.  Their simulations predict $L_{\rm [C{\sc II}]} \sim (0.3 -
1)\times 10^8\,{\rm \lsun}$ for their brightest LBGs (${\rm SFR} \gs 30\,{\rm
\msun\,yr^{-1}}$) which is slightly below the prediction of our integrated ${\rm
\cii-SFR}$ relation ($L_{\rm [C{\sc II}]} \sim 6\times 10^8\,{\rm \lsun}$; Fig.\
\ref{cii_sfr}).

Employing a semi-analytical model of galaxy formation, \cite{popping14b} made
predictions of the integrated \cii luminosity from galaxies at $z=2$ with
$M_{\rm *}\sim 10^8-10^{12}\,{\rm \msun}$ and ${\rm SFR \sim 0.1-100\,{\rm
\msun\,yr^{-1}}}$ \citep[see also][]{popping14a}. The galaxies are assumed to
have exponential disk gas density profiles, with randomly placed
`over-densities' mimicking GMCs (constituting $\sim 1\%$ of the total volume).
The gas is embedded in a background UV radiation field of 1\,Habing, with local
variations in the radiation field set to scale with the local SFR surface
density. The C$^+$ abundance is set to scale with the abundance of carbon in the cold gas. 
The excitation of \cii\ occurs via collisions with $e^-$, H, and H$_2$, and the
line emission is calculated with a 3D radiative transfer code that takes into
account the kinematics and optical depth effects of the gas.  For galaxies with
SFRs similar to our model galaxies ($\sim 5 - 50\,{\rm \msun\,yr^{-1}}$) their
simulations predict ensemble-median \cii\ luminosities of $\ls (1-6)\times
10^7\,{\rm \lsun}$.\footnote{Since \citet{popping14b} plots $L_{\rm [C{\sc
II}]}$ against $L_{\rm IR}$ and not SFR, we have converted our SFRs to $L_{\rm
IR}$ in order to crudely estimate their predicted \cii\ luminosities (see their
Fig.\ 11). For the ${\rm SFR}\rightarrow L_{\rm IR}$ conversion we have used
\citep{bell03}.  Not all the star formation will be obscured and so the IR
luminosities, and thereby the \cii\ luminosities given here will be upper
limits.} This is lower than the \cii\ luminosities predicted by our simulations
and also somewhat on the low side of the observed $z\sim 0$ $L_{\rm [C{\sc
II}]}-{\rm SFR}$ relation (Fig.\ \ref{cii_sfr}). Allowing for the $+$2$\sigma$
deviation from the median of the \citet{popping14b} models results in $L_{\rm
[C{\sc II}]}\ls (2-30)\times 10^7\,{\rm \lsun}$ for ${\rm SFR \sim 5
- 50\,{\rm \msun\,yr^{-1}}}$, which matches the observed $L_{\rm [C{\sc
  II}]}-{\rm SFR}$ relation.

The remaining \cii simulation studies in the literature focus on $z \gs 6$
galaxies \citep{vallini13,vallini15,munoz14}.  \citet{vallini13} uses a
GADGET-2 cosmological SPH simulation with a mass resolution $\delta m_{\rm SPH}
= 1.32\times 10^5\,{\rm \msun}$ and gravitational softening length $\sim
2h^{-1}\,{\rm kpc}$.  They adopt a two-phased ISM model (CNM$+$WNM), and
heating and cooling mechanisms similar to that of \citet{nagamine06}, in order
to predict the \cii\ emission from a $z=6.6$ Lyman alpha emitter (LAE) with
${\rm SFR \simeq 10\,{\rm \msun\,yr^{-1}}}$.  They investigated two cases of
fixed metallicity: $\Z=1$ and $\Z=0.02$.  In their simulations, the CNM ($T
\sim 250\,{\rm K}$ and $n\sim 50\,{\rm cm^{-3}}$) is found to be responsible
for $\sim 95\%$ of the total \cii emission, with the WNM phase ($T\sim
500\,{\rm K}$ and $n\sim 1\,{\rm cm^{-3}}$) contributing the remaining 5\%.
\cite{vallini15} presents an update to their 2013 model, in which the same
$z=6.6$ SPH simulation as in \cite{vallini13} is considered but now with the
implementation of a density-dependent prescription for the metallicity of the
gas, and the inclusion of \cii contributions from PDRs (in addition to their
previous two-phased CNM$+$WNM ISM model), and accounting for the effect of the
CMB on the \cii emission.  As a result of these updates, it is found that the
\cii emission is now dominated by the PDRs, with $<10\%$ coming from the CNM.
This is qualitatively consistent with the results from our simulations at $z=2$
where the PDR gas dominates the total \cii emission at least at the low SFR end
($\ls 10\,\sfru$).  However, our simulations do not incorporate the CNM to
the same extent as that of \cite{vallini15}, as the lowest densities found in
the neutral gas in our simulations is of order $\sim100\,{\rm cm^{-3}}$, i.e.,
above typical CNM densities of $\sim 20-50\,{\rm cm^{-3}}$.  It is therefore
reassuring that \cite{vallini15} find the CNM contribution to the total \cii
emission to be benign.  Assuming that $\Sigma_{\rm SFR} \propto \Sigma_{\rm H_2}$ 
and $\Sigma_{\rm H_2} \propto \CIIsd$, \cite{vallini15} scale the \cii
luminosity of their fiducial LAE model (${\rm SFR} = 10\,{\rm \msun\,yr^{-1}}$)
in order to generate a $L_{\rm [CII]} - {\rm SFR}$ relation.

\citet{munoz14} make analytical predictions of the \cii luminosities for a
range of galaxy types at $z\gs 6$ (e.g., Lyman-alpha emitters, starburst
galaxies and quasars, spanning a range in SFR from tens of ${\rm
\msun\,yr^{-1}}$ to several thousand) as part of their efforts to develop an
analytical framework for disk galaxy formation and evolution at these early
epochs.  In their study, the \cii emitting gas is assumed to come from
photo-dissociation regions only. Throughout their models, the metallicity is
kept fixed at solar.  By tuning their models, i.e., either increasing the star
formation efficiency at high redshifts or lowering the depletion of metals onto
dust grains, they arrive at the \cii$-$SFR relation: $L_{\rm [CII]}/\lsun =
5\times 10^8 \left ( {\rm SFR}/100\,{\rm \msun\,yr^{-1}}\right )^{0.9}$. While
this is essentially a linear relation, it struggles to match the expected \cii
luminosities based on observations due to the somewhat lower normalization.

\section{Conclusion}
We have adapted \sigame (described in previous Chapter) to include simulations of the \cii\ emission from
star-forming galaxies. 
\sigame was applied to SPH simulations of seven star-forming galaxies at $z =
2$ with stellar masses in the range $\sim (0.4 - 6.6)\times 10^{10}\,{\rm
\msun}$ and SFRs $\sim 5-60\,{\rm \msun\,yr^{-1}}$ in order to
make predictions of the \cii\ line emission from MS galaxies during
the peak of the cosmic star formation history.

A key result of our simulations is that the total \cii\ emission budget from our
galaxies is dominated by the molecular gas phase ($\gs 70\%$) in the central
regions ($R\ls 1\,{\rm kpc}$) where the bulk of the star formation occurs and is
most intense, and by PDR regions further out ($R\gs 1-2\,{\rm kpc}$) where the
molecular \cii emission has dropped by at least an order of magnitude compared
to their central values.  The PDR gas phase,
while rarely able to produce \cii emission as intense as the molecular gas in
the central regions, is nonetheless able to maintain significant levels of \cii
emission from $R\sim 2\,{\rm kpc}$ all the way out to $\sim 8\,{\rm kpc}$ from
the center. The net effect of this is that on global scales the PDR gas
can produce between 8\% and 67\% of the total \cii luminosity with the
molecular gas responsible for the remaining emission. We see a trend in which
galaxies with higher SFRs also have a higher fraction of their
total \cii luminosity coming from the molecular phase.  Our simulations
consistently show that the ionized gas contribution to the \cii\ luminosity is
negligible ($\ls 3\%$), despite the fact that this phase dominates the ISM
mass budget (see Fig.\ \ref{m_cii}). Therefore, the ionized gas phase is an
inefficient \cii line emitter in our simulations.

The integrated \cii\ luminosities of our simulated galaxies strongly correlate
with their SFRs, and in a manner that agrees well (both in terms of slope and
overall normalization) with the observed $L_{\rm [CII]} - {\rm SFR}$ relations
for normal star-forming galaxies at both low and high redshifts.  We have also
examined the relationship between the $1\,{\rm kpc}$-averaged surface densities
of $L_{\rm [C{\sc II}]}$ and ${\rm SFR}$ across our simulated galaxies. The
resulting $\CIIsd-\SFRsd$ relation spans six orders of magnitude in \SFRsd
($\sim 10^{-5}-10\,{\rm \msun\,yr^{-1}\,kpc^{-2}}$), extending beyond the
observed ranges at both the low and high end of the relation.  In the
\SFRsd-range where a direct comparison with observations can be made ($\sim
0.001-1\,{\rm \msun\,yr^{-1}\,kpc^{-2}}$) we find excellent agreement with our
simulations.

Our simulations suggest that the correlation between \cii and SFR -- both the
integrated and the resolved versions -- is determined by the combined
\cii-contribution from the molecular and PDR phases (the ionized gas
makes a negligible contribution), with the former exhibiting the steepest slope
and dominating the \cii emission at the high-SFR-end.  We argue that this is
due to the fact that the \cii luminosity scales with the amount of molecular
gas present in our simulated galaxies.  A similar luminosity-mass scaling is
not seen for the other phases.  Our work therefore suggest that the observed
$\cii-{\rm SFR}$ relation is a combination of the line predominantly tracing
the molecular gas (i.e., the star formation 'fuel') at high SFR levels/surface
densities, while at low SFRs/surface densities the line is tracing PDR
gas being exposed to a weaker, interstellar UV-field.  As a consequence, we
hypothesize that galaxies with large mid-plane pressures and large molecular
gas fractions will display a steeper ${\rm \cii - SFR}$ relationship than
galaxies where a larger fraction of the ISM is atomic/ionized gas. In the
future we will extend this study to a larger sample of model galaxies, in
particular with a larger spread in SFRs and metallicities.

In the future we feel strongly encouraged to extend this study to a larger sample of
galaxies, in particular with a larger spread in SFRs.

\chapter{Outlook}

\section{Improvements on \textnormal{\sigame}}
The aim of \sigame is to model line emission from the ISM of galaxies at any redshift. 
Such a code has to be as true as possible to the observations of gas and dust, on 
small and large scale. 
At high-$z$ (and assuming that physical laws do not change in time), this turns into 
a limitation in how well we can resolve the ISM spatially or spectroscopically, 
sometimes leading to rather crude assumptions based on what can be observed locally.
As a recently developed code, \sigame has several such assumptions with room for improvement, 
of which three are listed below.

\subsection{Dust temperatures with a radiative transfer code}
With \sigame we have demonstrated a new method that takes into account local FUV fields and CR 
intensities in galaxies. 
However, our method of scaling these fields to the local SFR and total stellar mass, 
can be made much more realistic with the inclusion of an actual dust radiative transfer (RT) code. 
One such code, developed recently, is Hyperion\footnote{\url{http://www.hyperion-rt.org/}} \citep{robitaille11}
for which Powderday\footnote{\url{https://bitbucket.org/desika/powderday}} can be used 
as an interface between Hyperion and galaxy formation simulations.
And many alternatives exist as evident from the review by \cite{steinacker13}.
Implementing one of these would allow us to self-consistently derive dust temperatures, taking into account anisotropic scattering, 
absorption and (re-)emission by interstellar dust, all based on the distribution of stars and dust in the galaxy.


\subsection{Modeling of asymmetric GMCs}
At present, there exists a gap between simulations of the ISM on galaxy-scales and those on 
scales relevant for the 
formation of individual stars. On galaxy-scales, simulations are typically restricted to model star formation 
using empirical correlations between gas density and star formation rate. 
However, more detailed simulations are emerging, following dense gas from kpc-scales down to 
$\sim0.1\,$pc where star clusters are formed \citep{butler14,dobbs15}.
With modifications, \sigame could be applied to resolved simulations of single star formation such as 
those of \cite{vazquez10b,vazquez11} and \cite{dale13}, in order to make a more precise grid in CO line intensities as a function of 
more global parameters. 
The move from spherically symmetric GMC models towards the clumpy, filamentary clouds observed is an important step 
forward for \sigame.

\subsection{Including heating by X-rays and turbulent dissipation}
Particularly towards the Galactic center, observations demand the presence of non-photon driven heating such 
as cosmic rays and turbulent dissipation \citep{ao14}.
But also in normal star-forming GMCs, turbulent dissipation has been suggested as a cause for most 
of the observed excess in CO lines at $J_{\rm upper}\geq 6$ \citep{pon12,pon14}.
In AGNs, the circum-nuclear disks can act like an X-ray Dominated Region (XDR), and X-ray heating is 
therefore important when modeling these central regions \citep[e.g.][]{aalto07,garcia-burillo10}.
We have focused our research projects on normal star-forming galaxies, but if \sigame is to work more 
generally on starburst and AGNs, turbulent dissipation and X-ray heating must be included in the model.

\section{Going to higher redshift}

\subsection{The evolution of \xco with redshift}
Gas mass estimates depend crucially on the dust-to-gas mass ratio or the X-factor, \xco. 
At $z\sim2$, \sigame predicts \xco factors about half that of the MW for massive star-forming 
galaxies, and applying the same method 
to simulations at higher redshift, we could investigate how \xco changes with redshift. 
Such predictions will be important for future observational campaigns trying to pin down the 
evolution of $f_{\rm gas}$ with redshift. 
\xco could also readily be applied to observations of any CO line, if we knew 
the shape of the CO ladder beforehand. 
\cite{narayanan12} developed a parametrization of the CO ladder with \SFRsd that \sigame 
can now improve on by seeing how it changes at $z>2$ for normal star-forming galaxies.

\subsection{The full calibration to normal galaxies at $z\sim2$}
As evident from the study of CO and \cii line emission in normal star-forming $z\sim2$ galaxies, 
we are in need of more observations, in particular of the full CO SLED, in order 
to test models, such as \sigame, and use them to their fullest. 
I plan to continue my involvement in the HELLO ({\it Herschel} Extreme Lensing 
Line Observations) project which takes advantage of powerful lensing 
(by e.g. foreground groups of galaxies) in order to detect \cii line emission from normal galaxies 
at $z\sim2$. With the high spectral resolution of {\it Herschel}, kinematic studies of the line profiles 
are used to estimate rotation speed and gas velocity dispertion \citep{rhoads14}.
Follow-up observations of CO lines with JVLA and ALMA make these galaxies excellent callibrators for \sigame, 
which itself will help in the interpretation of these future line observations.

\subsection{Galaxies during the epoch of re-ionization}
In the early universe, at redshifts above $7$, spectroscopic obervations in the FIR might be our only way 
to characterize the modest MS galaxies responsible for most star formation, 
that are otherwise proving difficult to capture in rest-frame UV with existing spectrographs.
Observing and modeling the gas conditions during the Epoch of Re-ionization (EoR) will 
enable us to answer questions such as:
{\it How did galaxies first aquire their gas masses? 
When did the gas become enriched by the first supernovae?
What role did the gas play in setting the SFR?}

Observing with band 7 of ALMA, \cite{capak15} found $>3\sigma$ detections of \cii in a sample of 9 normal 
(SFR$\,<3$ to $\sim170$\,\sfru) galaxies at $z\sim5-6$, of which only 4 are detected in dust continuum, underlining 
the importance of \cii as a tracer of the ISM at high redshift.
However, many attempts with ALMA (and PdBI) pointing at $z\sim7$ star-forming MS galaxies for hours, 
have only resulted in upper limits on both \cii line 
emission and dust continuum \citep{ouchi13,maiolino15,ota14,schaerer15,gonzalez-lopez14}.
One preferred explanation is low metallicity because the galaxies might not have had enough time 
to enrich the primordial gas with metallicities to a degree that their line emission is 
strong enough for detection \citep{ouchi13,ota14,gonzalez-lopez14,schaerer15}. 
Another explanation based on recent simulations, is that molecular clouds in these 
primeval systems are destroyed and dispersed by strong stellar feedback, pushing 
most of the carbon to higher ionization states \citep{maiolino15,vallini13}. 
A difference in PDR structure or physical states of the ISM between star-forming galaxies 
at high-redshift and local ones, might also be enough to explain the low luminosity in 
\cii and FIR \citep{ota14}. 

\begin{wrapfigure}[23]{l}{7cm}
\centering
\includegraphics[width=7cm]{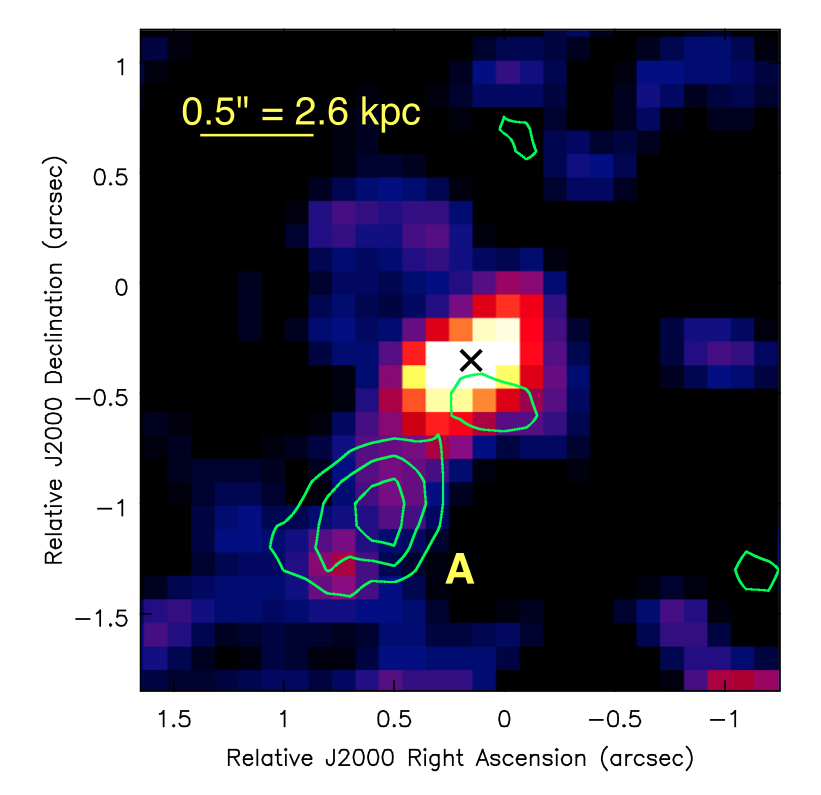}
\caption{\footnotesize{Image in rest frame UV stellar continnum and Ly$\alpha$ emission of 
the $z=7.109$ galaxy BDF3299. Green contours show the levels of 2, 3 and 4 times the noise 
per beam in the \cii map. The detected clump at 4\,kpc distance from the galaxy center 
(black cross) is marked 'A' \citep{maiolino15}.}}
\label{f:maio}
\end{wrapfigure}

Interestingly, \cite{maiolino15} detect a strong ($7\sigma$) \cii signal from 
a position which is offset from the actual galaxy, BDF3299 at $z=7.109$, as shown in Fig.\,\ref{f:maio}. This is interpreted as an accreting/satellite 
gas clump of neutral gas about $4$\,kpc from the galaxy itself.  
At slightly higher redshift, \cite{watson15} managed to detect dust continuum emission from the galaxy 
A1689-zD1 of SFR$\sim10\,\sfru$. 
Such observations nourish the hope that we 
are but few steps away from using \cii observations to charaterise the gas in normal galaxies 
and/or the mass assembly from the Circumgalactic Medium (CGM) in the early universe.

\begin{figure}[htbp] 
\centering
  \includegraphics[width=0.48\columnwidth]{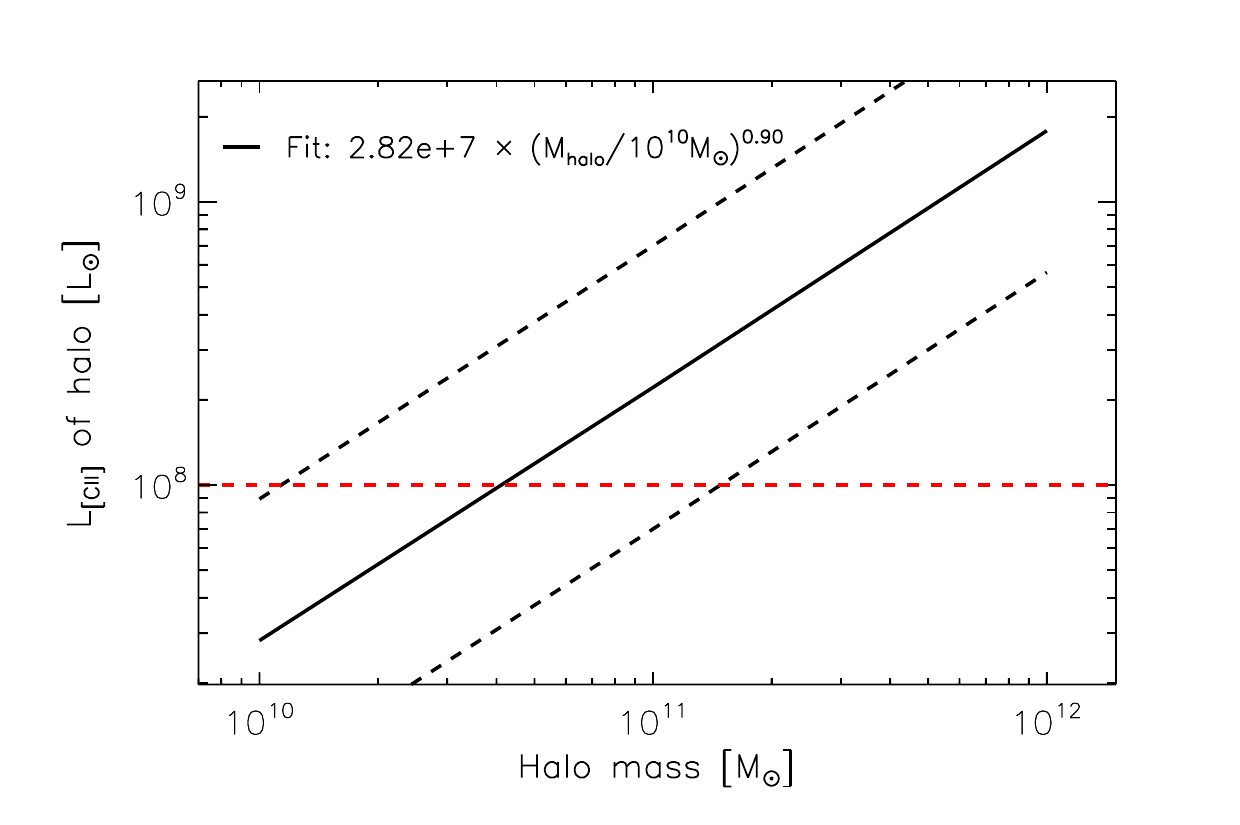}
  \includegraphics[width=0.48\columnwidth]{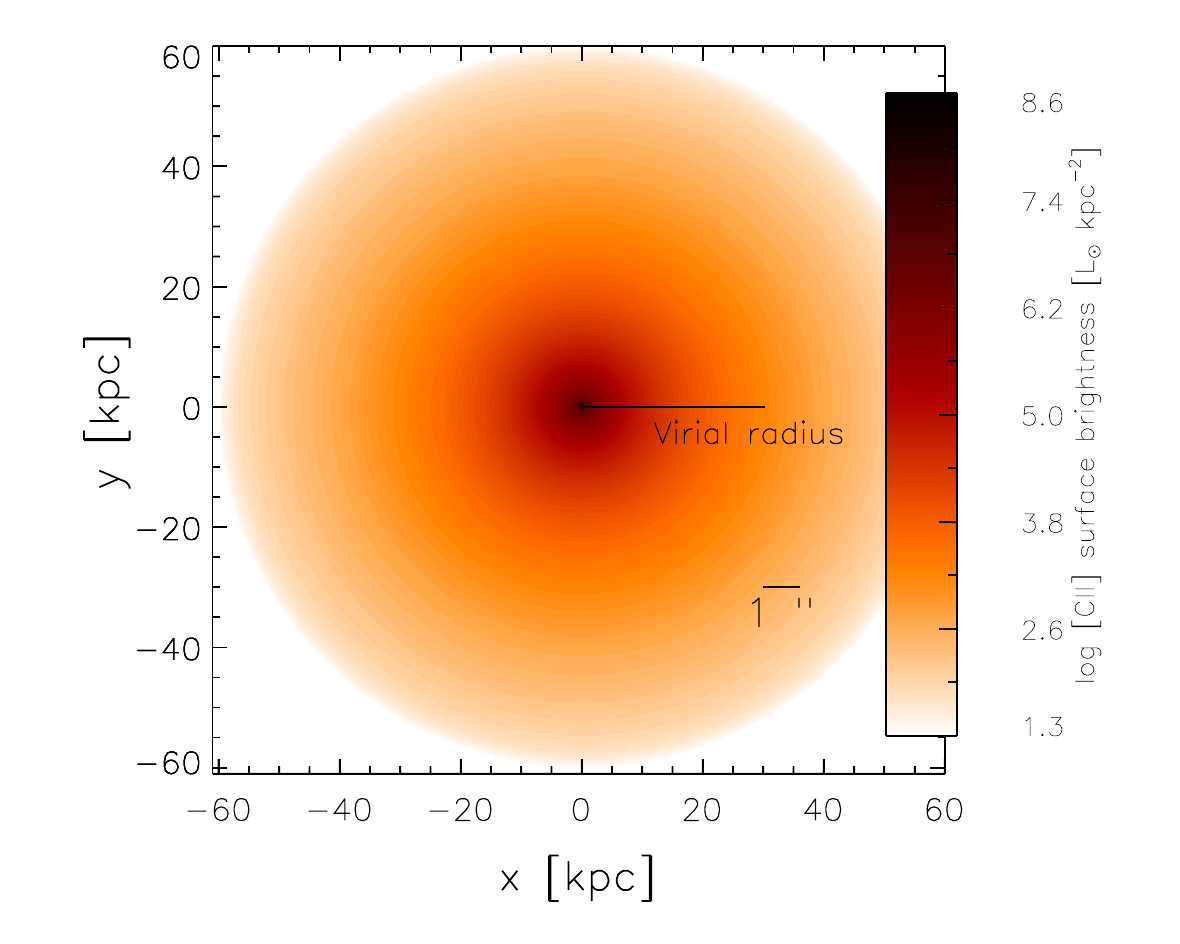}
 \caption{\footnotesize{{\it Left: } \cii luminosity as a function of stellar mass for our fiducial case (solid line) 
 and the optimistic and pessimistic cases (dashed lines). 
 {\it Right: } Map of \cii surface brightness from a galaxy of stellar mass $5\e{9}\,$\msun at $z=6$ 
 (corresponding to a halo mass of $2.7\e{11}\,\msun$ within the virial radius, marked on the figure.}}
 \label{ciihalo}
\end{figure}

With this motivation, I have recently made preliminary attempts at simulating \cii emission from 
halos at high redshift, in collaboration with Kristian Finlator, here at the Dark Cosmology Centre. 
As a simple test, we adopted a NFW profile for the radial halo density profile, 
and, since little is known about the metal abundance at these redshifts, 
assumed three values for the abundance of carbon, going from 
pessimistic to optimistic: $\Z_{\rm C}=[10^{-4.5},10^{-4},10^{-3.5}]$. 
The resulting rough estimate of \cii luminosity as a function of halo mass at $z=6$ 
is shown in Fig.\,\ref{ciihalo} together with a surface brightness map for a halo mass of $5\e{9}\,\msun$.
With the use of actual cosmological simulations, rather than simple spherically symmetric halo models or 
constant metallicity and temperature, we can make much more precise predictions for the detectability of the CGM.

\begingroup
\section{References}
\def\chapter*#1{}
\bibliographystyle{apj} 
	\setlength{\bibsep}{1pt}
	\setstretch{1}
\bibliography{bibsCII}
\endgroup

%% file: Part_II.tex

\part{The AGN-galaxy co-evolution at $z\sim2$}
\label{part2}

\chapter{How to detect an AGN}
A general concern when estimating the SFR by fitting the broadband SED of a galaxy with 
stellar population synthesis models, is the relative dominance of AGN 
versus star formation. The dominant contributor to the mid-infrared (MIR) light is from UV-light reprocessed by dust, 
but the UV light can be emitted by either young stars or an AGN. 
At $z\sim2$ it is not clear, without the aid of high spatial and spectral resolution, what is causing the observed 
MIR emission from massive galaxies; is dust in the central regions being heated by AGN activity, 
is dust across a larger region of the galaxy being heated by star formation, or is a combination of the two scenarios taking place?

X-ray emission, on the other hand, does not suffer from strong dust obscuration. 
Observing in X-ray can thus lift the apparent degeneracy in interpreting the origin of the MIR light, 
and thereby help to give the true AGN fraction of a sample of galaxies. A galaxy dominated by AGN activity 
can be distinguished from a galaxy dominated by star formation by having a harder X-ray spectrum if the AGN 
is obscured by dust, or by simply having a very high X-ray luminosity, $L_{0.5\text{-}8\text{\,keV}}$. 
The most heavily obscured `Compton-thick' AGNs, with column densities $N_\text{H}>10^{24}\:\text{cm}^{-2}$, 
might be missed by X-ray selection, but can instead be identified via an excess of MIR emission over that expected 
from purely star-forming galaxies \cite[e.g.,][]{daddi07b,treister09}.

The shape of the MIR spectrum can also reveal AGNs, as the intense nuclear emission re-emitted by 
dust leads to a power-law spectrum in the MIR. For this purpose, color cuts have been devised using the 
{\it Spitzer} data \citep[e.g.,][]{stern05,donley12}. While this technique has the capability of detecting 
even Compton-thick AGNs, otherwise missed in X-ray, it has a low efficiency for low- to moderate-luminosity 
AGNs and its robustness has yet to be verified at $z\gtrsim2$ \citep{cardamone08}. 

In a galaxy where star formation is dominant, the total hard band X-ray luminosity, 
$L_{2\text{-}10\text{\,keV}}$, can be used to estimate the SFR \citep[e.g.,][]{grimm03,ranalli03,lehmer10}. 
An AGN will reveal itself if the X-ray SFR inferred this way is much larger than the SFR derived from other 
tracers such as H$\alpha$, UV, or MIR luminosity. Also, if star formation dominates the radiation output, 
the $L_x\rightarrow\,$SFR conversion will give an upper limit on the SFR as has been obtained for 
sub-mm galaxies \citep{laird10} and for massive, star-forming galaxies at $1.2<z<3.2$ \citep{kurczynski12}.

At redshifts around $2$, high-ionization lines in the rest-frame UV can be used as AGN indicators, 
but X-ray observations remain a more efficient way of identifying AGNs \citep{lamareille10}. 

\section{Dissecting the X-ray emission from an AGN} \label{2:xray}
I will focus this section on X-ray emission generated by the accretion of matter onto a SMBH, as 
described in Section \ref{intro:agn}, since it is this characteristic that I use for detecting AGN 
in my study of the presence of AGN at high-$z$ (Part \ref{part2}).

\begin{figure}[!htbp] 
\centering
\includegraphics[width=0.6\columnwidth]{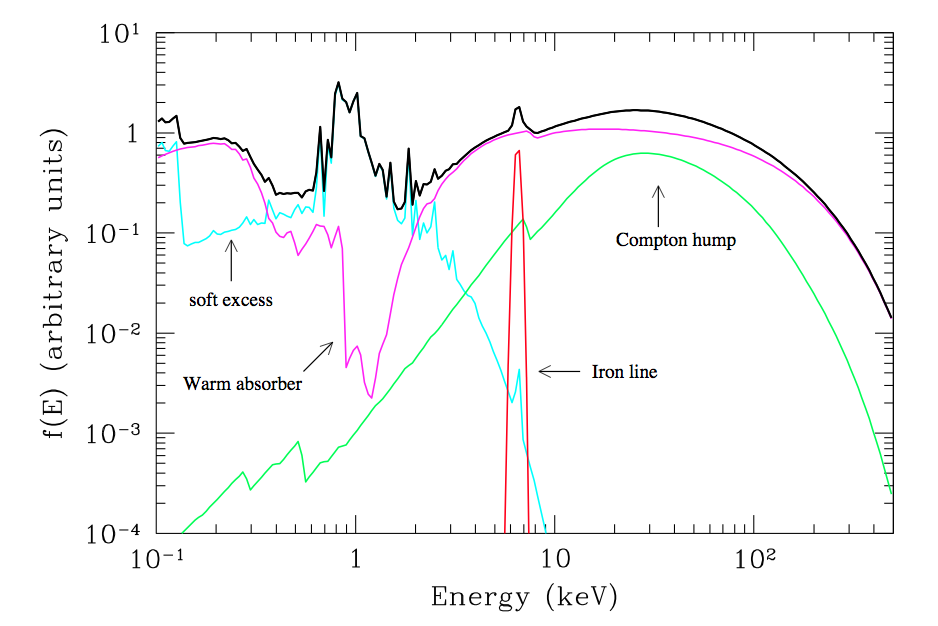}
\caption{\footnotesize{The X-ray spectrum (black) of a typical type I AGN, with curves showing the 
contrution from a primary continuum component (pink) resembling a power law but aborbed at soft 
energies by warm gas, a cold reflection component (green; Compton hump) and 
a soft excess (cyan). 
Also shown is iron K$\alpha$ emission -- the most relevant narrow feature (red), 
arising in the inner parts of the accretion disk. Figure from the review by \cite{risaliti04}.}}
\label{f:AGNspec}
\end{figure}

Fig.\,\ref{f:AGNspec} shows the X-ray part of the SED of a type I AGN, together with individual contributions 
from its components, the nature of which is elaborated on below. 

At high X-ray energies ($\sim30$\,keV), the SED is dominated by reflection component (`Compton hump') consisting of 
primary emission from the accretion disk being scattered by ionized gas, i.e. electrons. 
This component will be largest if the reflector is Compton-thick but smaller in the Compton-thin case 
in which case some of the incident radiation escapes without interaction \citep{risaliti04}. 
An additional hard component is sometimes present, most likely caused by inverse-Compton scattering of 
the radio-synchroton photons off of electrons in the relaticistic jet itself. 
At lower energies ($\lesssim2$\,keV) with an almost universal effective temperature of $\sim0.1-0.2\,$keV, 
a `soft excess' is often observed but less understood. 
Models suggest that the strongest soft excesses are found in AGNs with low mass accretion rates \cite{done12}, 
and typically atomic processes are invoked to explain the soft excess via either ionized reflectioni with light 
bending or the remaining part of a high-energy power law subject to ionized absorption \citep{vasudevan14}. 
But also comptonization models have been proposed, in which the soft excess consists of 
UV/soft-X-ray photons by a population of hot electrons \cite{matt14}. 
The joint NuSTAR/XMM program as well as future X-ray missions such as ASTRO-H and ASTROSAT (to be launched this year) 
are likely to break the degeneracy between these models (see Section \ref{intro:tele}).
Finally a warm, ionized absorber exists that reflects the incident continuum without changing 
its spectral shape. 

\begin{figure}[!htbp] 
\centering
\includegraphics[width=0.8\columnwidth]{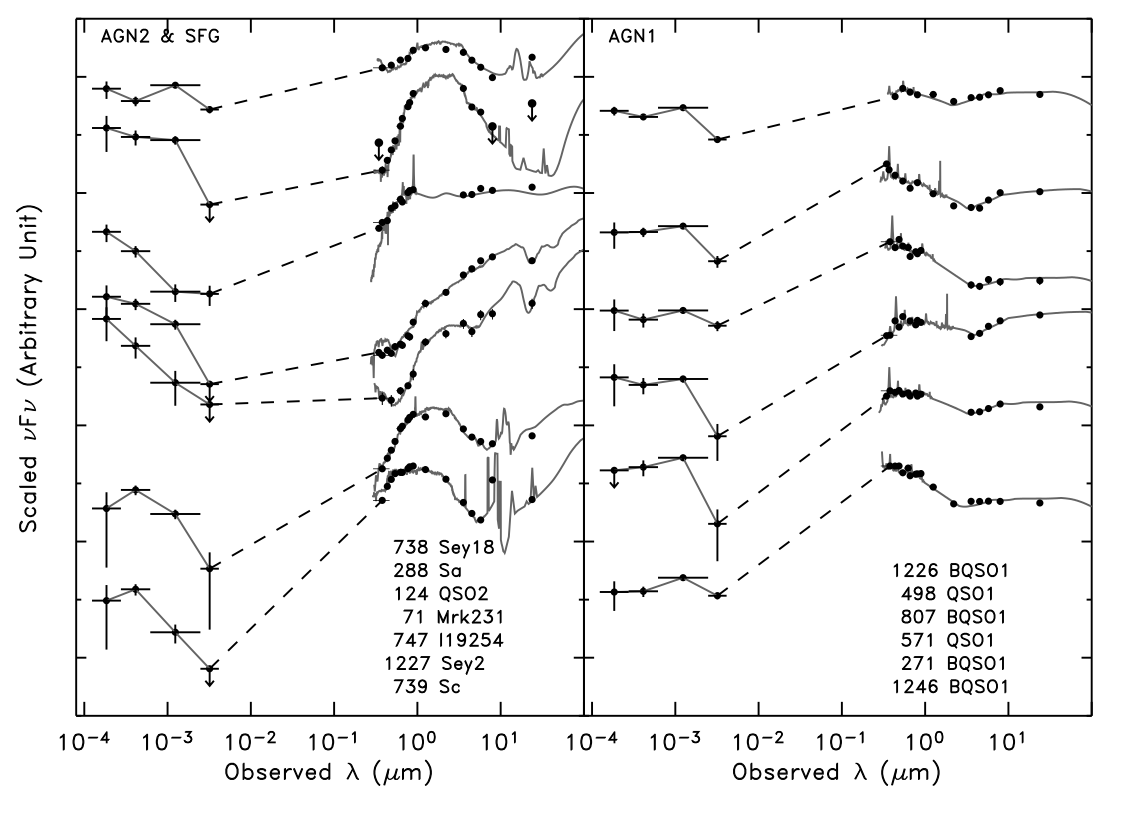}
\caption{\footnotesize{X-ray spectra of 13 galaxies from the work of \cite{tajer07}. 
{\it Left:} Top two spectra are consistent with star-forming template SEDs shown, 
while the following 4 galaxies have type 2 AGN SEDs. {\it Right:} Type 1 AGN SEDs. }}
\label{f:AGNsed}
\end{figure}

\begin{figure}[!htbp] 
\centering
\includegraphics[width=0.4\columnwidth]{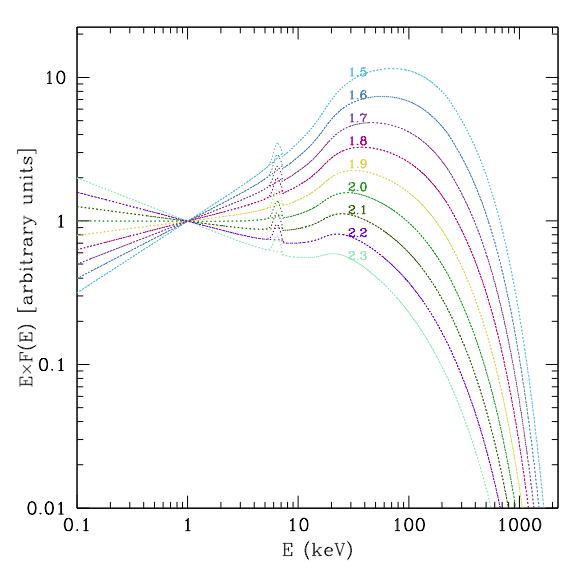}
\includegraphics[width=0.4\columnwidth]{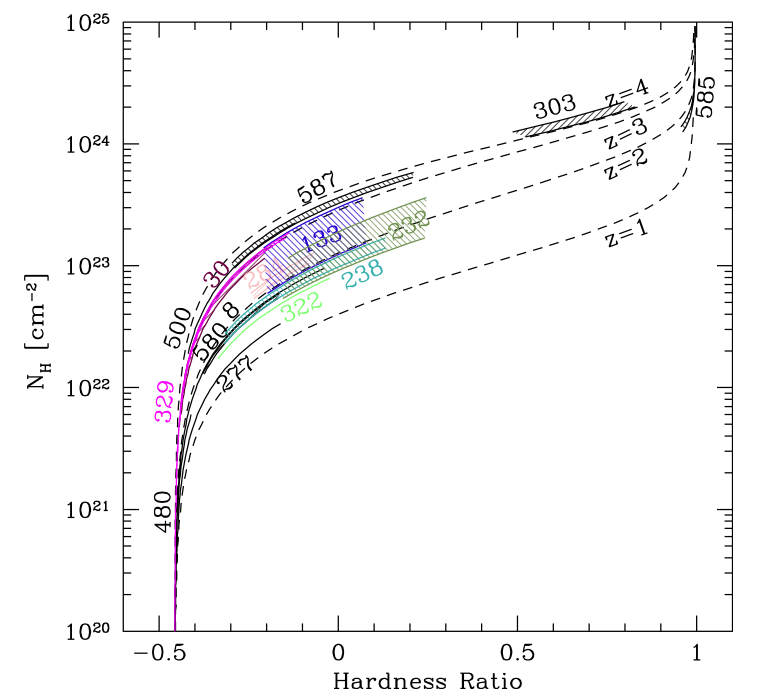}
\caption{\footnotesize{Measures of intrinsic power law slope in AGN. 
{\it Left:} SED templates for unobscured Seyferts normalized at $1\,$keV \citep{gilli07}. 
{\it Right:} Column density of obscuring material as a function of hardness ratio, HR \citep{treister09}.}}
\label{f:hr}
\end{figure}

To first order though, the intrinsic or `primary' emission from the central engine is a power law extending from 
1 to 100\,keV, with the slope determined by the different components described. 
The slope of the power law can be used to determine the amount of obscuration in the AGN, as 
more obscuration will result in a steeper power law (harder). 
In order to estimate the column density of obscuring material, one has to take a guess at 
the spectral slope, $\Gamma$, of the intrinsic emission from the obscured source:
\begin{align}
	dN/dE [\text{photons cm}^{-2}\text{ s}^{-1}\text{ keV}^{-1}] \propto E^{-\Gamma}
\end{align}
where $dN$ is the number of photons with energy in the $dE$ energy range. Note that a harder spectrum 
has {\it lower} $\Gamma$ as shown in the left-hand plot of Fig.\,\ref{f:hr}.
A value of $\Gamma=1.9$ is usually adopted, based on observations showing that $\Gamma$ 
is roughly constant for large samples of low-luminosity Seyfert galaxies and bright QSOs for redshifts 
up to $\sim5$ \citep[e.g.][]{nandra94,reeves00,piconcelli05}.

Fig.\,\ref{f:AGNsed} shows 13 SEDs from X-ray to FIR ($100\,\mu$m) of type I and type II AGN with best-fit galaxy 
templates. The X-ray part consists of data from XMM (see Section \ref{intro:tele}) sampled with 4 bands at 
0.3-0.5, 0.5-2, 2-4.5, 4.5-10 and 2-10\,keV.

The high angular resolution ($\sim0.5\arcsec$) of {\it Chandra} combined 
with the high sensitivity in the deep fields CDF-N and CDF-S (see Section \ref{intro:obs}), 
makes these fields advategous for searches of 
AGN at low and high redshift \citep{bauer04,xue11,treister09,alexander11}. 
For those studies, the observed X-ray spectra of {\it Chandra} are typically divided into 
a full ($0.5-8$\,keV), a soft ($0.5-2$\,keV) and a hard ($2-8$\,keV) band. 
A proxy for the powerlaw slope is the `hardness ratio', HR, defined as:
\begin{align}
	{\rm HR} = (H-S) / (H+S)
	\label{2:hr}
\end{align}
where $H$ and $S$ are the counts in hard and soft band respectively. 
Hardness ratio has the advantage, in comparison to $\Gamma$, 
of avoiding any assumptions regarding the shape of 
the spectrum.

Assuming an intrinsic obscured power law with $\Gamma=1.9$, HR converts into a column density 
that increases with the `hardness' of the spectrum as shown in the right-hand plot of Fig.\,\ref{f:hr}.
The criteria for finding an AGN are a high full-band X-ray luminosity ($\gtrsim2\e{42}\,$ergs/s) 
of a hard spectrum with HR~$>-0.2$ for sources of lower luminosity 
These criteria are meant to rule out galaxies with SEDs consisten with star formation processes only. 
However, a third possible origin of a bright X-ray component in the area-integrated SED, 
is a hot gas halo embracing the galaxy \citep{mulchaey10}. 
Souch a component can be ruled out, if the observed X-ray source is predominantly point-like, 
rather than extended.

\subsection{The AGN fraction at $z\gtrsim2$ from observations in X-ray}
Several studies of massive $z\sim2$ galaxies have therefore been made with the aim of uncovering AGN fractions, 
using the {\it Chandra} X-ray observatory. \cite{rubin04} performed a study of $40$ massive 
($M_{\ast}=(1$-$5)\times 10^{11}M_{\odot}$) red ($J_s-K_s\geq2.3$) galaxies at $z\gtrsim 2$ by analyzing a 
$91$ ks {\it Chandra} exposure. Roughly $5\%$ of these were found to host an AGN with intrinsic 
$L_{2\text{-}10\,\text{keV}}>1.2\times 10^{43}\,\text{erg\,s}^{-1}$. Assuming that the stacked 
X-ray signal from the remaining X-ray undetected galaxies in X-ray comes from star formation alone, 
they derived a mean SFR broadly consistent with the typical mean SFRs estimated from SED fits. 
\cite{alexander11} analyzed the 4Ms {\it Chandra} observations of $222$ $z\approx2$ star-forming BzK galaxies 
($M_{\ast}\sim10^{10}\text{-}10^{11}M_\odot$, \citealp{daddi07b}) in the Chandra Deep Field-South 
(CDF-S). $10\%$(23) showed X-ray emission consistent with AGN dominance, of which $5\%$(11) were 
found to contain heavily obscured AGNs, $4\%$(9) to have luminous, unobscured AGNs, and $3$ out of $27$ 
low-luminosity systems showed excess variability over that expected from star formation processes, 
indicating that at least some low-luminosity (rest-frame $L_{2\text{-}10\text{keV}}<10^{43}\text{\,erg\,s}^{-1}$) 
systems may contain AGNs.

\chapter{On the prevalence of AGN at $z\sim2$ (Paper III)}

\section{Aim of this project}
The aim of this project was to determine the AGN fraction in massive $z\sim2$ galaxies, 
and reveal any differences between quiescent and star-forming galaxies. 

\section{Our method and galaxy sample}
We addressed the matter by analyzing the X-ray emission from a mass-complete ($M_{\ast}>5\times10^{10}M_\odot$) 
sample of $1.5\leq z\leq 2.5$ galaxies residing in the CDF-S. The CDF-S, observed for $4\,$Ms, 
is currently the deepest X-ray view of the universe and so provides the best opportunity to study 
high-$z$ galaxies across a relatively large area ($464.5\,\text{arcmin}^2$).

We selected our galaxies from the FIREWORKS\footnote{\url{http://www.strw.leidenuniv.nl/fireworks/}} 
catalog \citep{wuyts08}, which covers a field situated within the CDF-S and contains photometry in 
$17$ bands from the $U$ band to the MIR. For this study we extracted a mass-complete sample of 
$M_{\ast}>5\times10^{10}M_\odot$ galaxies at $1.5\leq z\leq 2.5$ in a way similar to that of 
\cite{franx08} and \cite{toft09}. We used spectroscopic redshifts when available \citep{vanzella08,xue11}, 
and photometric redshifts from the FIREWORKS catalog otherwise. In order to maximize the 
signal-to-noise (S/N) on results from X-ray stacking \citep{zheng12} and ensure a relatively 
homogeneous PSF across the {\it Chandra} field employed, we considered only galaxies that lie within 
$6\arcmin$ of the average {\it Chandra} aimpoint. 

We adopted galaxy stellar masses from the SED fitting results of \cite{franx08} and \cite{toft09}. 
In short, the SED fits were carried out with models by \cite{bruzual03}, choosing the best fit from adopting 
three different star formation histories (a single stellar population with no dust, an exponentially 
declining star formation history of timescale $300\,\text{Myr}$ with dust, and a constant star formation history with dust). 
In cases where the spectroscopic redshift differed by more than $0.1$ from the original FIREWORKS 
photometric redshift, we redid the SED fits in FAST\footnote{\url{http://astro.berkeley.edu/~mariska/FAST.html}} 
using an exponentially declining star formation history with a range of possible timescales from $10\,\text{Myr}$ 
to $\sim1\,\text{Gyr}$. As a quality parameter of the SED modeling, we demanded an upper limit on the 
reduced $\chi^2_{\nu}$ of $10$ on the best-fit model. The SED fits provide SFR estimates, but we 
used SFRs derived from rest-frame UV+IR luminosities (see Section \ref{aha1}) as these include 
obscured star formation and are subject to less assumptions. From the observed photometry, rest-frame 
fluxes in $U$, $V$, $J$ band and at $2800\text{\,\AA}$ have been derived using 
InterRest\footnote{\url{http://www.strw.leidenuniv.nl/~ent/InterRest}} \citep{taylor09}. 

We divided the resulting sample of $123$ galaxies into quiescent and star-forming galaxies using 
the rest-frame $U$, $V$ and $J$ (falling roughly into the observed $J$, $K$ and IRAC 4.5$\,\mu\text{m}$ 
bands at $z\sim2$) colors. Dust-free but quiescent galaxies differ from dust-obscured starburst 
galaxies in that they obey the following criteria by \cite{williams09}: 
\begin{align}
U-V&>1.3  \\
V-J&<1.6 \\
(U-V)&>0.88\times(V-J)+0.49 \label{uvj2}
\end{align}
The fraction of quiescent galaxies identified within our sample using this method is $22\%\pm5\%(27/123)$. 
This is rather low compared to the $30$--$50\%$ found by \cite{toft09} for the same redshift and mass limit, 
but under the requirement that the sSFR (=SFR/$M_\ast$) from SED fitting is less than $0.03\,\text{Gyr}^{-1}$. 
If applying this criterion, we would have arrived at a fraction of $42\%\pm7\%(51/123)$. 
A possible reason for the discrepancy between the two methods may be that we were using the $UVJ$ criterion 
in the limits of the redshift range in which it has so far been established. As the redshift approaches $2$, 
the quiescent population moves to bluer $U-V$ colors, possibly crossing the boundaries of Equation (\ref{uvj2}). 
However, we preferred the $UVJ$ selection technique in contrast to a cut in sSFR, because rest-frame colors 
are more robustly determined than star formation rates from SED fits.

\subsection{X-ray data and stacking analysis}
The raw X-ray data from the CDF-S survey consist of 54 individual observations taken between 
1999 October and 2010 July. We build our analysis on the work of \cite{xue11}, who combined the observations 
and reprojected all images. They did so in observed full band ($0.5$--$8\,$keV), soft band ($0.5$--$2\,$keV) 
and hard band ($2$--$8\,$keV), and the resulting images and exposure maps are publicly 
available.\footnote{\url{http://www2.astro.psu.edu/users/niel/cdfs/cdfs-chandra.html}}
\begin{figure}[t!]
\centering
\hspace{.0em}\raisebox{0.cm}{\includegraphics[width=0.8\columnwidth]{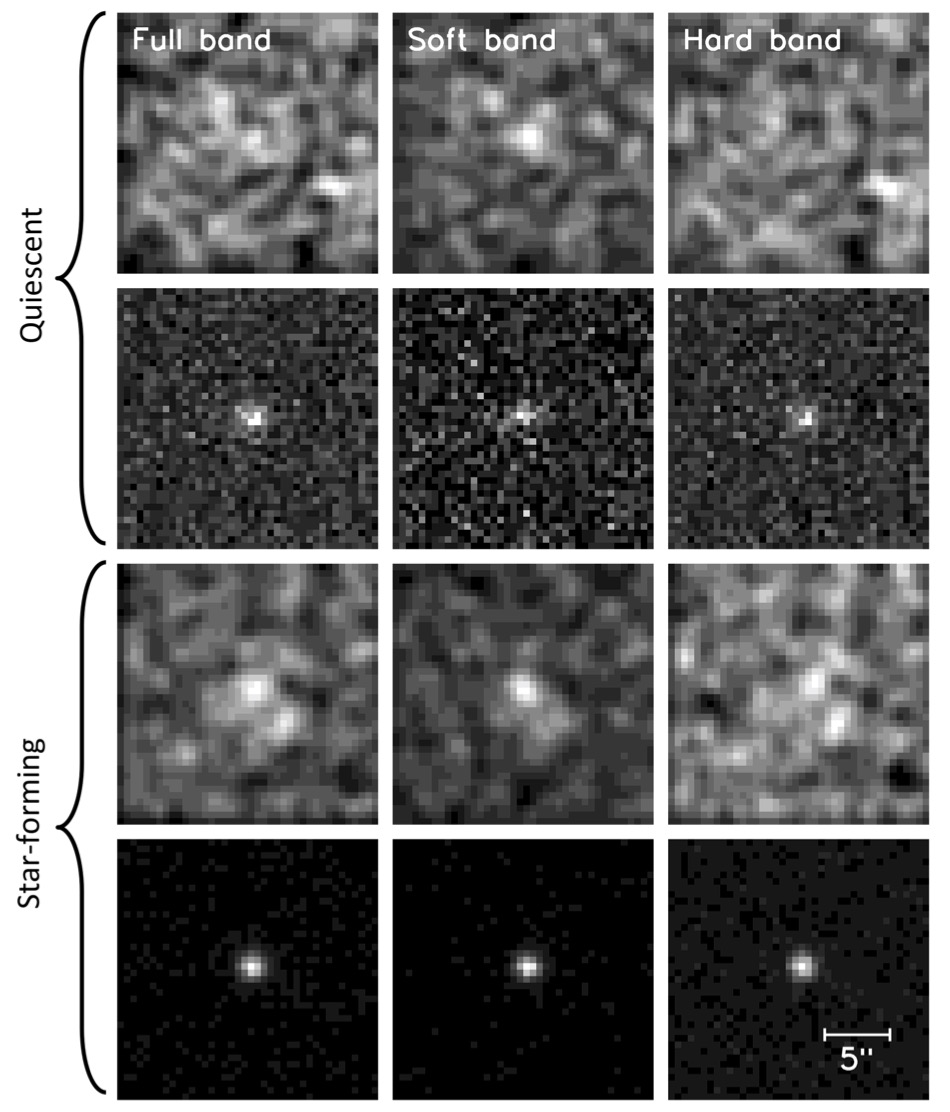}}
\caption{\footnotesize{Stacked, background-subtracted and exposure-corrected $20\arcsec\times20\arcsec$ images in the three energy bands 
(columns) for the individually non-detected (top row) and detected (bottom row) quiescent and star-forming galaxies. 
Images of the non-detections have been smoothed with a Gaussian of width 2 pixels.}}
\label{st_im}
\end{figure}
 \\
\noindent
We extracted source and background count rates in the X-ray maps for all galaxies using the method of \cite{cowie12}. 
Source counts were determined within a circular aperture of fixed radii: $0\arcsec.75$ and $1\arcsec.25$ at 
off-axis angles of $\theta\leq3\arcmin$ and $\theta>3\arcmin$ respectively. Background counts were estimated 
within an annulus $8\arcsec$--$22\arcsec$ from the source, excluding nearby X-ray sources from the catalog of \cite{xue11}.

Each galaxy was classified as `X-ray detected' if detected in at least one band at $\geq 3\sigma$ significance, 
thereby creating  $4$ subsamples containing $8$ quiescent and detected, $19$ quiescent and undetected, 
$43$ star-forming and detected, and $53$ star-forming and undetected galaxies. 

While the X-ray detected galaxies could be analyzed individually, we stacked the X-ray non-detected galaxies in order 
to constrain the typical X-ray flux from these. Stacked X-ray images of the $4$ subsamples are shown 
in Figure \ref{st_im} with all galaxies aligned to their $K_s$ band galaxy center positions from FIREWORKS. 
Representative count rates and associated errors for these stacks were calculated using the optimally 
weighted mean procedure of \cite{cowie12} and tabulated in Table \ref{counts_da} together with S/N values. 

\begin{table}[htbp]
\centering
\begin{tabular*}{0.7\columnwidth}{p{0.2cm}p{2.0cm}p{2.0cm}p{2.0cm}} \toprule
		& \multicolumn{3}{c}{Count Rate $\pm1\sigma$ ($10^{-6}\,\text{s}^{-1}$) in } 	\\
 		& Full Band	 		& Soft Band 			& Hard Band 	\\ \hline
$Q$		& $0.71\pm0.96$  	&  $2.01\pm0.41$   	&  $-1.67\pm0.86$\\ 
		& $(0.7)$  				&  $(4.9)$   			&  $(-1.9)$\\ 
SF		& $2.51\pm0.55$  	&  $1.08\pm0.24$    	&  $1.29\pm0.49$\\
		& $(4.5)$  				&  $(4.5)$    			&  $(2.7)$\\ \hline
\end{tabular*}
\caption{Stacking results for Quiescent ($Q$) and Star-forming (SF) galaxies not detected individually in 
X-Rays (in Parentheses the corresponding S/N)}
\label{counts_da}
\end{table}

The reliability of our chosen method for X-ray stacking and source count extraction was tested with Monte Carlo (MC) simulations. 
With $500$ MC simulations of $53$ (corresponding to the number of star-forming galaxies not detected in X-rays) randomly selected 
positions having no X-ray detections nearby, we obtained a histogram of stacked S/N values that is well fitted by a normal 
distribution with a center at $-0.02\pm0.8$, that is, consistent with no detection. 
Similar results were obtained using a sample size of $19$, the number of quiescent non-detections.

\begin{table*}[htbp]
\centering
\begin{tabular*}{\columnwidth}{p{1.5cm}p{3cm}p{7cm}} \toprule 
$L_{0.5\text{--}8\text{\,keV}}$ ($\text{\,erg\,s}^{-1}$)		& 	HR					&	Classification \\ \hline
$>3\times10^{42}$			&	$<-0.2$					& Unobscured AGNs ($N_\text{H}<10^{22}\:\text{cm}^{-2}$)		\\
$>3\times10^{42}$			&	$>-0.2\text{ and }<0.8$		& Moderately obscured AGNs ($10^{22}<N_\text{H}<10^{24}\:\text{cm}^{-2}$)		\\
$>3\times10^{42}$			&	$>0.8$					& Compton-thick AGNs ($N_\text{H}>10^{24}\:\text{cm}^{-2}$)		\\
$<3\times10^{42}$ 			&	$<-0.2$					& Star-forming galaxy \\
$<3\times10^{42}$ 			&	$>-0.2$					& Low-luminosity obscured AGNs or star-forming galaxy\\ \hline
\end{tabular*}
\caption{Classification scheme used in this work, with limits from \cite{szokoly04}, \cite{wang04}, and 
\cite{treister09}}
\label{class}
\end{table*}

We quantified the hardness of the X-ray spectra using the hardness ratio (see eq. \ref{2:hr}). 
Assuming a power-law spectrum, we also derived a photon index, $\Gamma$, with the online mission 
count rate simulator WebPIMMS\footnote{\url{http://heasarc.nasa.gov/Tools/w3pimms.html}} using a Galactic \ion{H}{1} column density of 
$8.8\times10^{19}\text{\,cm}^{-2}$ \citep{stark92} and not including intrinsic absorption. 
Whenever the S/N in either soft or hard band was 
below $2$, we used the corresponding $2\sigma$ upper limit on the count rate to calculate a limit on 
both HR and $\Gamma$, leading to a limit on the luminosity as well. When neither a hard- nor a soft-band flux 
was detected with an S/N above $2$, a typical faint source value of $\Gamma=1.4$ \citep{xue11} was assumed, 
corresponding to HR$\,\sim-0.3$ for an intrinsic powerlaw spectrum of $\Gamma=1.9$ \citep{wang04}.

We derived the unabsorbed flux from the count rate by again using the WebPIMMS tool, now with the $\Gamma$ tied 
to the observed value of either the individually detected galaxy or the stack in question. 
With this method, a count rate of $\sim 10^{-5}\text{\,counts\,s}^{-1}$ corresponds to a flux of nearly 
$10^{-16}\text{\,erg\,cm}^{-2}\text{\,s}^{-1}$ in full band at a typical value of $\Gamma=0.8$. 
Finally, the flux was converted into luminosity in the desired rest-frame bands using XSPEC version 
12.0.7 \citep{arnaud96}.

\subsection{Luminous AGN Identification} 
\label{aha}

\begin{figure}[htbp] 
\begin{center} $
\begin{array}{cc}
\includegraphics[width=0.57\columnwidth]{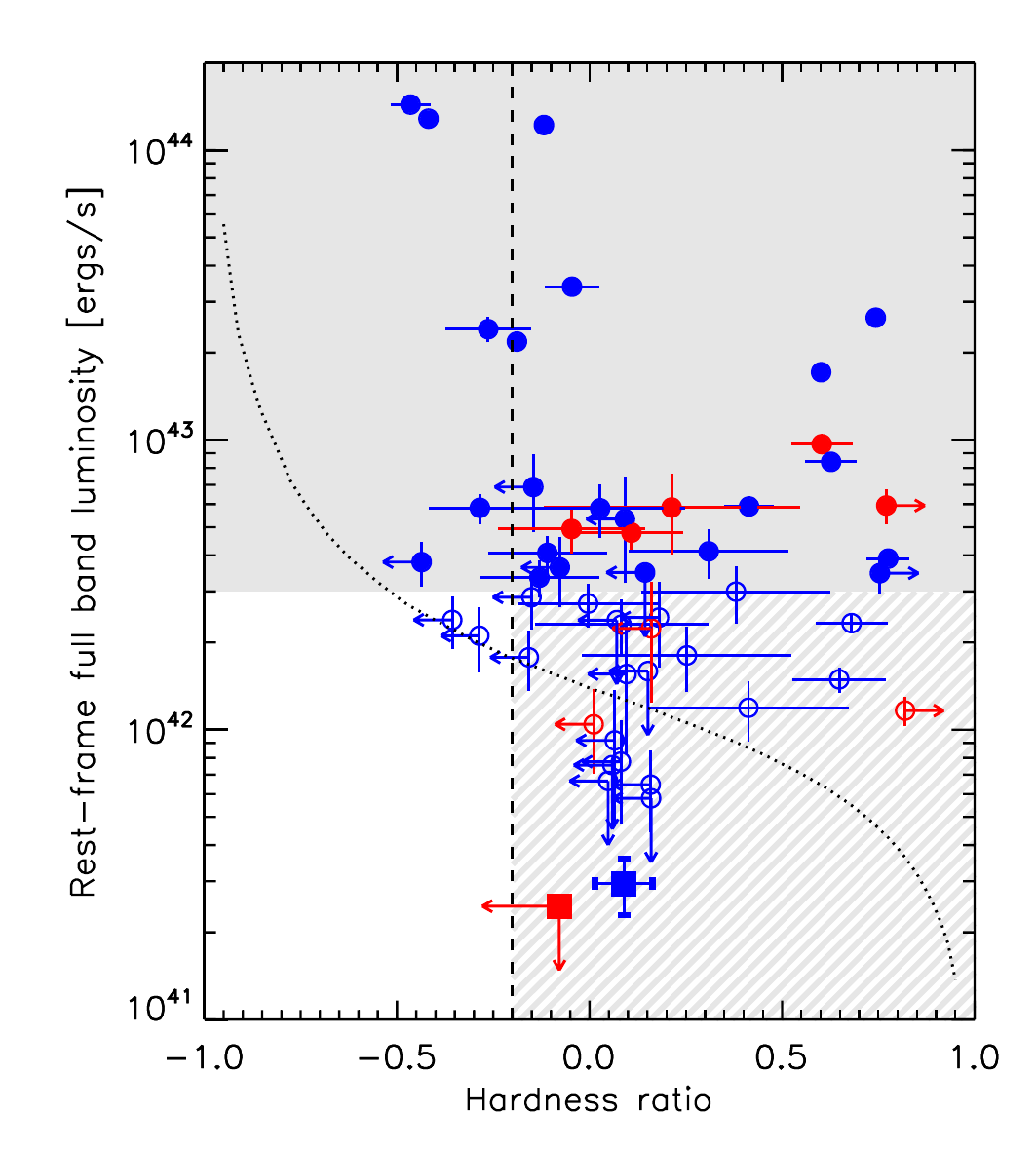} \\
\includegraphics[width=0.57\columnwidth]{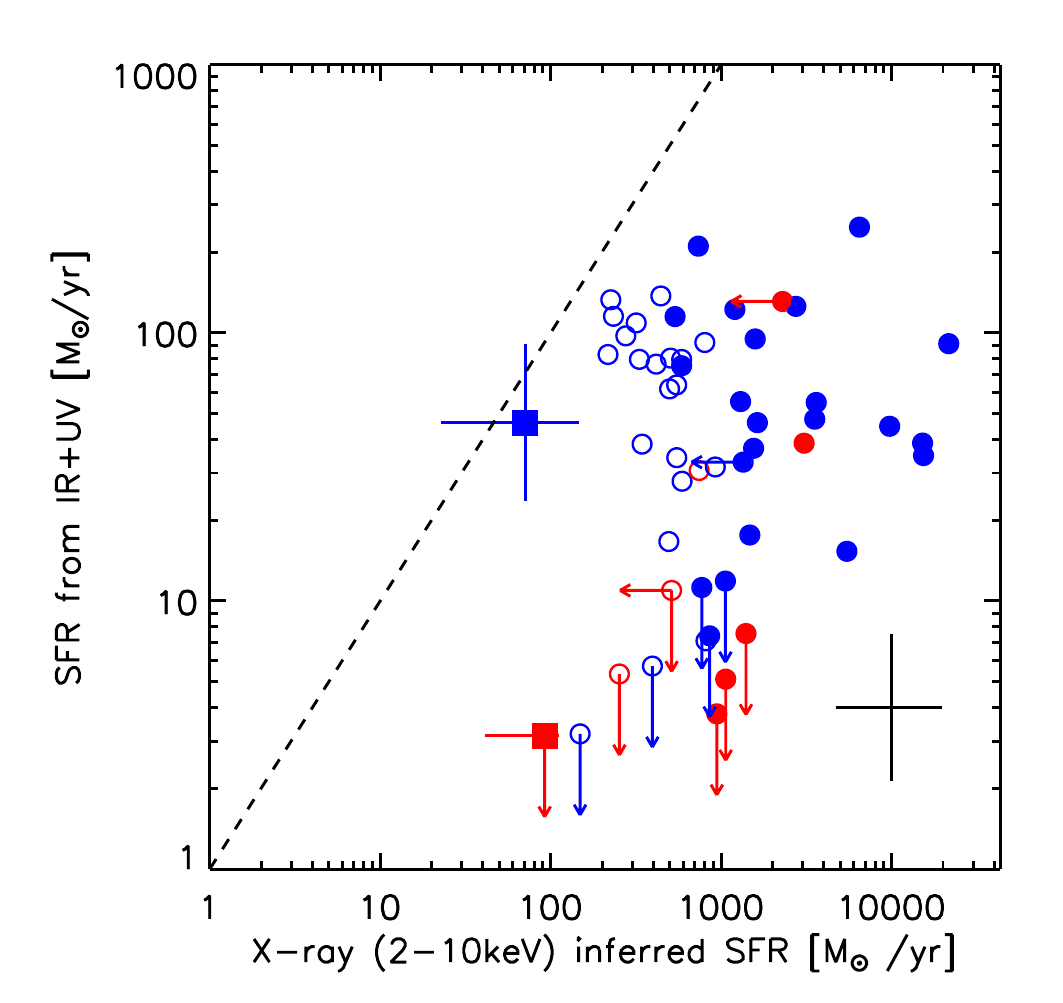}
\end{array} $
\end{center}
\caption{\footnotesize{Top: $L_{0.5\text{--}8\,\text{keV}}$ vs. HR for all galaxies. Red: quiescent galaxies. Blue: 
star-forming galaxies. Squares: stacks of non-detected samples. Filled circles: luminous AGNs (selected 
with the X-ray criteria indicated by a shaded area). Open circles: galaxies dominated by low-luminosity 
AGNs or star formation processes alone (selected with X-ray criteria indicated by a hatched area). 
Dashed line: separating obscured ($N_\text{H}\gtrsim10^{22}\:\text{cm}^{-2}$) from unobscured AGNs. 
Dotted line: typical detection limit. Errorbars display 2$\sigma$ errors and limits are indicated with 
arrows. Bottom: SFR$_\text{UV+IR}$ vs. SFR$_{2\text{--}10\text{keV}}$ (see the text). Dashed line: equality. 
Typical errors on X-ray detected galaxies are shown in the lower right corner. Symbols are as above.}}
\label{hr_lum}
\end{figure}

In the top panel of Figure \ref{hr_lum} are shown the rest-frame X-ray full band luminosities, 
$L_{0.5\text{--}8\,\text{keV}}$, not corrected for intrinsic absorption, versus observed hardness 
ratio, HR, for all X-ray detected galaxies as well as stacked non-detections. A typical detection 
limit on $L_{0.5\text{--}8\,\text{keV}}$ has been indicated with a dotted line in Figure \ref{hr_lum}, 
calculated as two times the background noise averaged over all source positions. A few detected galaxies 
were found below this limit due to these residing at relatively low $z$ and/or in regions of lower-than-average 
background. Galaxies with $L_{\text{0.5\text{--}8\,keV}}>3\times10^{42}\text{\,erg\,s}^{-1}$ were selected 
as luminous AGN, since star-forming galaxies are rarely found at these luminosities \citep{bauer04}. 
In addition, we adopted the HR criteria in Table \ref{class} in order to identify obscured and unobscured AGNs.

About $53\%(27/51)$ of the X-ray detected galaxies were identified as luminous AGNs, with $22$ moderately 
to heavily obscured ($-0.2<\;$HR$\;<0.8$) and $5$ unobscured (HR$\,<-0.2$) AGNs. The rest had X-ray 
emission consistent with either low-luminosity AGNs or star formation. We did not identify any Compton-thick 
AGNs directly, but three galaxies had lower limits on their hardness ratios just below HR$\;=0.8$, 
thus potentially consistent with Compton-thick emission (see Table \ref{class}). In total, we detected 
a luminous AGN fraction of $22\%\pm5\%(27/123)$, a fraction which is $23\%\pm5\%(22/96)$ for the 
star-forming and $19\%\pm9\%(5/27)$ for the quiescent galaxies.

Stacking the non-detections resulted in average X-ray source properties that exclude high luminosity AGNs. 
However, the inferred limits on their HR were consistent with a contribution from low-luminosity AGNs, the 
importance of which cannot be determined from this plot alone.

\subsection{X-ray Inferred SFR}
\label{aha1}

From the rest-frame hard band luminosity, $L_{2\text{--}10\,\text{keV}}$ it is possible to derive estimates of the SFR, 
as the number of high-mass X-ray binaries (HMXBs) is proportional to the SFR \citep{ranalli03}. \cite{kurczynski12} 
made a comparison of different SFR indicators, by applying three different $L_x\rightarrow\,$SFR conversions to $510$ 
star-forming BzK galaxies at $1.2<z<3.2$, selected as having $L_{\text{2\text{--}10\,keV}}<10^{43}\text{\,erg\,s}^{-1}$. 
While relations by \cite{persic04} and \cite{lehmer10} overestimated the true SFR (from rest-frame UV and IR light) 
by a factor of $\sim5$, the relation by \cite{ranalli03} provided a good agreement at $1.5<z<3.2$. But, as 
pointed out by \cite{kurczynski12}, all relations might lead to an overestimation due to contamination by 
obscured AGNs in the sample of SF galaxies. As we did not know the exact amount of obscured AGN contamination 
in the stacks, we chose to use the following relation by \cite{ranalli03} as a conservative estimate of the SFR: 
\begin{align}
\text{SFR}_{2\text{--}10\,\text{keV}}=2.0\times 10^{-40} \,L_{2\text{--}10\,\text{keV}}
\end{align}
with SFR measured in $M_\odot \text{\,yr}^{-1}$ and $L_{2\text{--}10\,\text{keV}}$ in $\text{erg\,s}^{-1}$. 
Another reason for not using the more recent relation by \cite{lehmer10} is that this relation was constructed 
for galaxies with SFR $>9\,M_{\odot}\text{\,yr}^{-1}$ only, whereas a large part of our sample had very low SFRs 
as inferred from SED fitting ($<1\,M_{\odot}\text{\,yr}^{-1}$). 
Following \cite{kurczynski12}, we used the observed soft band flux to probe the rest-frame hard band luminosity. 
The uncertainty on the SFR is estimated from the error on the observed soft-band flux together with a systematic 
error in the relation itself of $0.29\text{\,dex}$, as given by \cite{ranalli03}.

For comparison, the `true' SFR is inferred from rest-frame UV and IR light, SFR$_\text{UV+IR}$, following the 
method of \cite{papovich06}:
\begin{align}
\text{SFR}_{\text{UV+IR}}=1.8\times 10^{-10} (L_{\text{IR}}+3.3L_{2800})/L_\odot
\end{align}
where $L_{\text{IR}}$ is the total infrared luminosity and $L_{2800}$ is the monochromatic luminosity at rest-frame 
$2800\,\text{\AA}$. $L_{2800}$ comes from the rest-frame UV flux, $f_{2800\text{\AA}}$, 
and in this context, the errors on $f_{2800\text{\AA}}$ are negligible. We derived $L_{\text{IR}}$ from the observed 
$24\,\mu\text{m}$ flux, $f_{24\mu\text{m}}$, using redshift-dependent multiplicative factors, $a(z)$, from the work 
of \cite{wuyts08}, Section 8.2 in that paper):
\begin{align}
L_{\text{IR}}[L_{\odot,8\text{--}1000\mu \text{m}}]=10^{a(z)}\cdot f_{24\mu\text{m}} 
\end{align}
The errorbars on $L_{\text{IR}}$ derive from the errors on $f_{24\mu\text{m}}$ and we further assume an uncertainty 
of $0.3$ dex in the relation of \cite{papovich06} as found by \cite{bell03} when comparing to H$\alpha$- and radio-derived 
SFRs.  The X-ray undetected quiescent galaxies were not detected with S/N$\,>3$ in $24\,\mu\text{m}$ either 
(except in one case) leading to a mean flux of $1.2\pm1.7\mu\text{Jy}$, of which we adopted a 2$\sigma$ upper 
limit for the remaining analysis.

In the bottom panel of Figure \ref{hr_lum}, SFR$_{2\text{--}10\,\text{keV}}$ is compared to SFR$_\text{UV+IR}$ 
with the dashed line indicating equality. It is no surprise that nearly all of the individual X-ray detections 
have very high SFR$_{2\text{--}10\,\text{keV}}$ as compared to SFR$_\text{UV+IR}$, as we were largely insensitive 
to individual purely star-forming galaxies, given the detection limits in the top panel of Figure \ref{hr_lum} 
and the criteria in Table \ref{class}. 

For the star-forming stack, the X-ray inferred SFR of $71\pm51M_{\odot}\text{\,yr}^{-1}$ is consistent with the 
IR+UV inferred SFR of $46\pm33M_{\odot}\text{\,yr}^{-1}$, whereas the quiescent stack shows an SFR$_{2\text{--}10\,\text{keV}}$ 
of $92\pm65M_{\odot}\text{\,yr}^{-1}$ ($89\pm17M_{\odot}\text{\,yr}^{-1}$ when bootstrapping this sample $200$ times), 
well exceeding SFR$_\text{UV+IR}\leq3M_{\odot}\text{\,yr}^{-1}$.

\section{An overwhelmingly large AGN population}
\label{result}
\subsection{Luminous AGN Fraction}
In total, $22$ X-ray detected galaxies have emission consistent with containing a luminous 
(rest-frame $L_{0.5\text{--}8\text{keV}}>3\times10^{42}\text{\,erg\,s}^{-1}$) and 
obscured ($10^{22}<N_\text{H}<10^{24}\:\text{cm}^{-2}$) AGN, and a further five have emission 
consistent with a luminous unobscured ($N_\text{H}<10^{22}\:\text{cm}^{-2}$) AGN. 
This leads to a luminous AGN fraction of the full sample of $22\%pm5\%(27/123)$ and of the detected galaxies only, 
$53\%\pm13\%(27/51)$. The AGN fraction among both quiescent and star-forming galaxies, according to their X-ray spectra, 
is measured to be around $20\%$ as Table \ref{numbers} shows, meaning that AGNs in massive $z\sim2$ galaxies, 
even quiescent ones, are common, as proposed by \cite{kriek09} who studied the near-IR spectrum of one quiescent galaxy. 

\begin{table}[htbp]
\centering
\begin{tabular}{l|c|c} \toprule
							& 		Quiescent (27)			& Star-forming (96)			\\ \hline 
Luminous AGNs					&			$5$				& $22$		\\
\multirow{2}{*}{Low-luminosity AGNs}	&			$2$ (det)			& $19$ (det)			\\
							&			$12$--$19$ (non-det)	& $0$--$21$ (non-det)			\\ \hline
Luminous AGN fraction			&			$19\%\pm9\%$		& $23\%\pm5\%$				\\ 
Total AGN fraction				&			$70\%$--$100\%\,$		& $43\%$--$65\%\,$				\\ \hline
\end{tabular}
\caption{X-Ray derived AGN Numbers and Fractions for Quiescent and Star-forming Galaxies, 
Divided into Luminous AGNs and Detected and Non-detected Low-luminosity AGNs}
\label{numbers}
\end{table}

This luminous AGN fraction is high when compared to the $5\%$ found by \cite{rubin04} (see the introduction), 
but this is likely a consequence of their 4$\sigma$ detection limit of $1.2\times 10^{43}\,\text{erg\,s}^{-1}$ 
in rest-frame $2$--$10\,\text{keV}$ being about twice as high as our limit of $5.5\times 10^{42}\,\text{erg\,s}^{-1}$ 
(at $\Gamma=1.4$), as calculated from the average background noise in the observed soft band. Adopting the detection 
limit of \cite{rubin04} and requiring an S/N of at least $4$, we reduce our fraction of luminous AGN to $8\%\pm3\%(10/123)$, 
consistent with the results of \cite{rubin04}. \cite{alexander11}, using also the $4\,$Ms CDF-S data, found a much lower 
X-ray detection fraction of $21\%\pm3\%$ as compared to ours ($53\%\pm7\%$), and a luminous AGN fraction of only $9\%\pm2\%(20/222)$. 
We believe that the discrepancies have several reasons, the main ones being: (1) our use of a mass-complete sample, 
whereas the BzK selection technique used by \cite{alexander11} includes galaxies down to $M_{\ast}\sim10^{10}M_{\odot}$ 
for which the total AGN fraction, assuming a fixed Eddington ratio, is expected to be lower above our detection 
limit,\footnote{For a fixed Eddington ratio, and assuming that galaxy bulge mass increases with total stellar mass, 
the AGN X-ray luminosity is expected to scale with galaxy stellar mass according to the $M_{\text{bh}}$--$M_{\text{bulge}}$ 
relation \citep{haring04}} (2) our updated source count extraction and stacking method leading to higher S/N, and (3) the 
use of $\Gamma$ instead of HR in the AGN identification conducted by \cite{alexander11}. For comparison, \cite{tanaka12} 
recently discovered a group of quiescent galaxies at $z=1.6$ with only one main-sequence star-forming galaxy. 
This group differed from local groups in having a remarkably high AGN fraction of $38^{+23}_{-20}\%$, consistent with 
our result, and which they interpret as possible evidence for AGN activity quenching star formation. 

\subsection{Importance of Low-luminosity AGNs}
As seen in Figure \ref{hr_lum}, the X-ray-identified luminous AGNs in general show an excess in SFR compared to that 
inferred from IR+UV emission. Among the galaxies classified as being dominated by either low-luminosity AGN or star 
formation, about $\sim90\%(21/24)$ have SFR$_{2\text{--}10\,\text{keV}}$ more than 1$\sigma$ above SFR$_\text{UV+IR}$. 

Surprisingly, the quiescent stack also has a much larger SFR$_{2\text{--}10\,\text{keV}}$ than SFR$_{\text{IR+UV}}$. 
Even when removing the marginally undetected galaxies with $2<$S/N$<3$, the resulting 
SFR$_{2\text{--}10\,\text{keV}}=62\pm19M_{\odot}\text{\,yr}^{-1}$ is still more than 3$\sigma$ above SFR$_{\text{IR+UV}}$. 
This discrepancy is only further aggravated if instead assuming the SFR$-L_{X}$ relation of \cite{lehmer10}. If caused by 
obscured star formation, we would have expected an average $24\,\mu\text{m}$ flux of $90\,\mu\text{Jy}$ for the individual 
galaxies, in order to match the lower limit on SFR$_{2\text{--}10\,\text{keV}}$. This is far above the upper limit of 
$3.4\,\mu\text{Jy}$ for the stack, suggesting that the X-ray flux of this stack is instead dominated by low-luminosity AGNs, 
but that their contribution to the $24\,\mu\text{m}$ flux remains undetectable in the {\it Spitzer}--MIPS data. 

We derived a lower limit on the low-luminosity AGN contribution for this stack of $19$ objects by constructing a mock sample 
of the same size and with the same redshifts, but containing {\sc n} low-luminosity AGNs with luminosities 
$L_{0.5\text{--}8\,\text{keV}}=10^{42}\text{\,erg\,s}^{-1}$ just below the detection limit and $19-\text{{\sc n}}$ galaxies 
with $L_{0.5\text{--}8\text{\,keV}}=10^{41}\text{\,erg\,s}^{-1}$ (all of them with $\Gamma=1.4$). From random realizations 
of this mock sample we found that at least $\text{{\sc n}}=12$ low-luminosity AGNs are required to match the observed 
rest-frame $2$--$10$ keV luminosity of the stack. Similarly, we found that by removing at least $12$ randomly selected 
galaxies it is possible to match the low SFR predicted by IR+UV emission, though only with a small probability 
($\sim1\%$ of $200$ bootstrappings). We therefore adopted $63\%(=12/19)$ as a conservative lower limit 
on the low-luminosity AGN fraction among quiescent X-ray non-detections, but we note that the data were consistent 
with all the quiescent galaxies hosting low-luminosity AGNs if the luminosity of these was assumed to be only 
$L_{0.5\text{--}8\,\text{keV}}=7\times10^{41}\text{\,erg\,s}^{-1}$. 

Interestingly, the quiescent stack was only detected in the soft band, cf. Table \ref{counts_da}, whereas one might expect 
a significant contribution to the hard band flux from a low-luminosity AGN population as the one proposed above. The lack 
of a hard band detection can be explained by the fact that the sensitivity of the CDF-S observations drops about a factor 
of six from the soft to the hard band \citep{xue11}, meaning that the low-luminosity AGN population must be relatively 
unobscured ($\Gamma>1$, consistent with the value of $1.4$ assumed here), as it would otherwise have been detected in 
the hard band with an S/N$\,>2$.

The SFR$_{2\text{--}10\,\text{keV}}$ of the star-forming stack was consistent with its SFR$_\text{UV+IR}$, meaning that a 
strict lower limit to the low-luminosity AGN fraction here, is zero. However, performing the test above with the same 
model parameters, a maximum of $40\%(21/53)$ low-luminosity AGNs is possible, before the X-ray inferred SFR exceeds 
the upper limit on SFR$_\text{UV+IR}$.

It should be mentioned that the $24\,\mu\text{m}$ flux, especially at these high redshifts, is an uncertain estimator of 
the total rest-frame infrared luminosity, i.e., the entire dust peak, used in the conversion to SFR.  As shown by 
\cite{bell03}, one should ideally use the entire $8$--$1000\,\mu\text{m}$ range, e.g., by taking advantage of Herschel data, 
which can lead to systematic downward correction factors up to $\sim2.5$ for galaxies with  $L_\text{IR}\approx10^{11}L_{\odot}$ 
(similar to the inferred IR luminosities of our sample galaxies detected in $24\,\mu\text{m}$, showing a median of 
$L_\text{IR}=10^{11.5}L_{\odot}$) as demonstrated by \cite{elbaz10}. However, using the same conversion from $f_{24\mu\text{m}}$ 
to $L_\text{IR}$ as the one implemented in this study, \cite{wuyts11} showed that the resulting $L_\text{IR}$ for galaxies out 
to $z\sim3$ are consistent with those derived from PACS photometry with a scatter of $0.25$ dex. Hence, we do not expect the 
inclusion of Herschel photometry in the this study to significantly impact any of our results, and we leave any such 
analysis for future work.

Table \ref{numbers} gives an overview of the derived AGN fractions, both at high and low X-ray luminosity. Adding the numbers of 
luminous AGN, X-ray-detected low-luminosity AGNs as well as the estimated lower limit on the low-luminosity AGN fraction among 
non-detections, we arrive at a lower limit on the total AGN fraction of
\begin{align}
f_{\text{AGN}}\geq\frac{27+21+0.6\cdot19+0\cdot53}{123}=0.48
\end{align}
for all massive galaxies at $z\sim2$. While for the star-forming galaxies this fraction lies in the range from $43\%$--$65\%$, 
it must be $70\%$, and potentially $100\%$, for the quiescent galaxies. Using the upper limits on these numbers, a tentative 
upper limit on the total AGN fraction is $0.72$.\\

\subsection{Contribution from Hot Gas Halos}
We have so far considered star formation and AGN activity as causes of the X-ray emission observed, but a third possibility 
is an extended hot gas halo as seen around many nearby early-type galaxies in the same mass-range as our sample galaxies 
\citep{mulchaey10} and predicted/observed around similar spirals \citep{toft02,rasmussen09,anderson11,dai12}. AGN X-ray 
emission is expected to come from a very small, $R<1\,\text{pc}$, accretion disk surrounding the central black hole of 
the host galaxy \citep{lobanov07}, whereas very extended star formation in the galaxy or a hot gas halo surrounding it 
would lead to more extended emission. We investigated the possibilities for these latter cases by comparing radial surface 
brightness profiles of the $51$ individually X-ray detected galaxies out to a radius of $8\arcsec$ in both the stacked 
observed image and a correspondingly stacked PSF image. 

\begin{figure}[htbp] 
\centering
\includegraphics[width=0.7\columnwidth]{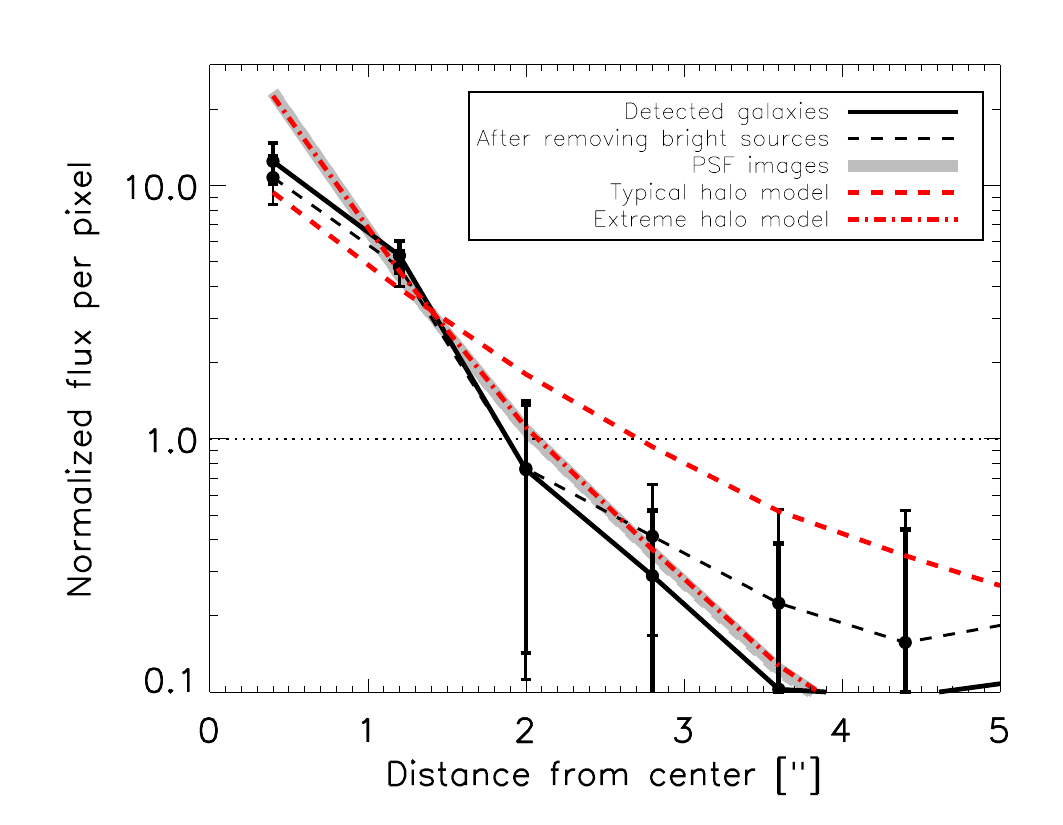}
\caption{\footnotesize{Comparison of the radial surface brightness profiles in PSF images vs. observed images in full band for all 
detected galaxies. Black: stacked observed image centered on $K$-band positions from FIREWORKS for all detected galaxies 
(solid line) and if excluding the three X-ray brightest sources (dashed line). Grey line: stacked PSF image. Red: stack 
of halo model images, with $\beta=0.5$, $R_c=2.5\,\text{kpc}$ (dashed) and $\beta=1$, $R_c=1\,\text{kpc}$ (dot-dashed). }}
\label{psf_im_is13}
\end{figure}

The profiles were calculated in full band only, because of the high S/N here as compared to the other bands (cf. Figure \ref{st_im}). 
For each galaxy, we extracted the background subtracted source count per pixel within $10$ concentric rings of width $0\arcsec.8$ around 
the galaxy center positions from the FIREWORKS catalog, and each profile was normalized to the mean count rate in all rings. 
The same procedure was applied to the corresponding PSF images, extracted with the library and tools that come with 
CIAO,\footnote{\url{http://cxc.cfa.harvard.edu/ciao4.3/ahelp/mkpsf.html}} allowing for extraction at the exact source 
positions on the detector. We verified the robustness of using PSF models from the CIAO calibration database for this 
purpose, by repeating the above procedure for $51$ known X-ray bright point-like sources from the catalog of \cite{xue11}, 
and confirming that the resulting mean profile was fully consistent with that of the corresponding model PSFs.

As can be observed in Figure \ref{psf_im_is13}, the combined radial profile of our X-ray detected galaxies in the full band 
is consistent with the PSF of point sources stacked at the same detector locations. A Kolmogorov-Smirnov (K-S) test yields 
a statistic of $0.2$ and a probability of $99.96\%$ that the two profiles are drawn from the same parent sample.

Omitting the three sources with $L_{0.5\text{--}8\text{\,keV}}>10^{44}\text{\,erg\,s}^{-1}$ (cf. Figure \ref{hr_lum}) leads 
to a more extended profile (dashed black line in Figure \ref{psf_im_is13}), but the result still shows a high corresponding 
K-S probability of $70\%$. We also compared the profiles while recentering the images on the center of the X-ray emission as 
derived using WAVDETECT\footnote{WAVDETECT is part of the CIAO software.} instead of the $K$-band center positions listed 
in the FIREWORKS catalog, in order to test for the impact of any off-set between the X-ray and optical centroids. However, 
the recentering only resulted in small variations within the errorbars on the original profile. The same conclusions applied 
to the subsample of X-ray detected star-forming galaxies only, whereas the S/N was too low to perform a similar study on the 
quiescent, X-ray detected sample alone.

For comparison, we also simulated the stacked profile of extended hot gas halos, using for the surface brightness profile 
the $\beta$-model first introduced by \cite{cavaliere76}. With default parameters of $\beta=0.5$ and core radius 
$R_c=2.5\,\text{kpc}$, both taken from the study of $z\lesssim 0.05$ early-type galaxies by \cite{osullivan03}, 
the stacked profile of the halos, convolved with the corresponding PSFs, is well outside the errorbars of the measured profile. 
Only if assuming an extremely compact case of $\beta=1$ and $R_c=1\,\text{kpc}$, does the halo emission become sufficiently 
compact to mimic that of the PSF profile (see dash-dotted line in Figure \ref{psf_im_is13}), but the halo model then overpredicts 
the observed surface brightness on scales smaller than $1\arcsec$. In conclusion, a hot gas halo alone, described by a $\beta$-model, 
cannot explain the emission, unless one chooses model parameters that render the profile indistinguishable from a point-like source.

\subsection{Quenching of Star Formation by AGNs?} \label{2:quench}
It remains debated whether the presence of AGN is connected with internal galaxy properties associated with secular evolution or, 
to a higher degree, with external processes \citep{darg10}. For example, the cosmic star formation rate density and the number density 
of AGN share a common growth history with a peak in activity around $z\sim 2$--$3$ \citep{treister12a,dunlop11}, hinting at a 
co-evolution between SFR and supermassive black hole (SMBH) accretion \citep{schawinski11}. 
We investigated the correlation, if any, between the presence of AGNs and internal/external processes, treating luminous and low-luminosity AGNs 
separately. As typical internal properties governing secular evolution we focused on stellar mass, $M_\ast$, and SFR \citep{peng10}, 
while major mergers were taken as a likely case of external processes and as a phenomenon often associated with luminous AGNs \citep{treister12}. 

Starting with our X-ray identified luminous AGNs, we found that the fraction of these does not correlate with star formation in 
our sample (cf. Table \ref{numbers}). In addition, the distribution of luminous AGNs and that of the rest with regards to their 
SFR$_{\text{IR+UV}}$ are similar, with a K-S test yielding a statistic of $0.21$ with a probability of $59\%$. Similarly, 
\cite{harrison12} found no correlation between AGN luminosity and quenching of star formation, albeit at a higher luminosity 
($L_{2-8\,\text{keV}}>10^{44} \text{erg\,s}^{-1}$) than probed here. Dividing the sample according to $M_\ast$ instead, 
by constructing two bins around the median mass of $1.1\times10^{11}\,M_\odot$, we arrived at similar luminous AGN fractions above and 
below this mass limit, namely $21\%\pm7\%(13/61)$ and $23\%\pm7\%(14/62)$, respectively. We thus find no clear evidence for the 
luminous AGN fraction of our sample to correlate with internal properties, suggesting that an external factor is of larger importance. 

An alternative is that luminous AGNs are primarily triggered by non-secular processes such as major mergers. \cite{treister12} 
found that major mergers are the only processes capable of triggering luminous ($L_{\text{bol}}\gtrsim10^{45}\,\text{erg\,s}^{-1}$) 
AGN at $0<z<3$. Our luminous AGN fraction is consistent with the major merger fraction of massive ($M_{\ast}>10^{11}M_{\odot}$) 
galaxies at $1.7<z<3$ found by \cite{man12} to be $15\%\pm8\%$ (compared to a luminous AGN fraction of $11\pm3\%(13/123)$ when 
excluding galaxies with $M_{\ast}<10^{11}M_{\odot}$). This is consistent with the idea that our luminous AGNs are triggered by 
major mergers, but a more direct test of this scenario would be to search for major mergers using the imaging data available in 
CDF-S. Indeed, \cite{newman12} used {\it HST} CANDELS imaging in the GOODS-South field residing within CDF-S (along with the UKIRT 
Ultra Deep Survey) to arrive at roughly equal pair fractions when comparing massive ($M_{\ast}>10^{10.7}M_{\odot}$) quiescent to 
massive star-forming galaxies at $0.4<z<2$. Again, this is in agreement with our result that the luminous AGN fraction does not 
vary between quiescent and star-forming galaxies. A detailed morphological study of our X-ray identified luminous AGN using the 
{\it HST}/WFC3 data to search for evidence of merging is an interesting extension to the work presented here, 
which we leave for a future study. 

Turning toward our low-luminosity AGNs, we {\it do} see X-ray evidence for an enhanced population among the quiescent galaxies when 
compared to their star-forming equivalents. The mean masses of our non-detected samples of quiescent and star-forming galaxies are 
similar ($14.6\pm1.6$ and $12.7\pm1.0\times10^{10}\,M_\odot$ respectively), suggesting that the relevant factor here is SFR. At 
$0.01<z<0.07$, \cite{schawinski09} found that massive ($M_{\ast}\gtrsim10^{10}M_{\odot}$) host galaxies of low-luminosity AGNs 
all lie in the green valley, that is, at some intermediate state after star formation quenching has taken place and before the 
SED is truly dominated by old stellar populations. The observation that these host galaxies had been quiescent for $\sim100\text{\,Myr}$, 
ruled out the possibility that one short-lived, luminous AGN suppressed star formation and, at the same time, made it unlikely 
that the same AGN quenching star formation was still active, given current AGN lifetime estimates of $\sim10^7$--$10^8$\text{\,yr} 
\citep{dimatteo05}. Rather, the authors favored a scenario in which a low-luminosity AGN already shut down star formation, 
followed by a rise in luminosity, making the AGN detectable. At $z\sim 2$, SED fits of massive ($M_{\ast}>10^{11}M_{\odot}$) 
quiescent galaxies show that the quenching typically took place $\sim1\;$Gyr before the time of observation 
(\citealp{toft12}; Krogager, J.-K. et al., in preparation), demanding an even longer delay {\it or} an episodic AGN activity 
as frequently applied in models \citep{croton06}. Episodic AGN activity could explain why we see evidence for a higher 
low-luminosity AGN fraction among quiescent as compared to star-forming galaxies, but it would also require the low-luminosity 
AGN phase in quiescent galaxies to last at least as long as the dormant phase. Future modeling and observations will show whether 
this is in fact possible. 

We conclude that our data are consistent with a scenario in which luminous AGNs in massive galaxies at $z\sim2$ are connected 
with major mergers or other non-secular processes, while the presence of low-luminosity AGNs in the majority of quiescent galaxies 
suggests that these AGNs present an important mechanism for quenching star formation and keeping it at a low level. Ultimately 
what happens at $z\sim2$ has to agree with the subsequent evolution that changes size and morphology of quiescent galaxies \citep{barro13}.

\section{Conclusions}
\label{sum}
Our main conclusions are on the following two topics: 
\begin{enumerate}
\item {\it Luminous AGN fraction}\\
We find a luminous AGN fraction of $22\%\pm5\%$ among massive ($M_{\ast}>5\times10^{10}M_\odot$) galaxies at redshifts 
$1.5 \leq z \leq 2.5$, using their X-ray properties extracted from the 4 Ms Chandra Deep Field South observations. 
Among the X-ray detected galaxies, $53\%\pm13\%$ harbor high-luminosity AGNs, while stacking the galaxies not detected in X-ray, 
leads to mean detections consistent with low-luminosity AGNs or pure star formation processes. The luminous AGN fraction 
among quiescent and star-forming galaxies is similar ($19\%\pm9\%$ and $23\%\pm5\%$, respectively) and does not depend 
on galaxy $M_{\ast}$. \\
We confirmed that extended X-ray emission from a hot gaseous halo is not a viable explanation for the observed X-ray 
emission of the X-ray detected galaxies.
\item {\it Limits on total AGN fraction}\\
We convert the rest-frame hard band X-ray luminosity into an upper limit on the star formation rate, 
SFR$_{2\text{--}10\,\text{keV}}$ and compare to that derived from the rest-frame IR+UV emission, SFR$_{\text{IR+UV}}$. 
All luminous AGNs show an excess in SFR$_{2\text{--}10\,\text{keV}}$ as expected, and so does a large fraction ($\sim90\%$) 
of the remaining detected galaxies. While the star-forming galaxies not detected in X-ray have a mean X-ray inferred SFR of 
$71\pm51M_{\odot}\text{\,yr}^{-1}$, consistent with their SFR$_{\text{IR+UV}}$, the stack of quiescent galaxies shows an 
excess in SFR$_{2\text{--}10\,\text{keV}}$ of a factor $>10$ above the upper limit on SFR$_{\text{IR+UV}}$. For these 
galaxies, we find that a minimum fraction of $\sim60\%$ must contain low-luminosity 
($L_{0.5\text{--}8\,\text{keV}}\approx10^{42}\text{\,erg\,s}^{-1}$) AGNs if the SFR estimates from X-ray are to be explained, 
and that low-luminosity AGNs might be present in all of them. On the other hand, for the star-forming stack, we derive a 
low-luminosity AGN fraction of $0$--$40\%$. \\
Gathering all low- and high-luminosity AGNs, we derive a lower limit to the total AGN fraction of $48\%$, with a tentative 
upper limit of $72\%$.
\end{enumerate}

Our study was the first to present observational evidence that, at $z\sim2$, the majority of quiescent galaxies host a 
low- to a high-luminosity AGN, while the AGN fraction is significantly lower in star-forming galaxies. These findings 
are consistent with an evolutionary scenario in which low-luminosity AGN quench star formation via the energetic output 
from SMBH accretion, which, if believed to continue in an episodic fashion as often invoked by models, would need to 
have `dormant' phases at least as long as `active' phases.

We find that the high-luminosity AGNs are likely related to non-secular processes such as major mergers. 
In the future, examining the X-ray properties of galaxies in a larger sample, cross-correlated with signs of major mergers, 
may shed further light on the co-evolution of AGNs and host galaxy. 

\chapter{Outlook}
Since we published these results in 2013, many additional observational studies have been carried out 
with the aim of shedding light on AGN-galaxy co-evolution at $z\sim2$. 
A short summary is provided here together with future directions in this field.

\section{The AGN-morphology connection}
First of all, we found no clear evidence for a correlation between the presence of a luminous AGN 
and the internal properties of massive galaxies at $z\sim2$. 
This was interpreted as an argument for the need of external factors 
triggering the luminous AGN phase rather than events associated with secular evolution.
However, we only considered stellar mass and SFR, 
whereas more recent observations have established a clear correlation 
between the presence (and strength) of AGN and the morphology of its host galaxy.

In this regard, \cite{barro13} analysed the morphologies and SFRs of $\Mstar>10^{10}\,\msun$ 
galaxies at $z=1.4-3$ in the GOODS-S and CANDELS UDS fields, for comparison with AGN activity as a function of redshift. 
\cite{barro13} found a population of compact star-forming galaxies (cSFGs; defined as having 
$\log\left( \Mstar/r_e^{1.5}\right)$ below $10.3\,\msun\,{\rm kpc}^{-1.5}$) 
at $z=2.6-3.0$ that dissappears before the compact quiescent galaxies (CQGs) of similar structural properties appear 
at lower redshifts. At the same time, the AGN fraction among cSFGs is higher at $z>2$ than at lower redshifts and 
X-ray luminous ($L_{\rm 2-8\,keV}>10^{43}\,$erg\,\ps) AGNs are found $\sim30\,\%$ more 
frequently in cSFGs than in non-compact ones at $z>2$.
This was interpreted as a sign that at least some cSFGs evolve directly into the CQGs observed 
at $z\sim2$ via the evolutionary path described in Section \ref{intro:dots}, whereas 
non-compact quiescent galaxies, formed later on, are the result of normal SFGs that are quenched by 
other mechanisms which partially preserve the structural properties.

Deriving the total IR luminosity directly from the $24\,\mu$m flux only, is a rough approximation that 
could possibly affect our results and those of \cite{barro13}, as the possible contribution from an AGN to 
the FIR luminosity can lead to overestimates of the SFR. 
\cite{mancini15} did a more careful estimate of \Lir with ({\it Spitzer}+{\it Herschel}) 
photometry from 24\,$\mu$m to 250$\,\mu$m 
of 56 $\Mstar\geq 10^{11}\,$\msun galaxies at $1.4\leq z \leq 2$ 
in the GOODS-S field, for a better estimate of SFR$_{\rm IR+UV}$. Defining AGN as those 
with excess X-ray and/or radio (VLA 1.4\,GHz) luminosity with respect to the expected SFR$_{\rm IR+UV}$ or SFR$_{\rm SED}$, 
and classifying galaxies as star-forming/quiescent based on sSFR as well as $UVJ$ color, 
the authors found relatively high AGN fractions of $\sim40\pm10\,\%$ and $\sim22\pm7\,\%$ 
for star-forming and quiescent galaxies, respectively 
(though fractions very close to ours when adopting the same criteria). 
In agreement with \cite{barro13}, \cite{mancini15} also found that the presence of an AGN correlates with 
compactness as probed by a steep radial surface brightness profile in rest-frame optical.
Over a wider redshift range, $1.4<z<3$ but with a similar method, \cite{rangel14} concluded 
that compact galaxies, both quiscent and star-forming, 
display higher intrinsic $2-8$\,keV luminosities and levels of obscuration that in extended galaxies.

{\it These studies show that of out massive galaxies at $z\sim2$, compact ones contain the most powerful AGN, and 
that when studying quiescent/star-forming galaxy samples across redshift, observations seem to promote a picture in which 
cSFGs evolve directly into CQGs via a relatively obscured  AGN phase, albeit of high (absorption-corrected) 
X-ray luminosity.}

\section{Direct observations of AGN feedback}
Direct evidence for the radiative and quasar mode of AGN feedback is now existing for 
a few, mostly nearby, objects \citep[see review by][]{fabian12}. 
For example, in a study of the UV absorption line profiles of 226 $\Mstar\geq 10^{11}\,$\msun galaxies at $1<z<3$, 
\cite{cimatti13} found possible gas outflow velocities of up to $\sim500\,$km\,\ps, 
but only in galaxies classified as luminous AGNs (with $L_{\rm 2-8\,keV}>10^{42.3}$\,ergs\,\ps). 
This indicates that the gas here is moving faster than the escape velocity of active galaxies, resulting in the 
ejection of part of the ISM possible decrease in SFR.
With the X-ray satellite Nustar, sensitive to high energy X-rays of $3-79\,$keV, it is also possible to see 
such outflows in X-ray absorption and emission lines as demonstrated by \cite{nardini15}.
Likewise but at high redshift, \cite{nesvadba11} observed evidence for outflows of turbulent gas traced by 
narrow-line emission in the rest-frame UV and optical of two quasars at $z\sim3.4$ and $\sim3.9$. 
In addition to observations in UV, AGN feedback has been confirmed with radio and sub-mm observations, 
in particular in the form of strong molecular outflows \citep[e.g.][]{feruglio10,cicone14,tombesi15}.
Comparing the mass outflow (or inflow) rates of ionized or molecular gas to the accretion rate to the AGN, 
hints at the effect that these might have on the global SFR, as summarized by e.g. \cite{bergmann12} and \cite{storchi-bergmann14}.
However, {\it more observations of both jets and outflows, 
observationally confirmed in at least some AGNs at redshifts out to almost $z\sim4$, 
are needed in order to settle how they affect the ISM of their host galaxies as a whole.}


\section{Future directions: Discerning variability with time}
In particular the arrival of new oberving facilities, will open up this field towards high redshift
by studying both individual cases and global properties of large samples.

Seeing the strong connection between compactness and AGN activity at high-$z$, as revealed over recent years, 
this field will benefit tremendously from the arrival of JWST, that will enable rest-frame optical imaging out to $z\sim10$ 
with 3 times better angular resolution than HST at $1.6\mu$m \citep{beichman12}. 
That means being able to see whether AGNs are associated with compact galaxies all the way out to Epoch of Re-ionization (EoR). 
With the high NIR spectral resolution of JWST, the absence of poly-aromatic hydrocarbon (PAH) emission 
at $\sim6-11\,\mu$m can also be used to select possibile AGN candidates 
without the need of X-ray observations \citep{windhorst09,hanami11}. 
The combined forces of JVLA, JWST and ALMA will also allow detections of molecular outflows at high-$z$ 
by measuring gas kinematics as traced by molecular and fine structure lines \citep{fabian12}.

\begin{figure}[!htbp]
\centering
\includegraphics[width=8cm]{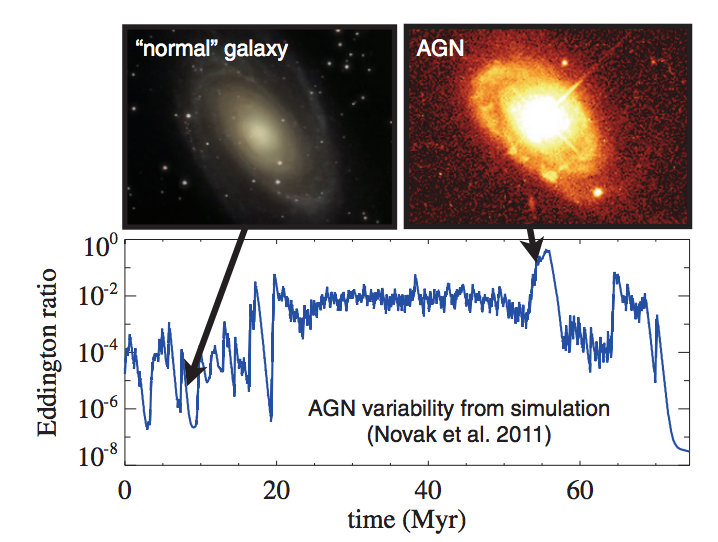}
\caption{\footnotesize{Illustration of AGN variability from \cite{hickox14} who 
took the Eddington ratio, a measure of BHAR, as a function of time from a hydrodynamic simulation presented 
by \cite{novak11} ({\it bottom}). Depending on the time of observation, the same galaxy 
can happen to be in an inactive state or a bright AGN state.}}
\label{f:novak11}
\end{figure}

A potentially very important factor, but not completely revealed in any of the above studies, 
is time variation.  
As mentioned in Section \ref{2:quench}, major mergers at $0<z<3$ have been associated with the most luminous AGNs \citep{treister12}, 
whereas moderate-luminosity ($L_{0.5\text{-}8\text{\,keV}}\sim10^{42-44}$\,ergs\,\ps) AGNs at $z\sim2$ are 
not associated with disturbed morphologies, indicative of recent mergers, any more than non-active galaxies are \citep{kocevski12}.
This said, the apparent lack of merger signatures in AGNs cannot rule out a connection between AGNs and mergers. 
First of all, if mergers in the initial phase generally produce obscured AGNs, 
then these will be harder to detect in X-rays, unless one has access to the hard X-rays 
that penetrate the obscuring gas better than the soft ones. 
One would then expect moderate AGNs to be primarily observed in post-starburst galaxies, 
in agreement with our stacking of un-detected X-ray sources, that revealed a 
{\it higher} signal from quiescent galaxies than from SFGs. 
Another exception would be if there exists a time delay between the merger and the actual onset of the luminous AGN phase. 
With JWST and upcoming X-ray observatories (see Section\,\ref{intro:tele}), we can start to piece together the redshift distribution of 
AGNs and mergers/pairs to look for a possible delay.
A second time-related issue, possibly blurring the observable relationship between AGNs and their hosts, is variability of the AGN 
luminosity as shown in Fig.\,\ref{f:novak11}. 
Models show, that when reasonable variability for the AGN luminosity is assumed, the weak correlations between SFR and $L_{\rm AGN}$ can be 
reproduced, even if there is a tight underlying correlation between SFR and BHAR \citep{novak12,hickox14}.

In conclusion, new observatories can explore the presence of AGNs at $z>2$, with particular focus on compactness, jets and 
in/outflows that are crucial for the overall SFR of the host galaxy. 
We also need to understand the time domain of AGNs in more detail, and 
as Fig.\,\ref{f:novak11} illustrates, such observations of AGN variability will be aided by a wealth of emerging simulations 
attempting to model the AGN-host co-evolution 
\citep[e.g.][]{vogelsberger13,gabor14,degraf14}.

\clearpage

\begingroup
\section{References}
\def\chapter*#1{}
\bibliographystyle{apj} 
	\setlength{\bibsep}{1pt}
	\setstretch{1}
\bibliography{bibsX}
\endgroup

%% file: summary.tex
\chapter*{Summary}


Piecing together the puzzle of galaxy evolution 
precise knowledge about {\it the shape of the pieces to the puzzle}, 
i.e. the physical conditions of the galaxies, not just of the stellar component, 
but of the interstellar medium (ISM) and black hole as well. 
This thesis focuses on deriving the actual shape of those puzzle pieces, with particular focus on 
characterising the ISM as well as the gas accretion onto the central black hole, in order to 
better understand their effects on the SFR and general evolution of massive galaxies.
With this aim in mind, the following three studies were carried out:

\subsubsection*{Amount and distribution of molecular gas in the ISM}
A combination of density and temperature is the main agent shaping the CO 
spectral line energy distribution (SLED), and observations thereof can consequently be used to 
probe the conditions of the star-forming, molecular gas in the ISM. 
However, detailed modeling is needed in order to interpret observations. 
For that purpose, a new method is constructed and presented here; 
SImulator of GAlaxy Millimeter/submillimeter Emission (\sigame). 
The method is combined with the radiative transfer code \lime in order to simulate CO 
line emission, and by applying \sigame to three galaxies from cosmological simulations, the following 
conclusions are drawn for simulated massive galaxies at $z\sim2$:
\begin{itemize}[label=\raisebox{0.25ex}{\tiny$\bullet$}]
\item The CO SLED resembles the low-excitation one of the Milky Way (MW), in that it peaks at CO$(3-2)$, 
but also with significant line intensity at higher transitions, not seen in the MW.
\item Global \aco factors range from 1.4 to 1.6\,\msun\,pc$^{-2}$\,(K\,km\,s$^{-1}$)$^{-1}$ or about 
one third of the typically assumed MW value.
\item Total CO$(3-2)$ luminosities are within the range of 
corresponding observed samples at reshifts $z\sim1-2.5$, 
however on the low side, most likely due to relatively low molecular gas masses 
in our model galaxies.
\item Radial profiles of line ratios within each galaxy reveal more excited gas towards the center, 
in agreement with observations of nearby galaxies, and suggesting ULIRG-like environments 
in the central ($R<5\,$kpc) regions.
\item The \aco factor displays a decrease towards the center of each model galaxy, however, 
by a factor that is lower than what is observed in local spiral galaxies.
\end{itemize}

\subsubsection*{Tracing SFR on local and global scales}
\sigame is expanded and adapted in order to model \cii ~-- the fine-structure line of singly ionized 
carbon (\cplus), thereby being the first code to simulate the \cii emission 
reliably on kpc-scales in normal star-forming galaxies. 
\cii is typically the strongest cooling line in neutral ISM, and correlates strongly with SFR on 
global and local scales, but its origin is so far unclear. 
These are the major findings from applying \sigame to seven $z=2$ star-forming galaxies from another simulation: 
\begin{itemize}[label=\raisebox{0.25ex}{\tiny$\bullet$}]
\item \sigame is able to reproduce the observed \Lcii-SFR relation of normal galaxies at $z>0.5$. 
\item The \cii emission that originates in molecular gas tends to dominate the total \cii luminosity in 
the central regions ($R\lesssim1\,$kpc), but 
contributions from PDRs dominate further out ($R\gtrsim1-2\,$kpc). 
The more diffuse \hii regions always contribute with a negligible amount.
\item On resolved (1\,kpc) scales, an expression is provided for the \CIIsd-\SFRsd relation.
\item The low scatter around the modelled \Lcii-SFR relation in molecular gas owes to a tight correlation between 
molecular gas amount and SFR. On resolved scales, \cii contributions from all ISM phases are increased at 
high metallicity, but whether GMCs of PDRs dominate the net \cii output depends on local pressure as well. 
PDRs can maintain high \cii efficiency at low as well as high pressures, while GMCs require high pressures 
as typically present in the central regions of galaxies.
\item A weak trend is found between increasing metallicity and decreasing 
global SFR/\Lcii ratio, but the study is hampered by a limited range in metallicity for 
our sample of galaxies. 
On resolved scales, SFR/\Lcii does increase towards the center of each galaxy 
together with metallicity, as also observed 
in local spiral galaxies.
\end{itemize}

\subsubsection*{Importance of AGNs in massive galaxies at $z\sim2$}
The deep {\it Chandra} X-ray survey CDF-S is analysed, and AGNs of high and low X-ray luminosity are extracted 
via AGN classification methods, and stacking techniques of non-detections, in X-ray. 
The galaxies investigated come from a mass-complete (at $\Mstar>5\e{10}\,\msun$) 
sample at $1.5\leq z\leq 2.5$ with stellar masses from fits of stellar population synthesis models to their 
spectral energy distribution. 
By statistical analysis, the following results are obtained:
\begin{itemize}[label=\raisebox{0.25ex}{\tiny$\bullet$}]
\item Among both star-forming and quiescent galaxies, a similar fraction of $22\pm5\,\%$ 
is found to have rest-frame X-ray luminosities ($L_{0.5\text{-}8\text{\,keV}}>3\e{42}\,$erg\,\ps) 
consistent with hosting luminous AGNs. 
\item An excess in X-ray-inferred SFR compared to that from infrared and ultraviolet emission for the 
stacked non-detections, converts into even higher fractions of low- or high-luminosity AGNs among these. 
This fraction is higher ($70-100\,$\%) among quiescent galaxies than among star-forming ones ($43-65\,$\%). 
\item Hot gas halos are rejected as potential sources for the strong X-ray emission in the detected 
galaxies, based on the very limited spatial extent of their X-ray emission.
\item The fraction of luminous AGNs does not depend on SFR nor stellar mass, suggesting that external effects 
are triggering the AGNs.
\end{itemize}

\chapter*{Sammenfatning}


Det er et sandt puslespil at stykke galakse-udviklingen sammen ud fra observationer ved 
forskellige r\o dforskydninger. 
For at l\o se puslespillet er det f\o rst of fremmest afg\o rende 
at kende formen af brikkerne, dvs. den fysiske tilstand af galakserne 
-- ikke bare af deres stellare del men ogs\aa ~af deres interstellare medium (ISM) og 
centrale sorte hul. 
Med dette m\aa l for \o je, best\aa r denne afhandling af f\o lgende tre projekter:

\subsubsection*{M\ae ngde og fordeling af molekyl\ae r gas i ISM af galakser}
Den spektrale linie energi fordeling (SLED; Spectral Line Energy Distribution) for CO molekylet 
er bestemt af hovedsagelig t\ae thed og temperatur af den gas hvor linien udsendes fra, og 
observationer af CO SLEDs kan derfor bruges til at bestemme tilstanden af den stjernedannende, 
molekyl\ae re gas i ISM af galakser. 
For at kunne overs\ae tte observationer af SLED'en til s\aa danne gas tilstande, 
er detaljeret modellering n\o dvendig. 
Til det form\aa l bliver en ny metode skabt og pr\ae senteret her; 
SImulator of GAlaxy Millimeter/submillimeter Emission (\sigame). 
Denne metode kombineres med en kode for str\aa lingstransport af millimeter of infrar\o d str\aa ling 
(\lime) for at kunne simulere CO linie emission, og ved at anvende \sigame p\aa ~tre 
galakser fra kosmologiske simulationer, bliver de f\o lgende konklusioner draget for 
massive galakser ved $z\sim2$:

\begin{itemize}[label=\raisebox{0.25ex}{\tiny$\bullet$}]
\item CO SLED'en minder om den lavt eksiterede SLED i M\ae lkevejen, i og med at den har maksimum 
i CO$(3-2)$, men de har ogs\aa ~signifikant linie intensitet ved h\o jere overgange, hvilket ikke ses 
i M\ae lkevejen.
\item Globale \aco faktorer ligger p\aa ~omkring $1/3$-del af M\ae lkevejens, nemlig fra $1.4$ til 
$1.6$\,\msun\,pc$^{-2}$\,(K\,km\,s$^{-1}$)$^{-1}$.
\item De totale CO$(3-2)$ luminositeter er alle indenfor spredningen af sammenlignelige 
observerede stjernedannende galakser ved r\o dforskydning $z\sim1-2.5$, 
men til den lave side, h\o jst sandsynlig pga. relativt lave 
molekyl\ae re gas masser i vores model galakser.
\item Radielle profiler af CO linie forhold viser at hver model galakse indeholder mere eksiteret gas i 
de centrale dele sammenlignet med l\ae ngere ude i disken. Dette er i overensstemmelse med 
observationer af lokale galakser og antyder at de centrale ($R<5\,$kpc) omr\aa der minder om de forhold man 
finder i ULIRGs (Ultra-Lumin\o se Infrar\o de Galakser).
\item \aco faktoren aftager mod centrum i galakserne som ogs\aa ~observeret, dog med et mindre 
relativt fald ift. til det man typisk oberverer i lokale spiral-galakser.
\end{itemize}

\subsubsection*{Afd\ae kning af SFR p\aa ~sm\aa ~og store skalaer}
\sigame udvides og adapteres til at modellere \cii -- fin-struktur linien fra enkelt-ioniseret 
carbon (\cplus), og er dermed den f\o rste kode til trov\ae rdigt at simulere \cii emission p\aa 
~kpc-skala i normale stjernedannende galakser. 
\cii er typisk den linie der kraftigst k\o ler neutral ISM, og den skalerer med SFR p\aa ~globale 
s\aa vel som p\aa ~lokale skalaer. 
Men dens fysiske oprindelse i gassen er stadig uklar. 
Disse er de f\o r ste konklusioner, som \sigame indtil nu har f\o rt til p\aa ~det omr\aa de:
\begin{itemize}[label=\raisebox{0.25ex}{\tiny$\bullet$}]
\item \sigame kan reproducere den observerede relation mellem \Lcii og SFR for normale galakser 
ved $z>0.5$.
\item \cii emission udspringer hovedsageligt fra molekyl\ae r gas i de centrale ($R\lesssim1\,$kpc) 
dele af galakserne, men bidrag fra PDRs (Photon Dominated Regions) dominerer l\ae ngere ude 
($R\gtrsim1-2\,$kpc). Den mere diffuse \hii gas bidrager med neglicibel \cii intensitet.
\item P\aa ~opl\o ste (1\,kpc) skalaer, opstilles et udtryk for \CIIsd som funktion af \SFRsd.
\item Den lave spredning i \Lcii-SFR relationen for molekyl\ae r gas skyldes en t\ae t sammenh\ae ng mellem 
molekyl\ae r gas masse of total SFR. 
P\aa ~opl\o ste (1\,kpc) skalaer er \cii emissionen st\o rre i omr\aa der med relativ h\o j 
metallicitet, men hvorvidt GMCs eller PDRs dominerer \cii budgettet afh\ae nger ogs\aa ~af 
det lokale tryk. PDRs kan opretholde h\o j \cii `effekticitet' ved lavt s\aa vel som h\o jt 
tryk, hvorimod GMCs kr\ae ver h\o jt tryk som typisk kun er at finde i de centrale omr\aa der.
\item En svag korrelation findes mellem stigende metallicitet of aftagende global 
SFR/\Lcii rate, men det er sv\ae rt at konkluderer noget da vores galakser ikke sp\ae nder vidt nok i 
metallicitet. Dog ser vi p\aa ~mindre skalaer at SFR/\Lcii raten stiger imod centrum af 
galakserne sammen med metalliciteten, hvilket ogs\aa 
~er blevet observeret i lokale spiral-galakser.
\end{itemize}

\subsubsection*{En vigtig rolle af AGNs i massive galakser ved $s\sim2$}
Aktive galaksekerner (AGNs; Active Galactic Nucleui) af h\o j og lav X-ray luminositet afd\ae kkes i 
feltet CDF-S, der er blevet studeret i r\o ntgen med lang eksponering ($4\,$Ms) af {\it Chandra} rum-observatoriet. 
Gruppen af galakser som unders\o ges d\ae kker fuldst\ae ndigt stellar masser over $5\e{10}$\,\msun ved 
$1.5\leq z\leq 2.5$. Ved brug af AGN klassificerings-teknikker (og opsummerings-metoder for de galakser 
der ikke detekteres individuelt i r\o ntgen), opn\aa es de f\o l gende resultater:
\begin{itemize}[label=\raisebox{0.25ex}{\tiny$\bullet$}]
\item En br\o kdel p\aa ~$22\pm5\,\%$ blandt stjerne-dannende s\aa vel som blandt passive galakser, 
har X-ray lystyrker svarende til lysst\ae rke AGNs.
\item En endnu st\o rre br\o kdel blandt de ikke-detekterede galakser, viser tegn p\aa ~at indeholde 
AGNs af lav til h\o j luminositet, med h\o jere br\o kdel ($70-100\,\%$) blandt passive galakser 
sammenlignet med de stjerne-dannende ($43-65$\,\%).
\item Grundet den rumligt koncentrerede X-ray emission fra de unders\o gte galakser, kan 
en ekstremt varm gas halo udelukkes som \aa rsag til X-ray emissionen i de detekterede galakser.
\item Br\o kdelen af lysst\ae rke AGNs afh\ae nger ikke af SFR eller stellar masse, 
hvilket tyder p\aa ~at eksterne faktorer spiller ind p\aa ~ant\ae ndelsen af deres AGNs.
\end{itemize}

%% file: akn.tex

\chapter*{Acknowledgments}
My passion for astronomy was sparked during long, cold nights in the observatory of my high 
school, Alssundgymnasiet S\o nderborg, in Southern Denmark. 
In that way, I feel fortunate for having benefitted from an educational system that incorporates astronomy.
But more so, I am grateful to my first astronomy teacher, Mogens Winther, for his incredible enthusiasm 
and to those classmates that made it fun to stay up at night.
Since then, so many inspirational people have opened up the sky for me, 
both in Aarhus, where I finished my bachelors in physics, and later on in Copenhagen for my master and PhD. 
These past 4 years, that journey has been eased and guided by two amazing people, my supervisors Thomas and Sune. 
In your own very separate ways, you have taught me how to believe in myself as a researcher, 
and no words can describe my respect for you. 

From the moment I hand in my thesis, I will miss life at Dark. 
Dark Cosmology Centre is nothing less of a paradise for any astronomer; daily provisions of coffee and fruit 
and a non-tiring staff of administration and support. 
Thank you for patiently answering all of my stupid questions and fixing my mistakes (even on weekends). 
And to the rest of my `darklings'; thank you for all those great moments; excursions, 
movie nights, alleviating lunch breaks, parties or laughs at the coffee machine.

I would not have gotten very far, in astronomy or anything else, without my friends. 
First `tak' to my old friends in Copenhagen that I occasionally forget how lucky 
I am to have; Tomatito, Lotte (x2), Helene. 
Thanks to my dear friends spread across the world -- your warmth, generosity and wisdom still reaches and 
amazes me; Marty, Cat, Urs and papa-J, Sam, Chang, Rosa, Scott (the dad and the pilot), 
Fan-Fan, Hannah, Carol, Lila, Henry, 
Charlotte, Britta and Kaare to name a few, but certainly not all! 
I will also not forget to thank my flamenco girls; Rut, Rebecka, Elisabeth and all those crazy, lovely people I 
meet doing capoeira or aerial acrobatics, in Denmark and accross the world.

Towards the end of my PhD, I was blessed by meeting my own `curandero'. 
From the other side of the ocean, you support me in everything I do and make me laugh on my most solemn days. 
I will never know how many bruises and dangers you saved me from.

Finally, I can't wait to thank my family (yes, that includes you guys, 
t\'ia m\'ia and Erk who I always wish to see more). 
You're there with love and encouragement (and sund fornuft), even when Als seems light years away. 
In particular, thank you mamita for taking those field trips to 
Copenhagen and reminding me how to go shopping.

\bigskip
{\it Tak, gracias, obrigada, thank you!}

\bigskip
Karen

%% file: appendix.tex
\begin{appendices}

\chapter{Appendix to Chapter \ref{paper1}}

\section{Thermal balance of the atomic gas phase}
\label{apB}
As explained in \S\,\ref{WCNM}, we cool the initial hot SPH gas by 
requiring:
\begin{equation}
\Hcrhi = \Cions + \Crec + \Cff. \nonumber
\end{equation}

\bigskip

\Hcrhi is the heating rate of the atomic gas due to cosmic ray ionizations
\citep{draine11}:
\begin{equation}
\begin{aligned}
	\Hcrhi = &1.03\e{-27}\nhi \left( \frac{\crihi}{10^{-16}} \right) \\
	&\times\left[ 1+4.06\left( \frac{\xe}{\xe+0.07}\right)^{1/2}\right]~~{\rm erg\,cm^{-3}\,s^{-1}}, 
	\label{test}
\end{aligned}
\end{equation}
where \crihi is the primary CR ionization rate of \hi atoms (determined locally
in our simulations according to eq.\ \ref{equation:CR}), and \xe is the
hydrogen ionization fraction calculated with a procedure kindly provided by I.
Pelupessy; see also Pelupessy (2005).  The term containing \xe in
eq.\,\ref{test} accounts for the fact that in highly ionised gas, electrons
created by primary CR ionization have a high probability of transferring their
kinetic energy into heat via long-range Coulomb scattering off free electrons.
For low ionization gas, this term becomes insignificant as a higher fraction of
the energy of the primary electrons goes to secondary ionizations or excitation
of bound states instead of heating.  

\Cions is the total cooling rate due to line emission from H, He, C, N, O, Ne, 
Mg, Si, S, Ca, and Fe, calculated using the publically available code of 
\cite{wiersma09} which takes \Tk, \nH 
and the abundances of the above elements as input. 
\cite{wiersma09} compute the cooling rates with the
photoionization package \texttt{CLOUDY} assuming CIE. They also adopt a value 
for the meta-galactic UV and X-ray field equal to that expected at $z\sim 2$ \citep{haardt01}.
At $z\sim2$, the emission rate of \hi ionizing radiation is higher by a factor of 
about $\sim 30$ than at $z=0$ \citep{puchwein15}, and thus plays an important role
in metal line cooling calculations.

\Crec is the cooling rate due to hydrogen recombination emission \citep{draine11}:
\begin{equation}
	\Crec = \alpha_B\ne\nHplus\langle E_{rr} \rangle~~{\rm ergs\,cm^{-3}\,s^{-1}},
\end{equation}
where $\alpha_B$ is the radiative recombination rate for hydrogen in the case of
optically thick gas in which ionizing photons emitted during recombination are
immediately re-absorbed. We adopt the approximation for $\alpha_B$ given by
\cite{draine11}: 
\begin{equation}
	\alpha_{\rm B} = 2.54\e{-13}T_4^{\left(-0.8163-0.0208\ln T_4 \right)}~~{\rm cm}^3\,\ps,
\end{equation}
where $T_4$ is defined as $\Tk/10^4$\,K. The density of ionised hydrogen,
\nHplus, is set equal to the electron density, \ne, and $E_{rr}$ is the
corresponding mean kinetic energy of the recombining electrons:
\begin{equation}
	\langle E_{rr}\rangle = \left[0.684-0.0416\,\ln T_4 \right]\,k_{\rm B}\Tk~~{\rm ergs}.	
\end{equation}

\Cff is the cooling rate due to free-free emission from electrons in a pure H
plasma (i.e., free electrons scattering off H$^+$), and is given by
\citep{draine11}:
\begin{equation}
	\Cff = 0.54 T_4^{0.37} k_B\Tk\ne\nHplus\alpha_{\rm B}~~\rm{ergs\,cm^{-3}\,s^{-1}},
\end{equation}
where the recombination rate, $\alpha_{\rm B}$, is calculated in the same way as for \Crec.

\begin{figure*}
\centering
\includegraphics[width=0.5\textwidth]{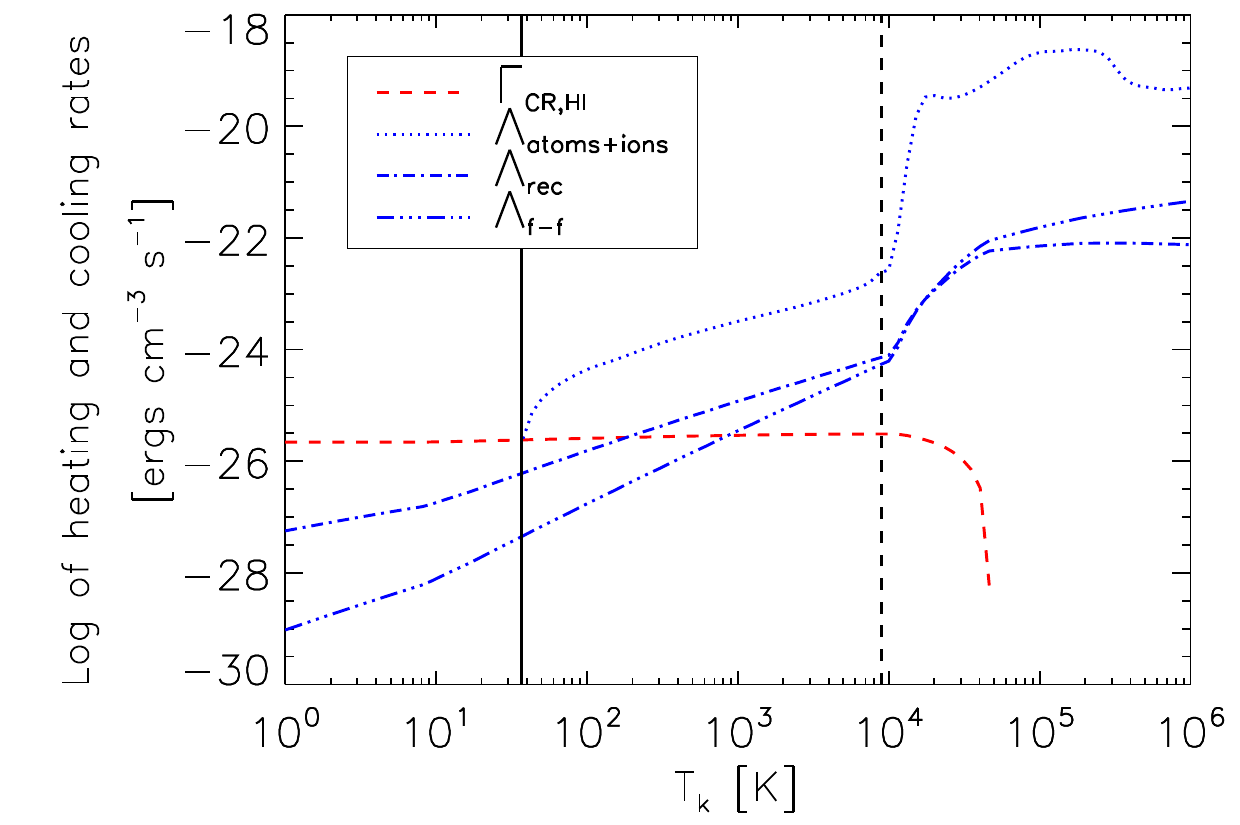}\includegraphics[width=0.5\textwidth]{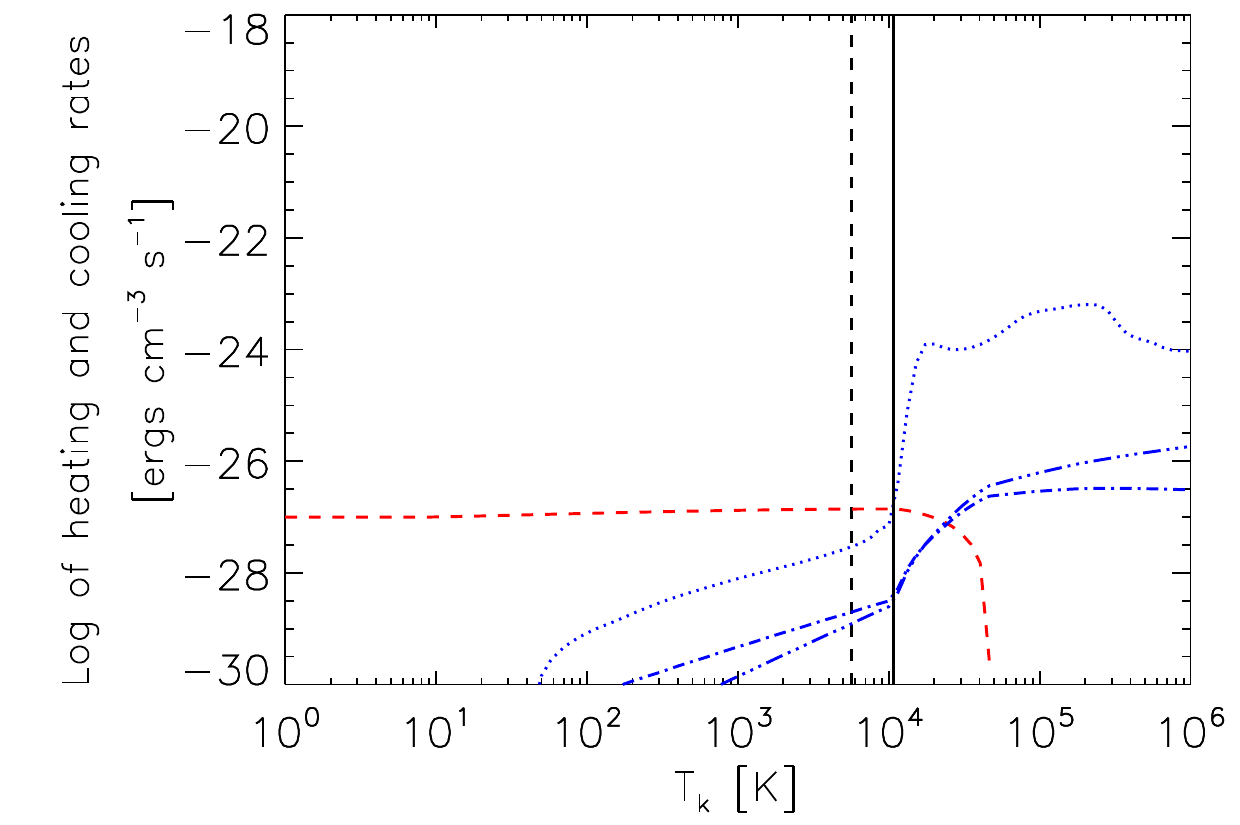}
\caption{\footnotesize{Heating and cooling rates as functions of temperature for two SPH gas 
particles with [$\nH=14.62$\,\cmpc, $\cri=4.02\e{-16}$\,\ps, \xe($T_{\rm 
k,SPH}=8892\,K)=0.006$] (left) and [$\nH=0.09$\,\cmpc, $\cri=3.92\e{-16}$\,\ps, 
\xe($T_{\rm k,SPH}=5649\,K)=0.001$] (right). In both cases the metal line emission
is the main cooling agent. The left-hand side plot 
illustrates the case of an SPH particle of high gas density and metallicity, 
leading to relatively efficient metal line cooling and in a low equilibrium temperature 
of $\Tk=37$\,K.  The right-hand side plot shows a case of lower density and 
metallicity causing less cooling by metal lines, hence a higher equilibrium 
temperature despite a slightly lower CR heating rate. The dashed vertical lines in the two
panels indicate the original SPH gas temperatures of the SPH particles, and the
solid vertical lines mark their final equilibrium temperatures.
}}
\label{apB1}
\end{figure*}

Figure \ref{apB1} shows the above heating and cooling rates pertaining to two
example SPH particles with similar initial temperatures ($\sim 10^4\,{\rm K}$).
Because of different ambient conditions (i.e., \nhi, \xe, \Z, and \cri) the
equilibrium temperature solutions for the two gas particles end up being
significantly different.

\section{Thermal balance of the molecular gas phase}
\label{apC}
As described in Section \ref{structure} \texttt{S\'IGAME} assumes that the molecular
gas resides exclusively in giant molecular clouds that have Plummer radial
density profiles (i.e., given by eq.\ \ref{equation:plummer-profile}).
Throughout the clouds the gas temperature is solved for according to the heating
and cooling equilibrium requirement $\Hpe+\Hcrh2 = \Ch2+\Cco+\Ccii+\Coi$+\Cgd
(eq.\ \ref{Tk_GMC_e1}).

\Hpe is the heating rate of the gas due to photo-electric ejection of electrons 
from dust grains by FUV photons, and is given by \citep{bakes94}:
\begin{equation}
	\Hpe = 10^{-24}\epsilon \ga \nH~~{\rm \hu},
\end{equation}
where \ga is the local attenuated FUV field in Habing units, derived following eq.\,\ref{ga}, 
and $\epsilon$ is the heating efficiency:
\begin{align}
	\epsilon = &\frac{4.87\times10^{-2}}{1+4\times10^{-3}(\ga T^{0.5}/n_e)^{0.73}}\\
		&+\frac{3.65\times10^{-2}}{1+4\times10^{-3}(\ga T^{0.5}/n_e)^{0.73}}, \nonumber
\end{align}
where \ne is the electron density, calculated as $\xe \nH$, with \xe again calculated using the procedure of 
I. Pelupessy.

\Hcrh2 is the heating rate by cosmic rays traveling through molecular gas \citep{stahler05}:
\begin{equation}
	\Hcrh2 = 1.068\e{-24}\left (\frac{\crimol}{10^{-16}}\right )\left( \frac{\nh2}{10^3\,\cmpc}\right)~~{\rm ergs\,cm^{-3}\,s^{-1}}
\end{equation}
where \crimol is the local CR primary ionization rate of \h2 molecules, which
is approximately $1.6\times$ higher that of \hi atoms \citep{stahler05}.

\Ch2 is the \h2 line cooling rate, and we use the parameterization made by
\cite{papa14} that includes the two lowest \h2 rotational lines (S(0) and S(1),
the only lines excited for  $\Tk\lesssim1000\,{\rm K}$):
\begin{align}
	\Ch2 = &2.06\times10^{-24} \frac{n_{\rm H_2}}{1+{\rm r}_{\rm op}} \left[ 1+\frac{1}{5}e^{510{\rm K}/T_{\rm k}}\left( 1+\frac{n_{0}}{n_{\rm H_2}} \right) \right]^{-1} \\
	&\times(1+R_{10})\rm{\,ergs\,cm^{-3}\,s^{-1}}, \nonumber
\end{align}
where ${\rm R}_{10}$ is defined as:
\begin{equation}
 	{\rm R}_{10} = 26.8\,{\rm r}_{\rm op}\left[ \frac{1+(1/5)e^{510{\rm K}/T_{\rm k}}\left( 1+\frac{n_0}{n_{\rm H_2}}\right)}{1+(3/7)e^{845\,{\rm K}/T_{\rm k}}\left( 1+\frac{n_1}{n_{\rm H_2}}\right)} \right],
\end{equation}
and $n_{\rm 0}\sim54$\,\cmpc and $n_{\rm 1}\sim10^3$\,\cmpc are the critical
densities of the S(0):2-0 and S(1):3-1 rotational lines. ${\rm r}_{\rm op}$ is
the ortho-\h2/para-\h2 ratio (set to 3 which is the equilibrium value).

For the cooling rates due to the [C\,{\sc ii}]$158\,{\rm \mu m}$ and [O\,{\sc
i}]$63\,{\rm \mu m}$+$146\,{\rm \mu m}$ fine-structure lines we adopt the
parameterizations by \citet{rollig06}. The C\,{\sc ii} cooling rate (\Ccii) is:
\begin{align}
	\Ccii = &2.02\e{-24}n\Z \\ 
	&\times \left[ 1+\frac{1}{2}e^{92{\rm K}/T_{\rm k}}\left(1+1300/\nH \right) \right]^{-1}~~{\rm ergs\,cm^{-3}\,s^{-1}}, \nonumber
\end{align}
where a carbon to hydrogen abundance ratio that scales with metallicity
according to $\chi_{\rm [C]}=1.4\e{-4}$\,\Z is assumed. For the parameterization
of the O\,{\sc i} cooling rate ($\Coi = \Lambda_{63\mu \rm{m}}+\Lambda_{146\mu
\rm{m}}$) we refer to eqs.\ A.5 and A.6 in \citet{rollig06} and simply note that
we adopt \citep[in accordance with][]{rollig06} an oxygen to hydrogen abundance
ratio of $\chi_{\rm [O]}=3\e{-4}\,\Z$.

\Cco is the cooling rate due to CO rotational transitions. We use the parameterization provided by \cite{papa13}:
\begin{equation}
	\Cco = 4.4\times10^{-24} \left(\frac{n_{\rm H_2}}{10^4} \right)^{3/2} \left( \frac{T_{\rm k}}{10\,{\rm K}} \right)^2 \left( \frac{\chi_{\rm CO}}{\chi_{\rm [C]}} \right)\,{\rm ergs\,cm^{-3}\,s^{-1}}, 
\end{equation} 
where $ \chi_{\rm CO}/\chi_{\rm [C]}$ is the relative CO to neutral carbon
abundance ratio, the value of which we determine by interpolation, assuming that
$\chi_{\rm CO}/\chi_{\rm [C]}=(0.97,0.98,0.99,1.0)$ for
$\nh2=5\times10^3,10^4,10^5,10^6$\,\cmpc, respectively \citep{papa13}.  

\Cgd is the cooling rate due to gas-dust interactions and is given by \citep{papa11}:
\begin{equation}
	\Cgd = 3.47\times10^{-33}\nH^2\sqrt{\Tk}(\Tk-T_{\rm dust})\,{\rm ergs\,cm^{-3}\,s^{-1}}, 
        \label{equation:dust-gas-cooling}
\end{equation}
where the dust temperature ($T_{\rm dust}$) is calculated using
eq.\,\ref{Tk_GMC_e2} (Section \ref{Tk_GMC}). 

\begin{figure*} 
\centering
\includegraphics[width=0.5\textwidth]{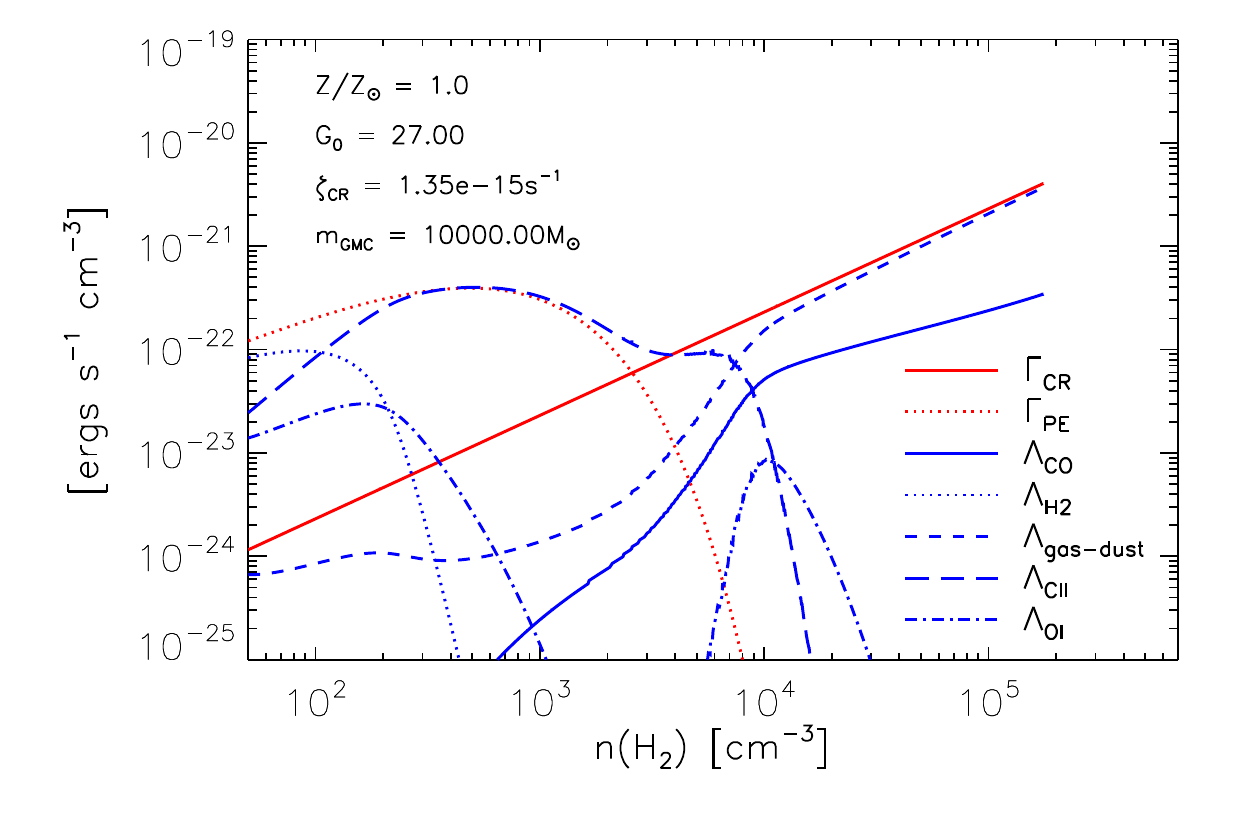}\includegraphics[width=0.5\textwidth]{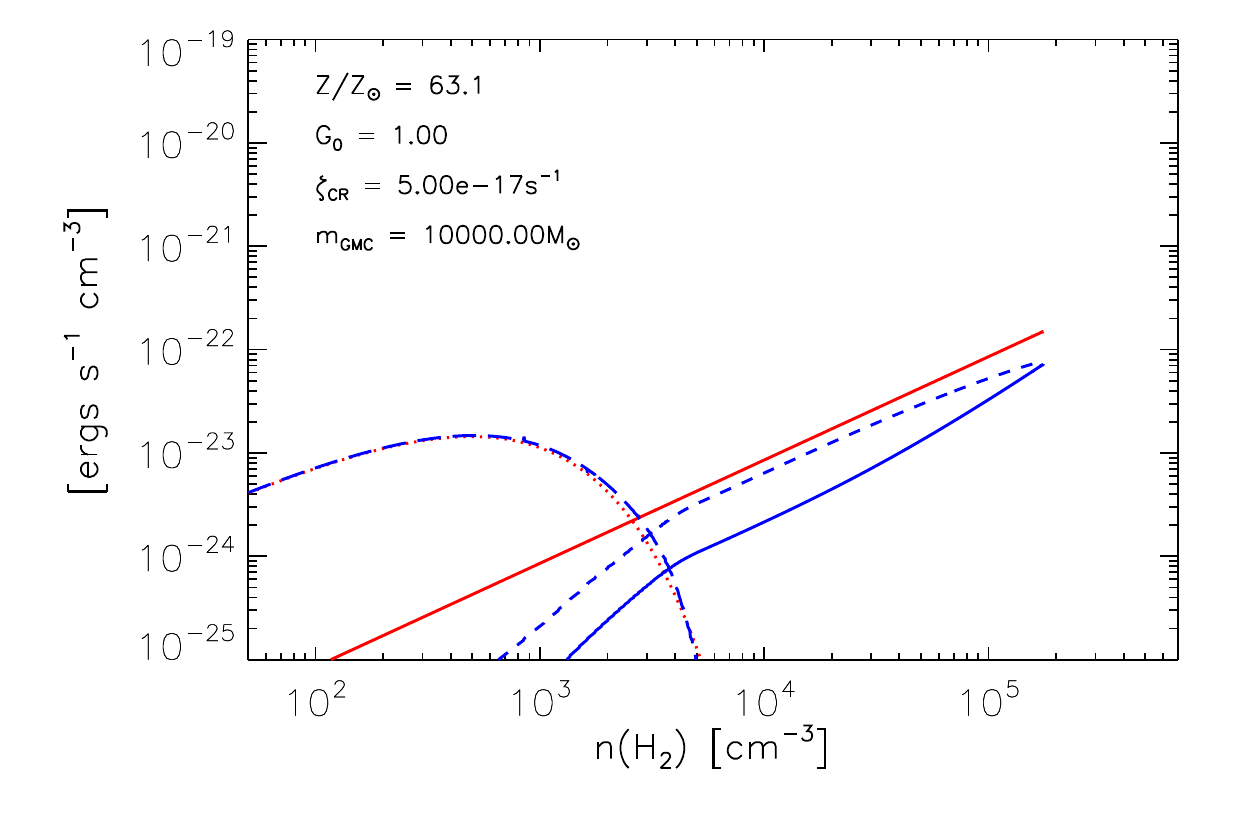}
\caption{\footnotesize{Equilibrium heating (red curves) and cooling (blue curves) rates of the gas as
functions of \h2 density for two different GMC models with 
[$\Mgmc=10^4$\,\msun, $\Z=1$, $\g0=27$, $\cri=1.35\e{-15}\,\ps$, $\Pe=10^4\,$K\,\cmpc] (left) and 
[$\Mgmc=10^4$\,\msun, $\Z=63$, $\g0=1$, $\cri=5\e{-17}\,\ps$, $\Pe=10^4\,$K\,\cmpc] (right). 
In the case of high FUV and CR radiation fields but low metallicity (left), the heating 
is dominated by cosmic ray heating (red solid) in the inner region ($\nh2\gtrsim1500$\,\cmpc) and 
photoelectric heating (red dotted) in the outer region.
Cooling is dominated by gas-dust interactions (blue dashed) 
in the inner region ($\nh2\gtrsim10000$\,\cmpc) and by \cii as well as \h2 line cooling 
(blue long-dashed and dotted) in the outer region.
In the opposite case of low FUV and CR radiation fields but high metallicity (right), 
the same heating and cooling mechanisms are dominating the energy balance throughout the cloud, except 
in the inner region, where cooling by CO line emission (blue solid) is more important than it is in the 
case of low metallicity.}}
\label{apC1}
\end{figure*}

\section{GMC models}
\label{apD}
\begin{figure*}
\hspace{-1cm}
\includegraphics[width=1\columnwidth]{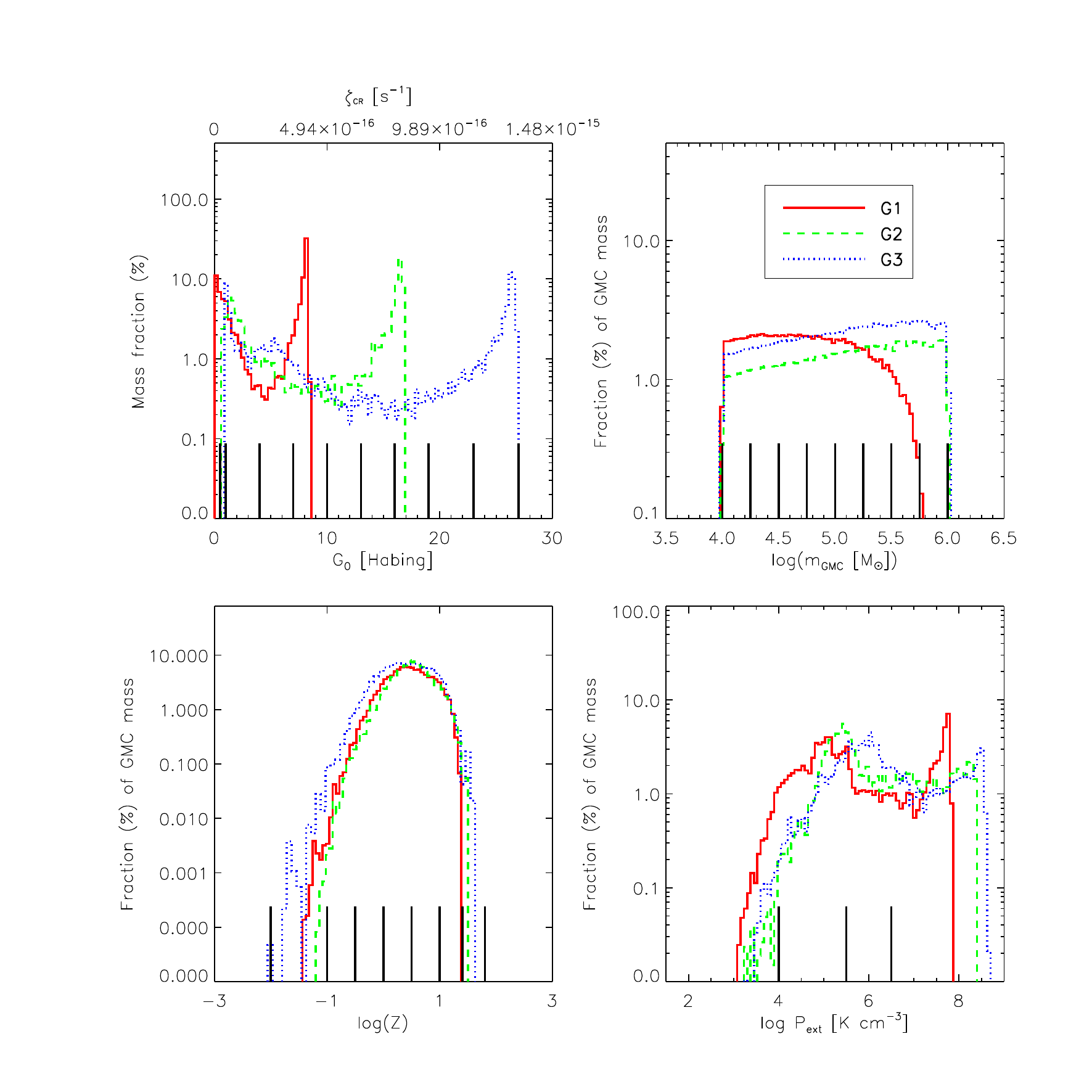}
\caption{\footnotesize{Mass-weighted histograms of the basic parameters of the GMCs in G1 (red
solid), G2 (green dashed), and G3 (blue dotted). From top left and
clockwise: the local far-UV field (\g0) (and CR ionization rate since
$\cri\propto\g0$), GMC mass (\Mgmc), metallicity (\Z) and external pressure
(\Pe).  Black vertical lines indicate the \g0, \Mgmc, \Z, \Pe-values for which
$\Tk-\nh2$ curves were calculated (see Figure \ref{apD3}) -- a total of 630 GMCs
which make up our grid GMC models. Each GMC in the galaxies is assigned the
$\Tk-\nh2$ curve of the GMC model at the closest grid point.}}
\label{apD1}
\end{figure*}

\begin{figure*} 
\centering
\hspace{-1cm}
\includegraphics[width=1\columnwidth]{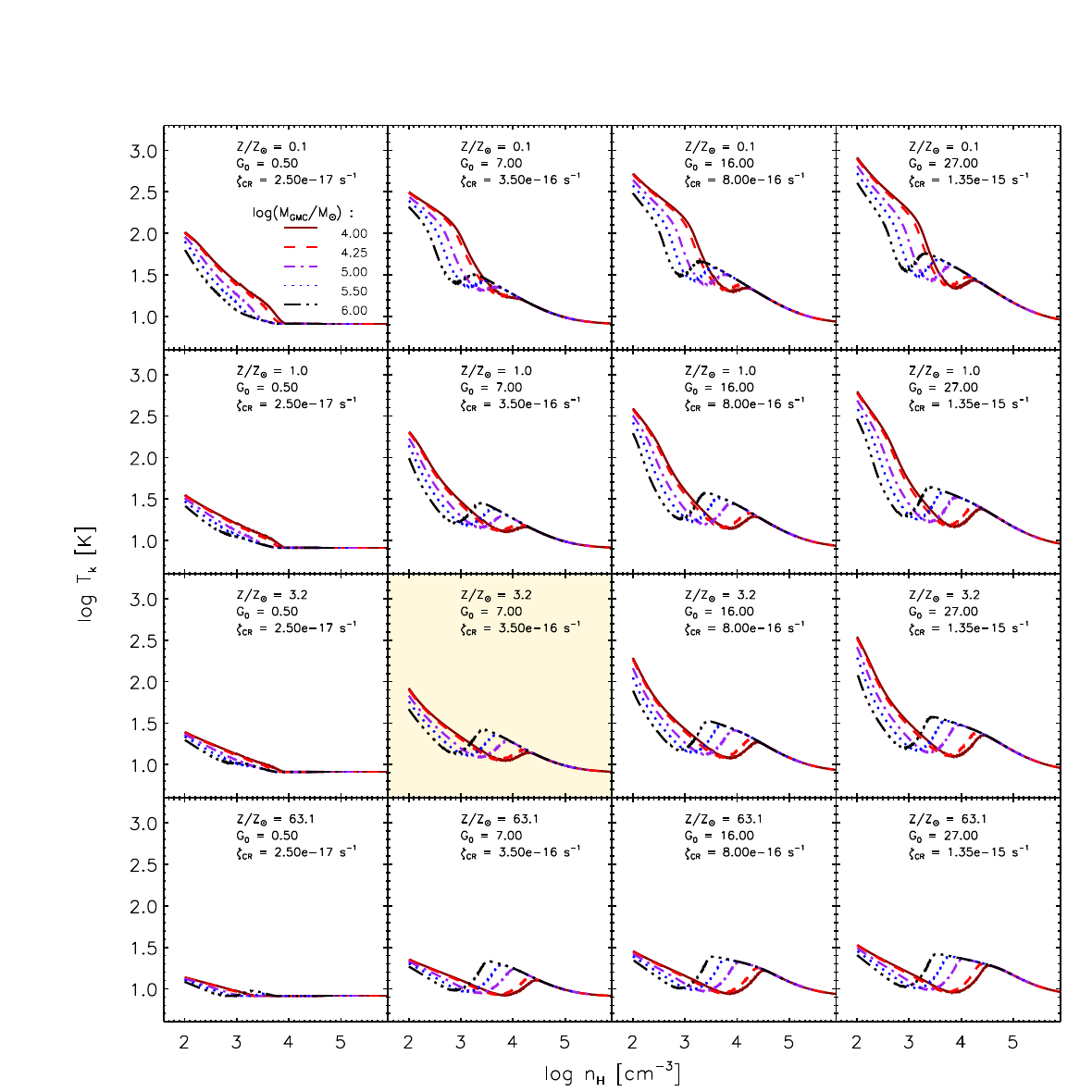} 
\caption{\footnotesize{Kinetic temperature versus \h2 density curves for 80 out of the 630
grid model GMCs that span the full (\g0, $M_{\rm GMC}$, \Z) parameter space set
by and marked on top of the distributions in Figure \ref{apD1} (see also
Section \ref{gmcgrid}) for a pressure of $\Pe=10^4$\,K\,\cmpc.  The grid model
most often assigned to GMCs in G1 is indicated by the red dashed curve in the
highlighted panel and corresponds to $\g0=7.0$ ($\cri=3.5\e{-16}$\,\ps), $\log
\Mgmc/\msun =4.25$, and $\Z=3.2$.  In general, higher metallicity (from top to
bottom) leads to more cooling via emission lines of ions, atoms and molecules,
and hence lower temperatures.  On the other hand, higher UV and CR fields (from
left to right) cause more heating and therefore higher \Tk. The decreasing
trend of \Tk with higher values of \nh2 is mainly caused by the gradual
attenuation of the UV field as one moves into the cloud.  The `bump' at
$\nh2\sim10^3-10^4$\,\cmpc corresponds to the transition from CII line cooling
to the less efficient CO line cooling at higher densities.  At densities above
$\nh2=10^4$\,\cmpc, gas-dust interactions set in and eventually cools the gas
down to the CMB temperature in all GMC cores.}}
\label{apD3}
\end{figure*}

Each SPH particle is divided into several GMCs as described in \S\,\ref{split},
and we derive the molecular gas density and temperature within each from three
basic parameters which are SFR density, GMC mass, \Mgmc, and metallicity,
$\Z/Z_{\odot}$.  Derived from these basic parameters are the far-UV and cosmic
ray field strengths, the \h2 gas mass fraction of each SPH particles, as well as
the GMC properties used to derive the CO excitation and emission; \h2 density
and temperature.  Histograms of the basic parameters are shown in
Figure\,\ref{apD1}, while properties derived thereof can be found in Figure\,\ref{apD3}.

\begin{figure*} 
\centering
\hspace{-1cm}
\includegraphics[width=1\columnwidth]{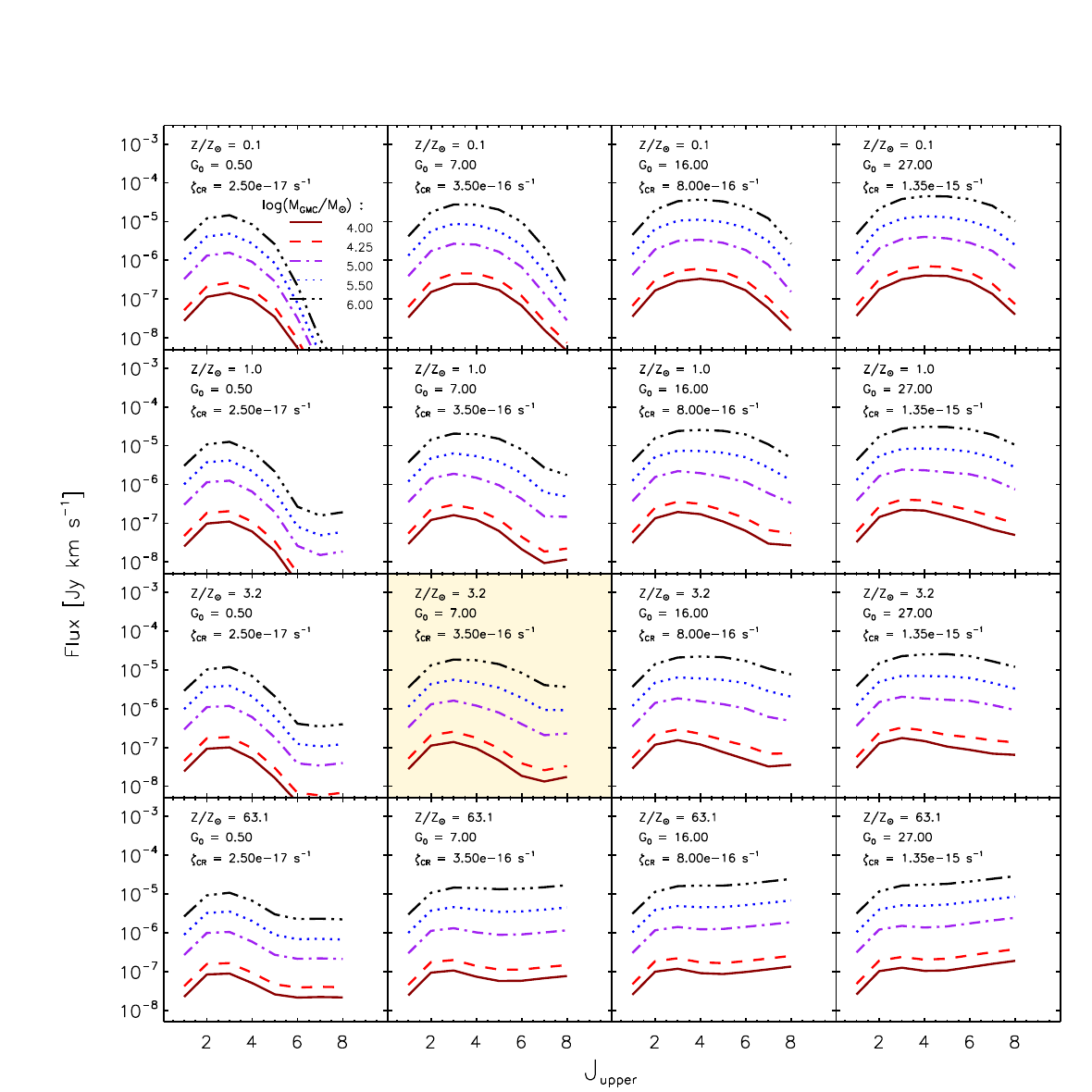}   
\caption{\footnotesize{CO SLEDs obtained with \lime for the 80 model GMCs whose
$\Tk-\nh2$ curves are shown in Figure \ref{apD3}. The CO SLED used most often in
G1 is shown as the red dashed curve in the highlighted panel.  These CO
SLEDs were made for a fixed external pressure of $\Pe/k_{\rm B}=10^4\,{\rm
\cmpc\,K}$ and a default Plummer density profile.}}
\label{apD4}
\end{figure*}

\begin{figure*} 
\centering
\hspace{-1cm}
\includegraphics[width=1\columnwidth]{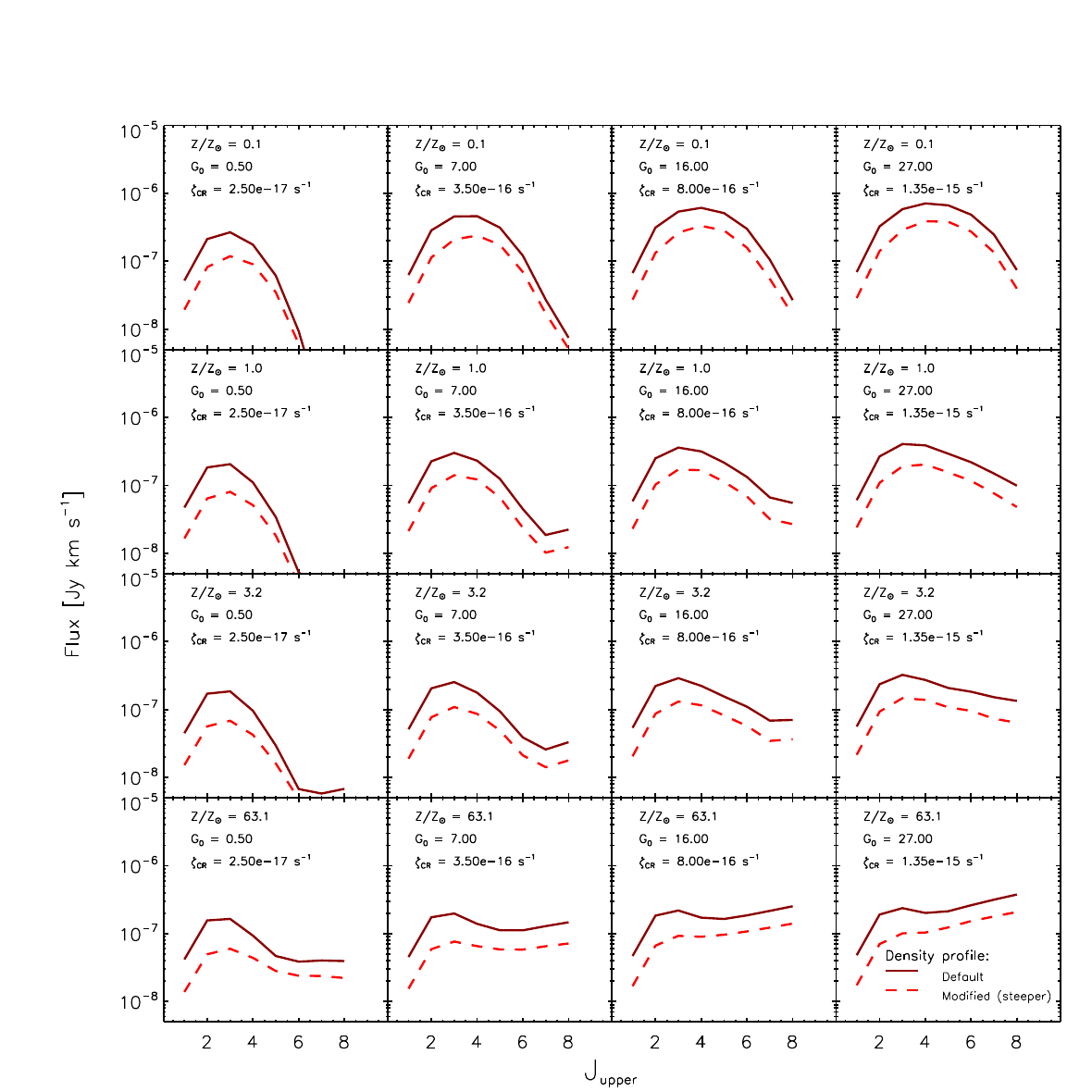}   
\caption{\footnotesize{CO SLEDs obtained with \lime for the same [\g0,\Zn] values as 
shown in the panels of Figure \ref{apD4} for Plummer density profiles with power-law
index $-5/2$ (solid curve) and $-7/2$ (dashed curve).  In all panels, the
external pressure has been fixed to $\Pe/k_{\rm B}=10^4\,{\rm \cmpc\,K}$ and the GMC mass 
to $\Mgmc=10^{4.25}\,{\rm \msun}$.}}
\label{apD5}
\end{figure*}

\begin{figure*} 
\centering
\hspace{-1cm}
\includegraphics[width=1\columnwidth]{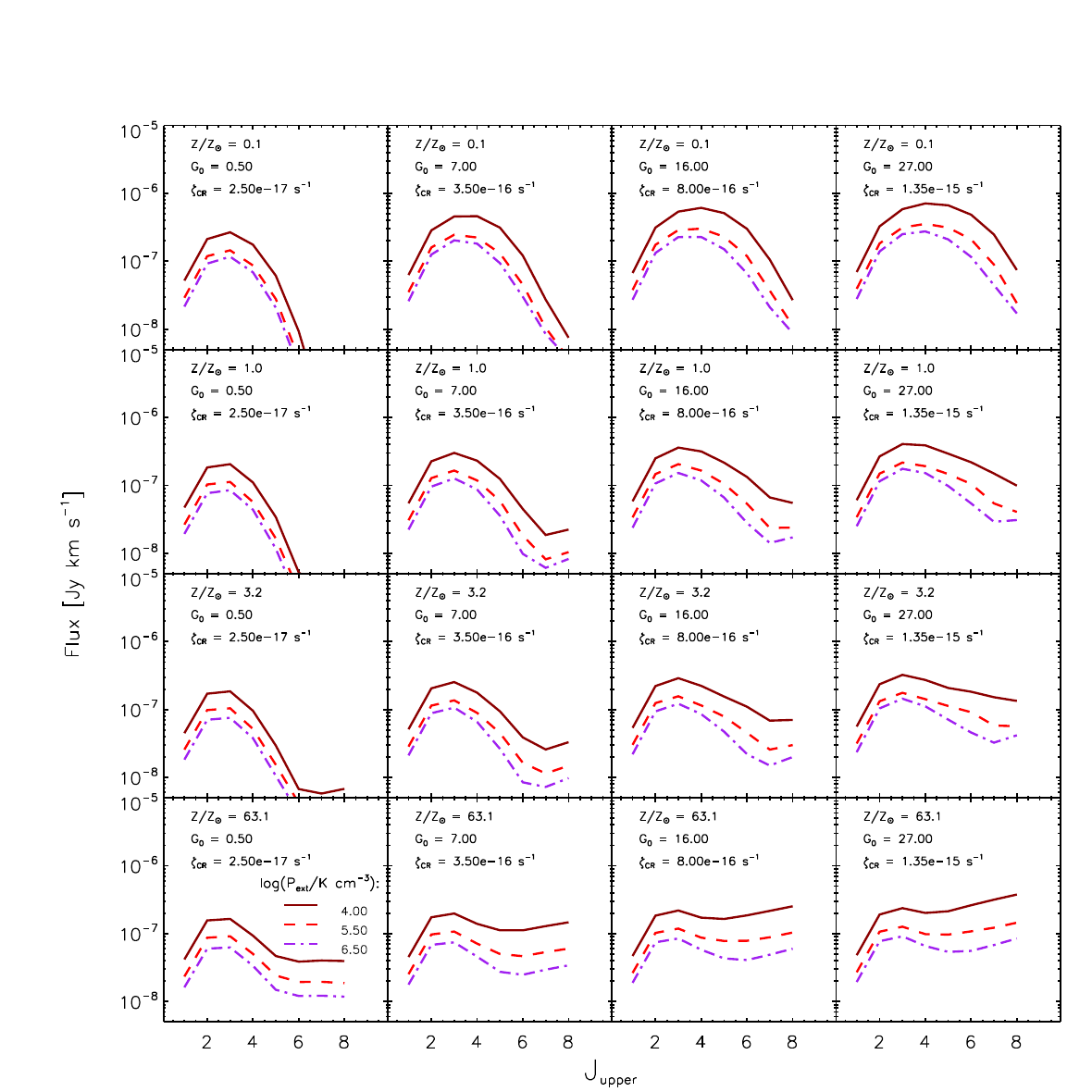}   
\caption{\footnotesize{CO SLEDs obtained with \lime for the same [\g0,\Zn] values as
shown in the panels of Figure \ref{apD4} for $\Pe/k_{\rm B} = 10^4\,{\rm
cm^{-3}\,K}$ (solid curves), $10^{5.5}\,{\rm cm^{-3}\,K}$ (dashed curves), and
$10^{6.5}\,{\rm cm^{-3}\,K}$ (dot-dashed curves).  In all panels, the GMC mass
is fixed to $10^{4.25}\,\msun$.  Higher pressure environments are seen to lead
to a decrease in luminosity for all transitions.}}
\label{apD6}
\end{figure*}

\clearpage


\section{Testing \sigame on MW-like galaxies}
\label{apMW}

As a benchmark test, \sigame was applied to galaxy simulations at $z=0$ with
properties similar to those of the MW. The reasoning here was that since much of
the sub-grid physics adopted by \sigame relies on empirical relations observed in
the MW (e.g., the GMC mass spectrum index), the method ought to come
close to re-producing the observed CO properties of spiral galaxies, such as our
MW, in the local universe.

The three galaxies (hereafter denoted MW1, MW2, and MW3 in order of increasing
SFR) were selected from a more recent version of the SPH simulation described
in Section\,\ref{cosmological_simulations}. The properties of MW1, MW2, and MW3
are listed in Table\,\ref{MWgal} and are seen to be within a factor $\sim 2-6$
of the stellar mass \citep[${\rm M}_{*,\rm{MW}}=6\e{10}$\,\msun;][]{mcmillan11}
and SFR \citep[$1.9\pm0.4$\,\sfru; ][]{chomiuk11} of the MW.

The steps outlined in Section \ref{subsection:methodology} (and further described in
Sections \ref{WCNM} to \ref{Tk_GMC}) were followed, but with two minor modifications: 1) the
CMB temperature at $z=0$ was set to 2.725\,K rather than 8.175\,K used at
$z=2$, and 2) the resolved FUV fields of our MW-like model galaxies span a
range below that seen in our $z=2$ model galaxies (cf. Figure\,\ref{apD1})
meaning that the \g0 parameter values for our GMC grid had to be adjusted
correspondingly. The following \g0 grid points were chosen: $[0.05,
0.1, 0.15, 0.1, 0.2, 0.4, 0.6, 0.8, 1.0, 1.2]$\,Habing.

\begin{figure*} 
\centering
\hspace{-1cm}
\includegraphics[width=1\columnwidth]{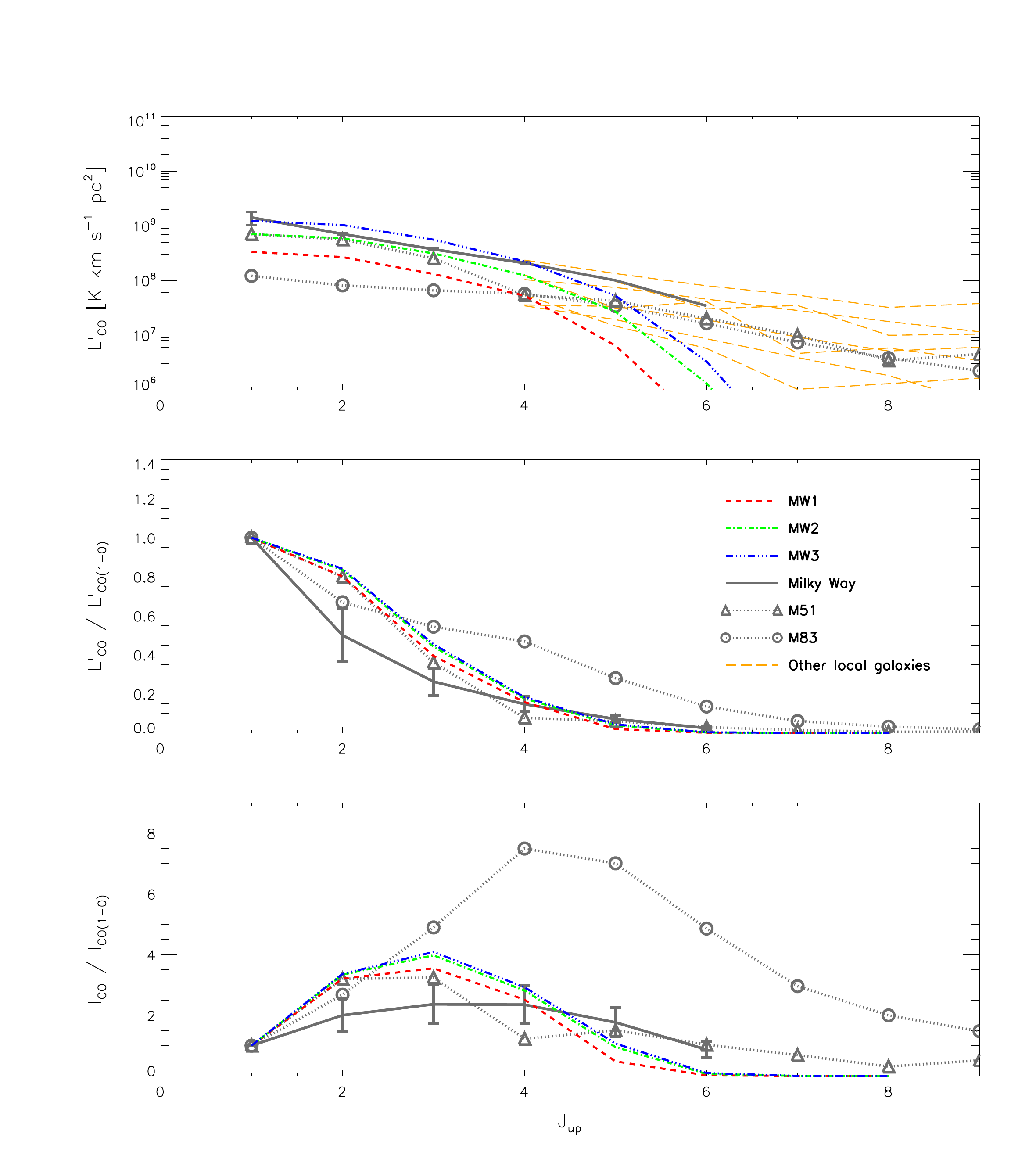}   
\caption{\footnotesize{Global CO SLEDs of our three model galaxies MW1, MW2 and MW3 shown as
red (dashed), green (dash-dot) and blue (dash-dot-dot-dot) curves,
respectively, selected to represent the MW in terms of stellar mass and SFR.
The SLEDs are given as absolute line luminosities in units of
K\,\kms\,pc$^{-2}$ (top panel), as brightness temperature ratios normalised to
the CO(1$-$0) transition (middle panel), and as velocity-integrated intensity
ratios normalised to CO(1$-$0) (bottom panel).  The observed global CO SLED of
the MW is shown in grey \citep{fixsen99}, and global CO measurements of the
local star-forming galaxy M\,51 are displayed with triangles. 
The CO observations for M\,51 are from
\citet{vlahakis13} and \citet{hughes13}.  Also shown, with
long-dashed orange lines, are observed CO SLEDs of local galaxies with IR luminosities within $20\%$
of that of the MW \citep{kamenetzky15}.}}
\label{apD7}
\end{figure*}

In Figure\,\ref{apD7} we compare the resulting CO
SLEDs with that of the MW from \cite{fixsen99}, as well as with those of M\,51
($M_{\ast}=4.7\e{10}\,\msun;$ \cite{cooper12}, ${\rm SFR}=2.6$\,\sfru;
\citealt{schuster07}) from \cite{vlahakis13}, \cite{hughes13} and 
\cite{kamenetzky15}, M\,83 
($M_{\ast}=7.9\e{10}\,\msun$ and ${\rm SFR}=3.2$\,\sfru;
\citealt{jarrett13}) from \cite{wu15} 
and other local galaxies of IR luminosity within 20\% of that of 
the MW ($1.8\e{10}$\,\lsun; \citealt{wright91}) from \cite{kamenetzky15}. 
M\,51 is a nearby galaxy slightly smaller in size and stellar mass than the MW, 
whereas M\,83 is more massive and star-forming than the MW. 
The dispersion in observed CO SLEDs among the MW, M\,83, M\,51 
and other local galaxies, allows for 
a large range in normalization and shape of the CO SLED within 
which our model galaxies find themselves.

The CO line luminosities of the MW lie roughly in the range spanned by 
the CO SLEDs of MW2 and MW3 up to and including $J_{\rm up} = 4$, 
but significantly above that of MW1.
For the higher $J$ transitions the line luminosities of our simulations 
drop off more rapidly than the MW, 
meaning that our model galaxies have lower luminosities compared to
the MW (and other local galaxies of IR luminosities within 20\% from that
of the MW; long-dashed orange lines) at $J_{\rm up} > 4$.
This suggests that our simulations are missing a warm/dense component which is
required to significantly excite these high-$J$ lines. 
The agreement between the CO SLED of MW2 and that of M\,51 is good at 
low transitions ($J_{\rm up} < 6$), which is encouraging given that the two galaxies
have nearly identical SFR and $M_{\ast}$. 
At higher $J$ values, the CO SLED of M\,51 has an excess of CO emission when compared with MW2, 
corresponding to a warm gas phase not captured by our models. 
This warm phase is more pronounced in signature when looking at the CO SLED of M\,83, 
the line ratios of which are above our model galaxies for all $J_{\rm up} > 2$. 
Our models MW1, MW2 and MW3 only agree in CO luminosity with M\,51 at the 
$J_{\rm up} = 4$ and $5$ transitions. 

As for brightness temperature ratios (see middle panel) of our simulated MW-like 
galaxies only agree with the MW within the observational errors at CO(4$-$3), 
and other wise display a CO SLED shape noticably different from the MW. 
Where our model galaxies peak at the CO(3$-$2) transition in 
velocity-integrated intensity (see bottom panel), similar to M\,51, 
the MW and M\,83 peak at CO(4$-$3).
However, nowhere do the simulated line ratios differ by more than a factor of 2. 

The molecular gas masses of MW1, MW2 and MW3 are $4.6$, $5.4$ and 
$8.5\e{9}\,\msun$, respectively, significantly above $(1.0\pm0.3)\e{9}\,\msun$ 
for the MW \citep{heyer15}. 
The corresponding global ($R<20\,$kpc) \aCO factors are 
$13.6$, $7.7$ and $7.0\,\msun\,\rm{pc}^{-2}$\,(K\,km\,s$^{-1}$)$^{-1}$. 
For the inner disk ($R<10\,\rm{kpc}$) of the MW, $\alpha_{\rm CO,MW} 	
\simeq 4.3\pm 0.1\,{\rm \msun\,pc^{-2}\,(K\,km\,s^{-1})^{-1}}$ 
\citep[e.g.,][]{strong96,dame01,pineda08,bolatto13}.
However, typically the measurements of \aCO in the MW do not probe the central 
($<1\,$kpc) region. 
Excluding the central $R<1\,$kpc region and going out to $R=10\,$kpc, 
the \aCO factors of MW1, MW2 and MW3 are 
$11.6$, $6.9$ and $6.2\,\msun\,\rm{pc}^{-2}$\,(K\,km\,s$^{-1}$)$^{-1}$, 
i.e. factors of $2.7$, $1.6$ and $1.4$ above $\alpha_{\rm CO,MW}$, respectively.

\begin{center}
\begin{table}
\centering
\caption{\footnotesize{Properties of the three $z=0$ MW-like galaxies (MW1, MW2, and MW3)
used to benchmark \sigame.}}
\begin{tabular}{p{1cm}p{2cm}p{2cm}p{2cm}p{1cm}p{1cm}p{1cm}p{1cm}}
\hline
\hline
 	 			&	SFR 		& 	$M_{\ast}$	 		& $M_{\rm SPH}$ 		 &	$f_{\rm SPH}$	&	$\Z$	& 	$R_{\rm cut}$  \\ 
			 	&	[\sfru]		& 	[$10^{10}$\,\msun]	& [$10^{10}$\,\msun]	 &					&		& 	[kpc]  \\ 
\hline
MW1				&	2.2			&	$0.97$				&	$1.17$				 &	55\%			&	2.07	&	$20$		\\ 
MW2				& 	4.1			&	$3.12$				&	$1.27$				 &	29\%			&	3.31	&	$20$		\\ 
MW3 			& 	6.2			&	$3.83$				&	$1.69$				 &	31\%			&	3.90	&	$20$		\\ 
\hline
\end{tabular}
\\
	{\bf Notes.} \footnotesize{All quantities are determined within a radius $R_{\rm cut}$,
	which is the radius where the cumulative radial stellar mass function of
	each galaxy becomes flat. The gas mass ($M_{\rm SPH}$) is the total SPH gas
	mass within $R_{\rm cut}$.  The metallicity ($\Z=Z/Z_{\odot}$) is the mean
	of all SPH gas particles within $R_{\rm cut}$.}
\label{MWgal}
\end{table}
\end{center}

\chapter{Appendix to Chapter \ref{paper2}}

\section{Density of singly ionized carbon} \label{ap:fcii}

The strength of the \cii emission from any gas phase depends on the fraction of total carbon atoms 
that are singly ionized, \fcii. 
For calculating \fcii, we chose to employ the microphysics code \cloudy v13.03, 
that simultaneously solves the equations for statistical and thermal equilibrium for a 
wide range of optional elements. 
We set up \cloudy to work on a small parcel of gas with a single temperature and a constant density across its volume. 
The incident radiation field on the gas parcel is set to the local interstellar radiation field, 
as provided by \cloudy, but scaled by a factor (in practice, our \g0). 
In addition, a cosmic ray ionization rate can be specified in units of \ps.

\begin{figure*} 
\centering
\includegraphics[width=0.40\columnwidth]{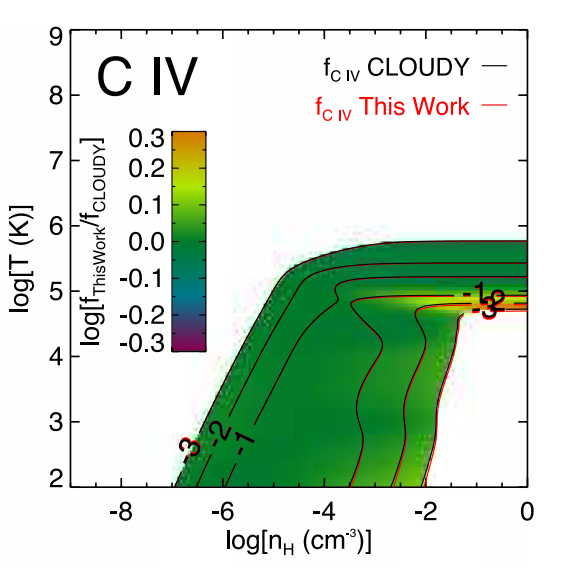}
\includegraphics[width=0.50\columnwidth]{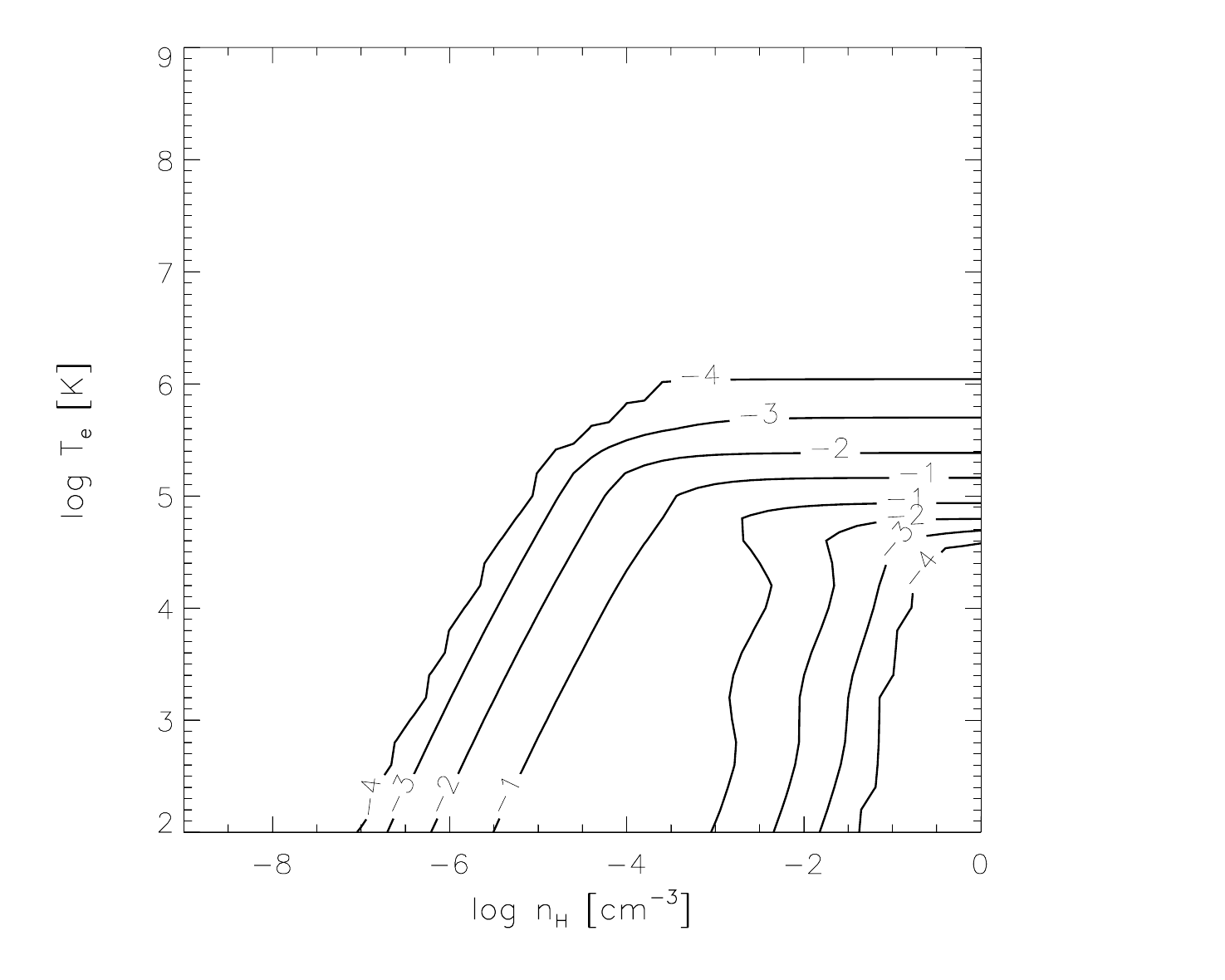}
\caption{\footnotesize{A comparison of $f_{\rm{\rm{C}\textsc{iv}}}$ as calculated by \cite{oppenheimer13} using \cloudy v10.00 (left) 
and by \sigame using \cloudy v13.03 (right). In both cases a Haardt \& Madau 2005 background radiation field (embedded in \cloudy) 
at $z=1$ was used together with solar abundances. A good agreement is seen, in particular with decreasing 
$f_{\rm{\rm{C}\textsc{iv}}}$ towards high densities and low temperatures, where carbon exists in lower ionizational states 
such as \rm{C}\textsc{ii}, see Fig.\,\ref{f:CII} below.}}
\label{f:CIV}
\end{figure*}

As a test to judge whether we are using CLOUDY correctly, the fraction of \rm{C}\textsc{iv}, $f_{\rm{\rm{C}\textsc{iv}}}$, 
as a function of density and temperature 
was compared to the work of \cite{oppenheimer13} by reproducing 
their CIV fractions at $z=1$ for a \cite{haardt01} background radiation field. 
This comparison is shown in Fig.\,\ref{f:CIV} with plots of the CIV abundance (relative to the total number of carbon atoms) 
as a function of hydrogen density and kinetic gas temperature. 
A good agreement with the figure of \cite{oppenheimer13} is achieved when using the build-in Haardt \& Madau 2005 background 
of \cloudy (`HM05' command in \cloudy, including galaxies and quasars but not the CMB) at $z=1$ together with default solar abundances. 

\begin{figure*} 
\centering
\includegraphics[width=0.9\columnwidth]{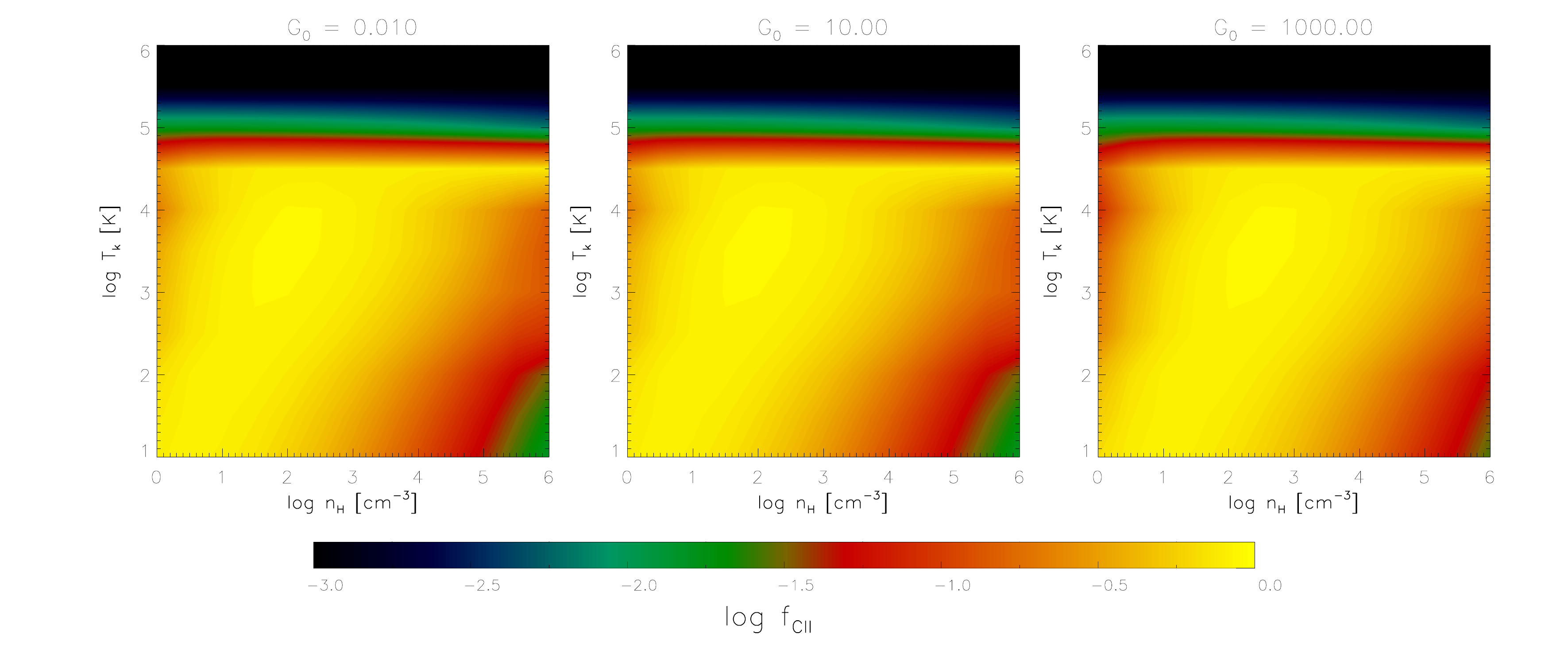}\\
\centering
\includegraphics[width=0.9\columnwidth]{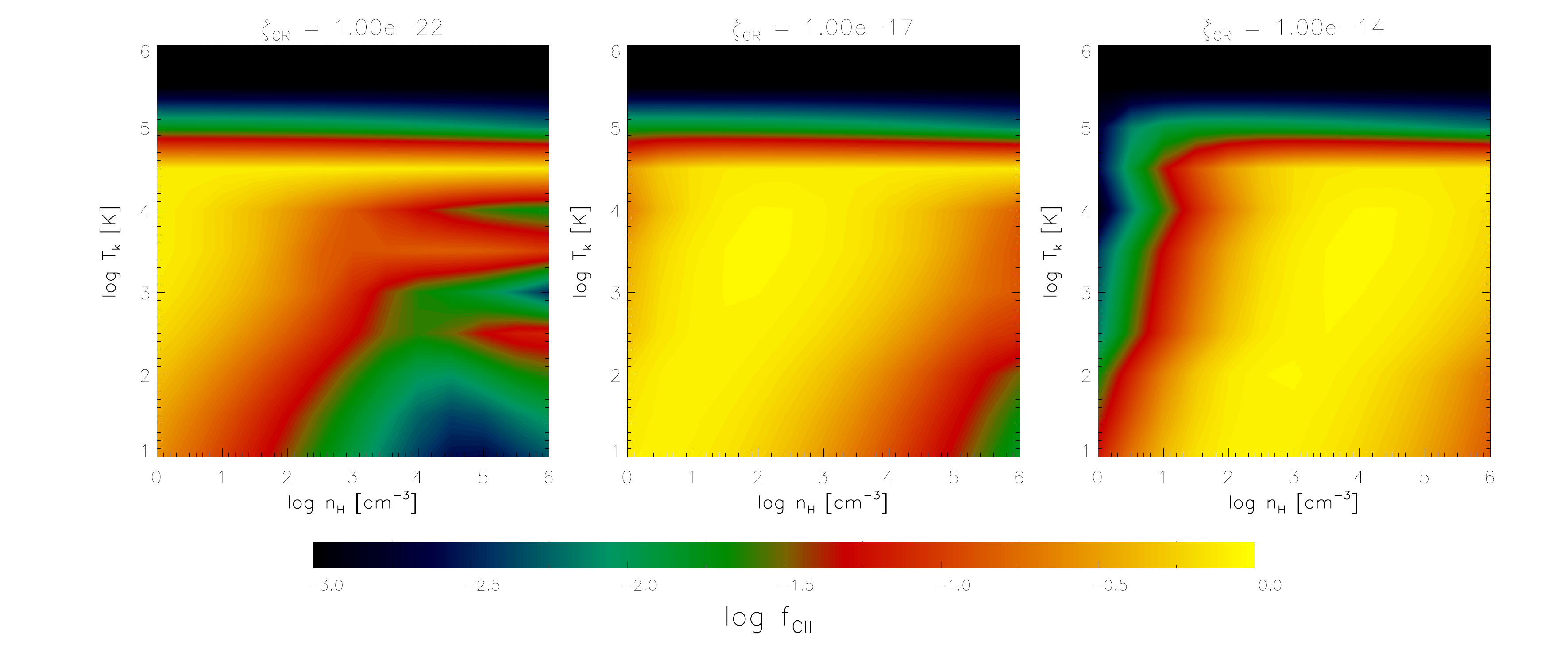}
\caption{\footnotesize{Maps of \fcii as function of gas temperature and hydrogen density made with 
\cloudy v13.03 as part of the grid used in \sigame to derive \xe and \fcii in the ISM. 
{\it Top row:} For three values of \g0 (and \cri fixed at $10^{-17}\,$\ps). 
{\it Bottom row:} For three values of \cri (with \g0 fixed to 1\,Habing). 
Note how \fcii drops at $\Tk\gtrsim10^5\,$K in all cases, and depends on \cri while being largely insensitive 
to changes in \g0.}}
\label{f:CII}
\end{figure*}

In order to keep the computing time of \sigame low, we set up a grid in [\nH, \Tk, \cri, \g0] and calculated \fcii with 
\cloudy for each grid point. 
Though metallicity and element abundances (of C, O, Si and Fe) are followed in our simulated galaxies, 
we chose the default solar composition that comes with \cloudy for the sake of having fewer parameters in the grid. 
Visualizations of this grid are shown in Fig.\,\ref{f:CII} for three values of \g0 (and \cri fixed at $10^{-17}\,$\ps, close to the 
adopted $\crimw=10^{-17}\,$\ps) in the top row and 
three values of \cri (with \g0 fixed to 1\,Habing, close to 
$G_{0{\rm,MW}}=0.6\,$Habing) in the bottom row.
As can be seen, \fcii does not seem to depend at all on \g0, but much more on the intensity of cosmic rays. 

\section{Electron fraction} \label{ap:xe}

\begin{figure*} 
\centering
\hspace{-1cm}
\includegraphics[width=1\columnwidth]{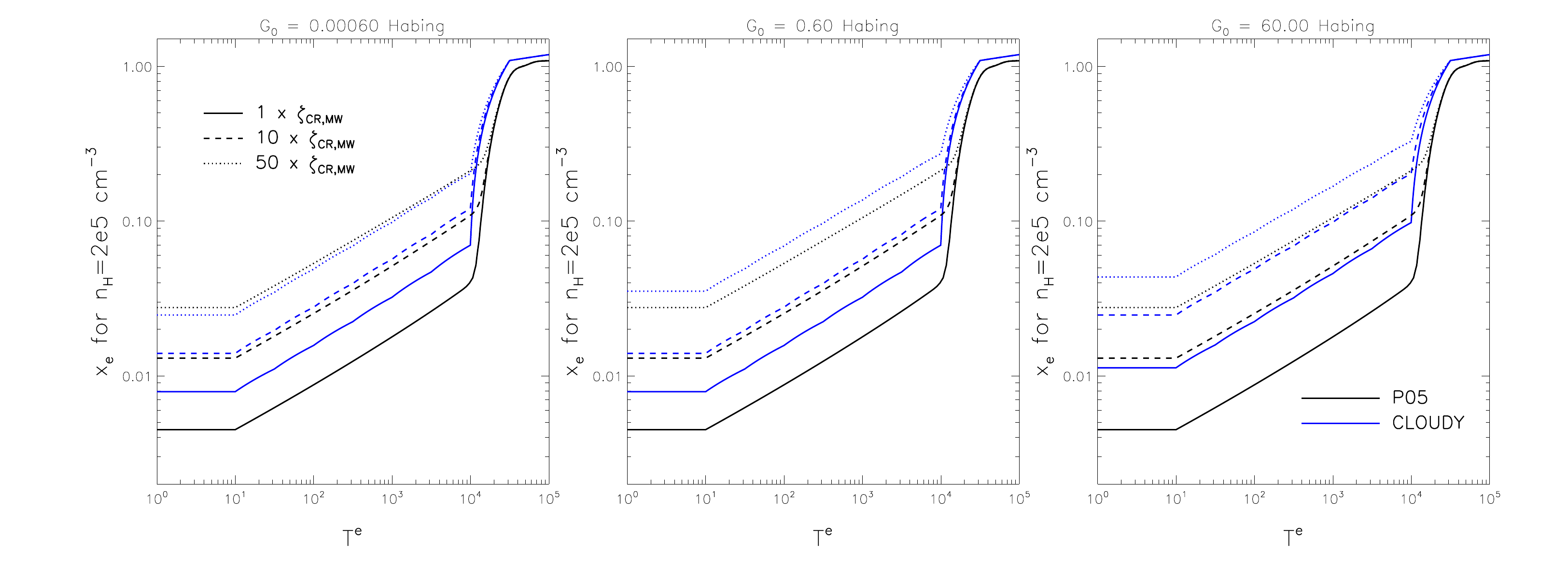}
\caption{\footnotesize{A comparison of \cloudy and the method described in \cite{pelupessy05} for 
deriving \xe as a function of electron temperature, \Te, at a density of $\nH=2\e{5}\,\cmpc$. 
The two methods agree about the shape of increasing \xe with \Te, including the 
sharp rise at $\Tk\sim10^4$\,K due to Balmer lines. 
The P05 method does not include ionization by the FUV radiation field, meaning that it takes the same shape in each panel. 
Note that \xe rises above 1 due to the inclusion of other elements than hydrogen, that can contribute to the electron number.}}
\label{f:xe}
\end{figure*}

In order to be consistent with the calculation of \fcii, we also decided to calculate the electron fraction, \xe, 
with \cloudy v13.03 rather than the method of \cite{pelupessy05} (P05) used in the previous Chapter \ref{paper1}. 
Fig.\,\ref{f:xe} shows how \cloudy and the P05 method compare as a function of electron temperature at a density of $\nH=2\e{5}\,\cmpc$, 
for different combinations of interstellar radiation field (\g0) and cosmic ray ionization rate (\cri). 
The P05 method is aimed for application in the interior of GMCs and therefore only considers 
ionization by cosmic rays that are expected to dominate the ionization of hydrogen, 
whereas \cloudy takes \g0 and \cri as separate variables. 

\begin{figure*} 
\centering
\hspace{-1cm}
\includegraphics[width=1\columnwidth]{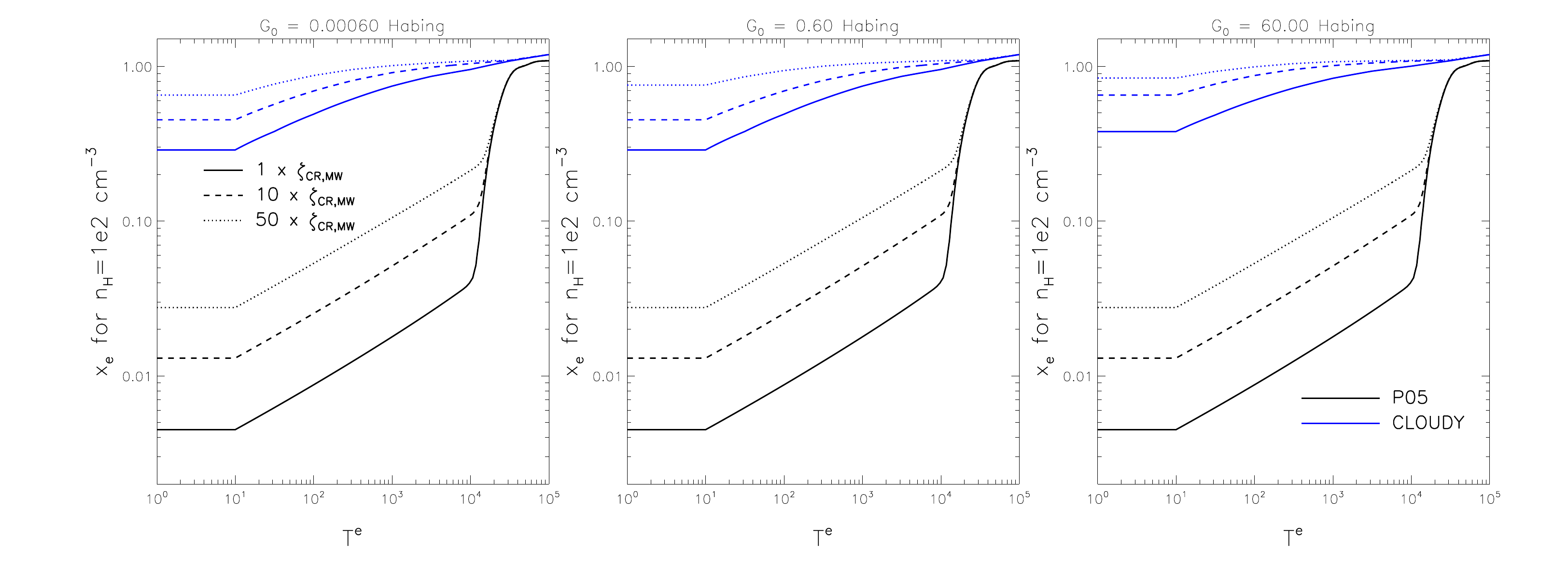}
\caption{\footnotesize{Same as Fig.\,\ref{f:xe}, but for a density of $\nH=100\,\cmpc$. 
}}
\label{f:xe2}
\end{figure*}

Furthermore, P05 method does not depend on density, whereas the calculation with \cloudy does. 
Fig.\,\ref{f:xe2} gives the same comparison as the previous figure, but for a lower density of $\nH=100\,\cmpc$, 
where \cloudy derives much larger electron fractions than the P05 model.
This is expected, since at low densities, the gas is less shielded from the incident radiation field. 
We therefore consider the inclusion of the \cloudy, with its sensitivity towards \g0 and \nH, 
an improvement of \sigame with respect to the electron fraction.

\section{\cii\ excitation and emission} \label{expr}
The cooling and heating mechanisms relevant for the generation of \cii emission 
are the same as those employed in Chapter\,\ref{gas1} 
and described in Appendix\,\ref{apB}-\ref{apC}, but we summarize them in Table\,\ref{lookup}.

\begin{table}[htbp]
\centering
\caption{\footnotesize{Provides a summary of the various heating and
cooling mechanisms adopted in the molecular and atomic gas regions of our GMC models,
along with the physical quantities on which they depend.}}
\begin{threeparttable}
\begin{tabular*}{\columnwidth}{p{1cm}p{5cm}p{3cm}p{6cm}} 
\hline
\hline\\ 
\multicolumn{4}{c}{Cooling and Heating Rates in GMC Models} 	\\ [0.1cm]
Process			&												& Parameters			&	Reference		\\
\toprule 
\Hpe 			& Photo-electric heating 						& \g0, \Tk, \ne, \nH	&  	\cite{bakes94}		\\
\Hcrhi 			& Cosmic ray heating in atomic gas				& \cri, \nhi, \xe		&	\cite{draine11}		\\ 
\Hcrh2 			& Cosmic ray heating in molecular gas			& \cri, \nh2			&	\cite{stahler05}		\\ 
\Ch2 			& H2 line cooling								& \nh2, \Tk				&	\cite{papa14} \\
\Coi 			& \oi ~line cooling								& \nH, \Tk				&	\cite{rollig06} \\
\Ccii	 		& \cii\ line cooling in ionized gas				& \g0, \Tk, \nH, \Xc, $\sigma_v$ 	&	\cite{goldsmith12} \\
\bottomrule
\end{tabular*}
\footnotesize{{\bf Note.} The most important
references are given, and we also refer to \cite{olsen15} for a
detailed description.}
\end{threeparttable}
\label{lookup}
\end{table}

For a two-level system such as \cii embedded in a radiation field with energy
density $U$ at the transition frequency, the general rate equation governing the
population levels can be written as: 
\begin{align}
	\frac{n_u}{n_l}	=	\frac{B_{lu}U+C_{lu}}{A_{ul}+B_{ul}U+C_{ul}} 
					=	\frac{gB_{ul}U+gKC_{ul}}{A_{ul}+B_{ul}U+C_{ul}}, 
\end{align}
where $A_{ul}$ ($=2.3\e{-6}$\,\ps) is the spontaneous emission rate, and
$B_{ul}$ and $B_{lu}$ are the stimulated emission and absorption rate
coefficients, respectively \citep{goldsmith12}. Here we have invoked detailed
balance, i.e.\ $B_{lu}U=gB_{ul}U$ and $C_{lu}=gKC_{ul}$, where $g$ ($=g_u/g_l$)
is the ratio of the statistical weights and $K = e^{-h\nu/k_{\rm B}T_{\rm k}} =
e^{-91.25\,{\rm K}/T_{\rm k}}$.  We can then write the fraction of C$^+$ in the
upper level, $f_u$, as:
\begin{align}
	f_u				&=	\frac{n_u}{n_l+n_u} =	\frac{1}{\frac{n_l}{n_u}+1}  \nonumber\\
					&=	\frac{gB_{ul}U+gKC_{ul}}{A_{ul}+(1+g)B_{ul}U+C_{ul}+gKC_{ul}}  \nonumber\\	
					&=	\frac{gK+gB_{ul}U/C_{ul}}{1+gK+A_{ul}/C_{ul}+(1+g)B_{ul}U/C_{ul}}.
					\label{eq:f_u}
\end{align}
In general, $U = (1-\beta)U(T_{\rm ex}) + \beta U(T_{\rm bg})$, where $U(T_{\rm
bg})$ is the energy density from a background field (e.g.\ CMB or radiation from
dust), and $\beta$ is the escape probability fraction at the \cii frequency
\citep[see also][]{goldsmith12}.  We shall ignore any background, however, in which case we
can write the stimulated downward rate as:
\begin{align}
	B_{ul}U 		= 	\frac{(1-\beta)A_{ul}}{e^{-91.25\,{\rm K}/T_{\rm ex}}-1},
	\label{eq:BulU}
\end{align}
and the excitation temperature as:
\begin{align}
	e^{-91.25\,{\rm K}/T^{\rm ex}}	=	K\left( 1+\frac{\beta A_{ul}}{C_{ul}} \right),
	\label{eq:Tex}
\end{align}
see \cite{goldsmith12}. For the escape probability we assume a spherical
geometry with a radial velocity gradient proportional to radius, such that \
$\beta = (1 - \exp(-\tau))/\tau$.  In calculating the integral in eq.\
\ref{eq:CII-integral} we split the integral up into 100 radial bins of thickness
$\Delta R$ (see Section \ref{cii_em}), and we approximate the optical depth in
each such bin with that of a homogeneous static slab of gas of thickness $\Delta
R$ \citep{draine11}:
\begin{align}
	\tau	=	\frac{g_u}{g_l}\frac{A_{ul}c^3}{4(2\pi)^{3/2}\nu^3\sigma_v}n_l\Delta R
				\left( 1-\frac{n_ug_l}{n_lg_u} \right), 
				\label{eq:tau}
\end{align}
where $\sigma_v$ is the gas velocity dispersion, which is calculated according
to eq.\,\ref{sigma_v_Pe} for the PDR and molecular gas regions in the GMCs, and
is set to the local velocity dispersion in the SPH simulation in the case of the
ionized clouds.  We use eqs.\ \ref{eq:f_u}, \ref{eq:BulU}, \ref{eq:Tex}, and
\ref{eq:tau} to iteratively solve for consistent values of $f_u$ and $\beta$ in
each $\Delta R$ bin. This is done by first assuming optically thin emission ($\beta=1$) 
in order to get an initial estimate of $f_u$ (eq.\
\ref{eq:f_u}), which is subsequently used to calculate $\tau$ and $\beta$ (eq.\
\ref{eq:tau}) and from that $T_{\rm ex}$ and $B_{ul}U$ (eqs.\ \ref{eq:Tex} and
\ref{eq:BulU}), etc.  Once consistent values for $f_u$ and $\tau$ have been
reached, we calculate the total cooling rate according to eq.\,\ref{eq:ccii}.
This cooling rate is used to determine the thermal balance at various points
within the GMCs as well as the \cii emission from the ionized clouds as
described in Section \ref{cii_em}.  We emphasize that our methods assumes that
the \cii emission from the different $\Delta R$ bins within a cloud is
radiatively de-coupled, and that the total \cii emission from a cloud is
therefore the sum of the emission from each bin.

\bigskip

\begin{figure}[htbp] 
\centering
\includegraphics[width=0.8\columnwidth]{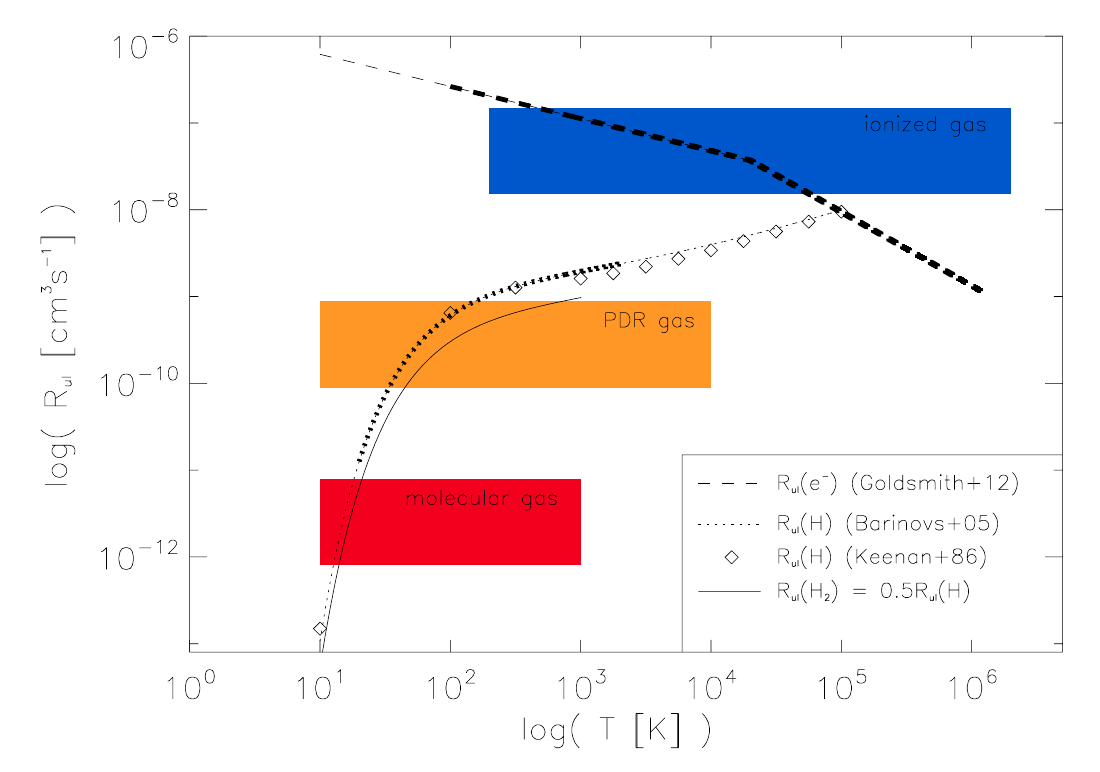}
\caption{\footnotesize{\cii deexcitation coefficients ($R_{ul}$) as a function of temperature
for collisions with $e^-$ (dashed line), H\,{\sc i} (dotted line), and H$_2$
(solid line and diamonds). The curves are thicker over the temperature ranges where 
they are formally applicable. The blue, orange, and red rectangles indicate the
temperature range encountered in the ionized, atomic, and molecular phase in our
simulated galaxies, respectively (see sections \ref{split12} and \ref{split2}).}}
\label{figure:rate-coefficients-vs-T}
\end{figure}

As explained in Section \ref{cii_em} \cii is collisionally (de)excited by H$_2$
in the molecular phase, by $e^-$ and H{\sc i} in the PDR region, and by $e^-$ in
the ionized gas. For a single collision partner the collision rates are equal to
the density ($n$) of the collision partner times the rate coefficients, i.e., 
$C_{ul} = n R_{ul}$ and $C_{lu} = n R_{lu}$. In case of two collision partners
we have $C_{ul} = n_1 R_{ul,1} + n_2 R_{ul,2}$ and $C_{lu} = n_1 R_{lu,1} + n_2
R_{lu,2}$. Fig.\ \ref{figure:rate-coefficients-vs-T} shows the \cii\ deexcitation
rate coefficients that we have adopted in our work for collisions with $e^-$
H\,{\sc i}, and H$_2$ as a function of temperature.  For collisions with
electrons, we adopt the following expression for the deexcitation rate
coefficient as a function of electron temperature ($T_e$):
\begin{align}
	R_{ul}(e^-)		=	8.7\e{-8}(T_{e}/2000\,{\rm K})^{-0.37}\,{\rm cm^3\,s^{-1}},
\label{equation:rate_coefficient_e}
\end{align}
which is applicable for temperatures from $\simeq 100$\,K to $20{,}000$\,K
\citep{goldsmith12}. At temperatures $>20{,}000$\,K we set $R_{ul}(e^-) \sim
T_{e}^{-0.85}$ \citep{wilson02,langer15}. For the PDR gas, the electron
density is calculated using CLOUDY and the electron temperature is set to the
kinetic temperature of the gas (calculated according to eq.\ \ref{Tk_rH2}). For
the ionized gas we assume $n_{e}=n_{\rm H\textsc{ii}}$, i.e., $x_{e}=1$, and
the electron temperature gas is set to the SPH gas kinetic temperature (see
Section \ref{split2}).  For the deexcitation rate coefficient for collisions
with H\,{\sc i} we use the analytical expression provided by \cite{barinovs05}
and shown as the dotted curve in Fig.\ \ref{figure:rate-coefficients-vs-T}:
\begin{equation}
R_{ul}({\rm H\,\textsc{i}}) = 7.938\times 10^{11} \exp (-91.2/T_{\rm k}) \left ( 16 + 0.344\sqrt{T_{\rm k}} - 47.7/T_{\rm k}\right )\, {\rm cm^3\,s^{-1}}.
\label{equation:rate_coefficient_H}
\end{equation}
While \cite{barinovs05} cites an application range for the above expression of
$15\,{\rm K} < T_{\rm k} < 2000\,{\rm K}$, a comparison with $R_{ul}({\rm
H\,\textsc{i}})$-values found by \cite{keenan86} over the temperature range
$10\,{\rm K} < T_{\rm k} < 100,000\,{\rm K}$ shows that eq.\
\ref{equation:rate_coefficient_H} provides an excellent match over this larger
temperature range (Fig.\ \ref{figure:rate-coefficients-vs-T}).  For collisions
with H$_2$ we follow \cite{goldsmith12} and assume that the collision rate
coefficients are approximately half those for collisions with H over the
relevant temperature range for the molecular gas, i.e.\ $R_{ul}({\rm H}_2) =
0.5\times R_{ul}({\rm H})$.

\clearpage
\begingroup
\section{References}
\def\chapter*#1{}
\bibliographystyle{apj} 
	\setlength{\bibsep}{1pt}
	\setstretch{1}
\bibliography{bibsApp}
\endgroup

\end{appendices}